\documentclass[a4paper,12pt]{article}

\usepackage{amsmath,amsthm,amsbsy,amssymb,amsfonts,bm,accents,cite}
\usepackage[all]{xy}
\usepackage[unicode,bookmarksnumbered,bookmarksopen=true]{hyperref}

\makeatletter
\catcode`\@=11
\@addtoreset{equation}{section}

\makeatother

\renewcommand{\thefootnote}{\arabic{footnote}}

\newcommand{\sla}[1]{\setbox0=\hbox{$#1$} 
\dimen0=\wd0 \setbox1=\hbox{/} \dimen1=\wd1 
\ifdim\dimen0>\dimen1 \rlap{\hbox to \dimen0{\hfil/\hfil}} #1 
\else\rlap{\hbox to \dimen1{\hfil$#1$\hfil}} / \fi}

\newcommand{\Exp}[1]{\operatorname{e}^{#1}}
\newcommand{\abs}[1]{\lvert {#1} \rvert}
\newcommand{\rmd}{{\mathrm{d}}}
\newcommand{\ii}{i}
\newcommand{\nn}{\nonumber}
\newcommand{\Lie}{\pounds}
\newcommand{\gLie}{\hat{\pounds}}

\newcommand{\cA}{\mathcal A}
\newcommand{\cC}{\mathcal C}\newcommand{\cD}{\mathcal D}
\newcommand{\cF}{\mathcal F}
\newcommand{\cH}{\mathcal H}
\newcommand{\cI}{\mathcal I}
\newcommand{\cK}{\mathcal K}\newcommand{\cL}{\mathcal L}
\newcommand{\cM}{\mathcal M}
\newcommand{\cO}{\mathcal O}

\newcommand{\cT}{\mathcal T}

\newcommand{\rmc}{\text{c}}
\newcommand{\rmC}{\text{C}}
\newcommand{\rmD}{\text{D}}
\newcommand{\rmS}{\text{S}}
\newcommand{\TT}{\text{T}}

\makeatletter
\newcommand*{\rmT}{{\mathpalette\@transpose{}}}
\newcommand*{\@transpose}[2]{\raisebox{\depth}{$\m@th#1\intercal$}}
\makeatother

\newcommand{\OO}{\text{O}}
\newcommand{\SU}{\text{SU}}
\newcommand{\PSU}{\text{PSU}}
\newcommand{\SO}{\text{SO}}
\newcommand{\SL}{\text{SL}}
\newcommand{\diag}{\text{diag}}

\newcommand{\bmF}{{\bm{F}}}
\newcommand{\bmp}{{\bm{p}}}
\newcommand{\bmq}{{\bm{q}}}
\newcommand{\bmr}{{\bm{r}}}
\newcommand{\bms}{{\bm{s}}}

\newcommand{\bisC}{\hat{\bm{\cC}}}
\newcommand{\bisF}{\hat{\bm{\cF}}}

\newcommand\bre{\bar{e}}
\newcommand\breta{\bar{\eta}}
\newcommand\brgamma{\bar{\gamma}}
\newcommand\bromega{\bar{\omega}}
\newcommand\brPhi{\bar{\Phi}}

\newcommand\brC{\bar{C}}
\newcommand\brP{\bar{P}}
\newcommand\brGamma{\bar{\Gamma}}

\newcommand{\Loa}{a}
\newcommand{\Lob}{b}
\newcommand{\Loc}{c}
\newcommand{\Lod}{d}
\newcommand{\Loe}{e}
\newcommand{\Lof}{f}
\newcommand{\Log}{g}
\newcommand{\Loh}{h}

\newcommand{\Lobra}{\bar{a}}
\newcommand{\Lobrb}{\bar{b}}
\newcommand{\Lobrc}{\bar{c}}
\newcommand{\Lobrd}{\bar{d}}

\newcommand{\WSa}{\bar{\alpha}}
\newcommand{\WSb}{\bar{\beta}}

\newcommand{\SPa}{\alpha}
\newcommand{\SPb}{\beta}
\newcommand{\SPc}{\gamma}

\newcommand{\SPbra}{\bar{\SPa}}
\newcommand{\SPbrb}{\bar{\SPb}}
\newcommand{\SPbrc}{\bar{\SPc}}

\newcommand{\GV}{V}
\newcommand{\brGV}{\bar{V}}

\newcommand{\brTheta}{\bar{\Theta}}
\newcommand{\brbrTheta}{\bar{\bar{\Theta}}}
\newcommand{\brtheta}{\bar{\theta}}
\newcommand{\brOmega}{\bar{\Omega}}
\newcommand{\ODDeta}{\eta}
\newcommand{\OG}{G}
\newcommand{\CG}{g}
\newcommand{\Einv}{\mathcal{E}}
\newcommand{\AdS}[1]{\text{AdS}_{#1}}

\newcommand{\sfa}{{\mathsf{a}}}
\newcommand{\sfb}{{\mathsf{b}}}
\newcommand{\sfc}{{\mathsf{c}}}
\newcommand{\sfd}{{\mathsf{d}}}
\newcommand{\sfe}{{\mathsf{e}}}
\newcommand{\sff}{{\mathsf{f}}}
\newcommand{\sfi}{{\mathsf{i}}}
\newcommand{\sfj}{{\mathsf{j}}}
\newcommand{\sfk}{{\mathsf{k}}}

\newcommand{\sfD}{{\mathsf{D}}}
\newcommand{\sfP}{{\mathsf{P}}}
\newcommand{\sfJ}{{\mathsf{J}}}
\newcommand{\sfK}{{\mathsf{K}}}

\newcommand{\gP}{{\rm\bf P}}
\newcommand{\gJ}{{\rm\bf J}}
\newcommand{\gZ}{{\rm\bf Z}}
\newcommand{\gQ}{{\rm\bf Q}}
\newcommand{\dlT}{T}
\newcommand{\Pg}{P}
\newcommand{\gga}{\gamma}

\newcommand{\bos}{\text{b}}
\newcommand{\fer}{\text{f}}
\newcommand{\str}{{\rm Str}}
\newcommand{\ST}{{\rm st}}
\newcommand{\Tr}{{\rm Tr}}
\newcommand{\Ad}{{\rm Ad}}
\newcommand{\sgn}{{\rm sgn}}
\newcommand{\CYBE}{{\rm CYBE}}
\newcommand{\YB}{{\rm YB}}
\newcommand{\Pf}{{\rm Pf}}
\newcommand{\Span}{{\rm span}}
\newcommand{\alg}[1]{\mathfrak{#1}}

\newcommand\Atop[2]{\genfrac{}{}{0pt}{}{#1}{#2}}

\allowdisplaybreaks[3]

\newcommand{\Pskip}{6pt}

\begin{document}

\begin{titlepage}
\renewcommand{\thefootnote}{\fnsymbol{footnote}}

\begin{flushright}
\parbox{4cm}
{KUNS-2717}
\end{flushright}

\vspace*{1cm}

\begin{center}
\Large\textbf{Local $\beta$-deformations and Yang--Baxter sigma model}
\end{center}

\vspace{1.0cm}

\centerline{
{\large Jun-ichi Sakamoto$^a$%
\footnote{E-mail address: \texttt{sakajun@gauge.scphys.kyoto-u.ac.jp}}
and 
Yuho Sakatani$^b$}%
\footnote{E-mail address: \texttt{yuho@koto.kpu-m.ac.jp}}
}

\vspace{0.2cm}

\begin{center}
${}^a${\it Department of Physics, Kyoto University,}\\
{\it Kitashirakawa Oiwake-cho, Kyoto 606-8502, Japan}

\vspace*{1mm}

${}^b${\it Department of Physics, Kyoto Prefectural University of Medicine,}\\
{\it Kyoto 606-0823, Japan}

\end{center}

\vspace*{1mm}

\begin{abstract}
Homogeneous Yang--Baxter (YB) deformation of $\AdS{5}\times\rmS^5$ superstring is revisited. We calculate the YB sigma model action up to quadratic order in fermions and show that homogeneous YB deformations are equivalent to $\beta$-deformations of the $\AdS{5}\times\rmS^5$ background when the classical $r$-matrices consist of bosonic generators. In order to make our discussion clearer, we discuss YB deformations in terms of the double-vielbein formalism of double field theory. We further provide an $\OO(10,10)$-invariant string action that reproduces the Green--Schwarz type II superstring action up to quadratic order in fermions. When an AdS background contains a non-vanishing $H$-flux, it is not straightforward to perform homogeneous YB deformations. In order to get any hint for such YB deformations, we study $\beta$-deformations of $H$-fluxed AdS backgrounds and obtain various solutions of (generalized) type II supergravity. 
\end{abstract}

\thispagestyle{empty}
\end{titlepage}

\setcounter{footnote}{0}

\tableofcontents

\setlength{\parskip}{\Pskip}

\section{Introduction}

Yang--Baxter (YB) sigma model was originally introduced by Klim\v{c}\'ik \cite{Klimcik:2002zj} as a class of Poisson--Lie symmetric sigma models. 
It is characterized by a classical $r$-matrix that satisfies the modified classical YB equation (mCYBE). 
It was later shown to be integrable by constructing the Lax pair \cite{Klimcik:2008eq}. 
The original YB sigma model can be applied only to sigma models on group manifolds, but it was later generalized to coset sigma models in \cite{Delduc:2013fga} and to the case of the homogeneous classical YB equation (CYBE) in \cite{Matsumoto:2015jja}.

An interesting application of YB deformations is an integrable deformation of type IIB superstring theory on the $\AdS5\times \rmS^5$ background \cite{Delduc:2013qra,Delduc:2014kha,Kawaguchi:2014qwa}, that has been studied in the context of the AdS/CFT correspondence. 
Through various examples \cite{Matsumoto:2014nra,Matsumoto:2014gwa,Matsumoto:2015uja,vanTongeren:2015soa,vanTongeren:2015uha,Kyono:2016jqy}, it turned out that, when we employ an Abelian classical $r$-matrix, the YB-deformed $\AdS5\times \rmS^5$ superstring can be described as type IIB superstring on a TsT-transformed\footnote{A TsT transformation is a sequence of two Abelian $T$-dualities with a coordinate shift in between.} $\AdS5\times \rmS^5$ background \cite{Lunin:2005jy,Frolov:2005dj,Hashimoto:1999ut,Maldacena:1999mh,Herzog:2008wg,Maldacena:2008wh,Adams:2008wt} (see \cite{Osten:2016dvf} for a clear explanation and generalizations).
Namely, Abelian YB deformation was found to be equivalent to a TsT-transformation. 
For non-Abelian classical $r$-matrices, the deformations of the $\AdS5\times \rmS^5$ background have not been understood clearly; some deformed backgrounds were obtained through non-commuting TsT-transformations (see for example \cite{Borsato:2016ose}) and some were obtained through a combination of diffeomorphisms and $T$-dualities \cite{Orlando:2016qqu}, but it is not clear whether an arbitrary YB deformation can be realized as a combination of Abelian $T$-dualities and gauge symmetries of the supergravity (it was recently shown in \cite{Hoare:2016wsk,Borsato:2016pas,Hoare:2016wca,Borsato:2017qsx,Lust:2018jsx} that YB deformations can be also reproduced from non-Abelian $T$-dualities \cite{Fridling:1983ha,Fradkin:1984ai,delaOssa:1992vci,Gasperini:1993nz,Giveon:1993ai,Alvarez:1994np,Elitzur:1994ri,Sfetsos:2010uq,Lozano:2011kb,Itsios:2013wd}). 
As shown in a seminal paper \cite{Borsato:2016ose}, at least when an $r$-matrix satisfies a certain criterion called unimodularity, the deformed $\AdS5\times \rmS^5$ background are solutions of type IIB supergravity. 
Moreover, for a non-unimodular $r$-matrix, the deformed $\AdS5\times \rmS^5$ background was shown to satisfy the generalized supergravity equations of motion (GSE) \cite{Arutyunov:2015mqj,TW}, and a Killing vector $I^m$ appearing in the GSE was determined for a general $r$-matrix. 
In a recent paper \cite{Wulff:2018aku} (which appeared a few days after this manuscript was posted to the arXiv), a more tractable expression for $I^m$ has been given for $r$-matrices consisting only of bosonic generators (see Appendix \ref{app:conventions} for more details). 
On the same day, the second version of \cite{Hong:2018tlp} appeared on the arXiv, which has derived the GSE from the dual sigma model of \cite{Elitzur:1994ri} in a general NS-NS background, and has determined the Killing vector $I^m$ in the non-Abelian $T$-dualized backgrounds. 

Recently, in \cite{Sakatani:2016fvh,Baguet:2016prz,Sakamoto:2017wor,Sakamoto:2017cpu,Fernandez-Melgarejo:2017oyu}, the GSE and YB deformations were studied from a viewpoint of a manifestly $T$-duality covariant formulation of supergravity, called the double field theory (DFT) \cite{Siegel:1993xq,Siegel:1993th,Siegel:1993bj,Hull:2009mi,Hull:2009zb,Hohm:2010pp} and its extensions. 
Through various examples of non-Abelian YB deformations, it was noticed that YB deformations are equivalent to (local) $\beta$-transformations of the $\AdS5\times \rmS^5$ background \cite{Sakamoto:2017cpu}. 
The local $\beta$-transformations may be realized as gauge transformations in DFT, known as the generalized diffeomorphism, and for many examples of non-Abelian $r$-matrices, YB-deformed backgrounds were reproduced by acting generalized diffeomorphisms to the $\AdS5\times \rmS^5$ background. 
However, until now the equivalence between YB deformations and local $\beta$-transformations has not yet been proven. 

In this paper, we show the equivalence for YB deformations of the $\AdS5\times \rmS^5$ superstring. 
To be more precise, we show that, for a classical $r$-matrix consisting of the bosonic generators and satisfying the homogeneous CYBE, the YB deformed $\AdS5\times \rmS^5$ superstring action can be regarded as the Green--Schwarz (GS) type IIB superstring action \cite{Green:1983wt} defined in a $\beta$-transformed $\AdS5\times \rmS^5$ background. 
During the proof, we perform a suitable identification of the deformed vielbein and make a redefinition of the fermionic variable. 
These procedures can be clearly explained by using the double-vielbein formalism of DFT \cite{Siegel:1993xq,Jeon:2011cn,Jeon:2011vx,Jeon:2011sq,Jeon:2012kd,Jeon:2012hp}. 

We also find a manifestly $\OO(10,10)$-invariant string action that reproduces the conventional GS superstring action up to quadratic order in fermions. 
In the previous works, $T$-duality covariant string theories with the worldsheet supersymmetry were studied in \cite{Hull:2006va,Blair:2013noa,Driezen:2016tnz}. 
The GS-type string actions were also constructed in \cite{Bandos:2015cha,Park:2016sbw,Bandos:2016jez} but the target space was assumed to be flat and have no the R-R fluxes.
Our GS-type string action can apply to arbitrary curved backgrounds with the R--R fields and is a generalization of the previous ones (another $T$-duality manifest GS superstring action in a general background was proposed in \cite{Hatsuda:2014aza,Hatsuda:2015cia} although the relation to our action is unclear so far).

We expect that the equivalence between YB deformations and $\beta$-deformations will hold in more general backgrounds beyond the $\AdS5\times \rmS^5$ background. 
As a non-trivial example, we study local $\beta$-deformations of the $\AdS3\times \rmS^3\times \TT^4$ background that contains a non-vanishing $H$-flux.
In this case, due to the presence of $H$-flux, it is not straightforward to perform YB deformations.\footnote{There are several works \cite{Kawaguchi:2011mz,Kawaguchi:2013gma,Delduc:2014uaa,Delduc:2017fib,Demulder:2017zhz} where YB deformations of the WZ(N)W model based on the mCYBE have been studied.}
Therefore, we do not show the equivalence in this paper. 
However, thanks to the homogeneous CYBE for the local $\beta$-deformations, all examples of the $\beta$-deformed backgrounds are shown to satisfy the equations of motion of DFT, or the (generalized) supergravity. 

This paper is organized as follows. 
In section \ref{sec:DFT}, we review the double-vielbein formalism of DFT and find a simple $\beta$-transformation rule for the Ramond--Ramond (R--R) fields. 
We also find the action of the double sigma model for type II superstring that reproduces the conventional GS superstring action. 
In section \ref{sec:YB-AdS5xS5}, we concisely review YB deformations of $\AdS{5} \times \rmS^5$ superstring and show the equivalence of homogeneous YB deformations and local $\beta$-transformations. 
In section \ref{sec:AdS3}, we perform $\beta$-transformations of the $\AdS3\times \rmS^3\times \TT^4$ background and obtain various solutions. 
We also discuss a more general class of local $\OO(10,10)$ transformations that are based on the homogeneous CYBE. 
Section \ref{sec:discussion} is devoted to conclusions and discussions. 
Various technical computations are explained in the Appendices. 

\section{Local \texorpdfstring{$\beta$}{\textbeta}-deformations in DFT}
\label{sec:DFT}

In this section, we review the basics of the type II DFT and find a simple transformation rule for bosonic fields under local $\beta$-deformations. 
We also find a manifestly $\OO(10,10)$-invariant superstring action that reproduces the conventional GS type II superstring action. 

\vspace{-\Pskip}
\subsection{DFT fields and their parameterizations}

Bosonic fields in DFT are the generalized metric $\cH_{MN}$, the $T$-duality-invariant dilaton $d$, and the R--R potential $\bisC$, which is an $\OO(1,D-1)\times \OO(D-1,1)$ bispinor. 
In this subsection, we review their definitions and basic properties by turning off fermions (such as gravitino). 
Here, we employ the double-vielbein formalism developed in \cite{Siegel:1993xq,Jeon:2011cn,Jeon:2011vx,Jeon:2011sq,Jeon:2012kd,Jeon:2012hp} (see also \cite{Hohm:2010xe}), which is quite suitable for discussing YB deformations. 

\paragraph{NS--NS fields:}
The generalized metric $\cH_{MN}$ $(M,N=0,\dotsc,2D-1)$ is defined as
\begin{align}
\begin{split}
 \cH &\equiv (\cH_{MN}) \equiv E\,\mathsf{S}\, E^\rmT \,,\qquad E = (E_M{}^N) \in \OO(D,D)\,,
\\
 \mathsf{S}&\equiv (\mathsf{S}_{MN}) \equiv \diag(\underbrace{-1,\,+1,\cdots,+1}_{D},\,\underbrace{-1,\,+1,\dotsc,+1}_{D}) \,,
\end{split}
\end{align}
where the $\OO(D,D)$ property of the generalized vielbein $E$ is defined as
\begin{align}
 E \, \ODDeta\, E^\rmT = \ODDeta = E^\rmT\,\ODDeta\, E \,,\qquad 
 \ODDeta\equiv (\ODDeta_{MN})\equiv \begin{pmatrix} 0 & \delta_m^n \\ \delta^m_n & 0 \end{pmatrix} \qquad 
 (m,n=0,\dotsc,D-1). 
\end{align}
The familiar properties of the generalized metric
\begin{align}
 \cH^\rmT = \cH \,,\qquad \cH^\rmT\,\ODDeta\,\cH = \ODDeta\,,
\label{eq:H-properties}
\end{align}
follow from the above definitions. 
For an $\OO(D,D)$ matrix $h$ satisfying $h^\rmT\,\mathsf{S}\,h = \mathsf{S}$, namely an $\OO(1,D-1)\times \OO(D-1,1)$ matrix $h$, both $E$ and $E\,h$ give the same generalized metric $\cH$. 
Thus, the generalized metric $\cH$ can be regarded as a representative of a coset
\begin{align}
 \frac{\OO(D,D)}{\OO(1,D-1)\times \OO(D-1,1)} \,, 
\end{align}
where $\OO(1,D-1)\times \OO(D-1,1)$ is known as the double local Lorentz group \cite{Jeon:2011sq}. 
We raise or lower the $\OO(D,D)$ indices by using the $\OO(D,D)$-invariant metric $\ODDeta$ like $\cH_M{}^N \equiv \cH_{MP}\,\ODDeta^{PN}$\,, and then \eqref{eq:H-properties} indicates that the matrix $\cH_M{}^N$ has eigenvalues $\pm1$\,. 
Then, we introduce the double (inverse) vielbeins $\GV_{\Loa}{}^{M}$ and $\brGV_{\Lobra}{}^M$ ($\Loa,\,\Lobra=0,\dotsc,D-1$) as the eigenvectors
\begin{align}
 \cH^{M}{}_{N}\,\GV_{\Loa}{}^N =+\GV_{\Loa}{}^M\,,\qquad 
 \cH^{M}{}_{N}\,\brGV_{\Lobra}{}^N=-\brGV_{\Lobra}{}^M\,. 
\end{align}
Since the eigenvalues are different, they are orthogonal to each other
\begin{align}
 \cH_{MN}\,\GV_{\Loa}{}^M\,\brGV_{\Lobrb}{}^{N}=0 \,,\qquad 
 \ODDeta_{MN}\,\GV_{\Loa}{}^M\,\brGV_{\Lobrb}{}^{N}=0\,. 
\label{eq:H-eta-orthogonal}
\end{align}
Following \cite{Jeon:2011cn,Jeon:2011vx,Jeon:2011sq,Jeon:2012kd,Jeon:2012hp}, we normalize the double vielbeins as
\begin{align}
\begin{split}
 \eta_{\Loa\Lob}&=\ODDeta_{MN}\,\GV_{\Loa}{}^M \,\GV_{\Lob}{}^{N} = \cH_{MN}\,\GV_{\Loa}{}^M \,\GV_{\Lob}{}^{N} 
 = \diag(-1,\,+1,\dotsc,+1) \,,
\\
 \breta_{\Lobra\Lobrb}&=\ODDeta_{MN}\,\brGV_{\Lobra}{}^M \,\brGV_{\Lobrb}{}^{N} =-\cH_{MN}\,\brGV_{\Lobra}{}^M \,\brGV_{\Lobrb}{}^{N} 
 = \diag(+1,\,-1,\dotsc,-1) \,. 
\end{split}
\end{align}
By introducing $2D\times 2D$ matrices,
\begin{align}
 (\GV_A{}^M) \equiv \begin{pmatrix} \GV_{\Loa}{}^{M}\\ \brGV_{\Lobra}{}^{M} \end{pmatrix}\,,\qquad 
 (\eta_{AB}) \equiv \begin{pmatrix} \eta_{\Loa\Lob}& 0\\ 0& \breta_{\Lobra\Lobrb} \end{pmatrix} \,,\qquad 
 (\cH_{AB}) \equiv \begin{pmatrix} \eta_{\Loa\Lob}& 0\\ 0& -\breta_{\Lobra\Lobrb} \end{pmatrix} \,,
\end{align}
where $\{A\}\equiv\{\Loa,\,\Lobra\}$, the above orthonormal conditions are summarized as
\begin{align}
 \eta_{AB} = \GV_A{}^M\, \ODDeta_{MN}\, (\GV^\rmT)^N{}_B \,, \qquad 
 \cH_{AB} = \GV_A{}^M\, \cH_{MN}\, (\GV^\rmT)^N{}_B \,. 
\label{eq:DV-ortho-normal}
\end{align}
The matrix $\GV_A{}^M$ is always invertible and the inverse matrix is given by
\begin{align}
 (\GV^{-1})_M{}^A = \ODDeta_{MN}\,(\GV^\rmT)^N{}_B\,\eta^{BA} \,,
\end{align}
which indeed satisfies
\begin{align}
 \GV_A{}^M\,(\GV^{-1})_M{}^B = \ODDeta_{MN}\, \GV_A{}^M\,\GV_C{}^N \,\eta^{CB} = \delta_A^B \,. 
\end{align}
As long as we raise or lower the indices $M,\,N$ with $\ODDeta_{MN}$ and $A,\,B$ with $\eta_{AB}$ (namely, $\Loa,\,\Lob$ and $\Lobra,\,\Lobrb$ with $\eta_{\Loa\Lob}$ and $\eta_{\Lobra\Lobrb}$, respectively), there is no difference between $\GV_A{}^M$ and $(\GV^{-\rmT})_A{}^M\equiv \eta_{AB}\,\ODDeta^{MN}\,(\GV^{-\rmT})^B{}_N$\,. 
Thus, in the following, we may not show the inverse or the transpose explicitly. 

When $D\times D$ matrices, $\GV^m{}_{\Loa}$ and $\brGV^m{}_{\Lobra}$, are invertible, we can parameterize the double vielbeins as
\begin{align}
 (\GV^M{}_{\Loa}) =\frac{1}{\sqrt{2}} \begin{pmatrix} (e^{-\rmT})^m{}_{\Loa} \\ E_{mn}\,(e^{-\rmT})^n{}_{\Loa} \end{pmatrix} \,, \qquad 
 (\brGV^M{}_{\Lobra}) =\frac{1}{\sqrt{2}} \begin{pmatrix} (\bre^{-\rmT})^m{}_{\Lobra} \\ \bar{E}_{mn}\,(\bre^{-\rmT})^n{}_{\Lobra} \end{pmatrix} \,,
\end{align}
where we introduced matrix notations, $e\equiv (e_m{}^{\Loa})$ and $\bre\equiv (\bre_m{}^{\Lobra})$. 
From \eqref{eq:DV-ortho-normal}, we find
\begin{align}
 \bar{E}_{mn} = - E^\rmT_{mn}\,,\qquad 
 \CG_{mn} \equiv E_{(mn)} = (e \,\eta \, e^\rmT)_{mn} = - (\bre\,\breta \, \bre^\rmT)_{mn}\,.
\end{align}
By denoting $B_{mn}\equiv E_{[mn]}$, the parameterizations of the double vielbeins and the generalized metric become
\begin{align}
 &(\GV^M{}_{\Loa}) =\frac{1}{\sqrt{2}} \begin{pmatrix} (e^{-\rmT})^m{}_{\Loa} \\ (\CG+B)_{mn}\,(e^{-\rmT})^n{}_{\Loa} \end{pmatrix} \,, \qquad 
 (\brGV^M{}_{\Lobra}) =\frac{1}{\sqrt{2}} \begin{pmatrix} (\bre^{-\rmT})^m{}_{\Lobra} \\ -(\CG-B)_{mn}\,(\bre^{-\rmT})^n{}_{\Lobra} \end{pmatrix} \,,
\label{eq:dV-geometric}
\\
 &\cH = \begin{pmatrix} (\CG-B\,\CG^{-1}\,B)_{mn} & (B\,\CG^{-1})_m{}^n \\ -(\CG^{-1}\,B)^m{}_n & \CG^{mn} \end{pmatrix} 
 = \begin{pmatrix} \delta_m^p & B_{mp} \\ 0 & \delta^m_p \end{pmatrix} 
 \begin{pmatrix} \CG_{pq} & 0 \\ 0 & \CG^{pq} \end{pmatrix} 
 \begin{pmatrix} \delta^q_n & 0 \\ -B_{qn} & \delta^q_n \end{pmatrix} \,.
\label{eq:H-geometric}
\end{align}
The dual parameterization, that can be prescribed when $\GV_m{}^{\Loa}$ and $\brGV_m{}^{\Lobra}$ are invertible, is
\begin{align}
\begin{split}
 \GV_M{}^{\Loa} &=\frac{1}{\sqrt{2}} \begin{pmatrix} \tilde{e}_m{}^{\Loa} \\ (\OG^{-1}-\beta)^{mn}\, \tilde{e}_n{}^{\Loa} \end{pmatrix} \,, \qquad 
 \brGV_N{}^{\Lobra} =\frac{1}{\sqrt{2}} \begin{pmatrix} \tilde{\bre}_m{}^{\Lobra} \\ -(\OG^{-1}+\beta)^{mn}\, \tilde{\bre}_n{}^{\Lobra} \end{pmatrix} \,,
\\
 \OG_{mn}&\equiv (\tilde{e}\,\eta\, \tilde{e}^\rmT)_{mn} = - (\tilde{\bre} \,\breta \, \tilde{\bre}^\rmT)_{mn}\,,\qquad \beta^{mn}=-\beta^{nm} \,,
\end{split}
\end{align}
and it provides the dual parameterization of the generalized metric
\begin{align}
 \cH = \begin{pmatrix} \OG_{mn} & (\OG\, \beta)_m{}^{n} \\
 -(\beta\,\OG)^m{}_{n} & (\OG^{-1}-\beta\,\OG\,\beta)^{mn} 
 \end{pmatrix} 
 = \begin{pmatrix} \delta_m^p & 0 \\
 - \beta^{mp} & \delta^m_p 
 \end{pmatrix} \begin{pmatrix} \OG_{pq} & 0 \\ 0 & \OG^{pq} 
 \end{pmatrix} 
 \begin{pmatrix} \delta^p_n & \beta^{pn} \\
 0 & \delta_p^n 
 \end{pmatrix} \,. 
\label{eq:H-non-geometric}
\end{align}
When both parameterizations are possible, comparing \eqref{eq:H-geometric} and \eqref{eq:H-non-geometric}, we obtain
\begin{align}
\begin{split}
 E^{mn} &\equiv (E^{-1})^{mn} = \OG^{mn} - \beta^{mn} \qquad \bigl(E_{mn} \equiv \CG_{mn} + B_{mn}\bigr)\,,
\\
 \CG_{mn}&= E_{mp}\,E_{nq}\,\OG^{pq}\,,\qquad B_{mn}= E_{mp}\,E_{nq}\,\beta^{pq}\,. 
\end{split}
\label{eq:relation-open-closed}
\end{align}
In the following, we raise or lower the indices of $\{e_m{}^{\Loa},\,\bre_m{}^{\Lobra},\,\tilde{e}_m{}^{\Loa},\,\tilde{\bre}_m{}^{\Lobra}\}$ as
\begin{align}
\begin{alignedat}{2}
 e^m{}_{\Loa}&=\CG^{mn}\,e_n{}^{\Lob}\,\eta_{\Lob\Loa}\,,&\qquad 
 \bre^m{}_{\Lobra}&=\CG^{mn}\,\bre_n{}^{\Lobrb}\,\breta_{\Lobrb\Lobra}\,,
\\
 \tilde{e}^m{}_{\Loa}&=\OG^{mn}\,\tilde{e}_n{}^{\Lob}\,\eta_{\Lob\Loa}\,,&\qquad 
 \tilde{\bre}^m{}_{\Lobra}&=\OG^{mn}\,\tilde{\bre}_n{}^{\Lobrb}\,\breta_{\Lobrb\Lobra}\,, 
\end{alignedat}
\end{align}
and then we obtain relations like $(e^{-\rmT})^m{}_{\Loa}=e^m{}_{\Loa}$\,. 
We can then omit the inverse or the transpose without any confusions as long as the indices are shown explicitly. 
By using the two metrics, $\CG_{mn}$ and $\OG_{mn}$, we also introduce two parameterizations of the dilaton $d$,
\begin{align}
 \sqrt{\abs{\OG}}\,\Exp{-2\tilde{\phi}} = \Exp{-2d} = \sqrt{\abs{\CG}}\,\Exp{-2\Phi} \,. 
\label{eq:DFT-dilaton}
\end{align}

\paragraph{Ramond--Ramond fields:}
In order to study the ten-dimensional type II supergravity, let us consider the case $D=10$\,. 
Associated with the double local Lorentz group $\OO(1,9)\times\OO(9,1)$, we introduce two sets of gamma matrices, $(\Gamma^{\Loa})^{\SPa}{}_{\SPb}$ and $(\brGamma^{\Lobra})^{\SPbra}{}_{\SPbrb}$, satisfying
\begin{align}
\begin{split}
 &\{\Gamma^{\Loa},\,\Gamma^{\Lob}\} =2\,\eta^{\Loa\Lob}\,,\qquad 
  \{\brGamma^{\Lobra},\,\brGamma^{\Lobrb}\} = 2\,\breta^{\Lobra\Lobrb} \,,
\\
 &(\Gamma^{\Loa})^\dagger = - \Gamma^{0}\,\Gamma^{\Loa}\,(\Gamma^{0})^{-1} 
 = \mp \Gamma^{\Loa}\,, \qquad \Loa=\biggl\{\begin{array}{l} 0 \\[-1mm] 1,\dotsc,9 \end{array} ,
\\
 &(\brGamma^{\Lobra})^\dagger = + \brGamma^{0}\,\brGamma^{\Lobra}\,(\brGamma^{0})^{-1} 
 = \pm \brGamma^{\Lobra}\,, \qquad \Lobra=\biggl\{\begin{array}{l} 0 \\[-1mm] 1,\dotsc,9 \end{array} .
\end{split}
\end{align}
We also introduce the chirality operators
\begin{align}
\begin{split}
 &\Gamma^{11}\equiv \Gamma^{012\cdots 9}\,,\qquad 
 \brGamma^{11}\equiv \brGamma^{012\cdots 9}\,,\qquad (\Gamma^{11})^\dagger=\Gamma^{11}\,,\qquad
 (\brGamma^{11})^\dagger=\brGamma^{11}\,,
\\
 &\{\Gamma^{\Loa},\,\Gamma^{11}\} =0\,,\qquad \{\brGamma^{\Lobra},\,\brGamma^{11}\} = 0\,,\qquad
 (\Gamma^{11})^2 =1\,,\qquad (\brGamma^{11})^2=1\,,
\end{split}
\end{align}
and the charge conjugation matrices $C_{\SPa\SPb}$ and $\brC_{\SPbra\SPbrb}$ satisfying%
\footnote{In order to follow the convention of \cite{Arutyunov:2015qva}, we employ the charge conjugation matrices $C_-$ and $\bar{C}_-$ of \cite{Jeon:2012kd} rather than $C_+$ and $\bar{C}_+$. They are related as $C_-=C_+\,\Gamma^{11}$ and $\bar{C}_- = \bar{C}_+\,\brGamma^{11}$.}
\begin{align}
\begin{alignedat}{2}
 &C\,\Gamma^{\Loa}\,C^{-1} = -(\Gamma^{\Loa})^\rmT \,, &\qquad
  C &= - C^\rmT = - C^{-1}\,, \qquad C^*=C \,,
\\
 &\brC\,\brGamma^{\Lobra}\,\brC^{-1} = -(\brGamma^{\Lobra})^\rmT \,,&\qquad 
 \brC &= -\brC^\rmT = - \brC^{-1}\,, \qquad \brC^*=\brC \,.
\end{alignedat}
\label{eq:CC-properties}
\end{align}
We can show $C\,\Gamma^{11}\,C^{-1}=-\Gamma^{11}$ and $\brC\,\brGamma^{11}\,\brC^{-1}=-\brGamma^{11}$ by using
\begin{align}
 C\,\Gamma^{\Loa_1\cdots\Loa_n}\,C^{-1} = (-1)^{\frac{n(n+1)}{2}} \,(\Gamma^{\Loa_1\cdots\Loa_n})^\rmT \,, \qquad
 \brC\,\brGamma^{\Lobra_1\cdots\Lobra_n}\,\brC^{-1} = (-1)^{\frac{n(n+1)}{2}} \,(\brGamma^{\Lobra_1\cdots\Lobra_n})^\rmT \,. 
\end{align}
We raise or lower the spinor indices by using the charge conjugation matrices like
\begin{align}
\begin{alignedat}{2}
 (\Gamma^{\Loa})_{\SPa\SPb} &\equiv (\Gamma^{\Loa})^{\SPc}{}_{\SPb}\, C_{\SPc\SPa}\,,&\qquad 
 (\Gamma^{\Loa})^{\SPa\SPb} &\equiv C^{\SPb\SPc}\,(\Gamma^{\Loa})^{\SPa}{}_{\SPc} \,,
\\
 (\brGamma^{\Lobra})_{\SPbra\SPbrb} &\equiv (\brGamma^{\Lobra})^{\SPbrc}{}_{\SPbrb}\,\brC_{\SPbrc\SPbra}\,,&\qquad 
 (\brGamma^{\Lobra})^{\SPbra\SPbrb} &\equiv \brC^{\SPbrb\SPbrc}\,(\brGamma^{\Lobra})^{\SPbra}{}_{\SPbrc} \,,
\end{alignedat}
\end{align}
and then from \eqref{eq:CC-properties} we have
\begin{align}
 (\Gamma^{\Loa})_{\SPa\SPb} = (\Gamma^{\Loa})_{\SPb\SPa}\,,\qquad 
 (\brGamma^{\Lobra})^{\SPbra\SPbrb} = (\brGamma^{\Lobra})^{\SPbrb\SPbra} \,.
\end{align}

We define the R--R potential as a bispinor $\bisC^{\SPa}{}_{\SPbrb}$ with a definite chirality
\begin{align}
 \Gamma^{11}\,\bisC\,\brGamma^{11} = \pm\, \bisC \,,
\end{align}
where the sign is for type IIA/IIB supergravity. 
The R--R field strength is defined as
\begin{align}
\begin{split}
 &\bisF{}^{\SPa}{}_{\SPbrb} \equiv \cD_+ \bisC{}^{\SPa}{}_{\SPbrb} \equiv \frac{1}{\sqrt{2}}\,\bigl(\Gamma^M\,\cD_M \bisC + \Gamma^{11}\,\cD_M \bisC\,\brGamma^M\bigr){}^{\SPa}{}_{\SPbrb}\,,
\\
 &\Gamma^M\equiv \GV^M{}_{\Loa}\,\Gamma^{\Loa}\,,\qquad 
 \brGamma^M\equiv \brGV^M{}_{\Lobra}\,\brGamma^{\Lobra}\,,
\end{split}
\label{eq:F-C-relation}
\end{align}
where $\cD_+$ is a nilpotent operator introduced in \cite{Jeon:2012kd}, and the covariant derivative $\cD_M$ for a bispinor $\bm{\cT}^{\SPa}{}_{\SPbrb}$ and the spin connections are defined as \cite{Jeon:2011cn,Jeon:2011vx,Jeon:2012kd}
\begin{align}
\begin{split}
 &\cD_M \bm{\cT}^{\SPa}{}_{\SPbrb} \equiv \partial_M \bm{\cT}^{\SPa}{}_{\SPbrb} + \Phi_{M}{}^{\SPa}{}_{\SPc}\,\bm{\cT}^{\SPc}{}_{\SPbrb} - \bm{\cT}^{\SPa}{}_{\SPbrc}\,\brPhi_{M}{}^{\SPbrc}{}_{\SPbrb}\,,
\\
 &\Phi_{M}{}^{\SPa}{}_{\SPb} \equiv \frac{1}{4}\,\Phi_{M\Loc\Lod}\,(\Gamma^{\Loc\Lod})^{\SPa}{}_{\SPb}\,,\qquad 
 \brPhi_{M}{}^{\SPbra}{}_{\SPbrb}\equiv \frac{1}{4}\,\brPhi_{M\Lobrc\Lobrd}\,(\Gamma^{\Lobrc\Lobrd}){}^{\SPbra}{}_{\SPbrb}
\\
 &\Phi_{M\Loc\Lod} \equiv \GV^N{}_{\Loc}\,\nabla_M \GV_{N\Lod}
 = \GV^N{}_{\Loc}\,\bigl(\partial_M \GV_{N\Lod} - \Gamma_{M}{}^P{}_N\,\GV_{P\Lod} \bigr) \,,
\\
 &\brPhi_{M\Lobrc\Lobrd} \equiv \brGV^N{}_{\Lobrc}\,\nabla_M \brGV_{N\Lobrd}
 = \brGV^N{}_{\Lobrc}\,\bigl(\partial_M \brGV_{N\Lobrd} - \Gamma_{M}{}^P{}_N\,\brGV_{P\Lobrd} \bigr) \,,
\end{split}
\end{align}
where $\nabla_M$ is the (semi-)covariant derivative in DFT \cite{Jeon:2010rw,Jeon:2011cn,Hohm:2011si} (see also \cite{Sakatani:2016fvh} which employs the same convention as this paper). 
Since $\cD_+$ flips the chirality, $\bisF$ has the opposite chirality to $\bisC$ \cite{Jeon:2012kd}
\begin{align}
 \Gamma^{11}\,\bisF\,\brGamma^{11} = \mp\, \bisF \,. 
\end{align}
As it has been shown in \cite{Jeon:2012kd}, $\bisF$ transforms covariantly under the $\OO(1,9)\times\OO(9,1)$ double Lorentz transformations, and transforms as a scalar under generalized diffeomorphisms. 
Further, from the nilpotency of $\cD_+$, $\bisF$ is invariant under gauge transformations of R--R potential
\begin{align}
 \delta \bisC = \cD_+ \bm{\lambda} \,,\qquad \Gamma^{11}\,\bm{\lambda}\,\brGamma^{11} = \mp \bm{\lambda} \,,
\end{align}
and the Bianchi identity is given by
\begin{align}
 \cD_+ \bisF = 0 \,.
\end{align}
As in the case of the democratic formulation \cite{Fukuma:1999jt,Bergshoeff:2001pv}, the self-duality relation
\begin{align}
 \bisF = -\Gamma^{11}\,\bisF \ \bigl(= \pm\, \bisF \,\brGamma^{11}\bigr)\,,
\label{eq:F-self-dual}
\end{align}
for type IIA/IIB supergravity is imposed by hand at the level of the equations of motion. 

\paragraph{Section condition and gauge symmetry:}
In DFT, fields are defined on the doubled spacetime with the generalized coordinates $(x^M)=(x^m,\,\tilde{x}_m)$, where $x^m$ are the standard ``physical'' $D$-dimensional coordinates and $\tilde{x}_m$ are the dual coordinates. 
For the consistency of DFT, we require that arbitrary fields or gauge parameters $A(x)$ and $B(x)$ satisfy the so-called section condition \cite{Siegel:1993th,Hull:2009mi,Hull:2009zb},
\begin{align}
 \ODDeta^{MN}\,\partial_M A(x)\,\partial_N B(x) = 0\,,\qquad \ODDeta^{MN}\,\partial_M \partial_N A(x) = 0 \,. 
\end{align}
In general, under this condition, fields and gauge parameters can depend on at most $D$ coordinates out of the $2D$ coordinates $x^M$. 
We frequently choose the ``canonical solution'' where all fields and gauge parameters are independent of the dual coordinates; $\tilde{\partial}^m \equiv \frac{\partial}{\partial\tilde{x}_m}=0$\,. 
In this case, DFT reduces to the conventional supergravity. 
Instead, if all fields depend on $(D-1)$ coordinates $x^i$ and only the dilaton $d(x)$ has an additional linear dependence on a dual coordinates $\tilde{z}$, DFT reduces to the generalized supergravity as discussed in \cite{Sakatani:2016fvh,Sakamoto:2017wor}. 

When the section condition is satisfied, the gauge symmetry of DFT is generated by the generalized Lie derivative \cite{Siegel:1993th,Hull:2009zb}
\begin{align}
 \gLie_V W^M \equiv V^N\,\partial_N W^M - \bigl(\partial_N V^M -\partial^M V_N \bigr)\,W^N\,. 
\end{align}
This symmetry is interpreted as diffeomorphisms in the doubled spacetime, $x^M\to x^M + V^M(x)$\,. 
Indeed, under the canonical section $\tilde{\partial}^m=0$\,, this symmetry consists of the conventional diffeomorphisms and $B$-field gauge transformations. 
If we parameterize the diffeomorphism parameter as $(V^M)=(v^m,\,\tilde{v}_m)$, the vector $v^m$ corresponds to the $D$-dimensional diffeomorphism parameter while the 1-form $\tilde{v}_m$ corresponds to the gauge parameter of the $B$-field gauge transformation, $B_2\to B_2 + \rmd \tilde{v}_1$\,. 
Under the canonical section, this is the whole gauge symmetry, but if we choose a different section, the generalized diffeomorphism may generate other local $\OO(D,D)$ transformations, such as $\beta$-transformations. 
For more details, the reader may consult a concise review \cite{Hohm:2013bwa}. 

\vspace{-\Pskip}
\subsection{Diagonal gauge fixing}
\label{sec:diagonal-gauge}

In this subsection, we review the diagonal gauge fixing introduced in \cite{Jeon:2011cn,Jeon:2012kd}. 

\subsubsection{NS--NS fields}

In order to constrain the redundantly introduced two vielbeins $e_{m}{}^{\Loa}$ and $\bre_{m}{}^{\Lobra}$\,, we implement the diagonal gauge fixing 
\begin{align}
 e_{m}{}^{\Loa} = \bre_{m}{}^{\Lobra} \,, 
\label{eq:diagonal-gauge}
\end{align}
which is important to reproduce the conventional supergravity. 
Before the diagonal gauge fixing, the double vielbeins transform as
\begin{align}
 \GV_{\Loa}{}^M \to h^M{}_N\,\,\GV_{\Loa}{}^N\,,\qquad
 \brGV_{\Lobra}{}^M \to h^M{}_N\,\,\brGV_{\Lobra}{}^N \,,
\end{align}
under a global $\OO(10,10)$ rotation or a finite generalized diffeomorphism. 
We parameterize the $\OO(10,10)$ matrix $h^M{}_N$ as
\begin{align}
\begin{split}
 &h_M{}^N = \begin{pmatrix} \bmp_{m}{}^{n} & \bmq_{mn} \\ \bmr^{mn} & \bms^{m}{}_{n} \end{pmatrix} \,,\qquad 
 h^M{}_N = \begin{pmatrix} \bms^{m}{}_{n} & \bmr^{mn} \\ \bmq_{mn} & \bmp_{m}{}^{n} \end{pmatrix} 
\\
 &\bigl(\bmp\,\bms^\rmT + \bmq\,\bmr^\rmT =\bm{1}\,,\quad 
 \bmr\,\bms^\rmT + \bms\,\bmr^\rmT =0 \,,\quad 
 \bmp\,\bmq^\rmT + \bmq\,\bmp^\rmT =0 \bigr)\,, 
\end{split}
\end{align}
and then obtain the following transformation rule:
\begin{align}
\begin{alignedat}{2}
 e_m{}^{\Loa} &\to \bigl[\bigl(\bms^\rmT + E^\rmT\,\bmr^\rmT\bigr)^{-1}\bigr]_m{}^n\, e_n{}^{\Loa}\,,&\qquad 
 \bre_m{}^{\Lobra} &\to \bigl[\bigl(\bms^\rmT - E\,\bmr^\rmT\bigr)^{-1}\bigr]_m{}^n\, \bre_n{}^{\Lobra}\,,
\\
 \tilde{e}_m{}^{\Loa} &\to \bigl(\bmp + \bmq\, E^{-1}\bigr)_m{}^n\,\tilde{e}_n{}^{\Loa}\,,&\qquad 
 \tilde{\bre}_m{}^{\Lobra} &\to \bigl(\bmp - \bmq\, E^{-\rmT}\bigr)_m{}^n\,\tilde{\bre}_n{}^{\Lobra}\,, 
\\
 E_{mn} &\to [(\bmq+\bmp\,E)\,(\bms+\bmr\,E)^{-1}]_{mn} \,,&\qquad 
 E^{mn} &\to [(\bmr + \bms\,E^{-1})\,(\bmp + \bmq\,E^{-1})^{-1}]^{mn} \,. 
\label{eq:double-vielbein-transf}
\end{alignedat}
\end{align}
At the same time, the dilaton transforms as
\begin{align}
 \Exp{-2d} \to \abs{\det(\bmp_{m}{}^{n})}\,\Exp{-2d} \,, 
\end{align}
and the bispinors of R--R fields, $\bisC$ and $\bisF$, are invariant.

As we can see from \eqref{eq:double-vielbein-transf}, under a (geometric) subgroup (where $\bmr^{mn}=0$),
\begin{align}
 h_M{}^N = \begin{pmatrix} \bmp_{m}{}^{n} & \bmq_{mn} \\ 0 & (\bmp^{-\rmT})^{m}{}_{n} \end{pmatrix} \,,\qquad 
 h^M{}_N = \begin{pmatrix} (\bmp^{-\rmT})^{m}{}_{n} & 0 \\ \bmq_{mn} & \bmp_{m}{}^{n} \end{pmatrix} 
 \qquad 
 \bigl(\bmp\,\bmq^\rmT = - \bmq\,\bmp^\rmT \bigr)\,,
\end{align}
$e_m{}^{\Loa}$ and $\bre_m{}^{\Lobra}$ transform in the same manner. 
However, if we perform a general $\OO(10,10)$ transformation with $\bmr^{mn}\neq 0$, even if we choose the diagonal gauge in the original duality frame $(e_m{}^{\Loa}=\bre_m{}^{\Lobra})$, after the transformation, $e_m{}^{\Loa}\to e'_m{}^{\Loa}$ and $\bre_m{}^{\Lobra}\to \bre'_m{}^{\Lobra}$, we obtain
\begin{align}
 \bre'_m{}^{\Lobra} = (\Lambda^{-1})^{\Lobra}{}_{\Lob}\,e'_m{}^{\Lob} \,,\qquad 
 \Lambda^{\Loa}{}_{\Lobrb} \equiv \bigl[e^{\rmT}\,(\bms + \bmr\,E)^{-1}\,(\bms - \bmr\,E^\rmT)\,e^{-\rmT} \bigr]^{\Loa}{}_{\Lobrb} \in \OO(9,1) \,.
\label{eq:Lambda-def}
\end{align}
In order to maintain the diagonal gauge \eqref{eq:diagonal-gauge}, we shall simultaneously perform an $\OO(9,1)$ local Lorentz transformation for barred tensors that compensates the deviation of $\bre_m{}^{\Lobra}$ from $e_m{}^{\Loa}$. 
Namely, we modify the $\OO(10,10)$ transformation as \cite{Jeon:2012kd}
\begin{align}
 \GV_M{}^{\Loa} \to h_M{}^N\, \GV_{N}{}^{\Loa}\,,\qquad
 \brGV_M{}^{\Lobra} \to h_M{}^N\, \Lambda^{\Lobra}{}_{\Lobrb}\,\brGV_{N}{}^{\Lobrb} \,. 
\label{eq:modified-O(10-10)}
\end{align}
After the diagonal gauge fixing, since there is no more distinction between $\{\Loa,\,\SPa\}$ and $\{\Lobra,\,\SPbra\}$, we may simply replace $\{\Lobra,\,\SPbra\}$ by $\{\Loa,\,\SPa\}$. 
In this replacement, we should be careful about the signature
\begin{align}
 \breta_{\Loa\Lob} = -\eta_{\Loa\Lob}\,,\qquad \brC_{\SPa\SPb} = C_{\SPa\SPb} \,.
\end{align}
In addition, we relate the two sets of gamma matrices as
\begin{align}
 \brGamma^{\Loa}=\Gamma^{11}\,\Gamma^{\Loa}\qquad 
 \bigl(\,\{\brGamma^{\Loa},\,\brGamma^{\Lob}\} = - \{\Gamma^{\Loa},\,\Gamma^{\Lob}\} = 2\,\breta^{\Loa\Lob}\, \bigr)\,,\qquad 
 \brGamma^{11}=-\Gamma^{11}\,.
\end{align}

\subsubsection{R--R fields}

According to the diagonal gauge fixing, there is no distinction between the two spinor indices $\SPa$ and $\SPbra$, and we can convert the bispinors into polyforms:
\begin{align}
 \bisC^{\SPa}{}_{\SPb} = \sum_{n} \frac{1}{n!}\,\hat{\cC}_{\Loa_1\cdots \Loa_n}\,(\Gamma^{\Loa_1\cdots \Loa_n})^{\SPa}{}_{\SPb} \,,\qquad 
 \bisF^{\SPa}{}_{\SPb} = \sum_{n} \frac{1}{n!}\,\hat{\cF}_{\Loa_1\cdots \Loa_n}\,(\Gamma^{\Loa_1\cdots \Loa_n})^{\SPa}{}_{\SPb} \,. 
\end{align}
From the identity,
\begin{align}
 \Gamma^{11}\,\Gamma^{\Loa_1\cdots \Loa_p} = \frac{(-1)^{\frac{p(p+1)}{2}}}{(10-p)!}\,\epsilon^{\Loa_1\cdots \Loa_p \Lob_1\cdots \Lob_{10-p}}\,\Gamma_{\Lob_1\cdots \Lob_{10-p}} \,,
\end{align}
where $\epsilon_{0\cdots 9}=-\epsilon^{0\cdots 9}=1$\,, the self-duality relation \eqref{eq:F-self-dual} can be expressed as
\begin{align}
 \hat{\cF}_p = (-1)^{\frac{p(p-1)}{2}}\, * \hat{\cF}_{10-p} \,.
\end{align}
Here, we have defined
\begin{align}
\begin{alignedat}{2}
 \hat{\cF} &\equiv \sum_p \hat{\cF}_p\,,&\qquad 
 \hat{\cF}_p &\equiv \frac{1}{p!}\,\hat{\cF}_{m_1\cdots m_p}\,\rmd x^{m_1}\wedge \cdots \wedge\rmd x^{m_p}\,,
\\
 \hat{\cC} &\equiv \sum_p \hat{\cC}_p\,,&\qquad 
 \hat{\cC}_p &\equiv \frac{1}{p!}\,\hat{\cC}_{m_1\cdots m_p}\,\rmd x^{m_1}\wedge \cdots \wedge\rmd x^{m_p}\,,
\end{alignedat}
\end{align}
where the R--R fields with the curved indices are defined as
\begin{align}
 \hat{\cF}_{m_1\cdots m_p} \equiv e_{m_1}{}^{\Loa_1}\cdots e_{m_p}{}^{\Loa_p}\,\hat{\cF}_{\Loa_1\cdots \Loa_p}\,,\qquad
 \hat{\cC}_{m_1\cdots m_p} \equiv e_{m_1}{}^{\Loa_1}\cdots e_{m_p}{}^{\Loa_p}\,\hat{\cC}_{\Loa_1\cdots \Loa_p}\,.
\label{eq:B-curved-flat}
\end{align}
In addition, if we define the components of the spin connections as
\begin{align}
\begin{split}
 \Phi_{\Loa\Loc\Lod} &\equiv \GV^M{}_{\Loa}\,\Phi_{M\Loc\Lod} \,,\qquad 
 \brPhi_{\Lobra\Lobrc\Lobrd} \equiv \brGV^M{}_{\Lobra}\,\brPhi_{M\Lobrc\Lobrd} \,,
\\
 \Phi_{\Lobra\Loc\Lod} &\equiv \brGV^M{}_{\Lobra}\,\Phi_{M\Loc\Lod} \,,\qquad 
 \brPhi_{\Loa\Lobrc\Lobrd} \equiv \GV^M{}_{\Loa}\,\brPhi_{M\Lobrc\Lobrd} \,,
\end{split}
\end{align}
and compute their explicit forms under the canonical section $\tilde{\partial}^m=0$ as
\begin{align}
\begin{split}
 &\!\begin{alignedat}{2}
 \Phi_{\Lobra\Lob\Loc} &= \frac{1}{\sqrt{2}}\,\Bigl(\omega_{\Loa\Lob\Loc} + \frac{1}{2}\, H_{\Loa\Lob\Loc}\Bigr) \,,&\qquad 
 \brPhi_{\Loa\Lobrb\Lobrc} &= \frac{1}{\sqrt{2}}\,\Bigl(-\omega_{\Loa\Lob\Loc} + \frac{1}{2}\, H_{\Loa\Lob\Loc}\Bigr)\,,
\\
 \Phi_{[\Loa\Lob\Loc]} &= \frac{1}{\sqrt{2}}\,\Bigl(\omega_{[\Loa\Lob\Loc]} + \frac{1}{6}\,H_{\Loa\Lob\Loc}\Bigr)\,,&\qquad 
 \brPhi_{[\Lobra\Lobrb\Lobrc]} &= \frac{1}{\sqrt{2}}\,\Bigl(-\omega_{[\Loa\Lob\Loc]} + \frac{1}{6}\,H_{\Loa\Lob\Loc}\Bigr)\,,
\end{alignedat}
\\
 &\eta^{\Loa\Lob}\,\Phi_{\Loa\Lob\Loc} = \frac{1}{\sqrt{2}}\,\bigl(\eta^{\Loa\Lob}\,\omega_{\Loa\Lob\Loc} -2\,e_{\Loc}{}^m\,\partial_m\Phi\bigr) 
 = \breta^{\Lobra\Lobrb}\,\brPhi_{\Lobra\Lobrb\Lobrc} \,,
\\
 &\omega_{\Loa\Lob\Loc}\equiv e_{\Loa}{}^{m}\,\omega_{m\Lob\Loc} \,,\qquad 
  \omega_m{}^{\Loa\Lob} \equiv 2\,e^{n[\Loa}\,\partial_{[m} e_{n]}{}^{\Lob]} - e^{\Loa p}\,e^{\Lob q}\,\partial_{[p} e_{q]}{}^{\Loc}\,e_{m\Loc} \,,
\\
 &H_{\Loa\Lob\Loc}\equiv e_{\Loa}{}^{m}\,e_{\Lob}{}^{n}\,e_{\Loc}{}^{p}\,H_{mnp} \,, \qquad 
  H_{mnp}\equiv 3\,\partial_{[m}B_{np]}\,,
\end{split}
\label{eq:spin-connections}
\end{align}
we can show that the relation \eqref{eq:F-C-relation} between $\bisF$ and $\bisC$ can be expressed as \cite{Jeon:2012kd}\footnote{Here, we have used the following identities for type IIA/IIB theory:
\begin{align*}
\begin{split}
 &\frac{1}{2}\, \bigl(\Gamma^{m}\,\partial_{m}\bisC \mp \partial_m\bisC\,\Gamma^{m}\bigr) = \sum_n \frac{1}{n!}\,(\rmd \cC)_{\Loa_1\cdots\Loa_n}\,\Gamma^{\Loa_1\cdots\Loa_n} \,,
\\
 &\frac{1}{2}\, \partial_m\Phi\,\bigl(\Gamma^{m}\,\bisC \mp \bisC\,\Gamma^{m}\bigr) = \sum_n \frac{1}{n!}\,(\rmd\Phi \wedge \cC)_{\Loa_1\cdots\Loa_n}\,\Gamma^{\Loa_1\cdots\Loa_n} \,,
\\
 &\frac{1}{8}\, \omega_{m\Loa\Lob}\,\bigl[\Gamma^{\Loa}\,(\Gamma^{\Lob\Loc}\,\bisC-\bisC\,\Gamma^{\Lob\Loc})\mp (\Gamma^{\Lob\Loc}\,\bisC-\bisC\,\Gamma^{\Lob\Loc})\,\Gamma^{\Loa}\bigr] = -\sum_n \frac{\omega_{[\Loa_1}{}^{\Lob}{}_{\Loa_2}\,\cC_{|\Lob|\Loa_3\cdots\Loa_n]}}{2!\,(n-2)!} \,\Gamma^{\Loa_1\cdots\Loa_n} \,,
\\
 &\frac{1}{16}\,H_{\Loa\Lob\Loc}\,\Bigl[\frac{1}{3}\, \bigl(\Gamma^{\Loa\Lob\Loc}\,\bisC \mp \bisC\,\Gamma^{\Loa\Lob\Loc}\bigr) 
 + \bigl(\Gamma^{\Loa}\,\bisC\,\Gamma^{\Lob\Loc} \mp \Gamma^{\Lob\Loc}\,\bisC\,\Gamma^{\Loa}\bigr) \Bigr]
 = \sum_n \frac{1}{n!}\,(H_3\wedge \cC)_{\Loa_1\cdots\Loa_n}\,\Gamma^{\Loa_1\cdots\Loa_n} \,. 
\end{split}
\end{align*}
}
\begin{align}
 \hat{\cF} = \rmd \hat{\cC} - \rmd \Phi\wedge \hat{\cC} + H_3\wedge \hat{\cC} \,. 
\label{eq:cF-cC-relation}
\end{align}

Originally, the R--R fields were invariant under global $\OO(10,10)$ transformations or generalized diffeomorphisms, but after the diagonal gauge fixing, according to the modified transformation rule \eqref{eq:modified-O(10-10)}, they transform as
\begin{align}
 \bisC \to \bisC \, \Omega^{-1} \,,\qquad 
 \bisF \to \bisF \, \Omega^{-1} \,,
\label{eq:R-R-beta-transf}
\end{align}
where $\Omega$ is the spinor representation of the local Lorentz transformation \eqref{eq:Lambda-def},
\begin{align}
 \Omega^{-1}\,\brGamma^{\Loa}\,\Omega = \Lambda^{\Loa}{}_{\Lob}\, \brGamma^{\Lob} \qquad 
 \bigl(\Lambda^{\Loa}{}_{\Lob} = \bigl[e^{\rmT}\,(\bms + \bmr\,E)^{-1}\,(\bms - \bmr\,E^\rmT)\,e^{-\rmT} \bigr]^{\Loa}{}_{\Lob} \bigr)\,.
\label{eq:Omega-def}
\end{align}

For later convenience, we here introduce several definitions of R--R fields that can be summarized as follows:
\begin{align}
\vcenter{
\xymatrix@R=35pt{
 (\check{C},\,\check{F};\,\check{\bm{C}},\,\check{\bm{F}}) \ar@/_/[rr]_{\Atop{\beta\text{-twist}}{(\text{for polyform} \Exp{-\beta\vee})}} \ar@/_/[d]_{\Atop{\tilde{\phi}\text{-untwist}}{\Exp{\tilde{\phi}}}} & \qquad\quad & \ar@/_/[ll]_{\Atop{\beta\text{-untwist}}{(\text{for polyform} \Exp{\beta\vee})}} (A,\,F;\,\sla{A},\,\sla{F}) \ar@/^/[rr]^{\Atop{B\text{-untwist}}{(\text{for polyform} \Exp{-B_2\wedge})}} &\qquad & (\hat{C},\,\hat{F};\,\hat{\bm{C}},\,\hat{\bm{F}}) \ar@/^/[ll]^{\Atop{B\text{-twist}}{(\text{for polyform} \Exp{B_2\wedge})}} \ar@/_/[d]_{\Atop{\Phi\text{-untwist}}{\Exp{\Phi}}}
\\
 (\check{\cC},\,\check{\cF};\,\check{\bm{\cC}},\,\check{\bm{\cF}}) \ar@/_/[u]_{\Atop{\tilde{\phi}\text{-twist}}{\Exp{-\tilde{\phi}}}} \ar@{<->}[rrrr]^{\text{Eqs.~\eqref{eq:R-R-relation1} and \eqref{eq:R-R-relation2}}}_{\text{(local Lorentz transformation)}} & \qquad & &\qquad\quad & (\hat{\cC},\,\hat{\cF};\,\bisC,\,\bisF) \ar@/_/[u]_{\Atop{\Phi\text{-twist}}{\Exp{-\Phi}}}
}
}
\label{eq:R-R-diagram}
\end{align}
The quantities at the lower right, polyforms $(\hat{\cC},\,\hat{\cF})$ and bispinors $(\bisC,\,\bisF)$, are already defined, which we call $(B,\,\Phi)$-untwisted fields. 
There, the curved indices and flat indices are interchanged by using the usual vielbein $e_m{}^{\Loa}$ like \eqref{eq:B-curved-flat}. 
The quantities at the upper right, which we call the $B$-untwisted fields, are defined as
\begin{align}
 \hat{\bm{C}} \equiv \Exp{-\Phi}\bisC\,,\qquad 
 \hat{\bm{F}} \equiv \Exp{-\Phi}\bisF\,,\qquad 
 \hat{C}\equiv \Exp{-\Phi}\hat{\cC}\,,\qquad 
 \hat{F}\equiv \Exp{-\Phi}\hat{\cF}\,. 
\end{align}
The curved and flat indices are again related as
\begin{align}
 \hat{C}_{m_1\cdots m_n}\equiv e_{m_1}{}^{\Loa_1}\cdots e_{m_n}{}^{\Loa_n}\, \hat{C}_{\Loa_1\cdots \Loa_n}\,,\qquad
 \hat{F}_{m_1\cdots m_n}\equiv e_{m_1}{}^{\Loa_1}\cdots e_{m_n}{}^{\Loa_n}\, \hat{F}_{\Loa_1\cdots \Loa_n}\,. 
\end{align}
The $B$-untwisted fields are rather familiar R--R fields satisfying
\begin{align}
 \hat{F} = \rmd \hat{C} + H_3\wedge \hat{C} \,,
\end{align}
which can be shown from \eqref{eq:cF-cC-relation}. 
We also define a polyform $A$ and its field strength $F$ as
\begin{align}
 A= \Exp{-\Phi} \Exp{B_2\wedge} \hat{\cC} = \Exp{B_2\wedge} \hat{C} \,,\qquad 
 F= \Exp{-\Phi} \Exp{B_2\wedge} \hat{\cF} = \Exp{B_2\wedge} \hat{F} \,. 
\end{align}
These are utilized in \cite{Fukuma:1999jt,Hassan:1999mm,Hohm:2011dv} to define R--R fields as $\OO(D,D)$ spinors (see also \cite{Sakamoto:2017wor})
\begin{align}
 \sla{A} \equiv \sum_n \frac{1}{n!}\, A_{m_1\cdots m_n}\,\gamma^{m_1\cdots m_n}\vert0\rangle\,,\qquad 
 \sla{F} \equiv \sum_n \frac{1}{n!}\, F_{m_1\cdots m_n}\,\gamma^{m_1\cdots m_n}\vert0\rangle\,. 
\end{align}
By using the dual fields $(\tilde{e}_m{}^{\Loa},\,\beta^{mn},\,\tilde{\phi})$\,, we can also introduce the dual R--R fields,
\begin{center}
\begin{tabular}{ll}
 \underline{$\bullet$ $\beta$-untwisted fields:} & polyforms $(\check{C},\,\check{F})$ and bispinors $(\check{\bm{C}},\,\check{\bm{F}})$\,,
\\[2mm]
 \underline{$\bullet$ $(\beta,\,\tilde{\phi})$-untwisted fields:} \qquad\quad& polyforms $(\check{\cC},\,\check{\cF})$ and bispinors $(\check{\bm{\cC}},\,\check{\bm{\cF}})$\,.
\end{tabular}
\end{center}
By introducing an operator $\beta\vee F\equiv \frac{1}{2}\,\beta^{mn}\,\iota_m\,\iota_n F$, we define these polyforms as
\begin{align}
 \check{C} \equiv \Exp{\beta\vee} A \,,\qquad 
 \check{F} \equiv \Exp{\beta\vee} F \,. \qquad
 \check{\cC} \equiv \Exp{\tilde{\phi}} \Exp{\beta\vee} A \,,\qquad 
 \check{\cF} \equiv \Exp{\tilde{\phi}} \Exp{\beta\vee} F \,,
\end{align}
and their flat components as
\begin{align}
\begin{split}
 \check{C}_{\Loa_1\cdots \Loa_p} &\equiv \tilde{e}_{\Loa_1}{}^{m_1}\cdots \tilde{e}_{\Loa_p}{}^{m_p}\,\check{C}_{m_1\cdots m_p}\,,\qquad 
 \check{F}_{\Loa_1\cdots \Loa_p} \equiv \tilde{e}_{\Loa_1}{}^{m_1}\cdots \tilde{e}_{\Loa_p}{}^{m_p}\,\check{F}_{m_1\cdots m_p}\,,
\\
 \check{\cC}_{\Loa_1\cdots \Loa_p} &\equiv \tilde{e}_{\Loa_1}{}^{m_1}\cdots \tilde{e}_{\Loa_p}{}^{m_p}\,\check{\cC}_{m_1\cdots m_p}\,,\qquad 
 \check{\cF}_{\Loa_1\cdots \Loa_p} \equiv \tilde{e}_{\Loa_1}{}^{m_1}\cdots \tilde{e}_{\Loa_p}{}^{m_p}\,\check{\cF}_{m_1\cdots m_p}\,,
\end{split}
\label{eq:dual-RR-flat-components}
\end{align}
by using the dual vielbein $\tilde{e}_m{}^{\Loa}$\,. 
Their corresponding bispinors are defined as
\begin{align}
\begin{split}
 \check{\bm{C}} &\equiv \sum_n \frac{1}{n!}\,\check{C}_{\Loa_1\cdots \Loa_n}\,\Gamma^{\Loa_1\cdots \Loa_n} \,,\qquad
 \check{\bm{F}} \equiv \sum_n \frac{1}{n!}\,\check{F}_{\Loa_1\cdots \Loa_n}\,\Gamma^{\Loa_1\cdots \Loa_n} \,,
\\
 \check{\bm{\cC}} &\equiv \sum_n \frac{1}{n!}\,\check{\cC}_{\Loa_1\cdots \Loa_n}\,\Gamma^{\Loa_1\cdots \Loa_n} \,,\qquad
 \check{\bm{\cF}} \equiv \sum_n \frac{1}{n!}\,\check{\cF}_{\Loa_1\cdots \Loa_n}\,\Gamma^{\Loa_1\cdots \Loa_n} \,.
\end{split}
\end{align}

\subsubsection{Single \texorpdfstring{$T$}{T}-duality}

As a simple application of the formula \eqref{eq:R-R-beta-transf}, let us explain how the R--R fields transform under a single $T$-duality along the $x^z$-direction,
\begin{align}
 (h^M{}_N) = \begin{pmatrix} \bm{{1_{10}}} - \bm{e}_z & \bm{e}_z \\ \bm{e}_z & \bm{{1_{10}}} - \bm{e}_z \end{pmatrix}\,, \qquad 
 \bm{e}_z \equiv \diag (0,\dotsc,0,\, \overset{z\text{-th}}{1},\,0,\dotsc,0)\,. 
\end{align}
In this case, the vielbein and the dilaton transform as
\begin{align}
 e' = \bigl[\bm{{1_{10}}} - (\bm{{1_{10}}}-E^\rmT)\, \bm{e}_z \bigr]^{-1} \, e
 = \bigl[\bm{{1_{10}}} + \CG_{zz}^{-1}\, (\bm{{1_{10}}}-E^\rmT)\, \bm{e}_z \bigr] \, e \,, \qquad 
 \Exp{\Phi'} = \frac{1}{\sqrt{\CG_{zz}}} \Exp{\Phi} \,,
\label{eq:abelian-NS-NS}
\end{align}
and the Lorentz transformation matrix is
\begin{align}
 \Lambda \equiv (\Lambda^{\Loa}{}_{\Lob}) = e^{\rmT}\,\bigl[\bm{{1_{10}}} - \bm{e}_z \,(\bm{{1_{10}}} - E)\bigr]^{-1}\,\bigl[\bm{{1_{10}}} - \bm{e}_z \,(\bm{{1_{10}}}+ E^\rmT)\bigr]\,e^{-\rmT} \,.
\end{align}
This can be simplified as
\begin{align}
 \Lambda^{\Loa}{}_{\Lob} = \delta^{\Loa}_{\Lob} - 2\,\frac{e_z{}^{\Loa} \,e_{z\Lob}}{\CG_{zz}} \,,
\end{align}
and we can easily see that the R--R field transforms under the $T$-duality as \cite{Hassan:1999mm}
\begin{align}
 \bisC' = \bisC\, \Omega_z^{-1} \,,\qquad 
 \Omega_z \equiv \frac{e_{z \Loa}}{\sqrt{\CG_{zz}}}\,\brGamma^{\Loa}\,\brGamma^{11} = \frac{1}{\sqrt{\CG_{zz}}}\,\Gamma_z = \Omega_z^{-1} \quad (\Gamma_m\equiv e_m{}^{\Loa}\,\Gamma_{\Loa})\,,
\end{align}
where we have supposed $\CG_{zz}\geq 0$. 

From the identity \eqref{eq:Gamma-n-1}, we obtain
\begin{align}
 \bisC' = \bisC\, \Omega_z^{-1}
 = \frac{1}{\sqrt{\CG_{zz}}}\sum_n \frac{1}{n!}\biggl( n\,\hat{\cC}_{[\Loa_1\cdots\Loa_{n-1}}\,e_{\Loa_n]z} + \hat{\cC}_{\Loa_1\cdots\Loa_n\Lob}\,e_z{}^{\Lob} \biggr)\,\Gamma^{\Loa_1\cdots\Loa_n}\,.
\label{eq:C-transformation-single}
\end{align}
By using the $B$-untwisted R--R potentials, $\hat{\bm{C}} = \Exp{-\Phi}\bisC$ and $\hat{C}=\Exp{-\Phi}\hat{\cC}$, \eqref{eq:C-transformation-single} is expressed as
\begin{align}
 \hat{\bm{C}}' = \sum_n \frac{1}{n!}\biggl( n\,\hat{C}_{[\Loa_1\cdots\Loa_{n-1}}\,e_{\Loa_n]z} + \hat{C}_{\Loa_1\cdots\Loa_n\Lob}\,e_z{}^{\Lob} \biggr)\,\Gamma^{\Loa_1\cdots\Loa_n}\,,
\end{align}
where we have used \eqref{eq:abelian-NS-NS}. 
For the curved components, using the transformation rule of the vielbein \eqref{eq:abelian-NS-NS}, we obtain
\begin{align}
 \hat{C}'_{m_1\cdots m_n} 
 &= e'_{m_1}{}^{\Loa_1}\cdots e'_{m_n}{}^{\Loa_n}\, \bigl(n\,\hat{C}_{[\Loa_1\cdots\Loa_{n-1}}\,e_{\Loa_n]z} + \hat{C}_{\Loa_1\cdots\Loa_n\Lob}\,e_z{}^{\Lob} \bigr)
\nn\\
 &=n\,\hat{C}_{[m_1\cdots m_{n-1}}\,\CG_{m_n]z} + \hat{C}_{m_1\cdots m_{n}z}
\nn\\
 &\quad +n\,\CG_{zz}^{-1}\,
 \bigl[\hat{C}_{[m_1\cdots m_{n-1}}\,\CG_{zz}-(n-1)\,\hat{C}_{[m_1\cdots m_{n-2}|z|}\,\CG_{m_{n-1}|z|}\bigr]\,\bigl(\delta^z_{m_n]}-E^{\rmT}_{m_n]z}\bigr)
\nn\\
 &= \hat{C}_{m_1\cdots m_{n}z} +n\, \biggl[\hat{C}_{[m_1\cdots m_{n-1}} -(n-1)\,\frac{\hat{C}_{[m_1\cdots m_{n-2}|z|}\,\CG_{m_{n-1}|z|}}{\CG_{zz}}\biggr]\,\bigl(\delta^z_{m_n]}+B_{m_n]z}\bigr)\,.
\end{align}
This reproduces the famous transformation rule,
\begin{align}
\begin{split}
 &\hat{C}'_{i_1\cdots i_n} = \hat{C}_{i_1\cdots i_{n}z} +n\,\hat{C}_{[i_1\cdots i_{n-1}}\,B_{i_n]z} + n\,(n-1)\,\frac{\hat{C}_{[i_1\cdots i_{n-2}|z|}\,B_{i_{n-1}|z|}\,\CG_{i_n]z}}{\CG_{zz}} \,,
\\
 &\hat{C}'_{i_1\cdots i_{n-1}z} 
 = \hat{C}_{i_1\cdots i_{n-1}} - (n-1)\,\frac{\hat{C}_{[i_1\cdots i_{n-2}}\,\CG_{i_{n-1}]z}}{\CG_{zz}} \,,
\end{split}
\end{align}
where we have decomposed the coordinates as $\{x^m\}=\{x^i,\,x^z\}$. 

It is also noted that, under the single $T$-duality after taking the diagonal gauge, an arbitrary $\OO(1,9)$ spinor $\Psi_1^{\SPa}$ and an $\OO(9,1)$ spinor $\Psi_2^{\SPbra}$ transform as
\begin{align}
 \Psi_1 \ \to \ \Psi'_1=\Psi_1 \,,\qquad \Psi_2 \ \to \ \Psi'_2=\Omega \,\Psi_2 = \frac{e_{z \Loa}}{\sqrt{\CG_{zz}}}\, \brGamma^{\Loa}\,\brGamma^{11}\,\Psi_2 \,. 
\label{eq:Psi-Tdual}
\end{align}
When we consider a single $T$-duality connecting type IIA and type IIB superstring, these transformations are applied to the spacetime fermions $\Theta_1$ and $\Theta_2$ introduced later. 

\vspace{-\Pskip}
\subsection{\texorpdfstring{$\beta$}{\textbeta}-transformation of R--R fields}

In this subsection, we consider local $\beta$-transformations
\begin{align}
 h_M{}^N = \begin{pmatrix} \bm{{1_{10}}} & \bm{{0_{10}}} \\ \bmr^{mn}(x) & \bm{{1_{10}}} \end{pmatrix} \,,\qquad 
 h^M{}_N = \begin{pmatrix} \bm{{1_{10}}} & \bmr^{mn}(x) \\ \bm{{0_{10}}} & \bm{{1_{10}}} \end{pmatrix} \qquad \bigl(\bmr^{mn}=-\bmr^{nm}\bigr) \,. 
\end{align}
From the general transformation rule \eqref{eq:R-R-beta-transf}, the R--R fields should transform as $\bisC \to \bisC'=\bisC\, \Omega^{-1}$ and $\bisF \to \bisF' = \bisF\, \Omega^{-1}$\,. 
We here find an explicit form of $\Omega$ associated with $\beta$-transformations [the final result is obtained in \eqref{eq:Omega-general}]. 

\subsubsection{Gauge fixing for dual fields}

Let us first specify the dual vielbein $\tilde{e}_m{}^{\Loa}$ explicitly. 
As we can see from \eqref{eq:double-vielbein-transf}, under $\beta$-transformations, we have the following transformation rules:\footnote{The transformation rule of $e_m{}^{\Loa}$ given \eqref{eq:beta-rule-NS} make sense only when $(E^{-\rmT})^{mn}$ is not singular. When $(E^{-\rmT})^{mn}$ is singular, we should express it as $e_m{}^{\Loa} \to e'_m{}^{\Loa}=\bigl[\bigl(\bm{1} - E^\rmT\, \bmr\bigr)^{-1}\bigr]_m{}^n\, e_n{}^{\Loa}$\,. When both $E_{mn}$ and $E^{mn}$ are singular, we should choose another parameterization of the double vielbein, although we do not consider such cases in this paper.}
\begin{align}
\begin{split}
 e_m{}^{\Loa} &\ \to\ e'_m{}^{\Loa}=\bigl[\bigl(E^{-\rmT} - \bmr\bigr)^{-1}\,E^{-\rmT}\bigr]_m{}^n\, e_n{}^{\Loa}\,, 
\\
 \tilde{e}_m{}^{\Loa} &\ \to\ \tilde{e}'_m{}^{\Loa}=\tilde{e}_m{}^{\Loa} \,, \qquad
 E^{mn}\ \to\ E'^{mn} = E^{mn} + \bmr^{mn} \,. 
\end{split}
\label{eq:beta-rule-NS}
\end{align}
Then, we can consistently relate $e_m{}^{\Loa}$ and $\tilde{e}_m{}^{\Loa}$ as
\begin{align}
 \tilde{e}_{m}{}^{\Loa} = E_{mn}\,e^n{}_{\Lob}\,\eta^{\Lob\Loa} \,. 
\label{eq:tilde-e-gauge}
\end{align}
This is equivalent to a direct identification of two parameterizations,
\begin{align}
 \frac{1}{\sqrt{2}} \begin{pmatrix} e^m{}_{\Lob}\,\eta^{\Lob\Loa} \\ (\CG+B)_{mn}\,e^n{}_{\Lob}\,\eta^{\Lob\Loa} \end{pmatrix} = \GV^{M\Loa}= \frac{1}{\sqrt{2}} \begin{pmatrix} (\OG^{-1}-\beta)^{mn}\, \tilde{e}_n{}^{\Loa} \\ \tilde{e}_m{}^{\Loa} \end{pmatrix} \,, 
\end{align}
and consistent with the relation \eqref{eq:relation-open-closed}. 
If we introduce the flat components of $E^{mn}$ as
\begin{align}
 \Einv^{\Loa\Lob} \equiv \tilde{e}_m{}^{\Loa}\, \tilde{e}_n{}^{\Lob}\,E^{mn} \equiv \eta^{\Loa\Lob} - \beta^{\Loa\Lob} \,,
\label{eq:E-cEinv-cE}
\end{align}
we obtain
\begin{align}
 E_{mn} = \tilde{e}_m{}^{\Loa}\,\tilde{e}_n{}^{\Lob}\,(\Einv^{-1})_{\Loa\Lob} = e_m{}^{\Loa}\,e_n{}^{\Lob}\, (\Einv^{\rmT})_{\Loa\Lob}\,. 
\end{align}
Namely, we have simple expressions,
\begin{align}
\begin{alignedat}{2}
 \CG_{mn} &= e_{m}{}^{\Loa}\,e_{n}{}^{\Lob}\,\eta_{\Loa\Lob} \,,&\qquad 
 B_{mn} &= e_{m}{}^{\Loa}\,e_{n}{}^{\Lob}\,\beta_{\Loa\Lob}\,,
\\
 \OG_{mn} &= \tilde{e}_{m}{}^{\Loa}\,\tilde{e}_{n}{}^{\Lob}\,\eta_{\Loa\Lob} \,,&\qquad 
 \beta^{mn} &= \tilde{e}^m{}_{\Loa}\,\tilde{e}^n{}_{\Lob}\,\beta^{\Loa\Lob}\,. 
\end{alignedat}
\label{eq:g-G-B-beta}
\end{align}
In terms of $\Einv^{\Loa\Lob}$, the relation \eqref{eq:tilde-e-gauge} can also be expressed as
\begin{align}
 e_m{}^{\Loa} = \tilde{e}_{m}{}^{\Lob}\,(\Einv^{-\rmT})_{\Lob}{}^{\Loa}\,. 
\label{e-etilde-Einv}
\end{align}
From \eqref{e-etilde-Einv}, the relation \eqref{eq:DFT-dilaton} between the two dilatons, $\Phi$ and $\tilde{\phi}$, can be expressed as
\begin{align}
 \Exp{\Phi} = (\det\Einv_{\Loa}{}^{\Lob})^{-\frac{1}{2}} \Exp{\tilde{\phi}}\,. 
\label{eq:two-dilatons}
\end{align}

\subsubsection{Relation between untwisted R--R fields}

From \eqref{eq:R-R-diagram}, the relation between $(B,\,\Phi)$-untwisted R--R polyforms and the $(\beta,\,\tilde{\phi})$-untwisted R--R polyforms can be expressed as
\begin{align}
\begin{alignedat}{2}
 \check{\cF} &= \Exp{\tilde{\phi}-\Phi}\Exp{\beta\vee} \Exp{B_2\wedge} \hat{\cF} \,, \qquad& 
 \hat{\cF} &= \Exp{\Phi-\tilde{\phi}}\Exp{-B_2\wedge} \Exp{-\beta\vee} \check{\cF} \,, \\
 \check{\cC} &= \Exp{\tilde{\phi}-\Phi}\Exp{\beta\vee} \Exp{B_2\wedge} \hat{\cC} \,, \qquad& 
 \hat{\cC} &= \Exp{\Phi-\tilde{\phi}}\Exp{-B_2\wedge} \Exp{-\beta\vee} \check{\cC} \,. 
\end{alignedat}
\label{eq:R-R-relation1}
\end{align}
As we show in Appendix \ref{app:formula-proof} by a brute force calculation, if rephrased in terms of bispinors, these relations have quite simple forms
\begin{align}
\begin{alignedat}{2}
 &\bisF = \check{\bm{\cF}}\,\Omega_0^{-1}\,,\qquad 
 \bisC = \check{\bm{\cC}}\,\Omega_0^{-1}\,,\qquad &
 &\check{\bm{\cF}} = \bisF \,\Omega_0\,,\qquad 
 \check{\bm{\cC}} = \bisC \,\Omega_0\,, 
\label{eq:R-R-relation2}
\\
 &\Omega_0^{-1} =(\det \Einv_{\Loc}{}^{\Lod})^{-\frac{1}{2}} \text{\AE}\bigl(-\tfrac{1}{2}\,\beta^{\Loa\Lob}\,\Gamma_{\Loa\Lob}\bigr)\,,\qquad&
 &\Omega_0 =(\det \Einv_{\Loc}{}^{\Lod})^{-\frac{1}{2}} \text{\AE}\bigl(\tfrac{1}{2}\,\beta^{\Loa\Lob}\,\Gamma_{\Loa\Lob}\bigr)\,, 
\end{alignedat}
\end{align}
where $\text{\AE}$ is an exponential-like function with the gamma matrices totally antisymmetrized \cite{Hassan:1999mm}
\begin{align}
 \text{\AE}\bigl(\tfrac{1}{2}\,\beta^{\Loa\Lob}\,\Gamma_{\Loa\Lob}\bigr) \equiv \sum^5_{p=0}\frac{1}{2^{p}\,p!}\, \beta_{\Loa_1\Loa_2}\cdots\beta_{\Loa_{2p-1}\Loa_{2p}}\,\Gamma^{\Loa_1\cdots \Loa_{2p}}\,. 
\end{align}
In fact, this $\Omega_0$ is a spinor representation of a local Lorentz transformation,\footnote{Note that $\brGamma^{\Loa}=\Gamma^{11}\,\Gamma^{\Loa}$ also satisfies the same relation, $\Omega_0^{-1} \,\brGamma^{\Loa}\,\Omega_0 = \bigl(\Einv^{-1}\, \Einv^\rmT\bigr)^{\Loa}{}_{\Lob}\, \brGamma^{\Lob}$\,.}
\begin{align}
 \Omega_0^{-1} \,\Gamma^{\Loa}\,\Omega_0 = \bigl(\Einv^{-1}\, \Einv^\rmT\bigr)^{\Loa}{}_{\Lob}\, \Gamma^{\Lob} \,, 
\end{align}
as we can show by employing the formula provided below \cite{Hassan:1999mm} (see Appendix \ref{app:omega} for a proof). 
In this sense, the $(B,\,\Phi)$-untwisted fields and the $(\beta,\,\tilde{\phi})$-untwisted fields are related by a local Lorentz transformation. 

\noindent\textbf{\underline{Formula:}} For an arbitrary antisymmetric matrix $a_{\Loa\Lob}$, the spinor representation of a local Lorentz transformation
\begin{align}
 \Lambda^{\Loa}{}_{\Lob} \equiv \bigl[(\eta+a)^{-1}\,(\eta-a)\bigr]^{\Loa}{}_{\Lob} = \bigl[(\eta-a)\,(\eta+a)^{-1}\bigr]^{\Loa}{}_{\Lob}\ \in \OO(1,D-1)\,,
\end{align}
is given by
\begin{align}
\begin{split}
 \Omega_{(a)} &=\bigl[\det (\delta_{\Loc}^{\Lod}\pm a_{\Loc}{}^{\Lod})\bigr]^{-\frac{1}{2}} \text{\AE}\bigl(-\tfrac{1}{2}\,a_{\Loa\Lob}\,\Gamma^{\Loa\Lob}\bigr)\,,
\\
 \Omega_{(a)}^{-1} &=\bigl[\det (\delta_{\Loc}^{\Lod}\pm a_{\Loc}{}^{\Lod})\bigr]^{-\frac{1}{2}} \text{\AE}\bigl(\tfrac{1}{2}\,a_{\Loa\Lob}\,\Gamma^{\Loa\Lob}\bigr) \,, \qquad 
 \Omega_{(a)}^{-1} \,\Gamma^{\Loa}\,\Omega_{(a)} = \Lambda^{\Loa}{}_{\Lob}\, \Gamma^{\Lob} \,. 
\end{split}
\label{eq:Hassan-formula}
\end{align}

\subsubsection{General formula for \texorpdfstring{$\Omega$}{\textOmega}}

Now, let us find the explicit form of $\Omega$ for $\beta$-transformations [recall \eqref{eq:Omega-def}], satisfying
\begin{align}
 \Omega^{-1}\,\brGamma^{\Loa}\,\Omega = \Lambda^{\Loa}{}_{\Lob}\, \brGamma^{\Lob} \qquad 
 \bigl[\Lambda=e^{\rmT}\,(1+ \bmr\,E)^{-1}\,(1- \bmr\,E^\rmT)\,e^{-\rmT}\bigr]\,.
\end{align}
A key observation is that by using \eqref{eq:tilde-e-gauge}, \eqref{eq:E-cEinv-cE}, and \eqref{e-etilde-Einv}, $\Lambda^{\Loa}{}_{\Lob}$ can be decomposed into a product of two Lorentz transformations,
\begin{align}
\Lambda 
 = \Lambda'\, \Lambda^{-1}_{0} \,,\qquad
\Lambda' \equiv \eta^{-1}\,\Einv'^{-1}\, \Einv'^\rmT\,\eta \,,\qquad 
  \Lambda_{0}\equiv \eta^{-1}\,\Einv^{-1}\, \Einv^\rmT\,\eta\,,
\label{eq:Lambda-decomp}
\end{align}
where $\Einv'$ is defined by
\begin{align}
 \Einv'^{\Loa\Lob} \equiv \bigl[\tilde{e}^\rmT\,(E^{-1}+\bmr)\,\tilde{e}\bigr]^{\Loa\Lob} \equiv \eta^{\Loa\Lob} - \beta'^{\Loa\Lob} \qquad 
 \bigl(\beta'^{\Loa\Lob}\equiv \beta^{\Loa\Lob}- \bmr^{mn}\,\tilde{e}_m{}^{\Loa}\,\tilde{e}_n{}^{\Lob} \bigr)\,.
\end{align}
Then, we can check the following relations associated with $\Einv'^{\Loa\Lob}$:
\begin{align}
\begin{alignedat}{2}
 \Einv'^{\Loa\Lob}&= \tilde{e}_m{}^{\Loa}\,\tilde{e}_n{}^{\Lob}\, E'^{mn}
  = e'^{\Loa m}\,e'^{\Lob n}\, E'^\rmT_{mn} \,,&\qquad 
 E'^{mn}&=E^{mn}+\bmr^{mn} \equiv \OG'^{mn} - \beta'^{mn}\,,
\\
 e'_m{}^{\Loa} &\equiv E'^\rmT_{mn}\,\tilde{e}^{n\Loa} = (\Einv'^{-\rmT})_{\Lob}{}^{\Loa}\,\tilde{e}_{m}{}^{\Lob}\,,&\qquad 
 E'_{mn} &= \bigl[(E^{-1}+\bmr)^{-1}\bigr]_{mn} \equiv \CG'_{mn} + B'_{mn} \,,
\end{alignedat}
\end{align}
where $\CG'_{mn}$ and $B'_{mn}$ are the $\beta$-transformed metric and $B$-field, respectively. 
From the invariance of $d$, $\tilde{e}_m{}^{\Loa}$, and $\tilde{\phi}$ under $\beta$-transformations, the dilaton $\Phi$ in the $\beta$-transformed background becomes
\begin{align}
 \Exp{\Phi'} = (\det\Einv'_{\Loa}{}^{\Lob})^{-\frac{1}{2}} \Exp{\tilde{\phi}} 
 = \frac{(\det\Einv'_{\Loa}{}^{\Lob})^{-\frac{1}{2}}}{(\det\Einv_{\Loc}{}^{\Lod})^{-\frac{1}{2}}} \Exp{\Phi} \,. 
\label{eq:Phi-beta}
\end{align}

Corresponding to the decomposition \eqref{eq:Lambda-decomp}, we can also decompose $\Omega$ as
\begin{align}
\begin{split}
 \Omega &= \Omega' \,\Omega_{0}^{-1} = \bigl[\det (\Einv'\,\Einv)_{\Loe}{}^{\Lof}\bigr]^{-\frac{1}{2}} \, \text{\AE}\bigl(\tfrac{1}{2}\,\beta'^{\Loa\Lob}\,\Gamma_{\Loa\Lob}\bigr)\,\text{\AE}\bigl(-\tfrac{1}{2}\,\beta^{\Loc\Lod}\,\Gamma_{\Loc\Lod}\bigr)\,, 
\\
 \Omega^{-1} &= \Omega_0 \,\Omega'^{-1} = \bigl[\det (\Einv'\,\Einv)_{\Loe}{}^{\Lof}\bigr]^{-\frac{1}{2}} \,\text{\AE}\bigl(\tfrac{1}{2}\,\beta^{\Loa\Lob}\,\Gamma_{\Loa\Lob}\bigr)\, \text{\AE}\bigl(-\tfrac{1}{2}\,\beta'^{\Loc\Lod}\,\Gamma_{\Loc\Lod}\bigr)\,. 
\end{split}
\label{eq:Omega-general}
\end{align}
where we have defined
\begin{align}
 \Omega' \equiv (\det \Einv'_{\Loc}{}^{\Lod})^{-\frac{1}{2}} \text{\AE}\bigl(\tfrac{1}{2}\,\beta'^{\Loa\Lob}\,\Gamma_{\Loa\Lob}\bigr)\,,\qquad 
 \Omega'^{-1} = (\det \Einv'_{\Loc}{}^{\Lod})^{-\frac{1}{2}} \text{\AE}\bigl(-\tfrac{1}{2}\,\beta'^{\Loa\Lob}\,\Gamma_{\Loa\Lob}\bigr)\,.
\end{align}
This gives the desired local Lorentz transformation,
\begin{align}
 \Omega^{-1} \,\brGamma^{\Loa}\,\Omega = \Omega_{0}\,\Omega'^{-1} \,\brGamma^{\Loa}\,\Omega'\,\Omega_{0}^{-1} 
 = (\Lambda'\,\Lambda^{-1}_{0})^{\Loa}{}_{\Lob}\, \brGamma^{\Lob} = \Lambda^{\Loa}{}_{\Lob}\, \brGamma^{\Lob} \,. 
\end{align}
The $\beta$-transformed R--R field is then expressed as
\begin{align}
 \bisF' = \bisF\, \Omega^{-1} \,. 
\label{eq:beta-F-transf}
\end{align}
In terms of the differential form, we can express the same transformation rule as
\begin{align}
 \hat{\cF}' = \Exp{\Phi'-\Phi}\Exp{-B'_2\wedge} \Exp{\bmr\vee} \Exp{B_2\wedge} \hat{\cF} \qquad
 \bigl(\hat{\cF}'_{m_1\cdots m_n} \equiv e'_{m_1}{}^{\Loa_1}\cdots e'_{m_n}{}^{\Loa_n}\,\hat{\cF}'_{\Loa_1\cdots \Loa_n}\bigr)\,. 
\end{align}
In terms of the $B$-untwisted field $\hat{F}$, the $\beta$-untwisted field $\check{F}$, and the $(\beta,\,\tilde{\phi})$-untwisted field $\check{\cF}$, we can express the above formula as
\begin{align}
 \hat{F}' = \Exp{-B'_2\wedge} \Exp{\bmr\vee} \Exp{B_2\wedge} \hat{F} \,,\qquad 
 \check{F}' = \check{F}\,,\qquad \check{\cF}' = \check{\cF} \,. 
\end{align}
Namely, the $\beta$- or $(\beta,\,\tilde{\phi})$-untwisted field is invariant under $\beta$-transformations, which has been shown in \cite{Sakamoto:2017cpu} (see also \cite{Sakamoto:2017wor}) by treating the R--R fields, $A$ and $F$, as $\OO(D,D)$ spinors. 

Specifically, if the $B$-field and the dilaton $\Phi$ are absent before the $\beta$-transformation, we have $\beta^{\Loa\Lob}=0$\,, $\Einv_{\Loa}{}^{\Lob}=\delta_{\Loa}^{\Lob}$\,, and $\beta'^{\Loa\Lob}= - \bmr^{mn}\,\tilde{e}_m{}^{\Loa}\,\tilde{e}_n{}^{\Lob}$. 
Then, \eqref{eq:Omega-general} becomes
\begin{align}
 \Omega = (\det\Einv'_{\Loc}{}^{\Lod})^{-\frac{1}{2}}\,\text{\AE}\bigl(-\tfrac{1}{2}\,\bmr^{\Loa\Lob}\,\Gamma_{\Loa\Lob}\bigr) \qquad 
 (\bmr^{\Loa\Lob}\equiv \bmr^{mn}\,\tilde{e}_m{}^{\Loa}\,\tilde{e}_n{}^{\Lob}) \,. 
\end{align}
In section \ref{sec:YB-AdS5xS5}, we see that this $\Omega$ plays an important role in YB deformations of $\AdS{5}\times\rmS^5$ superstring [see Eq.~\eqref{eq:omega} where $2\,\eta\,\lambda^{\Loa\Lob}$ plays the same role as $\bmr^{\Loa\Lob}$ here]. 

\vspace{-\Pskip}
\subsection{\texorpdfstring{$T$}{T}-duality-invariant Green--Schwarz action}

In section \ref{sec:YB-AdS5xS5}, we study homogeneous YB deformations of the GS type IIB superstring action and show that YB deformations are equivalent to $\beta$-deformations of the target space. 
In order to show the equivalence, it will be useful to manifest the covariance of the GS superstring theory under $\beta$-transformations. 
In this section, we provide a manifestly $\OO(10,10)$ $T$-duality-covariant formulation of the GS type II superstring theory. 

A manifestly $T$-duality covariant formulations of string theory, the so-called double sigma model (DSM), has been developed in \cite{Duff:1989tf,Tseytlin:1990nb,Tseytlin:1990va,Hull:2004in,Hull:2006va,Copland:2011wx,Lee:2013hma} for the bosonic string. 
More recently, the DSM for the GS type II superstring theory was formulated in \cite{Park:2016sbw} (see also \cite{Hull:2006va,Blair:2013noa,Bandos:2015cha,Driezen:2016tnz,Bandos:2016jez} for other approaches to supersymmetric DSMs). 
The action by Park, in our convention, is given by
\begin{align}
 S &= \frac{1}{4\pi\alpha'}\int \Bigl[\,\frac{1}{2}\, \cH_{MN}\,\Pi^M\wedge *_{\gga} \Pi^N - D X^M\wedge \bigl(\cA_{M} + \Sigma_{M} \bigr)\Bigr]
\nn\\
 &= -\frac{1}{4\pi\alpha'}\int \sqrt{-\gga}\,\rmd^2\sigma\,
 \Bigl[\,\frac{1}{2}\,\gga^{\WSa\WSb}\,\cH_{MN}\,\Pi_{\WSa}^M\,\Pi_{\WSb}^N
  + \varepsilon^{\WSa\WSb}\, D_{\WSa}X^M\,\bigl(\cA_{\WSb M} + \Sigma_{\WSb M} \bigr)\Bigr] \,,
\label{eq:GS-DSM-Park}
\end{align}
where $\gga_{\WSa\WSb}$ is the intrinsic metric on the string worldsheet and
\begin{align}
\begin{split}
 &\Pi^M \equiv DX^M + \Sigma^M \,,\qquad DX^M \equiv \rmd X^M-\cA^M \,,\qquad 
 \varepsilon^{01}\equiv \frac{1}{\sqrt{-\gga}}\,,
\\
 &(X^M)\equiv \begin{pmatrix} X^m \\ \tilde{X}_m \end{pmatrix} \,,\qquad 
 \Sigma^M \equiv \begin{pmatrix} \Sigma^m \\ \tilde{\Sigma}_m \end{pmatrix} \equiv \frac{\ii}{\sqrt{2}}\,\bigl(\brTheta_1\,\Gamma^M\,\rmd \Theta_1 + \brbrTheta_2\,\brGamma^M\,\rmd \Theta_2\bigr) \,,
\end{split}
\end{align}
and a worldsheet 1-form $\cA^M(\sigma)$ is defined to satisfy,
\begin{align}
 \cA^M \, \partial_M T = 0\,,\qquad \cA^M\,\cA_M = 0\,,
\label{eq:cA-condition}
\end{align}
for arbitrary supergravity fields or gauge parameters $T(x)$\,. 
Here, the Dirac conjugates for the spacetime fermions $\Theta_1^{\SPa}$ and $\Theta_2^{\SPbra}$ are defined respectively as
\begin{align}
 \brTheta_1 \equiv \Theta_1^\dagger\,\Gamma^0\,,\qquad 
 \brbrTheta_2 \equiv -\Theta_2^\dagger\,\brGamma^0\,,
\end{align}
which indeed transform as
\begin{align}
 \brTheta_1 \ \to \ \brTheta_1 \, \Exp{-\frac{1}{4}\,\omega_{\Loa\Lob}\,\Gamma^{\Loa\Lob}}\,,\qquad
 \brbrTheta_2 \ \to \ \brbrTheta_2 \, \Exp{-\frac{1}{4}\,\bromega_{\Lobra\Lobrb}\,\brGamma^{\Lobra\Lobrb}} \,,
\end{align}
under a double Lorentz transformation $\Theta_1\to \Exp{\frac{1}{4}\,\omega_{\Loa\Lob}\,\Gamma^{\Loa\Lob}} \Theta_1$ and $\Theta_2\to \Exp{-\frac{1}{4}\,\bromega_{\Lobra\Lobrb}\,\brGamma^{\Lobra\Lobrb}} \Theta_2$\,. 
The Majorana--Weyl conditions are defined as\footnote{The non-standard factor $-\brGamma^{11}$ is introduced in the Majorana condition for $\Theta_2$ such that the condition becomes the standard Majorana condition after the diagonal gauge fixing; $\Theta_2 = C\, (\Gamma^0)^\rmT\,\Theta_2^*$\,.}
\begin{align}
\begin{alignedat}{2}
 &\Theta_1 = C\,(\Gamma^0)^\rmT\,\Theta_1^*\,,\qquad& 
 &\Theta_2 = -\brGamma^{11}\,\brC\,(\brGamma^0)^\rmT\,\Theta_2^* \,, 
\\
 &\Gamma^{11}\,\Theta_1 = \Theta_1\,,\qquad& 
 &\brGamma^{11}\,\Theta_2 = \pm \Theta_2\qquad (\text{IIA/IIB})\,,
\end{alignedat}
\end{align}
and then we obtain
\begin{align}
 \brTheta_1 = \Theta_1^\dagger\,\Gamma^0 = \Theta_1^\rmT\,C \,,\qquad 
 \brbrTheta_2 = -\Theta_2^\dagger\,\brGamma^0 = -\Theta_2^\rmT\,\brC\, \brGamma^{11} \,. 
\end{align}
In \cite{Park:2016sbw}, the target space was assumed to be flat, but here we generalize the action to arbitrary curved backgrounds. 

In order to consider the superstring action in the presence of fluxes, such as the $H$-flux and the R--R fluxes, we introduce generalized tensors,
\begin{align}
\begin{split}
 \cK_{MN}^{(1)}&\equiv -\frac{\ii}{\sqrt{2}}\,\GV_{(M}{}^{\Loa}\,\brGV_{N)}{}^{\Lobrb}\, \brTheta_1\,\Gamma_{\Loa}\,\Gamma^{\Loc\Lod}\,\Theta_1 \, \Phi_{\Lobrb\Loc\Lod} \,,
\\
 \cK_{MN}^{(2)}&\equiv -\frac{\ii}{\sqrt{2}}\,\brGV_{(M}{}^{\Lobra}\,\GV_{N)}{}^{\Lob}\, \brbrTheta_2\,\brGamma_{\Lobra}\,\brGamma^{\Lobrc\Lobrd}\,\Theta_2 \, \brPhi_{\Lob\Lobrc\Lobrd} \,,
\\
 \cK^{\text{\tiny(RR)}}_{MN} &\equiv \frac{\ii}{4}\,\GV_{(M}{}^{\Loa}\,\brGV_{N)}{}^{\Lobrb}\,\brTheta_{1}\, \Gamma_{\Loa}\, \bisF \, \brGamma_{\Lobrb} \Theta_{2} \,. 
\end{split}
\end{align}
Then, we add the following term to the DSM action \eqref{eq:GS-DSM-Park}:
\begin{align}
 \Delta S \equiv \frac{1}{8\pi\alpha'}\int \cK_{MN}\,\Pi^M\wedge *_{\gga} \Pi^N\,, \qquad 
 \cK_{MN}\equiv \cK_{MN}^{(1)}+\cK_{MN}^{(2)}+\cK_{MN}^{\text{\tiny(RR)}}\,. 
\end{align}
By choosing the diagonal gauge, the explicit form of $\cK_{MN}$ becomes
\begin{align}
 \cK_{MN} &= \begin{pmatrix}
  -(\CG\,\kappa^{\text{s}}\,\CG+B\,\kappa^{\text{s}}\,B+B\,\kappa^{\text{a}}\,\CG+\CG\,\kappa^{\text{a}}\,B)_{mn} & (B\,\kappa^{\text{s}} +\CG\,\kappa^{\text{a}})_m{}^n \\ - (\kappa^{\text{s}}\,B+\kappa^{\text{a}}\,\CG)^m{}_n & (\kappa^{\text{s}})^{mn}
 \end{pmatrix}
 \,,
\\
 \kappa_{mn} &\equiv -\frac{\ii}{4}\,\Bigl(\sqrt{2}\,\brTheta_1\,\Gamma_{m}\,\Gamma^{\Loa\Lob}\,\Theta_1\,\Phi_{n\Loa\Lob} 
 + \sqrt{2}\,\brbrTheta_2\,\brGamma_{n}\,\brGamma^{\Lobra\Lobrb}\,\Theta_2 \, \brPhi_{m\Lobra\Lobrb}
 - \frac{1}{2}\,\brTheta_{1}\, \Gamma_m\, \bisF \, \brGamma_n \Theta_{2}\Bigr)\,,
\end{align}
where we defined $\kappa^{\text{s}}_{mn}\equiv \kappa_{(mn)}$ and $\kappa^{\text{a}}_{mn}\equiv \kappa_{[mn]}$ and their indices are raised or lowered with the metric $\CG_{mn}$\,. 
Note that $\cK_{MN}$ is an $\OO(10,10)$ matrix up to quadratic order in $\Theta_I$ ($I=1,2$). 

The modification of the DSM action, $S\to S+\Delta S$, is equivalent to the replacement of the generalized metric
\begin{align}
 \cH_{MN} \ \to\ \cM_{MN}\equiv \cH_{MN} + \cK_{MN} \,. 
\label{eq:replacement}
\end{align}
The explicit form of $\cM_{MN}$ is given by
\begin{align}
 (\cM_{MN}) &= \begin{pmatrix}
  \delta_m^p & B_{mp} \\ 0 & \delta^m_p 
 \end{pmatrix}
 \begin{pmatrix}
  \CG_{pq}-\kappa^{\text{s}}_{pq} & (\kappa^{\text{a}})_{p}{}^q \\ -(\kappa^{\text{a}})^p{}_q & \CG^{pq}+(\kappa^{\text{s}})^{pq}
 \end{pmatrix}
 \begin{pmatrix}
  \delta^q_n & 0 \\ -B_{qn} & \delta_q^n
 \end{pmatrix} 
\nn\\
 &= \begin{pmatrix}
  \delta_m^p & \hat{B}_{mp} \\ 0 & \delta^m_p 
 \end{pmatrix}
 \begin{pmatrix}
  \hat{\CG}_{pq} & 0 \\ 0 & (\hat{\CG}^{-1})^{pq} 
 \end{pmatrix}
 \begin{pmatrix}
  \delta^q_n & 0 \\ -\hat{B}_{qn} & \delta_q^n
 \end{pmatrix} + \cO(\Theta^4)\,,
\end{align}
where we defined
\begin{align}
 \hat{\CG}_{mn}\equiv \CG_{mn}-\kappa^{\text{s}}_{mn}\,,\qquad 
 \hat{B}_{mn}\equiv B_{mn}+\kappa^{\text{a}}_{mn} \,. 
\end{align}
Then, we consider an action
\begin{align}
 S = \frac{1}{4\pi\alpha'}\int \Bigl[\,\frac{1}{2}\, \cM_{MN}\,\Pi^M\wedge *_{\gga} \Pi^N - D X^M\wedge \bigl(\cA_{M} + \Sigma_{M} \bigr)\Bigr]\,.
\label{eq:GS-DSM-ours}
\end{align}

Before we choose the diagonal gauge, spinors $\Theta_I$, R--R fields $\cF^{\SPa}{}_{\SPbrb}$, and the spin connections $\Phi_{\Lobra\Lob\Loc}$, $\brPhi_{\Loa\Lobrb\Lobrc}$ are invariant under global $\OO(10,10)$ transformations or (finite) generalized diffeomorphisms, while $\cH_{MN}$, $\cK_{MN}$, $\Pi^M$, and $\Sigma^M$ transform covariantly and the action is invariant.%
\footnote{More precisely, as discussed in \cite{Lee:2013hma,Park:2016sbw}, $\cA^M$ does not transform covariantly because $\rmd X^M$ does not transform covariantly, and $D X^M\wedge \cA_{M}$ is not invariant under generalized diffeomorphisms. However, the variation is only the total-derivative term and the action is invariant under generalized diffeomorphisms.}
The global double Lorentz symmetry is manifest but the local one is not manifest because $\Sigma^M$ contains non-covariant quantity $\rmd\Theta_I$ (or $\cK_{MN}$ contains the spin connection). 
The local symmetry becomes manifest only after eliminating the auxiliary fields. 
On the other hand, if we choose the diagonal gauge fixing, although the global $\OO(10,10)$ transformations are manifest, the covariance under generalized diffeomorphisms are lost, because the barred indices are rotated under the compensating local Lorentz transformation. 
Indeed, the transformation rule of fermionic fields after the diagonal gauge fixing is
\begin{align}
 \Theta_1 \ \to \ \Theta_1 \,,\qquad 
 \Theta_2 \ \to \ \Omega\, \Theta_2 \,, 
\label{eq:Theta-transf}
\end{align}
where $\Omega$ is the one given in \eqref{eq:Omega-def}, and in general, it is non-constant. 
Accordingly, $\rmd\Theta_2$ (and thus $\Sigma^M$ also) does not transform covariantly. 

It is interesting to note that all information on the curved background is contained in the generalized metric $\cM_{MN}$\,. 
The usual generalized metric $\cH_{MN}$ contains only the $P$-$P$ or $\brP$-$\brP$ components [see \eqref{eq:H-eta-orthogonal}] while other quantities such as the R--R fluxes are contained in the $P$-$\brP$ or $\brP$-$P$ components $\cK_{MN}$\,. 

In the following, we show that the action \eqref{eq:GS-DSM-ours} reproduces the conventional GS superstring action \cite{Cvetic:1999zs} up to quadratic order in fermions $\Theta_I$. 

\subsubsection{Classical equivalence to the type II GS action}

In order to reproduce the conventional action, we choose the canonical section $\tilde{\partial}^m =0$. 
Then, the condition \eqref{eq:cA-condition} for $\cA^M$ indicates that $\cA^M$ takes the form $(\cA^M)=(0,\,A_m)$ and $DX^M$ becomes
\begin{align}
 DX^M = \begin{pmatrix} \rmd X^m \\ \rmd \tilde{X}_m -A_m \end{pmatrix} \equiv \begin{pmatrix} \rmd X^m \\ P_m \end{pmatrix} \,,
\end{align}
where, for simplicity, we defined $P_m$ and treated it as a fundamental variable rather than $A_m$\,. 
The action then becomes
\begin{align}
 S &= \frac{1}{4\pi\alpha'}\int \Bigl[\,\frac{1}{2}\, \cM_{MN}\,\Pi^M\wedge *_{\gga} \Pi^N - P_{m}\wedge \bigl(\rmd X^m+\Sigma^{m}\bigr)\Bigr]
\nn\\
 &\quad - \frac{1}{4\pi\alpha'}\int \bigl(\rmd X^m\wedge \tilde{\Sigma}_m + \rmd X^m\wedge \rmd \tilde{X}_m \bigr) \,.
\end{align}
We can expand the first line as
\begin{align}
 &\frac{1}{2}\, \cM_{MN}\,\Pi^M\wedge *_{\gga} \Pi^N - P_{m}\wedge \bigl(\rmd X^m+\Sigma^{m}\bigr)
\nn\\
 &=\frac{1}{2}\, \hat{\CG}_{mn}\,\bigl(\rmd X^m+\Sigma^{m}\bigr)\wedge *_{\gga} \bigl(\rmd X^n+\Sigma^n\bigr)
\nn\\
 &\quad +\frac{1}{2}\, \hat{\CG}^{mn}\,\bigl[P_m+\tilde{\Sigma}_{m} -\hat{B}_{mp}\,(\rmd X^p+\Sigma^{p})\bigr]\wedge *_{\gga} \bigl[P_n+\tilde{\Sigma}_{n} -\hat{B}_{nq}\,(\rmd X^q+\Sigma^{q})\bigr]
\nn\\
 &\quad - P_{m}\wedge \bigl(\rmd X^m+\Sigma^{m}\bigr)
\nn\\
 &= \hat{\CG}_{mn}\,\bigl(\rmd X^m+\Sigma^{m}\bigr)\wedge *_{\gga} \bigl(\rmd X^n+\Sigma^n\bigr)
  + \bigl[\tilde{\Sigma}_n + \hat{B}_{mn}\, (\rmd X^m+\Sigma^{m})\bigr]\wedge \bigl(\rmd X^n+\Sigma^n\bigr) 
\nn\\
 &\quad +\frac{1}{2}\, \hat{\CG}^{mn}\,\bigl[P_m+\tilde{\Sigma}_{m} -\hat{B}_{mp}\,(\rmd X^p+\Sigma^{p})-\hat{\CG}_{mp}\,*_{\gga} (\rmd X^p+\Sigma^{p})\bigr]
\nn\\
 &\quad\qquad\qquad \wedge *_{\gga} \bigl[P_n+\tilde{\Sigma}_{n} -\hat{B}_{nq}\,(\rmd X^q+\Sigma^{q}) -\hat{\CG}_{nq}\,*_{\gga} (\rmd X^q+\Sigma^{q})\bigr]\,,
\end{align}
and eliminating the auxiliary fields $P_m$, we obtain
\begin{align}
 S &= \frac{1}{4\pi\alpha'}\int 
  \bigl[\,\hat{\CG}_{mn}\,\bigl(\rmd X^m+\Sigma^{m}\bigr)\wedge *_{\gga} \bigl(\rmd X^n+\Sigma^n\bigr)
  + \hat{B}_{mn}\,\bigl(\rmd X^m+\Sigma^{m}\bigr)\wedge \bigl(\rmd X^n+\Sigma^n\bigr) 
\nn\\
 &\qquad\qquad\quad - 2\, \rmd X^m\wedge \tilde{\Sigma}_{m}- \Sigma^{m} \wedge \tilde{\Sigma}_{m} - \rmd X^m\wedge \rmd \tilde{X}_m \bigr] \,.
\end{align}
By using the explicit expression for $\Sigma^M$,
\begin{align}
 \Sigma^M = \begin{pmatrix} \Sigma^m \\ 
 \tilde{\Sigma}_m 
\end{pmatrix}
 \equiv \begin{pmatrix} \Sigma^m \\ 
 \hat{\Sigma}_m + B_{mn}\, \Sigma^n
\end{pmatrix}\,,\quad
 \begin{pmatrix} \Sigma^m \\ \hat{\Sigma}_m \end{pmatrix}
 \equiv \frac{\ii}{2} \begin{pmatrix} \brTheta_1\,\Gamma^m\,\rmd \Theta_1 + \brbrTheta_2\,\brGamma^m\,\rmd \Theta_2 \\
 \brTheta_1\,\Gamma_m\,\rmd \Theta_1 - \brbrTheta_2\,\brGamma_m\,\rmd \Theta_2 \end{pmatrix} \,,
\end{align}
and neglecting quartic terms in $\Theta$ and the topological term, the action becomes
\begin{align}
\begin{split}
 S &= \frac{1}{2\pi\alpha'}\int \rmd^2\sigma \sqrt{-\gga}\,\cL \,,
\\
 \cL&=-\frac{1}{2}\,\bigl({\gga}^{\WSa\WSb}- \varepsilon^{\WSa\WSb}\bigr)\,(\hat{\CG}_{mn}+ \hat{B}_{mn})\, \partial_{\WSa}X^m \, \partial_{\WSb}X^n 
 - \CG_{mn}\, \partial_{\WSa}X^m \, \bigl(\gga^{\WSa\WSb}\,\Sigma_{\WSb}^n + \varepsilon^{\WSa\WSb}\, \hat{\Sigma}_{\WSb}^n \bigr) \,. 
\end{split}
\end{align}

In order to compare the obtained action with the conventional GS superstring action, let us further expand the Lagrangian as
\begin{align}
\begin{split}
 \cL &=- \Pg_{-}^{\WSa\WSb}\,(\CG_{mn}+ B_{mn})\, \partial_{\WSa}X^m \, \partial_{\WSb}X^n 
\\
 &\quad - \ii\,\Pg_{+}^{\WSa\WSb}\, \partial_{\WSa}X^m \, \brTheta_1\,\Gamma_m\,\Bigl(\partial_{\WSb} \Theta_1 +\frac{1}{4}\, \partial_{\WSb}X^n\,\omega_{+n\Loa\Lob} \, \Gamma^{\Loa\Lob} \,\Theta_1\Bigr)
\\
 &\quad - \ii\,\Pg_{-}^{\WSa\WSb}\, \partial_{\WSa}X^m \, 
  \brTheta_2\,\Gamma_m\,\Bigl(\partial_{\WSb} \Theta_2 - \frac{1}{4} \, \partial_{\WSb}X^n\,\omega_{-n\Loa\Lob} \, \brGamma^{\Loa\Lob}\,\Theta_2\Bigr)
\\
 &\quad + \frac{\ii}{8} \,\Pg_{+}^{\WSa\WSb}\, \brTheta_{1}\, \Gamma_m\, \bisF \, \brGamma_n \Theta_{2} \, \partial_{\WSa}X^m\,\partial_{\WSb}X^n \,. 
\end{split}
\end{align}
where we have defined
\begin{align}
 \Pg_{\pm\WSa\WSb}\equiv \frac{\gga_{\WSa\WSb}\pm \varepsilon_{\WSa\WSb}}{2}\,,\qquad 
 \brTheta_2\equiv \Theta_2^\dagger\,\Gamma^0 \qquad \bigl(\brbrTheta_2\,\brGamma_m = \brTheta_2\,\Gamma_m\bigr)\,,
\end{align}
and used the explicit form of the spin connection \eqref{eq:spin-connections},
\begin{align}
 \Phi_{m\Loa\Lob} = \frac{1}{\sqrt{2}}\,\omega_{+m\Loa\Lob}\,,\qquad 
 \brPhi_{m\Lobra\Lobrb}= -\frac{1}{\sqrt{2}}\,\omega_{-m\Loa\Lob}\,, \qquad 
 \omega_{\pm m\Loa\Lob}\equiv \omega_{m\Loa\Lob} \pm \frac{1}{2}\, e_m{}^{\Loc}\,H_{\Loc\Loa\Lob}\,.
\label{eq:Phi-omega-pm}
\end{align}
Further using
\begin{align}
 \brGamma^{\Loa\Lob}=-\Gamma^{\Loa\Lob}\,,\qquad 
 \bisF \, \brGamma_m = \mp\,\bisF \, \Gamma_m \quad (\text{IIA/IIB})\,,
\end{align}
we obtain the type II superstring action
\begin{align}
\begin{split}
 \cL_{\text{IIA/IIB}} &= - \Pg_{-}^{\WSa\WSb}\, (\CG_{mn}+B_{mn})\, \partial_{\WSa}X^m\,\partial_{\WSb}X^n
\\ 
 &\quad - \ii\, \bigl(\Pg_{+}^{\WSa\WSb} \,\partial_{\WSa} X^m\, \brTheta_{1}\, \Gamma_m\, D_{+\WSb}\Theta_{1} 
  + \Pg_{-}^{\WSa\WSb} \,\partial_{\WSa} X^m\, \brTheta_{2}\, \Gamma_m\, D_{-\WSb}\Theta_{2} \bigr)
\\
 &\quad \mp \frac{\ii}{8}\, \Pg_{+}^{\WSa\WSb}\, \brTheta_{1}\, \Gamma_m\, \bisF \, \Gamma_n\, \Theta_{2} \, \partial_{\WSa} X^m\, \partial_{\WSb} X^n \,,
\end{split}
\label{eq:GS-action-rewriting}
\end{align}
where we defined
\begin{align}
 D_{\pm \WSa} \equiv \partial_{\WSa} + \frac{1}{4}\, \partial_{\WSa} X^m\, \omega_{\pm m}{}^{\Loa\Lob}\, \Gamma_{\Loa\Lob} \,.
\end{align}

For type IIA superstring, defining $\Theta\equiv \Theta_1+\Theta_2$\,, we obtain a simple action
\begin{align}
\begin{split}
 \cL_{\text{IIA}} &=- \Pg_{-}^{\WSa\WSb}\,(\CG_{mn}+ B_{mn})\, \partial_{\WSa}X^m \, \partial_{\WSb}X^n 
\\
 &\quad - \frac{\ii}{2}\, \gga^{\WSa\WSb} \,\partial_{\WSa} X^m\,\brTheta\,\Gamma_m\, \bm{D}_{\WSb}\Theta 
  - \frac{\ii}{2}\, \varepsilon^{\WSa\WSb}\,\partial_{\WSa} X^m\,\brTheta\, \Gamma^{11} \, \Gamma_m\,\bm{D}_{\WSb}\Theta \,, 
\end{split}
\end{align}
where we defined
\begin{align}
 \bm{D}_{\WSa} \equiv \partial_{\WSa} + \frac{1}{4}\, \partial_{\WSa} X^m\, \omega_m{}^{\Loa\Lob}\, \Gamma_{\Loa\Lob} - \frac{1}{8}\,\partial_{\WSa} X^m\, H_m{}^{\Loa\Lob}\,\Gamma_{\Loa\Lob}\, \Gamma^{11}
 + \frac{1}{16} \, \partial_{\WSa}X^m\, \bisF \, \Gamma_m \,. 
\end{align}
On the other hand, for type IIB superstring, using the Pauli matrices $\sigma_i^{IJ}$ ($i=1,2,3$), we can rewrite the action in a familiar form
\begin{align}
\begin{split}
 \cL_{\text{IIB}} &= - \Pg_{-}^{\WSa\WSb}\, (\CG_{mn}+B_{mn})\, \partial_{\WSa}X^m\,\partial_{\WSb}X^n
\\
 &\quad - \frac{\ii}{2}\, \bigl(\gga^{\WSa\WSb}\,\delta^{IK}+\varepsilon^{\WSa\WSb}\,\sigma_3^{IK}\bigr) \,\brTheta_{I}\, \partial_{\WSa} X^m\, \Gamma_m\, D^{KJ}_{\WSb}\Theta_{J} \,, 
\end{split}
\label{eq:GS-action-conventional}
\end{align}
where we used \eqref{eq:bTGGGT} and defined
\begin{align}
\begin{split}
 D^{IJ}_{\WSa} &\equiv \delta^{IJ}\, \Bigl(\partial_{\WSa} + \frac{1}{4}\, \partial_{\WSa} X^m\, \omega_m{}^{\Loa\Lob}\, \Gamma_{\Loa\Lob} \Bigr) + \frac{1}{8}\,\sigma_3^{IJ}\, \partial_{\WSa} X^m\, H_{m\Loa\Lob}\,\Gamma^{\Loa\Lob}
\\
 &\quad - \frac{1}{8}\, \Big(\epsilon^{IJ}\, \bisF_1 + \sigma_1^{IJ}\, \bisF_3 +\frac{1}{2}\,\epsilon^{IJ}\, \bisF_5 \Big)\, \partial_{\WSa} X^n\, \Gamma_n \,,
\\
 \bisF_p&\equiv \frac{1}{p!}\,\hat{\cF}_{\Loa_1\cdots\Loa_p}\,\Gamma^{\Loa_1\cdots\Loa_p}\,,\qquad (\epsilon^{IJ}) \equiv \footnotesize{\begin{pmatrix} 0 & \,1\, \\ -1 & 0 \end{pmatrix}}\,. 
\end{split}
\end{align}

As discussed around \eqref{eq:Psi-Tdual}, under a single $T$-duality along the $x^z$-direction, the fermionic variables transform as
\begin{align}
 \Theta_1 \ \to \ \Theta_1 \,,\qquad 
 \Theta_2 \ \to \ \frac{1}{\sqrt{\CG_{zz}}}\,\Gamma_z\, \Theta_2 \,. 
\end{align}
Since it flips the chirality of $\Theta_2$, it maps type IIA and IIB superstring to each other. 

\section{YB deformations of \texorpdfstring{$\AdS{5} \times \rmS^5$}{AdS\textfiveinferior{x}S\textfivesuperior} superstring}
\label{sec:YB-AdS5xS5}

In this section, we revisit the homogeneous YB deformations of the $\AdS{5} \times \rmS^5$ superstring. 
After a concise review of the supercoset construction of $\AdS{5}\times\rmS^5$ superstring, we show that the action of the YB sigma model can be expressed as the GS superstring action in the $\beta$-deformed $\AdS{5} \times \rmS^5$ background. 

\vspace{-\Pskip}
\subsection{A supercoset construction of \texorpdfstring{$\AdS{5} \times \rmS^5$}{AdS\textfiveinferior{x}S\textfivesuperior} superstring}

\subsubsection{Metsaev--Tseytlin action}

Before considering YB deformations, we review a supercoset construction of the $\AdS{5} \times \rmS^5$ superstring action based on the supercoset
\begin{align}
 \frac{\PSU(2,2|4)}{\SO(1,4)\times \SO(5)}\,.
\end{align}
In order to perform the supercoset construction, we introduce a coset representative $g\in\SU(2,2|4)$\footnote{In order to obtain a matrix representation of the $\PSU(2,2|4)$ group (which is necessary in Appendix \ref{app:kappa}), we consider the $\SU(2,2|4)$ group and project out the central generator $\gZ$ as explained in Appendix \ref{app:psu-algebra}.} and define the left-invariant current $A$ as
\begin{align}
 A=g^{-1}\,\rmd g \,,
\end{align}
which satisfies the Maurer--Cartan equation
\begin{align}
 \rmd A + A\wedge A = 0\,. 
\end{align}
By using the projections $P^{(i)}$ $(i=0,1,2,3)$ to the $\mathbb{Z}_4$-graded components of $\alg{g}\equiv\alg{su}(2,2|4)$ [see \eqref{eq:P-i-projector}], we decompose the left-invariant current as
\begin{align}
 A = A^{(0)} +A^{(1)} +A^{(2)} + A^{(3)} \,. 
\end{align}
We also define projection operators $d_\pm$ as
\begin{align}
 d_{\pm} \equiv \mp P^{(1)}+2\,P^{(2)}\pm P^{(3)} \,,
\label{eq:dpm}
\end{align}
which satisfy
\begin{align}
 \str \bigl[ X\,d_\pm(Y) \bigr] = \str \bigl[ d_\mp (X)\, Y \bigr] \,.
\label{eq:dpm-transpose}
\end{align}

Now, we consider the sigma model action \cite{Metsaev:1998it}
\begin{align}
 S&=\frac{\dlT}{2}\int \str\bigl(A^{(2)}\wedge *_{\gga} A^{(2)} - A^{(1)}\wedge A^{(3)} \bigr) 
\nn\\
 &=-\frac{\dlT}{2}\int \rmd^2\sigma\,\sqrt{-\gga}\,\Pg_{-}^{\WSa\WSb} \, \str \bigl[\,A_{\WSa} \, d_-(A_{\WSb})\,\bigr]\,,
\label{AdS5S5-action}
\end{align}
where $\dlT \equiv R^2/2\pi\alpha'$ ($R$\,: the radius of $\AdS{5}$ and $\rmS^5$) is the dimensionless string tension. 
In order to relate the supercoset sigma model action to the $\AdS{5} \times \rmS^5$ superstring action, let us prescribe a concrete parameterization of $g$ and expand the action up the second order in fermions. 
We first decompose the group element into the bosonic and the fermionic parts,
\begin{align}
 g =g_{\bos}\cdot g_{\fer}\,,
\end{align}
and parameterize the bosonic part $g_{\bos}$ as (see Appendix \ref{app:psu-algebra} for the details)
\begin{align}
\begin{split}
 &g_{\bos} = g_{\AdS5}\cdot g_{\rmS^5}\,,\qquad 
 g_{\AdS5} \equiv \exp(x^\mu\,P_\mu)\cdot \exp(\ln z \,D)\,,
\\
 &g_{\rmS^5} \equiv \exp(\phi_1\, h_1+\phi_2\,h_2+\phi_3\,h_3)\cdot \exp(\xi\,\gJ_{56})\cdot \exp(r\,\gP_5)\,.
\end{split}
\label{eq:group-parameterization}
\end{align}
Here, $P_{\mu}$ ($\mu,\nu=0,\dotsc,3$) and $D$ are the translation and dilatation generators in the conformal algebra $\alg{so}(2,4)$, and $h_{i}\, (i=1,2,3)$ are Cartan generators of the $\alg{so}(6)$ algebra, given by
\begin{align}
 h_1 \equiv \gJ_{57}\,,\qquad 
 h_2 \equiv \gJ_{68}\,,\qquad 
 h_3 \equiv \gP_9\,. 
\end{align}
On the other hand, we parameterize the fermionic part $g_{\fer}$ as
\begin{align}
 g_{\fer} = \exp(\gQ^I\, \theta_I) \,, \qquad 
 \gQ^I\,\theta_I=(\gQ^I)^{\check{\SPa}\hat{\SPa}}\,\theta_{I\check{\SPa}\hat{\SPa}} \,,
\end{align}
where the supercharges $(\gQ^I)^{\check{\SPa}\hat{\SPa}}$ $(I=1,2)$ are labeled by two indices $(\check{\SPa}\,,\hat{\SPa}=1,\dotsc, 4)$ and $\theta_{I\check{\SPa}\hat{\SPa}}\,(I=1,2)$ are $16$-components Majorana--Weyl fermions.
Then, we can expand the left-invariant current $A$ as
\begin{align}
 A &= g_{\fer}^{-1}\,A_{(0)}\,g_{\fer} + \gQ^I\,\rmd\theta_I
\nn\\
 &= A_{(0)} + [A_{(0)},\,\gQ^I\,\theta_I] + \frac{1}{2}\,\bigl[[A_{(0)},\,\gQ^I\,\theta_I],\,\gQ^J\,\theta_J\bigr] + \gQ^I\,\rmd\theta_I + \cO(\theta^3)
\label{eq:A-AdS5xS5}
\\
 &= \Bigl(e^{\Loa}+\frac{\ii}{2}\,\brtheta_I\,\hat{\gamma}^{\Loa}\,D^{IJ}\theta_J\Bigr)\,\gP_{\Loa} -\frac{1}{2}\,\Bigl(\omega^{\Loa\Lob}-\frac{1}{4}\,\epsilon^{IK}\,\brtheta_I\,\gamma^{\Loc\Lod}\,R_{\Loc\Lod}{}^{\Loa\Lob}\,D^{KJ}\theta_J\Bigr)\,\gJ_{\Loa\Lob}
 +\gQ^I\,D^{IJ}\theta_J + \cO(\theta^3)\,,
\nn
\end{align}
where we have defined
\begin{align}
\begin{split}
 A_{(0)}&\equiv g_{\bos}^{-1}\,\rmd g_{\bos} = \Bigl(e_m{}^{\Loa}\,\gP_{\Loa} - \frac{1}{2}\, \omega_m{}^{\Loa\Lob}\,\gJ_{\Loa\Lob}\Bigr)\,\rmd X^m \,,
\\
 D^{IJ} &\equiv \delta^{IJ}\,\Bigl(\rmd + \frac{1}{4}\,\omega^{\Loa\Lob}\,\gamma_{\Loa\Lob}\Bigr)+\frac{\ii}{2}\,\epsilon^{IJ}\,e^{\Loa}\,\hat{\gamma}_{\Loa}\,, \qquad 
 (X^m)=(x^\mu,\,z,\,r,\,\xi,\,\phi_1,\,\phi_2,\,\phi_3)\,,
\end{split}
\end{align}
and used $\delta^{IJ}\,\brtheta_I\,\hat{\gamma}^{\Loa}\,\rmd \theta_J = 0$ and $\epsilon^{IJ}\,\brtheta_I\,\gamma^{\Loa\Lob}\,\rmd \theta_J = 0$.\footnote{%
We have also used
\begin{align*}
 &[A_{(0)},\,\gQ^I\,\theta_I]= \gQ^I\,\Bigl(\frac{1}{4}\,\delta^{IJ}\,\omega^{\Loa\Lob} \,\gamma_{\Loa\Lob} 
  + \frac{\ii}{2}\,\epsilon^{IJ}\, e^{\Loa}\, \hat{\gamma}_{\Loa} \Bigr)\,\theta_J \,,
\\
 &\bigl[[A_{(0)},\,\gQ^I\,\theta_I],\,\gQ^J\,\theta_J\bigr] 
 = \ii\,\brtheta_I\,\hat{\gamma}^{\Loa}\,\Bigl(\frac{1}{4}\,\delta^{IJ}\,\omega^{\Loc\Lod}\, \gamma_{\Loc\Lod} 
  + \frac{\ii}{2}\,\epsilon^{IJ}\, e^{\Lob}\, \hat{\gamma}_{\Lob} \Bigr)\,\theta_J\,\gP_{\Loa} 
\\
 &\qquad\qquad\qquad\qquad\qquad + \frac{1}{4}\,\epsilon^{IK}\, \brtheta_I\, \gamma^{\Loc\Lod}\,\Bigl(\frac{1}{4}\,\delta^{KJ}\,\omega^{\Loa\Lob}\, \gamma_{\Loa\Lob} 
  + \frac{\ii}{2}\,\epsilon^{KJ}\, e^{\Loa}\, \hat{\gamma}_{\Loa} \Bigr)\,\theta_J\,\rmd X^m\, R_{\Loc\Lod}{}^{\Loe\Lof}\,\gJ_{\Loe\Lof} 
\\
 &\qquad\qquad\qquad\qquad\qquad + \text{(irrelevant terms proportional to the central charge $\gZ$)}\,. 
\end{align*}
} 
The vielbein $e^{\Loa}=e_{m}{}^{\Loa}\,\rmd X^m$ takes the form
\begin{align}
 e^{\Loa} = \biggl(\frac{\rmd x^0}{z},\frac{\rmd x^1}{z} ,\frac{\rmd x^2}{z},\frac{\rmd x^3}{z}, \frac{\rmd z}{z},
\rmd r, \sin r\,\rmd \xi, \sin r\,\cos\xi\,\rmd\phi_1,\sin r\,\sin\xi\,\rmd\phi_2,\cos r\,\rmd\phi_3\biggr) \,,
\end{align}
and $\omega^{\Loa\Lob}=\omega_m{}^{\Loa\Lob}\,\rmd X^m$ and $R_{\Loa\Lob\Loc\Lod}$ are the corresponding spin connection and the Riemann curvature tensor (see Appendix \ref{app:conventions} for our conventions). 
Using the expansion \eqref{eq:A-AdS5xS5}, we can straightforwardly obtain
\begin{align}
 \frac{1}{2}\,\str\bigl[\,A_{\WSa}\,d_-(A_{\WSb})\,\bigr]
 &= \eta_{\Loa\Lob}\,e_{\WSa}{}^{\Loa}\,e_{\WSb}{}^{\Lob} 
  + \ii\, \bigl[\,e_{\WSb}{}^{\Loa}\, (\brtheta_1\,\hat{\gamma}_{\Loa}\, \partial_{\WSa} \theta_1) 
  + e_{\WSa}{}^{\Loa}\, (\brtheta_2\,\hat{\gamma}_{\Loa}\, \partial_{\WSb} \theta_2) \,\bigr]
\nn\\
 &\quad + \frac{\ii}{4}\,\Bigl[e_{\WSb}{}^{\Lob}\, e_{\WSa}{}^{\Loa}\,\omega_{\Loa}{}^{\Loc\Lod} \,(\brtheta_1\,\hat{\gamma}_{\Lob}\,\gamma_{\Loc\Lod} \,\theta_1)
 + e_{\WSa}{}^{\Loa}\, e_{\WSb}{}^{\Lob}\, \omega_{\Lob}{}^{\Loc\Lod} \, (\brtheta_2\,\hat{\gamma}_{\Loa}\,\gamma_{\Loc\Lod}\,\theta_2) \Bigr]
\nn\\
 &\quad - e_{\WSb}{}^{\Loa} \,e_{\WSa}{}^{\Lob}\, \brtheta_1\,\hat{\gamma}_{\Loa}\, \hat{\gamma}_{\Lob}\,\theta_2 \,,
\end{align}
where $e_{\WSa}{}^{\Loa}\equiv e_m{}^{\Loa}\,\partial_{\WSa}X^m$\,. 
Further using Eqs.~\eqref{eq:lift-32-AdS5-1}, \eqref{eq:lift-32-AdS5-2}, and \eqref{eq:lift-32-AdS5-3}, we obtain
\begin{align}
 \frac{1}{2}\,\str\bigl[\,A_{\WSa}\,d_-(A_{\WSb})\,\bigr]
 &= \CG_{mn}\,\partial_{\WSa} X^m\,\partial_{\WSb} X^n 
  + \ii\, \bigl[\,e_{\WSb}{}^{\Loa}\, \brTheta_1\,\Gamma_{\Loa}\, \partial_{\WSa} \Theta_1 
  + e_{\WSa}{}^{\Loa}\, \brTheta_2\,\Gamma_{\Loa}\, \partial_{\WSb} \Theta_2 \,\bigr]
\nn\\
 &\quad + \frac{\ii}{4}\,\Bigl[e_{\WSb}{}^{\Lob}\, e_{\WSa}{}^{\Loa}\,\omega_{\Loa}{}^{\Loc\Lod} \, \brTheta_1\,\Gamma_{\Lob}\,\Gamma_{\Loc\Lod} \,\Theta_1 
 + e_{\WSa}{}^{\Loa}\, e_{\WSb}{}^{\Lob}\, \omega_{\Lob}{}^{\Loc\Lod} \, \brTheta_2\,\Gamma_{\Loa}\,\Gamma_{\Loc\Lod}\,\Theta_2 \Bigr]
\nn\\
 &\quad - \frac{\ii}{8}\,e_{\WSb}{}^{\Loa} \,e_{\WSa}{}^{\Lob}\, \brTheta_I\,\Gamma_{\Loa}\,\bisF_5\,\Gamma_{\Lob}\,\Theta_J\,,
\end{align}
where $\bisF_5$ is a bispinor
\begin{align}
 \bisF_5 \equiv 4\,(\Gamma^{01234}+\Gamma^{56789})\,. 
\end{align}
Then, the action \eqref{AdS5S5-action} becomes the GS type IIB superstring action \eqref{eq:GS-action-conventional} with the target space given by the familiar $\AdS{5} \times \rmS^5$ background:
\footnote{Here, we have defined $(\eta_{\mu\nu})=\diag(-1,1,1,1)$ and
\begin{align*}
\begin{split}
 &\rmd s_{\rmS^5}^2\equiv \rmd r^2 + \sin^2 r\, \rmd\xi^2 + \cos^2\xi\,\sin^2 r\, \rmd\phi_1^2 + \sin^2r\,\sin^2\xi\, \rmd\phi_2^2 + \cos^2r\, \rmd\phi_3^2\,,
\\
 &\omega_{\AdS5} \equiv - \frac{\rmd x^0\wedge\rmd x^1\wedge \rmd x^2\wedge\rmd x^3\wedge\rmd z}{z^5}\,, 
\\
 &\omega_{\rmS^5} \equiv \sin^3r \cos r \sin\xi \cos\xi\,\rmd r\wedge \rmd\xi\wedge\rmd \phi_1\wedge \rmd\phi_2\wedge\rmd \phi_3\qquad (\omega_{\AdS{5}} = *_{10}\omega_{\rmS^5}) \,. 
\end{split}
\end{align*}
}
\begin{align}
 \rmd s^2 = \frac{\eta_{\mu\nu}\,\rmd x^\mu\,\rmd x^\nu + \rmd z^2}{z^2}+\rmd s_{\rmS^5}^2\,, \qquad
 B_{mn} =0\,,\qquad \Exp{\Phi}\hat{F}_5 = 4\,\bigl(\omega_{\AdS5}+\omega_{\rmS^5}\bigr) \,.
\end{align}
From the supergravity equations of motion (or the Weyl invariance of string theory), the dilaton is determined as $\Phi=0$\,. 

\subsubsection{Killing vectors}

For later convenience, let us calculate the Killing vectors $\hat{T}_i\equiv\hat{T}_i^m\,\partial_m$ associated with the bosonic symmetries $T_i$ of the $\AdS{5}$ background. 
From the general formula \eqref{eq:Killing-formula} explained in Appendix \ref{app:homogeneous-space}, the Killing vectors can be expressed as
\begin{align}
 \hat{T}_{i}=\hat{T}_{i}{}^{m}\,\partial_{m} 
 =\bigl[\Ad_{g_{\bos}^{-1}}\bigr]_{i}{}^{\Loa}\, e_{\Loa}{}^{m}\, \partial_{m}
 = \str\bigl(g_{\bos}^{-1}\,T_{i}\,g_{\bos}\,\gP_{\Loa}\bigr)\,e^{\Loa m}\, \partial_{m}\,,
\label{eq:Killing-Formula}
\end{align}
where we introduced a notation $g\,T_i\,g^{-1}\equiv [\Ad_{g}]_{i}{}^{j}\,T_j$\,. 
By using our parameterization \eqref{eq:group-parameterization}, the Killing vectors on the $\AdS{5}$ background are given by
\begin{align}
\begin{split}
 \hat{P}_\mu &\equiv \str\bigl(g_{\bos}^{-1}\,P_{\mu}\,g_{\bos}\, \gP_{\Loa} \bigr)\,e^{\Loa m}\,\partial_m = \partial_\mu \,,
\\
 \hat{K}_\mu &\equiv \str\bigl(g_{\bos}^{-1}\,K_{\mu}\,g_{\bos}\, \gP_{\Loa} \bigr)\,e^{\Loa m}\,\partial_m = \bigl(x^\nu\,x_\nu +z^2\bigr)\,\partial_\mu - 2\,x_\mu\,(x^\nu\,\partial_\nu+z\,\partial_z)\,,
\\
 \hat{M}_{\mu\nu} &\equiv \str\bigl(g_{\bos}^{-1}\,M_{\mu\nu}\,g_{\bos}\, \gP_{\Loa} \bigr)\,e^{\Loa m}\,\partial_m = x_\mu\,\partial_\nu -x_\nu\,\partial_\mu \,,
\\
 \hat{D} &\equiv \str\bigl(g_{\bos}^{-1}\,D\,g_{\bos}\, \gP_{\Loa} \bigr)\,e^{\Loa m}\,\partial_m = x^\mu\,\partial_\mu + z\,\partial_z \,.
\end{split}
\end{align}
The Lie brackets of these vector fields satisfy the same commutation relations \eqref{eq:so(2-4)-algebra} as the conformal algebra $\alg{so}(2,4)$ (with negative sign, $[\hat{T}_i,\,\hat{T}_j]=-f_{ij}{}^k\,\hat{T}_k$):
\begin{align}
\begin{split}
 &[\hat{P}_\mu,\, \hat{K}_\nu]= -2\,\bigl(\eta_{\mu\nu}\, \hat{D} - \hat{M}_{\mu\nu}\bigr)\,,\quad 
 [\hat{D},\, \hat{P}_{\mu}]= -\hat{P}_\mu\,,\quad [\hat{D},\,\hat{K}_\mu]= \hat{K}_\mu\,,
\\
 &[\hat{M}_{\mu\nu},\, \hat{P}_\rho] = -\eta_{\mu\rho}\, \hat{P}_\nu +\eta_{\nu\rho}\, \hat{P}_\mu \,,\quad 
 [\hat{M}_{\mu\nu},\, \hat{K}_\rho] = -\eta_{\mu\rho}\,\hat{K}_\nu + \eta_{\nu\rho}\,\hat{K}_\mu\,, 
\\
 &[\hat{M}_{\mu\nu},\,\hat{M}_{\rho\sigma}] = -\eta_{\mu\rho}\,\hat{M}_{\nu\sigma}+\eta_{\mu\sigma}\,\hat{M}_{\nu\rho} + \eta_{\nu\rho}\,\hat{M}_{\mu\sigma}-\eta_{\nu\sigma}\,\hat{M}_{\mu\rho}\,.
\end{split}
\end{align}

\vspace{-\Pskip}
\subsection{YB deformed \texorpdfstring{$\AdS{5} \times \rmS^5$}{AdS\textfiveinferior{x}S\textfivesuperior} backgrounds}
\label{sec:YB-sigma}

Let us now consider (homogeneous) YB deformations of the $\AdS{5} \times \rmS^5$ superstring action. 
A key ingredient that characterizes the YB deformation is an $R$-operator. 
It is a linear operator $R:\alg{g} \to \alg{g}$, that solves the homogeneous CYBE
\begin{align}
 \CYBE(X,Y) \equiv [R(X),\,R(Y)] - R([R(X),\,Y]+[X,\,R(Y)])=0\,,
\label{eq:CYBE}
\end{align}
where $X,\,Y \in\alg{g}$\,. 
We also define the dressed $R$-operator $R_g$ as
\begin{align}
 R_g(X)\equiv g^{-1}\, R(g\,X\,g^{-1})\,g \,,
\end{align}
which also satisfies the homogeneous CYBE \eqref{eq:CYBE},
\begin{align}
 \CYBE_g(X,Y) \equiv [R_g(X),\,R_g(Y)] - R_g([R_g(X),\,Y]+[X,\,R_g(Y)])=0\,, 
\label{eq:CYBE-g}
\end{align}
as long as $R$ satisfies the homogeneous CYBE. 
Then, the action of YB-deformed $\AdS{5} \times \rmS^5$ superstring is given by
\begin{align}
 S_{\YB}=-\frac{\dlT}{2}\int \rmd^2\sigma\,\sqrt{-\gga}\,\Pg_{-}^{\WSa\WSb}\, \str\bigl[A_{\WSa}\, d_-\circ\cO_-^{-1}(A_{\WSb})\bigr]\,, 
\label{eq:YBsM}
\end{align}
where we defined linear operators $\cO_{\pm}$ as
\begin{align}
 \cO_{\pm}=1\pm\eta\, R_g\circ d_\pm\,, 
\end{align}
and $\eta\in\mathbb{R}$ is a deformation parameter. 
This action reduces to the undeformed $\AdS{5}\times \rmS^5$ action \eqref{AdS5S5-action} by taking $\eta=0$\,. 

In this paper, we consider a class of $R$-operators that can be specified by using a skew-symmetric classical $r$-matrix. 
By introducing an $r$-matrix $r\in \alg{g}\otimes \alg{g}$ of the form
\begin{align}
 r=\frac{1}{2}\,r^{ij}\,T_{i}\wedge T_{j}\,, \qquad r^{ij}=-r^{ji}\,, \qquad T_{i}\in \alg{g}\,,
\end{align}
the $R$-operator is defined as
\begin{align}
 R(X) = r^{ij}\, T_{i}\,\str(T_{j}\,X) \,, \qquad X\in\alg{g} \,.
\label{eq:R-operator}
\end{align}
Then, the YB deformations are characterized only by the $r$-matrix. 
In terms of the $r$-matrix, the homogeneous CYBE \eqref{eq:CYBE} can be expressed as
\begin{align}
 f_{l_1l_2}{}^i\,r^{jl_1}\,r^{kl_2} + f_{l_1l_2}{}^j\,r^{kl_1}\,r^{il_2} + f_{l_1l_2}{}^k\,r^{il_1}\,r^{jl_2} =0\,,
\label{eq:CYBE-r}
\end{align}
where $f_{ij}{}^k$ is the structure constant, $[T_i,\,T_j]=f_{ij}{}^k\,T_k$\,. 

\subsubsection{Preparations}

In the following, we rewrite the YB-deformed action in the form of the conventional GS action, and show that the target space is a $\beta$-deformed $\AdS{5} \times \rmS^5$ background. 
In order to determine the deformed background, it is sufficient to expand the action up to quadratic order in fermions,
\begin{align}
 S_{\YB}=S_{(0)}+S_{(2)}+\cO(\theta^4)\,.
\end{align}
This kind of analysis has been performed in \cite{Arutyunov:2015qva} for the $q$-deformation of $\AdS{5}\times \rmS^5$ and in \cite{Kyono:2016jqy} for homogeneous YB deformations. 
However, the obtained deformed actions in the previous works are very complicated and it is not easy to read off the deformed background explicitly. 
In this paper, we provide a general formula for the deformed background for an arbitrary $r$-matrix satisfying homogeneous CYBE, though our analysis is limited to the cases where the $r$-matrices are composed only of the bosonic generators of the superalgebra $\alg{g}$\,.\footnote{Rewriting of the YB sigma model action to the standard GS form based on the $\kappa$-symmetry was done in \cite{Borsato:2016ose} to full order in fermionic variables, and there, the deformed background associated with a general $r$-matrix was determined.} 

In order to expand the YB sigma model action \eqref{eq:YBsM}, let us introduce some notations.
Since we are supposing that the $r$-matrices are composed of the bosonic generators, the dressed $R$-operator $R_{g_{\bos}}$ acts as
\begin{align}
 R_{g_{\bos}}(\gP_{\Loa}) =\lambda_{\Loa}{}^{\Lob}\,\gP_{\Lob}+\frac{1}{2}\,\lambda_{\Loa}{}^{\Lob\Loc}\,\gJ_{\Lob\Loc}\,, \quad
 R_{g_{\bos}}(\gJ_{\Loa\Lob}) =\lambda_{\Loa\Lob}{}^{\Loc}\,\gP_{\Loc}+\frac{1}{2}\,\lambda_{\Loa\Lob}{}^{\Loc\Lod}\,\gJ_{\Loc\Lod}\,, \quad
 R_{g_{\bos}}(\gQ^I) =0\,.
\label{eq:Rg-operation}
\end{align}
According to the definition \eqref{eq:R-operator}, the (dressed) $R$-operator is skew-symmetric
\begin{align}
 \str\bigl[R_{g_{\bos}}(X)\,Y\bigr] = r^{ij}\,\bigl[\Ad_{g_{\bos}^{-1}}\bigr]_i{}^k\,\bigl[\Ad_{g_{\bos}^{-1}}\bigr]_j{}^l\, \str(T_{k}\,Y)\,\str(T_{l}\, X) = - \str\bigl[X\,R_{g_{\bos}}(Y)\bigr] \,,
\end{align}
and by choosing $X$ and $Y$ as $\gP_{\Loa}$ or $\gJ_{\Loa\Lob}$, we obtain the following relations:
\begin{align}
 \lambda_{\Loa\Lob} \equiv \lambda_{\Loa}{}^{\Loc}\,\eta_{\Loc\Lob} = -\lambda_{\Lob\Loa}\,,\quad 
 \lambda_{\Loa\Lob}{}^{\Loc}=-\frac{1}{2}\,\eta^{\Loc\Lod}\,R_{\Loa\Lob\Loe\Lof}\,\lambda_{\Lod}{}^{\Loe\Lof}\,,\quad 
 \lambda_{\Loa\Lob}{}^{\Loe\Lof}\,R_{\Loe\Lof\Loc\Lod} = - \lambda_{\Loc\Lod}{}^{\Loe\Lof}\,R_{\Loe\Lof\Loa\Lob}\,. 
\label{eq:lambda-properties}
\end{align}

For later convenience, we introduce the deformed currents as
\begin{align}
 J_{\pm} \equiv \cO_{\pm}^{-1}\,A_{\pm}\,.
\label{eq:J-O-inv-A}
\end{align}
By using the results of Appendix \ref{app:expansion-O}, it can be expanded as
\begin{align}
 J_\pm&=\cO^{-1}_{\pm(0)}(A_{(0)})+\cO^{-1}_{\pm(0)}(A_{(1)})+\cO^{-1}_{\pm(1)}(A_{(0)})+\cO(\theta^2)
\nn\\
  &=e_{\pm}^{\Loa}\,\gP_{\Loa}-\frac{1}{2}\,W_{\pm}^{\Loa\Lob}\,\gJ_{\Loa\Lob} + \gQ^I\,D^{IJ}_{\pm}\theta_J +\cO(\theta^2)\,,
\label{eq:Jpm-expansion}
\end{align}
where we have defined
\begin{align}
 &e_{\pm}^{\Loa}\equiv e^{\Lob}\,k_{\pm \Lob}{}^{\Loa}\,,\qquad 
 k_{\pm \Loa}{}^{\Lob} \equiv \bigl[(1\pm 2\,\eta\,\lambda)^{-1}\bigr]{}_{\Loa}{}^{\Lob}\,, \qquad
 W_{\pm}^{\Loa\Lob} \equiv \omega^{\Loa\Lob}\pm 2\,\eta\,e_{\pm}^{\Loc}\,\lambda_{\Loc}{}^{\Loa\Lob}\,, 
\label{eq:e-torsionful-spin-pm}
\\
 &D^{IJ}_{\pm}\equiv \delta^{IJ}\,D_{\pm} +\frac{\ii}{2}\,\epsilon^{IJ}\,e_{\pm}^{\Loa}\,\hat{\gamma}_{\Loa}\,, \qquad 
 D_{\pm}\equiv \rmd+\frac{1}{4}\,W_{\pm}^{\Loa\Lob}\, \gamma_{\Loa\Lob}\,. 
\end{align}
As it turns out, $e_{\pm}^{\Loa}$ and $W_{\pm}^{\Loa\Lob}$\,, respectively, play the roles of the two vielbeins and the torsionful spin connections $\omega_{\pm}$ \eqref{eq:Phi-omega-pm} in the deformed background.\footnote{More precisely, we have $W_{\pm \Loa\Lob}\wedge e_{\mp}^{\Loa}\wedge e_{\mp}^{\Lob}=\bigl(\omega_{[\mp] \Loa\Lob} \pm\frac{1}{2}\,e_{\mp}^{\Loc}\,H'_{\Loc\Loa\Lob}\bigr)\wedge e_{\mp}^{\Loa}\wedge e_{\mp}^{\Lob}$\,, where $\omega_{[\pm]}$ represent the spin connections associated with the vielbeins $e_{\pm}$ and $H'_3$ represents the $H$-flux in the deformed background.} 

\subsubsection{NS--NS sector}
\label{sec:YB-NS-NS}

Let us first consider the NS--NS part of the YB sigma model action,
\begin{align}
 S_{(0)}=-\frac{\dlT}{2}\int \rmd^2\sigma\,\sqrt{-\gga}\,\Pg_{-}^{\WSa\WSb}\, \str\bigl[A_{\WSa (0)}\, d_-\circ\cO_{-(0)}^{-1}(A_{\WSb(0)})\bigr]\,. 
\end{align}
From \eqref{eq:Oinv-0}, we can easily see that it takes the form
\begin{align}
 S_{(0)}
 =-\dlT\int \rmd^2\sigma\,\sqrt{-\gga}\,\Pg_{-}^{\WSa\WSb}\, \eta_{\Loa\Lob}\,e_{\WSa}{}^{\Loa}\,e_{\WSb}{}^{\Loc}\,k_{-\Loc}{}^{\Lob}\,.
\label{eq:action-order0}
\end{align}
By comparing this with the NS--NS part of the GS action [i.e.~the first line of \eqref{eq:GS-action-conventional}], we can regard \eqref{eq:action-order0} as the NS--NS part of the string sigma model on a deformed background
\begin{align}
 \CG'_{mn} = e_{(m}{}^{\Loa}\,e_{n)}{}^{\Lob}\, k_{+\Loa\Lob} \,,\qquad 
 B'_{mn} = e_{[m}{}^{\Loa}\,e_{n]}{}^{\Lob}\, k_{+\Loa\Lob} \,. 
\label{eq:G-B-prime}
\end{align}
Then, we obtain
\begin{align}
 E'^{mn} \equiv \bigl[(\CG' + B')^{-1}\bigr]^{mn} = (k_+^{-1})^{\Loa\Lob}\, e_{\Loa}{}^{m}\,e_{\Lob}{}^{n} 
 = (\eta + 2\,\eta\,\lambda)^{\Loa\Lob}\, e_{\Loa}{}^{m}\,e_{\Lob}{}^{n} \,.
\end{align}
In the original $\AdS{5}\times\rmS^5$ background, the $B$-field is absent and we have
\begin{align}
 E_{mn}= \CG_{mn} = e_{m}{}^{\Loa}\,e_{n}{}^{\Lob}\,\eta_{\Loa\Lob}\,, \qquad 
 E^{mn}= \eta^{\Loa\Lob}\, e_{\Loa}{}^{m}\,e_{\Lob}{}^{n} \,. 
\end{align}
Therefore, the deformation can be summarized as
\begin{align}
 E^{mn} \ \to \ E'^{mn} = E^{mn} + 2\,\eta\,\lambda^{\Loa\Lob}\, e_{\Loa}{}^{m}\,e_{\Lob}{}^{n} \,.
\end{align}
By comparing this with the $\beta$-transformation rule \eqref{eq:beta-rule-NS}, we can regard the YB deformation as $\beta$-deformation with the parameter
\begin{align}
 \bmr^{mn} = 2\,\eta\,\lambda^{\Loa\Lob}\, e_{\Loa}{}^{m}\,e_{\Lob}{}^{n} \,. 
\label{eq:bmr-YB}
\end{align}
If we compute the dual field in the deformed background, we obtain
\begin{align}
 \OG'_{mn}&=\eta_{\Loa\Lob}\,e_m{}^{\Loa}\,e_n{}^{\Lob}\,, \qquad
 \beta'^{mn}=-2\,\eta\,\lambda^{\Loa\Lob}\,e_{\Loa}{}^m\,e_{\Lob}{}^n\,.
\label{eq:beta}
\end{align}
The dual metric is invariant under the deformation $\OG_{mn}\to \OG'_{mn} = \OG_{mn}$\,, while the $\beta$-field, which is absent in the undeformed background, is shifted as $\beta^{mn}=0\ \to\ \beta'^{mn}= -\bmr^{mn}$\,. 

In addition, the YB-deformed dilaton $\Phi'$ that is consistent with the kappa invariance (or supergravity equations of motion) has been proposed in \cite{Kyono:2016jqy,Borsato:2016ose} as
\begin{align}
 \Exp{\Phi'} = (\det k_{+})^{\frac{1}{2}}=(\det k_{-})^{\frac{1}{2}} \,. 
\label{eq:dilaton-YB}
\end{align}
In order to compare this with the $\beta$-transformation law of the dilation, we consider the two vielbeins $e_{\pm m}{}^{\Loa}=e^{\Lob}\,k_{\pm \Lob}{}^{\Loa}$ introduced in \eqref{eq:e-torsionful-spin-pm}. 
Here, we can rewrite $k_{\pm \Loa}{}^{\Lob}$ as
\begin{align}
 k_{\pm \Loa}{}^{\Lob} \equiv \bigl[(1\pm 2\,\eta\,\lambda)^{-1}\bigr]{}_{\Loa}{}^{\Lob}
 = e_{\Loa}{}^m\,\bigl[(1\pm \CG\,\bmr)^{-1}\bigr]_m{}^n\,e_n{}^{\Lob} 
 = e_{\Loa}{}^m\,\bigl[(1\pm E\,\bmr)^{-1}\bigr]_m{}^n\,e_n{}^{\Lob} \,,
\end{align}
by using $\bmr^{mn}$ of \eqref{eq:bmr-YB} and $B_{mn}=0$ in the undeformed background. 
Then, $e_{\pm m}{}^{\Loa}$ becomes
\begin{align}
 e_{\pm m}{}^{\Loa} = \bigl[(E^{-\rmT} \pm \bmr)^{-1}\,E^{\rmT}\bigr]{}_{m}{}^{n}\,e_n{}^{\Loa} \,. 
\end{align}
Comparing this with the $\beta$-transformation rule \eqref{eq:beta-rule-NS}, we can identify $e_{-m}{}^{\Loa}$ as the $\beta$-deformed vielbein $e'_m{}^{\Loa}$\,. 
Similarly, $e_{+m}{}^{\Loa}$ can be identified as the $\beta$-deformed barred vielbein $\bre'_m{}^{\Loa}$,
\begin{align}
 e_{-m}{}^{\Loa}\ \leftrightarrow\ e'_m{}^{\Loa}\,,\qquad 
 e_{+m}{}^{\Loa}\ \leftrightarrow\ \bre'_m{}^{\Loa}\,. 
\label{eq:e+-eebar}
\end{align}
Namely, we can express the deformed metric as
\begin{align}
 \CG'_{mn} = e_{+m}{}^{\Loa}\,e_{+n}{}^{\Lob}\,\eta_{\Loa\Lob}
    = e_{-m}{}^{\Loa}\,e_{-n}{}^{\Lob}\,\eta_{\Loa\Lob}\,,
\end{align}
and the invariance of $\Exp{-2d}=\Exp{-2\Phi}\sqrt{-\CG}$ under $\beta$-deformations shows
\begin{align}
 \Exp{-2\Phi'} = \frac{\sqrt{-\CG}}{\sqrt{-\CG'}} \Exp{-2\Phi} = \frac{\det (e_m{}^{\Loa})}{\det (e_{\pm m}{}^{\Loa})} \Exp{-2\Phi}
 = (\det k_{\pm})^{-1} \Exp{-2\Phi} \,. 
\end{align}
Recalling $\Phi=0$ in the undeformed background, the transformation rule \eqref{eq:dilaton-YB} can be understood as the $\beta$-transformation. 
Therefore, NS--NS fields are precisely $\beta$-deformed under the homogeneous YB deformation.

For later convenience, let us rewrite $\bmr^{mn}$ of \eqref{eq:bmr-YB} by using the $r$-matrix instead of $\lambda^{\Loa\Lob}$\,. 
From the definition, $\lambda^{\Loa\Lob}$ can be expressed as
\begin{align}
 \lambda^{\Loa\Lob} = \str\bigl[ R_{g_{\bos}}(\gP^{\Loa})\,\gP^{\Lob} \bigr] \,. 
\end{align}
By using the $r$-matrix $r=\frac{1}{2}\,r^{ij}\,T_i\wedge T_j$\,, this can be expressed as
\begin{align}
 \str\bigl[ R_{g_{\bos}}(\gP^{\Loa})\,\gP^{\Lob} \bigr]
 &= r^{ij}\, \str\bigl(g_{\bos}^{-1}\,T_{i}\,g_{\bos}\,\gP_{\Lob})\, \str(g_{\bos}^{-1}\,T_{j}\,g_{\bos}\,\gP_{\Loa}) 
 = -r^{ij}\,\bigl[\Ad_{g_{\bos}^{-1}}\bigr]_i{}^{\Loa}\,\bigl[\Ad_{g_{\bos}^{-1}}\bigr]_i{}^{\Lob}\,,
\end{align}
and we obtain
\begin{align}
 \bmr^{mn} = -2\,\eta\,r^{ij}\,\bigl[\Ad_{g_{\bos}^{-1}}\bigr]_i{}^{\Loa}\,\bigl[\Ad_{g_{\bos}^{-1}}\bigr]_j{}^{\Lob}\, e_{\Loa}{}^{m}\,e_{\Lob}{}^{n} \,. 
\end{align}
By using the Killing vectors \eqref{eq:Killing-Formula} we obtain a very simple expression
\begin{align}
 \bmr^{mn} = -2\,\eta\,r^{ij}\,\hat{T}_i^m\, \hat{T}_j^n \,. 
\label{eq:bmr-YB2}
\end{align}

The $\beta$-field after the deformation takes the form
\begin{align}
 \beta'^{mn} = - \bmr^{mn} = 2\,\eta\,r^{ij}\,\hat{T}_i^m\, \hat{T}_j^n \,,\qquad 
 \beta' \equiv \frac{1}{2}\,\beta'^{mn}\,\partial_m\wedge\partial_n 
    = 2\,\eta\,\biggl(\frac{1}{2}\,r^{ij}\,\hat{T}_i \wedge \hat{T}_j\biggr) \,,
\end{align}
and we can calculate the associated non-geometric $R$-flux
\begin{align}
 R^{mnp} \equiv 3\,\beta^{[m|q}\,\partial_q \beta^{|np]} \,. 
\end{align}
By using the Lie bracket for the Killing vector fields $[\hat{T}_i,\,\hat{T}_j]=-f_{ij}{}^k\,\hat{T}_k$\,, we obtain
\begin{align}
 R^{mnp} = -8\,\eta^2\,\bigl(f_{l_1l_2}{}^i\,r^{jl_1}\,r^{kl_2} + f_{l_1l_2}{}^j\,r^{kl_1}\,r^{il_2} + f_{l_1l_2}{}^k\,r^{il_1}\,r^{jl_2}\bigr)\,\hat{T}_i^m\,\hat{T}_j^n\,\hat{T}_k^p = 0\,, 
\end{align}
upon using the homogeneous CYBE \eqref{eq:CYBE-r}. 
This shows the absence of the $R$-flux in homogeneous YB-deformed backgrounds as noted in \cite{Sakamoto:2017cpu}. 

\subsubsection{R--R sector}
\label{sec:YB-RR}

Next, we determine the R--R fields from the quadratic part of the YB sigma model action $S_{(2)}$\,, and show that the R--R fields are also $\beta$-deformed with the $\bmr^{mn}$ given in \eqref{eq:bmr-YB2}. 

As noticed in \cite{Arutyunov:2015qva,Kyono:2016jqy}, the deformed action naively does not have the canonical form of the GS action \eqref{eq:GS-action-conventional}, and we need to choose the diagonal gauge and perform a suitable redefinition of the bosonic fields $X^m$\,. 
Since the analysis is considerably complicated, we relegate the details to Appendix \ref{app:deformed-Lagrangian}, and here we explain only the outline. 

The quadratic part of the deformed action $S_{(2)}$ can be decomposed into two parts
\begin{align}
 S_{(2)} = S_{(2)}^{\rmc} + \delta S_{(2)}\,.
\end{align}
For a while, we focus only on the first part $S_{(2)}^{\rmc}$ since the second part $\delta S_{(2)}$ is completely cancelled after some field redefinitions. 
The explicit expression of $S_{(2)}^{\rmc}$ is given by
\begin{align}
 S_{(2)}^{\rmc} &=-\ii\,\dlT\int \rmd^2\sigma\,\sqrt{-\gga}\,
  \biggl[ \Pg_+^{\WSa\WSb}\,e_{-\WSa}{}^{\Loa}\,\brtheta_1\,\hat{\gamma}_{\Loa}\,D_{+\WSb}\theta_1
       +\Pg_-^{\WSa\WSb}\,e_{+\WSa}{}^{\Loa}\,\brtheta_2\,\hat{\gamma}_{\Loa}\,D_{-\WSb}\theta_2
\nn\\
&\qquad\qquad\qquad\qquad\quad
       +\ii\, \Pg_+^{\WSa\WSb}\,\epsilon^{IJ}\,\brtheta_1\,e_{-\WSa}{}^{\Loa}\,\hat{\gamma}_{\Loa}\,e_{+\WSb}{}^{\Lob}\,\hat{\gamma}_{\Lob}\,\theta_2\biggr]\,.
\label{eq:L-order2}
\end{align}
This action contains the two deformed vielbeins $e_{\pm m}{}^{\Loa}$ similar to the DSM action \eqref{eq:GS-DSM-ours} prior to taking the diagonal gauge. 
As we observed in \eqref{eq:e+-eebar}, these vielbeins $e_{-m}{}^{\Loa}$ and $e_{+m}{}^{\Loa}$ correspond to the two vielbeins $e'_m{}^{\Loa}$ and $\bre'_m{}^{\Lobra}$ introduced in \eqref{eq:dV-geometric}, respectively. 
In order to rewrite the action into the canonical form of the GS action, we need to choose the diagonal gauge $e_m{}^{\Loa}=\bre_m{}^{\Lobra}$\,. 
For this purpose, we first rewrite the action \eqref{eq:L-order2} in terms of the $32\times 32$ gamma matrices. 
By using relations \eqref{eq:lift-32-AdS5-1}, \eqref{eq:lift-32-AdS5-2}, and \eqref{eq:lift-32-AdS5-3}, we obtain
\begin{align}
\begin{split}
 S^{\rmc}_{(2)}&=-\ii\,\dlT\int \rmd^2\sigma\,\sqrt{-\gga}\,\biggl[
 \Pg_+^{\WSa\WSb}\,\brTheta_1\,e_{-\WSa}{}^{\Loa}\,\Gamma_{\Loa}\,D_{+\WSb}\Theta_1
 +\Pg_-^{\WSa\WSb}\,\brTheta_2\,e_{+\WSa}{}^{\Loa}\,\Gamma_{\Loa}D_{-\WSb}\Theta_2
\\
 &\qquad\qquad\qquad\qquad\quad -\frac{1}{8}\,\Pg_+^{\WSa\WSb}\,\brTheta_1\,e_{-\WSa}{}^{\Loa}\,\Gamma_{\Loa} \,\bisF_5\, e_{+\WSb}{}^{\Lob}\,\Gamma_{\Lob}\,\Theta_2 \biggr]\,,
\end{split}
\end{align}
where $D_{\pm\WSa}\Theta_I\equiv \bigl(\partial_{\WSa}+\frac{1}{4}\,W_{\pm\WSa}{}^{\Loa\Lob}\, \Gamma_{\Loa\Lob}\bigr)\,\Theta_I$ and $\bisF_5$ is the undeformed R--R $5$-form field strength
\begin{align}
 \bisF_5 =\frac{1}{5!}\,\bisF_{\Loa_1\cdots \Loa_5}\,\Gamma^{\Loa_1\cdots \Loa_5} =4\,\bigl(\Gamma^{01234}+\Gamma^{56789}\bigr)\,. 
\end{align}
Next, we eliminate the barred vielbein $e_{+m}{}^{\Loa}$ by using
\begin{align}
 e_{+m}{}^{\Loa}=(\Lambda^{-1})^{\Loa}{}_{\Lob}\,e_{-m}{}^{\Lob}
 = \Lambda_{\Lob}{}^{\Loa}\,e_{-m}{}^{\Lob}\,,\qquad 
 \Lambda_{\Loa}{}^{\Lob} \equiv (k_-^{-1})_{\Loa}{}^{\Loc}\, k_{+\Loc}{}^{\Lob}\in \SO(1,9)\,,
\label{eq:e--e+-relation}
\end{align}
which follows from \eqref{eq:e-torsionful-spin-pm}. 
By further using the identity [recall the formula \eqref{eq:Hassan-formula}]
\begin{align}
 \Omega^{-1} \,\Gamma_{\Loa}\,\Omega = \Lambda_{\Loa}{}^{\Lob} \, \Gamma_{\Lob}\,, \qquad 
 \Omega =(\det k_-)^{\frac{1}{2}}\,\text{\AE}\bigl(-\eta\,\lambda^{\Loa\Lob}\,\Gamma_{\Loa\Lob}\bigr) \,,
\label{eq:omega}
\end{align}
the action becomes
\begin{align}
\begin{split}
 S^{\rmc}_{(2)}&=-\ii\,\dlT\int \rmd^2\sigma\,\sqrt{-\gga}\,\biggl[
 \Pg_+^{\WSa\WSb}\,\brTheta_1\,e'_{\WSa}{}^{\Loa}\,\Gamma_{\Loa}\,D_{+\WSb}\Theta_1
 +\Pg_-^{\WSa\WSb}\,\brTheta_2\,\Omega^{-1}\,e'_{\WSa}{}^{\Loa}\,\Gamma_{\Loa}\,\Omega\,D_{-\WSb}\Theta_2
\\
 &\qquad\qquad\qquad\qquad\quad -\frac{1}{8}\,\Pg_+^{\WSa\WSb}\,\brTheta_1\,e'_{\WSa}{}^{\Loa}\,\Gamma_{\Loa} \,\bisF_5\,\Omega^{-1}\, e'_{\WSb}{}^{\Loc}\,\Gamma_{\Loc}\,\Omega\,\Theta_2 \biggr]\,.
\end{split}
\end{align}
We then perform a redefinition of the fermionic variables $\Theta_I$\,,
\begin{align}
 \Theta'_1 \equiv \Theta_1\,,\qquad 
 \Theta'_2 \equiv \Omega \,\Theta_2 \,,
\label{eq:fermi-redef}
\end{align}
which corresponds to the transformation rule of fermions \eqref{eq:Theta-transf} under the diagonal gauge fixing. 
As the result of the redefinition, we obtain
\begin{align}
 S^{\rmc}_{(2)}&=-\dlT\int \rmd^2\sigma\,\sqrt{-\gga}\,\biggl[
 \Pg_+^{\WSa\WSb}\,\ii\,\brTheta'_1 \,e'_{\WSa}{}^{\Loa}\,\Gamma_{\Loa}\,D'_{+\WSb} \Theta'_1 
 +\Pg_-^{\WSa\WSb}\,\ii\,\brTheta_2\,e'_{\WSa}{}^{\Loa}\,\Gamma_{\Loa}\,D'_{-\WSb} \Theta'_2 
\nn\\
 &\qquad\qquad\qquad\qquad\quad -\frac{1}{8}\,\Pg_+^{\WSa\WSb}\,\ii\,\brTheta_1\,e'_{\WSa}{}^{\Loa}\,\Gamma_{\Loa}\,\bisF_5\,\Omega^{-1}\, e'_{\WSb}{}^{\Lob}\,\Gamma_{\Lob}\,\Theta'_2 \biggr]\,,
\label{eq:ScYB2-2}
\end{align}
where the derivatives $D'_{\pm}$ are defined as
\begin{align}
\begin{split}
 D'_+ &\equiv D_+ = \rmd + \frac{1}{4}\,W_+^{\Loa\Lob}\,\Gamma_{\Loa\Lob}\,,
\\
 D'_- &\equiv \Omega\circ D_-\circ \Omega^{-1} = \rmd + \frac{1}{4}\,W_-^{\Loa\Lob}\,\Omega\,\Gamma_{\Loa\Lob}\,\Omega^{-1} + \Omega\,\rmd \Omega^{-1} 
\\
 &= \rmd + \frac{1}{4}\,\bigl[\Lambda^{\Loa}{}_{\Loc}\,\Lambda^{\Lob}{}_{\Lod}\,W_-^{\Loc\Lod} +(\Lambda\,\rmd\Lambda^{-1})^{\Loa\Lob}\bigr] \,\Gamma_{\Loa\Lob} \,. 
\end{split}
\end{align}
As we show in Appendix \ref{app:torsionful-spin-connections}, the spin connection $\omega'^{\Loa\Lob}$ associated with the deformed vielbein $e'^{\Loa}$ and the deformed $H$-flux $H'_{\Loa\Lob\Loc}$ satisfy
\begin{align}
\begin{split}
 \omega'^{\Loa\Lob}+\frac{1}{2}\,e'_{\Loc}\,H'^{\Loc\Loa\Lob} &=W_{+}^{\Loa\Lob}\,,
\\
 \omega'^{\Loa\Lob}-\frac{1}{2}\,e'_{\Loc}\,H'^{\Loc\Loa\Lob} &= \Lambda^{\Loa}{}_{\Loc}\,\Lambda^{\Lob}{}_{\Lod}\,W_-^{\Loc\Lod} +(\Lambda\,\rmd\Lambda^{-1})^{\Loa\Lob} \,,
\end{split}
\label{eq:torsionful-spin}
\end{align}
and $D'_{\pm}$ can be expressed as
\begin{align}
 D'_{\pm} = \rmd + \frac{1}{4}\,\Bigl(\omega'^{\Loa\Lob}\pm\frac{1}{2}\,e'_{\Loc}\,H'^{\Loc\Loa\Lob}\Bigr)\,\Gamma_{\Loa\Lob}\,.
\end{align}
Then, the deformed action \eqref{eq:ScYB2-2} becomes the conventional GS action at order $\cO(\theta^2)$ by identifying the deformed R--R field strengths as
\begin{align}
 \bisF' =\bisF_5\,\Omega^{-1} \,. 
\label{eq:YBRR0}
\end{align}
The transformation rule \eqref{eq:YBRR0} has originally been given in \cite{Borsato:2016ose} but our expression of $\Omega$ \eqref{eq:omega} may be more useful to recognize the homogeneous YB deformations as $\beta$-twists. 
Indeed, \eqref{eq:YBRR0} is precisely the $\beta$-transformation rule of the R--R field strengths \eqref{eq:beta-F-transf}. 
Another evidence for the equivalence between YB deformations and local $\beta$-deformations, based on the $\kappa$-symmetry variations, is given in Appendix \ref{app:kappa}. 

Finally, let us consider the remaining part $\delta S_{(2)}$\,. 
This is completely canceled by redefining the bosonic fields $X^m$ \cite{Arutyunov:2015qva,Kyono:2016jqy},
\begin{align}
 X^m\ \to\ X^m + \frac{\eta}{4}\,\sigma_1^{IJ}\,e^{\Loc m}\,\lambda_{\Loc}{}^{\Loa\Lob}\,\brtheta_I\,\gamma_{\Loa\Lob}\,\theta_J + \cO(\theta^4)\,,
\label{eq:bosonic-shift}
\end{align}
as long as the $r$-matrix satisfies the homogeneous CYBE. 
Indeed, this redefinition gives a shift $S_{(0)} \to S_{(0)} + \delta S_{(0)}$\,, and as explained in Appendix \ref{app:bosonic-shift}, the deviation $\delta S_{(0)}$ satisfies a quite simple expression (the shift of $S_{(2)}$ is higher order in $\theta$)
\begin{align}
 &\delta S_{(0)}+\delta S_{(2)} 
\nn\\
 &=\frac{\eta^2\,\dlT}{2}\int \rmd^2\sigma\,\sqrt{-\gga}\, \Pg_-^{\WSa\WSb}\,\sigma_1^{IJ}\,
  \bigl[\CYBE^{(0)}_g\bigl(J_{+m}^{(2)},J_{-n}^{(2)}\bigr)\bigr]^{\Loa\Lob}\,\brtheta_I\, \gamma_{\Loa\Lob}\,\theta_J\,\partial_{\WSa}X^{m}\,\partial_{\WSb}X^n \,,
\end{align}
where $\CYBE^{(0)}_g (X,Y)$ represents the grade-$0$ component of $\CYBE_g (X,Y)$ defined in \eqref{eq:CYBE-g}. 
This shows that $\delta S_{(2)}$ is completely cancelled out by $\delta S_{(0)}$ when the $r$-matrix satisfies the homogeneous CYBE.

\section{\texorpdfstring{$\beta$}{\textbeta}-deformations with \texorpdfstring{$H$}{H}-flux: \texorpdfstring{$\AdS3\times \rmS^3\times \TT^4$}{AdS\textthreeinferior{x}S\textthreesuperior{x}T\textfoursuperior}}
\label{sec:AdS3}

In the previous section, we have shown that the YB sigma model on the $\AdS{5} \times \rmS^5$ background associated with an $r$-matrix $r=\frac{1}{2}\,r^{ij}\,T_i\wedge T_j$ can be regarded as the GS superstring theory defined on a $\beta$-deformed $\AdS{5} \times \rmS^5$ background with the $\beta$-deformation parameter $\bmr^{mn}=-2\,\eta\,r^{ij}\,\hat{T}_i^m\,\hat{T}_j^n$\,. 
The same conclusion will hold also for other backgrounds in string theory. 

In this section, we study deformations of an AdS background with $H$-flux. 
In the presence of $H$-flux, it is not straightforward to define the YB sigma model, and we shall concentrate only on $\beta$-deformations. 
As an example, we here consider the $\AdS3\times \rmS^3\times \TT^4$ solution
\begin{align}
\begin{split}
 &\rmd s^2 = \frac{-(\rmd x^0)^2+(\rmd x^1)^2+\rmd z^2}{z^2} + \rmd s^2_{\rmS^3} + \rmd s_{\TT^4}^2 \,,
\\
 &B_2= \frac{\rmd x^0\wedge \rmd x^1}{z^2} + \frac{1}{4}\,\cos\theta\,\rmd \phi \wedge \rmd \psi \,,\qquad \Phi=0\,,
\\
 &\rmd s^2_{\rmS^3} \equiv \frac{1}{4}\,\bigl[\rmd \theta^2 + \sin^2\theta\, \rmd \phi^2 + \bigl(\rmd \psi + \cos\theta\,\rmd \phi)^2\bigr] \,,
\end{split}
\label{eq:AdS3-S3-T4}
\end{align}
which contains the non-vanishing $H$-flux
\begin{align}
 H_3 =-2\,\frac{\rmd x^0\wedge \rmd x^1 \wedge \rmd z}{z^3}-\frac{1}{4}\,\sin\theta\rmd\phi \wedge \rmd \psi \wedge \rmd \theta\,. 
\end{align}
Using the Killing vectors $\hat{T}_i$ of the $\AdS3\times \rmS^3\times \TT^4$ background, we consider local $\beta$-deformations with deformation parameters of the form, $\bmr^{mn}=-2\,\eta\,r^{ij}\,\hat{T}_i^m\,\hat{T}_j^n$\,. 
We consider several $r$-matrices $r^{ij}$ satisfying the homogeneous CYBE, and show that all of the $\beta$-deformed backgrounds satisfy the equations of motion of (generalized) supergravity. 

In order to find the Killing vectors explicitly, we introduce a group parameterization for $\AdS3\times \rmS^3$ (for simplicity, we do not consider the trivial $\TT^4$ directions)
\begin{align}
\begin{split}
 &g =g_{\AdS{3}}\cdot g_{\rmS^3}\cdot g_{\TT^4}\,, \qquad
 g_{\AdS{3}} =\exp(x^{\mu} P_{\mu})\cdot\exp(\ln z\,D)\qquad (\mu=0,1)\,,
\\
 &g_{\rmS^3} =\exp(\phi\,T^L_4)\cdot\exp(\theta\,T^L_3)\cdot\exp(\psi\,T^R_4) \,. 
\end{split}
\end{align}
Here, similar to the $\AdS{5}\times\rmS^5$ case (see Appendix \ref{app:psu-algebra}), we have introduced the $\alg{so}(2,2)\times \alg{so}(4)$ generators $(\gP_{\check{\Loa}}\,, \gP_{\hat{\Loa}}\,,\gJ_{\check{\Loa}\check{\Lob}}\,, \gJ_{\hat{\Loa}\hat{\Lob}})$ $(\check{\Loa},\,\check{\Lob}=0,1,2;\,\hat{\Loa},\,\hat{\Lob}=3,4,5)$ as the following $8\times 8$ supermatrices:
\begin{align}
\begin{alignedat}{3}
 \gP_{\check{\Loa}} &= 
 \begin{pmatrix}
  \frac{1}{2}\,\bm{\gamma}_{\check{\Loa}} & \bm{0_4} \\ 
  \bm{0_4} & \bm{0_4} 
 \end{pmatrix},& \qquad 
  \gJ_{\check{\Loa}\check{\Lob}} &= 
 \begin{pmatrix}
  -\frac{1}{2}\,\bm{\gamma}_{\check{\Loa}\check{\Lob}} & \bm{0_4} \\ 
  \bm{0_4} & \bm{0_4} 
 \end{pmatrix},& \qquad
 \bm{\gamma}_{\check{\Loa}} &\equiv \begin{pmatrix}
  +\gamma_{\check{\Loa}}& \bm{0_2} \\
  \bm{0_2} &-\gamma_{\check{\Loa}}
 \end{pmatrix},
\\
 \gP_{\hat{\Loa}} &= 
 \begin{pmatrix}
  \bm{0_4} & \bm{0_4} \\ 
  \bm{0_4} & -\frac{\ii}{2}\,\bm{\gamma}_{\hat{\Loa}} 
 \end{pmatrix}& , \qquad 
 \gJ_{\hat{\Loa}\hat{\Lob}} &= 
 \begin{pmatrix}
  \bm{0_4} & \bm{0_4} \\
  \bm{0_4} & -\frac{1}{2}\,\bm{\gamma}_{\hat{\Loa}\hat{\Lob}} 
 \end{pmatrix}& , \qquad 
 \bm{\gamma}_{\hat{\Loa}}&\equiv
 \begin{pmatrix}
  -\gamma_{\hat{\Loa}} & \bm{0_2} \\
  \bm{0_2} &+\gamma_{\hat{\Loa}}
 \end{pmatrix}, 
\end{alignedat}
\end{align}
where $2\times 2$ gamma matrices $\gamma_{\check{\Loa}}$ and $\gamma_{\hat{\Loa}}$ are defined as
\begin{align}
 \{\gamma_0,\,\gamma_1,\,\gamma_2\} = \{\ii\,\sigma_3\,, \sigma_1\,, \sigma_2\}\,,\qquad
 \{\gamma_3,\,\gamma_4,\,\gamma_5\} = \{\sigma_1\,, \sigma_2\,, \sigma_3\} \,.
\end{align}
We have also defined the conformal basis $\{P_{\mu}\,, M_{\mu\nu}\,, D\,, K_{\mu}\}$ as
\begin{align}
 P_{\mu}\equiv \gP_{\mu}+\gJ_{\mu2}\,,\qquad 
 K_{\mu}\equiv \gP_{\mu}-\gJ_{\mu2}\,,\qquad
 M_{\mu\nu}\equiv \gJ_{\mu\nu}\,,\qquad 
 D\equiv \gP_{2}\,.
\end{align}
The generators of $\alg{su}(2)_L\times \alg{su}(2)_R\simeq \alg{so}(4)$ are defined as
\begin{align}
\begin{split}
 T^L_3&=\frac{1}{2}\,(\gP_3-\gJ_{4,5})\,,\qquad
 T^R_3=\frac{1}{2}\,(\gP_3+\gJ_{4,5})\,,
\\
 T^L_4&=\frac{1}{2}\,(\gP_4-\gJ_{5,3})\,,\qquad
 T^R_4=\frac{1}{2}\,(\gP_4+\gJ_{5,3})\,,
\\
 T^L_5&=\frac{1}{2}\,(\gP_5-\gJ_{3,4})\,,\qquad
 T^R_5=\frac{1}{2}\,(\gP_5+\gJ_{3,4})\,,
\end{split}
\end{align}
which satisfy the commutation relations,
\begin{align}
 [T^{L}_{i},\,T^{L}_{j}]=-\epsilon_{ijk}\,T^{L}_{k}\,,\qquad [T^{R}_{i},\,T^{R}_{j}]=\epsilon_{ijk}\,T^{R}_{k}\,,\qquad [T^{L}_{i},\,T^{R}_{j}]=0\quad 
 (\epsilon_{345}=1)\,.
\end{align}
By computing the Maurer--Cartan 1-form $A=g^{-1}\,\rmd g$, and using the supertrace formula
\begin{align}
\begin{alignedat}{2}
 &\str(\gP_{\Loa}\,\gP_{\Lob}) =\eta_{\Loa\Lob}\,,&\qquad &\str(\gJ_{\Loa\Lob}\,\gJ_{\Loc\Lod})=R_{\Loa\Lob\Loc\Lod} \,,
\\
 &R_{\check{\Loa}\check{\Lob}}{}^{\check{\Loc}\check{\Lod}} \equiv -2\, \delta_{[\check{\Loa}}^{[\check{\Loc}}\,\delta_{\check{\Lob}]}^{\check{\Lod}]}\,,& \qquad 
 &R_{\hat{\Loa}\hat{\Lob}}{}^{\hat{\Loc}\hat{\Lod}} \equiv 2\, \delta_{[\hat{\Loa}}^{[\hat{\Loc}}\,\delta_{\hat{\Lob}]}^{\hat{\Lod}]} \,,
\end{alignedat}
\end{align}
we can reproduce the above metric \eqref{eq:AdS3-S3-T4}. 

Then, we can find the Killing vectors $\hat{T}_i$ of this background associated with the generator $T_i$ by using the formula \eqref{eq:Killing-Formula}, or $\hat{T}_i^m = \str\bigl[ g^{-1}\,T^i\,g\,\gP_{\Loa} \bigr]\,e^{\Loa m}$\,. 
The result is summarized as
\begin{align}
\begin{split}
 \hat{P}_{\mu}&=\partial_{\mu}\,,\qquad
 \hat{M}_{\mu\nu}=x_{\mu}\,\partial_{\nu}-x_{\nu}\,\partial_{\mu}\,,\qquad
 \hat{D}=x^{\mu}\,\partial_{\mu}+z\,\partial_z\,,
\\
 \hat{K}_{\mu}& =(x^{\nu}\,x_{\nu}+z^2)\,\partial_{\mu}-2\,x_{\mu}\,(x^{\nu}\,\partial_{\nu}+z\,\partial_z)\,,
\\
 \hat{T}^L_3& =\cos\phi\,\partial_\theta+\sin\phi\,\Bigl(-\frac{1}{\tan\theta}\,\partial_\phi+\frac{1}{\sin\theta}\,\partial_{\psi}\Bigr) \,,
\\
 \hat{T}^R_3& =\cos\psi\,\partial_\theta+\sin\psi\,\Bigl(\frac{1}{\sin\theta}\,\partial_\phi-\frac{1}{\tan\theta}\,\partial_{\psi}\Bigr)\,,
\\
 \hat{T}^L_4&=\partial_{\phi}\,,\qquad
 \hat{T}^R_4=\partial_{\psi}\,,
\\
 \hat{T}^L_5& =\sin\phi\,\partial_\theta+\cos\phi\,\Bigl(\frac{1}{\tan\theta}\,\partial_\phi-\frac{1}{\sin\theta}\,\partial_{\psi}\Bigr) \,,
\\
 \hat{T}^R_5&=-\sin\psi\,\partial_\theta+\cos\psi\,\Bigl(\frac{1}{\sin\theta}\,\partial_\phi-\frac{1}{\tan\theta}\,\partial_{\psi}\Bigr) \,.
\end{split}
\end{align}

We note that among the AdS isometries, $\hat{P}_{\mu}$, $\hat{M}_{01}$, and $\hat{D}$ are symmetry of the $B$-field,
\begin{align}
 \Lie_{\hat{P}_{\mu}}B_2 =\Lie_{\hat{M}_{01}}B_2 =\Lie_{\hat{D}}B_2=0\,,
\end{align}
while the special conformal generators $\hat{K}_{\mu}$ change the $B$-field by closed forms,
\begin{align}
 \Lie_{\hat{K}_{0}}B_2 = -\frac{2\,\rmd x^1\wedge \rmd z}{z}\,,\qquad
 \Lie_{\hat{K}_{1}}B_2 = \frac{2\,\rmd x^0\wedge \rmd z}{z}\,.
\label{eq:K-B2-closed}
\end{align}
In the following, we first study $\beta$-deformations by using Killing vectors $\hat{P}_{\mu}$, $\hat{M}_{01}$, $\hat{D}$, and $T^R_4$\,. 
Then, non-trivial cases using the Killing vectors $\hat{K}_{\mu}$ are studied in section \ref{sec:generalized-isometries}. 

\vspace{-\Pskip}
\subsection{Abelian deformations}

Let us begin by studying simple examples associated with Abelian $r$-matrices. 
As it has been known well \cite{Matsumoto:2014nra,Matsumoto:2014gwa,Matsumoto:2015uja,vanTongeren:2015soa,vanTongeren:2015uha,Kyono:2016jqy,Osten:2016dvf}, YB deformations associated with Abelian $r$-matrices can be also realized as TsT-transformations. 

\subsubsection{\texorpdfstring{$r=\frac{1}{2}\,P_0\wedge P_1$}{r=P\textzeroinferior{\textwedge}P\textoneinferior}}

Let us first consider an Abelian $r$-matrix
\begin{align}
 r=\frac{1}{2}\,P_0\wedge P_1 \,.
\end{align}
From $\hat{P}_0=\partial_0$ and $\hat{P}_1=\partial_1$, the $\beta$-transformation parameter is $\bmr^{mn}= -\eta\,(\delta_0^{m}\,\delta_1^{n}-\delta_1^{m}\,\delta_0^{n})$\,. 
After the $\beta$-transformation, we obtain the background
\begin{align}
\begin{split}
 \rmd s^2&=\frac{-(\rmd x^0)^2+(\rmd x^1)^2}{z^2+2\,\eta}+\frac{\rmd z^2}{z^2} +\rmd s_{\rmS^3}^2+\rmd s_{\TT^4}^2\,,
\\
 B_2&=\frac{\rmd x^0\wedge \rmd x^1}{z^2+2\,\eta}+\frac{1}{4}\,\cos \theta\,\rmd \phi\wedge \rmd \psi\,,\qquad 
 \Exp{-2\Phi} = \frac{z^2+2\,\eta}{z^2} \,.
\end{split}
\end{align}
As we have mentioned above, we can also obtain the background by a TsT transformation from the background \eqref{eq:AdS3-S3-T4}; (1) T-dualize along the $x^1$-direction, (2) (active) shift $x^0\to x^0+\eta\,x^1$\,, (3) T-dualize along the $x^1$-direction.
This background is of course a solution of supergravity. 

As a side remark, noted that this background interpolates a linear dilaton background in the UV region ($z \sim 0$) and the undeformed $\AdS{3} \times \rmS^3\times \TT^4$ background in the IR region ($z \to \infty$). 
Indeed, by performing a coordinate transformation
\begin{align}
 x^{\pm}=\frac{x^0\pm x^1}{\sqrt{2}} \,,\qquad z=\Exp{\rho}\,,
\end{align}
the deformed background becomes
\begin{align}
\begin{split}
 \rmd s^2&=-\frac{2\Exp{-2\rho}}{1+2\,\eta\Exp{-2\rho}}\,\rmd x^+\,\rmd x^- + \rmd \rho^2 +\rmd s_{\rmS^3}^2+\rmd s_{\TT^4}^2\,,\qquad
 \Exp{-2\Phi} = 1+2\,\eta\,\Exp{-2\rho}\,,
\\
 B_2&=-\frac{\Exp{-2\rho}}{1+2\,\eta\Exp{-2\rho}}\,\rmd x^+\wedge\rmd x^- + \frac{1}{4}\,\cos \theta\,\rmd \phi\wedge \rmd \psi \,.
\label{eq:TT-AdS}
\end{split}
\end{align}
In the asymptotic region $\Exp{-2\rho} \gg \eta^{-1}$ (i.e.~$z\sim 0$), the background approaches to a solution that is independent of the deformation parameter $\eta$
\begin{align}
\begin{split}
 \rmd s^2&=-2\,\rmd x^+\rmd x^-+\rmd \rho^2 +\rmd s_{\rmS^3}^2+\rmd s_{\TT^4}^2\,,\qquad 
 \Phi =\rho\,,
\\
 B_2&=-\rmd x^+\wedge\rmd x^-+\frac{1}{4}\cos \theta\,\rmd \phi\wedge \rmd \psi \,,
\end{split}
\label{eq:TT-AdS-limit}
\end{align}
where we ignored the constant part of the dilaton and rescaled light-cone coordinates $x^{\pm}$ as
\begin{align}
x^{\pm}\to \sqrt{2\,\eta}\,x^{\pm}\,.
\end{align}
The $\AdS{3}$ part of the background \eqref{eq:TT-AdS} is precisely the geometry obtained via a null deformation of $\SL(2,\mathbb{R})$ WZW model \cite{Forste:1994wp} (see also \cite{Israel:2003ry}), which is an exactly marginal deformation of the WZW model (see \cite{Giveon:2017nie,Giveon:2017myj,Asrat:2017tzd,Giribet:2017imm} for recent studies). 
Note also that, under a formal $T$-duality along the $\rho$-direction, the solution \eqref{eq:TT-AdS-limit} becomes the following solution in DFT:
\begin{align}
\begin{split}
 \rmd s^2&=-2\,\rmd x^+\rmd x^-+\rmd \rho^2 +\rmd s_{\rmS^3}^2+\rmd s_{\TT^4}^2\,,\qquad 
 \Phi =\tilde{\rho}\,,
\\
 B_2&=-\rmd x^+\wedge\rmd x^-+\frac{1}{4}\cos \theta\,\rmd \phi\wedge \rmd \psi \,,
\end{split}
\end{align}
where the dilaton depends linearly on the dual coordinate $\tilde{\rho}$\,. 
This background can be also interpreted as the following solution of GSE:
\begin{align}
\begin{split}
 \rmd s^2&=-2\,\rmd x^+\rmd x^-+\rmd \rho^2 +\rmd s_{\rmS^3}^2+\rmd s_{\TT^4}^2\,,\qquad \Phi=0\,,
\\
 B_2&=-\rmd x^+\wedge\rmd x^-+\frac{1}{4}\cos \theta\,\rmd \phi\wedge \rmd \psi \,,\qquad I =\partial_\rho \,. 
\end{split}
\end{align}

\subsubsection{\texorpdfstring{$r=\frac{1}{2}\,P_+\wedge T^R_4$}{r=P\textplusinferior{\textwedge}T\textfourinferior R}}

As the second example, let us consider an Abelian $r$-matrix
\begin{align}
 r = \frac{1}{2}\,P_+\wedge T^R_4\qquad \Bigl(P_+ \equiv \frac{P_0 + P_1}{\sqrt{2}}\Bigr)\,.
\end{align}
For convenience, let us change the coordinates such that the background \eqref{eq:AdS3-S3-T4} becomes
\begin{align}
\begin{split}
 \rmd s^2&=-2\Exp{2\rho} \rmd x^+\,\rmd x^- + \rmd \rho^2 +\rmd s_{\rmS^3}^2+\rmd s_{\TT^4}^2\,,
 \qquad \Phi =0\,,
\\
 B_2&=-\Exp{2\rho}\rmd x^+\wedge\rmd x^- + \frac{1}{4}\,\cos \theta\,\rmd \phi\wedge \rmd \psi \,.
\end{split}
\end{align}
In this coordinate system, the Killing vectors take the form, $\hat{P}_+=\partial_+$ and $\hat{T}^R_4=\partial_\psi$. 
Then, the associated $\beta$-deformed (or TsT-transformed) background is given by
\begin{align}
\begin{split}
 \rmd s^2&=-2\Exp{2\rho} \rmd x^+\,\rmd x^- + \rmd \rho^2 +\frac{\eta}{2}\Exp{2\rho} \rmd x^-\,(\rmd\psi+2\,\cos\theta\,\rmd\phi) +\rmd s_{\rmS^3}^2+\rmd s_{\TT^4}^2\,,
 \quad \Phi =0\,,
\\
 B_2&=-\Exp{2\rho}\rmd x^+\wedge\rmd x^- + \frac{1}{4}\,\cos \theta\,\rmd \phi\wedge \rmd \psi -\frac{\eta}{4}\,\Exp{2\rho}\rmd x^-\wedge (\rmd \psi+2\,\cos\theta\,\rmd\phi) \,.
\end{split}
\end{align}
This background has been studied in \cite{Azeyanagi:2012zd}, where the twist was interpreted as a spectral flow transformation of the original model in the context of the NS--R formalism. 

\subsubsection{\texorpdfstring{$r = \frac{1}{2}\,D\wedge M_{01}$}{r=D{\textwedge}M\textzeroinferior\textoneinferior}}
\label{sec:AdS3-DwM}

Let us also consider a slightly non-trivial example $r = \frac{1}{2}\,D\wedge M_{01}$, which is also an Abelian $r$-matrix. 
The associated $\beta$-deformed background is given by
\begin{align}
\begin{split}
 \rmd s^2&= \frac{\eta_{\mu\nu}\,\rmd x^\mu\,\rmd x^\nu - 2\,\eta\,z^{-1}\,x_\mu\,\rmd x^\mu\, \rmd z +(1+\frac{2\,\eta\,x_\mu\,x^\mu}{z^2}) \,\rmd z^2}{z^2-\eta\,(\eta-2)\,x_\mu\,x^\mu} 
 + \rmd s^2_{\rmS^3} + \rmd s_{\TT^4}^2 \,,
\\
 \Exp{-2\Phi} &= \frac{z^2-\eta\,(\eta-2)\,x_\mu\,x^\mu}{z^2}\,,
\\
 B_2 &= \frac{\rmd x^0\wedge\rmd x^1-\eta\,z^{-1}\, (x^1\,\rmd x^0-x^0\,\rmd x^1)\wedge \rmd z}{z^2-\eta\,(\eta-2)\,x_\mu\,x^\mu} + \frac{1}{4}\,\cos\theta\,\rmd \phi \wedge \rmd \psi \,. 
\end{split}
\end{align}
We can easily check that this is a solution of the supergravity. 
In order to obtain the same background by performing a TsT transformation, we should first change the coordinates such that the Killing vectors $\hat{D}$ and $\hat{M}_{01}$ become constant, and perform a TsT transformation, and then go back to the original coordinates. 
The $\beta$-transformation is much easier in this case.

In order to describe the same $\beta$-deformation in the global coordinates
\begin{align}
\begin{split}
 \rmd s^2&=-\cosh^2\rho\,\rmd \tau^2+\sinh^2\rho\,\rmd \chi^2+\rmd\rho^2 +\rmd s_{\rmS^3}^2+\rmd s_{\TT^4}^2\,,\qquad \Phi=0\,,
\\
 B_2&=\cosh^2\rho\,\rmd\tau\wedge \rmd \chi +\frac{1}{4}\,\cos \theta\,\rmd \phi\wedge \rmd \psi\,,
\end{split}
\end{align}
we change the group parameterization as
\begin{align}
 g_{\AdS{3}}= \exp(\ii\,\tau\, D+\ii\,\chi\, M_{01})\cdot\exp(\rho\, \gP_1)\,.
\end{align}
In this case, we can compute the Killing vectors as
\begin{align}
 \hat{D}=-\ii\,\partial_{\tau}\,,\qquad \hat{M}_{01}=-\ii\,\partial_{\chi}\,.
\end{align}
Then, the $\beta$-deformed background becomes
\begin{align}
\begin{split}
 \rmd s^2 
 &=\frac{-\cosh^2\rho\,\rmd \tau^2+\sinh^2\rho\,\rmd \chi^2}{1+\eta\,(\eta-2)\,\cosh^2\rho}+\rmd\rho^2
 +\rmd s_{\rmS^3}^2+\rmd s_{\TT^4}^2\,,
\\
 \Phi&=\frac{1}{2}\,\ln\Bigl[\frac{1}{1+\eta(\eta-2)\cosh^2\rho}\Bigr] \,,
\\
 B_2&=(1-\eta)\,\frac{\cosh^2\rho\,\rmd\tau\wedge \rmd \chi}{1+\eta\,(\eta-2)\,\cosh^2\rho}
 +\frac{1}{4}\,\cos \theta\,\rmd \phi\wedge \rmd \psi \,.
\end{split}
\end{align}
If the deformation parameter $\eta$ and the angular coordinate $\chi$ are replaced as
\begin{align}
 \eta\to 1-\sqrt{\alpha}\,,\qquad \chi\to \sqrt{\alpha}\, \chi\,,
\end{align}
the AdS part of this background reproduce the background obtained in \cite{Giveon:1993ph,Israel:2003ry} through a current-current deformation of the $\SL(2,\mathbb{R})$ WZW model (see Eqs.~(5.1)--(5.3) in \cite{Israel:2003ry}). 

\vspace{-\Pskip}
\subsection{Non-unimodular deformations}
\label{sec:non-unimodular}

Let us next consider $\beta$-deformations associated with non-Abelian $r$-matrices. 
In particular, we consider non-unimodular $r$-matrices, namely non-Abelian $r$-matrices satisfying
\begin{align}
 \cI \equiv \eta\,r^{ij}\,[T_i,\,T_j] = \eta\,r^{ij}\,f_{ij}{}^k\,T_k \neq 0 \,. 
\label{eq:cI-formula}
\end{align}
In general, as was shown in \cite{Borsato:2016ose}, YB deformations associated with non-unimodular $r$-matrices give backgrounds that do not satisfy the usual supergravity equations but rather the GSE \cite{Arutyunov:2015mqj,TW,Sakatani:2016fvh,Baguet:2016prz,Sakamoto:2017wor}, which include non-dynamical Killing vector $I^m$ (see Appendix \ref{app:conventions}). 
As it was observed experimentally \cite{Araujo:2017jkb,Araujo:2017jap,Araujo:2017enj,Fernandez-Melgarejo:2017oyu}, the extra vector $I^m$ typically takes the form (see Appendix \ref{app:conventions} for a derivation in the case of the $\AdS{5}\times \rmS^5$ superstring)
\begin{align}
\begin{split}
 &I^m = \hat{\cI}^m \equiv - \eta\,r^{ij}\,[\hat{T}_i,\,\hat{T}_j]^m = \eta\,r^{ij}\,f_{ij}{}^k\,\hat{T}_k^m 
\\
 &\quad\bigl(\,[\hat{T}_i,\,\hat{T}_j]^m = \Lie_{\hat{T}_i}\hat{T}_j^m= - f_{ij}{}^k\,\hat{T}_k^m \,\bigr)\,. 
\end{split}
\label{eq:experimental}
\end{align}
Using $\bmr^{mn} = -2\,\eta\,r^{ij}\,\hat{T}_i^m\, \hat{T}_j^n$ obtained in \eqref{eq:bmr-YB2} and the Killing property of $\hat{T}_i$, we can also express the experimental formula as \cite{Araujo:2017jkb,Araujo:2017jap,Sakamoto:2017cpu,Araujo:2017enj,Fernandez-Melgarejo:2017oyu}
\begin{align}
 I^m = \sfD_n \bmr^{nm} \,,
\end{align}
where $\sfD_n$ is the usual covariant derivative associated with the undeformed $\AdS3\times \rmS^3\times \TT^4$ background. 

Interestingly, as we explain in section \ref{sec:AdS3-S3-T4MwedgeP}, in some examples, even for non-unimodular $r$-matrices, the $\beta$-deformed backgrounds satisfy the usual supergravity equations of motion. 
Such example has not been observed in the case of the $\AdS5\times \rmS^5$ background,
\footnote{See a recent paper \cite{Wulff:2018aku} for a general analysis of such backgrounds, called the ``trivial solutions'' of GSE.} 
and this is due to a particular property of the $\AdS3\times \rmS^3\times \TT^4$ background as explained below. 

\subsubsection{\texorpdfstring{$r = \frac{1}{2}\,\bar{c}^\mu\,D\wedge P_\mu$}{r=D{\textwedge}P}}

Let us consider the simplest non-unimodular $r$-matrix $r = \frac{1}{2}\,\bar{c}^\mu\,D\wedge P_\mu$\,, satisfying
\begin{align}
 \cI = \eta\,\bar{c}^\mu\,[D,\,P_\mu] = c^\mu P_\mu \neq 0 \qquad \bigl(c^\mu\equiv\eta\,\bar{c}^\mu\bigr)\,.
\end{align}
The $\beta$-deformed background becomes
\begin{align}
\begin{split}
 \rmd s^2&= \frac{\eta_{\mu\nu}\,\rmd x^\mu\,\rmd x^\nu + 2\,z^{-1}\,(c^0\,\rmd x^1 -c^1\,\rmd x^0)\,\rmd z + \bigl[1+2\,z^{-2}\,(c^1\,x^0-c^0\,x^1)\bigr]\,\rmd z^2}{z^2 + c_\mu\,c^\mu + 2\,(c^1\,x^0 - c^0 \,x^1)} 
\\
 &\quad + \rmd s^2_{\rmS^3} + \rmd s_{\TT^4}^2 \,,\qquad 
 \Exp{-2\Phi} = \frac{z^2 + c_\mu\,c^\mu + 2\,(c^1\,x^0 - c^0 \,x^1)}{z^2}\,,
\\
 B_2 &= \frac{\rmd x^0\wedge\rmd x^1 - z^{-1}\, c_\mu\,\rmd x^\mu\wedge \rmd z}{z^2 + c_\mu\,c^\mu + 2\,(c^1\,x^0-c^0\,x^1)} + \frac{1}{4}\,\cos\theta\,\rmd \phi \wedge \rmd \psi \,,
\end{split}
\end{align}
where $c_\mu \equiv \eta_{\mu\nu}\,c^\nu$\,. 
Although this is not a solution of the usual supergravity, by introducing a Killing vector,
\begin{align}
 I = c^\mu \hat{P}_\mu = c^\mu\,\partial_\mu \,,
\end{align}
it becomes a solution of the GSE.

\subsubsection{\texorpdfstring{$r = \frac{1}{2}\,\bar{c}^\mu\,M_{01}\wedge P_\mu$}{r=M\textzeroinferior\textoneinferior{\textwedge}P}}
\label{sec:AdS3-S3-T4MwedgeP}

The next example is a non-unimodular $r$-matrix $r = \frac{1}{2}\,\bar{c}^\mu\,M_{01}\wedge P_\mu$\,, satisfying
\begin{align}
 \cI = - c^\mu P_\mu \neq 0 \qquad \bigl(c^\mu\equiv\eta\,\bar{c}^\mu\bigr)\,.
\end{align}
The $\beta$-deformed background becomes
\begin{align}
\begin{split}
 \rmd s^2&= \frac{\eta_{\mu\nu}\,\rmd x^\mu\,\rmd x^\nu}{z^2 - 2\, c_\mu\,x^\mu} + \frac{\rmd z^2}{z^2} + \rmd s^2_{\rmS^3} + \rmd s_{\TT^4}^2 \,,\qquad 
 \Exp{-2\Phi} = \frac{z^2 - 2\, c_\mu\,x^\mu}{z^2}\,,
\\
 B_2 &= \frac{\rmd x^0\wedge \rmd x^1}{z^2 - 2\, c_\mu\,x^\mu} + \frac{1}{4}\,\cos\theta\,\rmd \phi \wedge \rmd \psi \,. 
\label{eq:AdS3_M^P}
\end{split}
\end{align}
where $c_\mu \equiv \eta_{\mu\nu}\,c^\nu$\,. 
As usual, by introducing
\begin{align}
 I = - c^\mu \hat{P}_\mu = -c^\mu\,\partial_\mu \,,
\end{align}
this background satisfies the GSE. 

Here, note that the defining properties of $I^m$,
\begin{align}
 \Lie_{I} \CG_{mn} = 0 \,,\qquad \Lie_{I} B_{mn} = 0 \,,
\end{align}
require that the parameters should satisfy $c^0=\pm c^1$\,. 
In terms of DFT, the above deformed background can be expressed as
\begin{align}
 (\cH_{MN})= \begin{pmatrix}
 (\CG-B\,\CG^{-1}\,B)_{mn} & (B\,\CG^{-1})_{m}{}^n \\
 -(\CG^{-1}\,B)^m{}_n & \CG^{mn} \end{pmatrix},\qquad 
 d = \Phi -\frac{1}{2}\,\ln \sqrt{-\CG} + I^\mu\,\tilde{x}_\mu \,. 
\end{align}
This solves the equations of motion of DFT for arbitrary parameters $c^\mu$\,, but they satisfy the strong constraint
\begin{align}
 \partial_P \cH_{MN}\,\partial^Pd=0\,,
\end{align}
only when $c^0=\pm c^1$\,. 
Therefore, we have to choose $c^0=\pm c^1$\,. 

In fact, this background has a distinctive feature that has not been observed before. 
According to the classification of \cite{Borsato:2016ose}, the condition for a YB-deformed background to be a standard supergravity background is the unimodularity condition. 
However, in this example, the background \eqref{eq:AdS3_M^P} satisfies the GSE even if we perform a rescaling $I^m \to \lambda\,I^m$ with arbitrary $\lambda\in\mathbb{R}$. 
In particular, by choosing $\lambda=0$, the background \eqref{eq:AdS3_M^P} without $I^m$ satisfies the usual supergravity equations of motion. 
As we explain below, the reason for the unusual behavior is closely related to the degeneracy of $(\CG \pm B)_{mn}$. 

According to \cite{Sakatani:2016fvh}, the condition for a solution of the GSE to be a standard supergravity background is given by
\begin{align}
 \gLie_{\bm{Y}} \cH_{MN} = 0\,, \qquad 
 \cH_{MN}\,\bm{Y}^M\,\bm{Y}^N= \nabla_M \bm{Y}^M \,,
\label{eq:condition-GSE-to-SUGRA}
\end{align}
where $\nabla_M$ is the (semi-)covariant derivative in DFT and
\begin{align}
 \bm{X}^M \equiv \begin{pmatrix} I^m \\ 0 \end{pmatrix},\qquad 
 \bm{Y}^M \equiv \cH^M{}_N\,\bm{X}^N = \begin{pmatrix} -(\CG^{-1}\,B)^m{}_n\,I^n \\ (\CG-B\,\CG^{-1}\,B)_{mn}\,I^n \end{pmatrix} \,. 
\end{align}
In our example with $c^0=\pm c^1$, $(\CG\pm B)_{mn}\,I^n=0$ is satisfied, and this leads to $\bm{Y}^M = \pm \bm{X}^M$\,. 
Then, from the null and generalized Killing properties of $\bm{X}^M$
\begin{align}
 \cH_{MN}\,\bm{X}^M\,\bm{X}^N=0\,,\qquad 
 \gLie_{\bm{X}} \cH_{MN} = 0\,, \qquad 
 \nabla_M \bm{X}^M =0 \,,
\end{align}
the condition \eqref{eq:condition-GSE-to-SUGRA} is automatically satisfied, and our GSE solution is also a solution of the standard supergravity. 
If we regard the background \eqref{eq:AdS3_M^P} as a solution of supergravity, the strong constraint is satisfied for an arbitrary $c_\mu$ and it is not necessary to require $c^0=\pm c^1$. 

\subsubsection{\texorpdfstring{$r = \frac{1}{2}\,\bigl(\bar{a}^\mu D\wedge P_\mu + \bar{b}^\mu M_{01}\wedge P_\mu\bigr)$}{r=a D{\textwedge}P+b M\textzeroinferior\textoneinferior{\textwedge}P}}

As a more general class of $r$-matrices, let us consider
\begin{align}
 r = \frac{1}{2}\,\bigl(\bar{a}^\mu D\wedge P_\mu + \bar{b}^\mu M_{01}\wedge P_\mu\bigr) \,. 
\end{align}
The homogeneous CYBE requires
\begin{align}
 \bar{a}^0\,\bar{b}^1 - \bar{a}^1\,\bar{b}^0 = 0 \,,\qquad -\bar{a}^0\,\bar{b}^0 + a^1\,\bar{b}^1 = 0 \,,
\end{align}
and we consider a non-trivial solution
\begin{align}
 r = \frac{1}{2}\,\bigl(\bar{c}\,D + \bar{d}\,M_{01}\bigr)\wedge (P_0\pm P_1) \,. 
\end{align}
The non-unimodularity becomes
\begin{align}
 \cI = (c-d)\,(P_0\pm P_1) \qquad \bigl(c \equiv\eta\, \bar{c}\,,\quad d\equiv \eta\,\bar{d}\,\bigr)\,.
\end{align}
The corresponding $\beta$-deformed background is given by
\begin{align}
\begin{split}
 \rmd s^2&= \frac{\eta_{\mu\nu}\,\rmd x^\mu\,\rmd x^\nu \mp 2\,c \,z^{-1}\,(\rmd x^0 \mp \rmd x^1)\,\rmd z + \bigl[1\pm 2\,z^{-2}\,(c\pm d)(x^0\mp x^1)\bigr]\,\rmd z^2}{z^2 \pm 2\,(c\pm d)(x^0\mp x^1)} 
\\
 &\quad
 + \rmd s^2_{\rmS^3} + \rmd s_{\TT^4}^2 \,,\qquad 
 \Exp{-2\Phi} = \frac{z^2 \pm 2\,(c\pm d)(x^0\mp x^1)}{z^2}\,,
\\
 B_2 &= \frac{\rmd x^0\wedge\rmd x^1 + c \, z^{-1}\,(\rmd x^0\mp \rmd x^1) \wedge \rmd z}{z^2\pm 2\,(c\pm d)(x^0\mp x^1)} + \frac{1}{4}\,\cos\theta\,\rmd \phi \wedge \rmd \psi \,. 
\label{eq:D+M^P-pm}
\end{split}
\end{align}
By introducing a Killing vector,
\begin{align}
 I = (c-d)\,(\hat{P}_0\pm \hat{P}_1) = (c-d)\,(\partial_0\pm \partial_1) \,,
\end{align}
this background becomes a solution of the GSE. 
In particular, when $c=d$, this becomes a supergravity background. 

Similar to the previous example, the Killing vector again satisfies $(\CG\pm B)_{mn}\,I^n=0$, and even if we rescale the Killing vector as $I^m\to \lambda\,I^m$, this is still a solution of the GSE. 
As a particular case $\lambda=0$, the background \eqref{eq:D+M^P-pm} becomes a solution of the usual supergravity. 

Let us also consider the case,
\begin{align}
 r=\frac{1}{2}\,\bar{a}^{\mu} D\wedge P_{\mu}\qquad \bigl(\text{i.e., }\bar{b}^\mu=0\bigr)\,. 
\end{align}
In this case, the $\beta$-deformed background becomes
\begin{align}
\begin{split}
 \rmd s^2&=\frac{\eta_{\mu\nu}\,\rmd x^\mu\,\rmd x^\nu +2\,z^{-1}\,(a^0\,\rmd x^1-a^1\,\rmd x^0)\,\rmd z +[1+2\,z^{-2}\,(a^1 x^0-a^0 x^1)]\,\rmd z^2}{z^2+a^{\mu} a_{\mu}+2(a^1 x^0-a^0 x^1)}\\
 &\quad + \rmd s^2_{\rmS^3} + \rmd s_{\TT^4}^2 \,,\qquad 
 \Exp{-2\Phi} = \frac{z^2+a^{\mu}\,a_{\mu}+2\,(a^1\,x^0-a^0\,x^1)}{z^2} \,, 
\\
 B_2&=\frac{\rmd x^0\wedge \rmd x^1+z^{-1}\rmd z\wedge (-a^0\,\rmd x^0+a^1\,\rmd x^1)}{z^2+a^{\mu}\,a_{\mu}+2\,(a^1\,x^0-a^0\,x^1)}+\frac{1}{4}\,\cos \theta\,\rmd \phi\wedge \rmd \psi \,,\quad 
 I =a^{\mu}\,\partial_{\mu}\,,
\end{split}
\end{align}
and this is a solution of the GSE for arbitrary $a^\mu \equiv \eta\,\bar{a}^{\mu}$\,. 
In this case, we can freely rescale the Killing vector $I^m$ only when $a^0=\pm a^1$\,. 

\vspace{-\Pskip}
\subsection{``\texorpdfstring{$\beta$}{\textbeta}-deformations'' with generalized isometries}
\label{sec:generalized-isometries}

In the previous subsections, we have not considered the special conformal generators $\hat{K}_{\mu}$\,. 
As in the case of $\AdS5\times\rmS^5$ background, if there is no $B$-field, we can obtain various solutions from $\beta$-deformations using $\hat{K}_{\mu}$. 
However, in the $\AdS3\times \rmS^3\times \TT^4$ background, we cannot naively use $\hat{K}_\mu$ according to $\Lie_{\hat{K}_{\mu}}B_2\neq 0$\,. 
Indeed, even for a simple Abelian $r$-matrix, such as $r=\frac{1}{2}\,K_0\wedge K_1$ or $r=\frac{1}{2}\,K_+\wedge P_+$, the $\beta$-deformed background does not satisfy the supergravity equations of motion. 
In this subsection, we explain how to utilize the special conformal generators, and obtain several solutions from (generalization of) $\beta$-deformations. 

In the canonical section $\tilde{\partial}^m=0$, if there exists a pair $(v^m,\,\tilde{v}_m)$ satisfying
\begin{align}
 \Lie_v \CG_{mn} = 0 \,,\qquad \Lie_v B_2 + \rmd \tilde{v}_1 = 0\,,\qquad \Lie_v \Phi =0 \,,
\end{align}
it means that the background admits a generalized Killing vector $(V^M)=(v^m,\,\tilde{v}_m)$ satisfying
\begin{align}
 \Lie_V \cH_{MN} = 0\,,\qquad \Lie_V d = 0\,. 
\end{align}
Then, the equation \eqref{eq:K-B2-closed} shows that there exist generalized Killing vectors $\hat{\mathsf{K}}_\mu^M$ associated with the Killing vectors $\hat{K}_\mu^m$\,. 
Since a generalized vector of the form $V^M=\partial^M f(x)$, which we call a trivial Killing vector, is always a generalized Killing vector, there is ambiguity in the definition of the generalized Killing vector. 
Using the ambiguity, we can find a set of generalized Killing vectors $\hat{\mathsf{T}}_i=(\hat{\mathsf{T}}_i^M)$ that satisfy
\begin{align}
 \gLie_{\hat{\mathsf{T}}_i}\cH_{MN} = \gLie_{\hat{\mathsf{T}}_i}d =0\,,\qquad 
 \gLie_{\hat{\mathsf{T}}_i}\hat{\mathsf{T}}_j^M + \gLie_{\hat{\mathsf{T}}_j}\hat{\mathsf{T}}_i^M =0 \,,
\end{align}
as well as the conformal algebra $\alg{so}(2,2)$ by means of the C-bracket
\begin{align}
 [V,\,W]_{\rmC}^M \equiv \frac{1}{2}\, (\gLie_V W - \gLie_W V)^M \,. 
\end{align}
Note that, according the requirement $\gLie_{\hat{\mathsf{T}}_i}\hat{\mathsf{T}}_j^M + \gLie_{\hat{\mathsf{T}}_j}\hat{\mathsf{T}}_i^M =0$, the C-bracket coincides with the D-bracket, $[V,\,W]_{\rmD}^M\equiv \gLie_V W^M$. 
We can find the following set of generalized Killing vectors:
\begin{align}
\begin{split}
 &\hat{\mathsf{D}} \equiv x^+\,\partial_+ + x^-\,\partial_- + z\,\partial_z \,, \qquad
 \hat{\mathsf{P}}_+ \equiv \partial_+\,, \qquad
 \hat{\mathsf{P}}_- \equiv \partial_-\,, 
\\
 &\hat{\mathsf{M}}_{+-} = x^+\,\partial_+ - x^-\,\partial_- + z^{-1}\,\tilde{\partial}^z \,,
\\
 &\hat{\mathsf{K}}_+ = z^2\,\partial_+ + 2\,(x^-)^2\,\partial_- + 2\,x^-\,z\,\partial_z 
           + 2\,\tilde{\partial}^- -\frac{2\, x^-}{z}\,\tilde{\partial}^z \,,
\\
 &\hat{\mathsf{K}}_- = 2\,(x^+)^2\,\partial_+ + z^2\,\partial_- + 2\, x^+\, z\,\partial_z 
           -2\,\tilde{\partial}^+ + \frac{2\,x^+}{z}\,\tilde{\partial}^z \,,
\end{split}
\end{align}
which satisfy
\begin{align}
 \eta_{MN}\,\hat{\mathsf{K}}_\pm^M\, \hat{\mathsf{P}}_\mp^N = \pm 2 \,, \qquad 
 \eta_{MN}\,\hat{\mathsf{D}}^M\, \hat{\mathsf{M}}_{+-}^N = 1\,, \qquad 
 \eta_{MN}\,\hat{\mathsf{T}}_i^M\,\hat{\mathsf{T}}_j^N = 0 \quad (\text{others}) \,. 
\end{align}
If we could find generators $\hat{\mathsf{T'}}_i$ which satisfy
\begin{align}
 \eta_{MN}\,\hat{\mathsf{T'}}_i^M\,\hat{\mathsf{T'}}_j^N = 0\,, 
\end{align}
they are on a common $D$-dimensional section, and we can find a duality frame where the generalized Killing vectors take the form $(\mathsf{T'}_i^M) =(\mathsf{T'}_i^m,\,0)$. 
If it is possible, the generalized Killing vectors reduces to the usual Killing vector and we can consider the usual $\beta$-deformations in such duality frame. 
However, it seems unlikely to be the case in the $\AdS3\times \rmS^3\times \TT^4$ background, and in the following, we employ the above set of generalized Killing vectors. 

\subsubsection{\texorpdfstring{$r=\frac{1}{8}\,\mathsf{K}_+\wedge \mathsf{P}_+$}{r=K\textplusinferior{\textwedge}P\textplusinferior}}

Let us first consider an Abelian $r$-matrix $r=\frac{1}{8}\,\mathsf{K}_+\wedge \mathsf{P}_+$ associated with the Abelian generalized isometries; $[\hat{\mathsf{K}}_+,\, \hat{\mathsf{P}}_+]_{\rmC}=0$. 
Since $\hat{\mathsf{K}}_+$ has the dual components, it is not clear how to perform a ``$\beta$-deformation.'' 
We thus change the generalized coordinates such that the dual components disappear. 

We here employ the simple coordinate transformation law by Hohm and Zwiebach \cite{Hohm:2012gk}. 
Namely, under a generalized coordinate transformation $x^M\to x'^M$, the generalized tensors are transformed as
\begin{align}
\begin{split}
 &\cH'_{MN}(x') = \cF_M{}^K(x',x)\,\cF_N{}^L(x',x)\,\cH_{KL}(x) \,,\qquad 
 \Exp{-2d'(x')} = \Bigl\lvert\det\frac{\partial x^M}{\partial x'^N}\Bigr\rvert\,\Exp{-2d'(x')}\,,
\\
 &\cF_M{}^N(x',x)\equiv \frac{1}{2}\,\biggl(\frac{\partial x'_M}{\partial x_P}\,\frac{\partial x^N}{\partial x'^P}+\frac{\partial x^P}{\partial x'^M}\,\frac{\partial x'_P}{\partial x_N}\biggr)\,. 
\end{split}
\end{align}
We can easily check that a generalized coordinate transformation
\begin{align}
 \tilde{z}' = \tilde{z} + \frac{\ln x^-}{z}\,, \qquad x'^M = x^M \quad (\text{others})\,,
\label{eq:gen-coord-trsf}
\end{align}
indeed removes the dual components; $(\mathsf{K'}_+^M)=(\hat{K'}_+^m,\,0)$ and $(\mathsf{P'}_+^M)=(\hat{P'}_+^m,\,0)$. 
In fact, this transformation $\cH'_{MN} = \cF_M{}^K\,\cF_N{}^L\,\cH_{KL}$ with
\begin{align}
 (\cF_M{}^N) = \begin{pmatrix} \bm{1_{10}} & \bmq_{mn} \\ 0 & \bm{1_{10}} \end{pmatrix}\,,\qquad \bmq_{-z}=-\bmq_{z-} = -\frac{1}{x^-\,z}\,,
\end{align}
is precisely a $B$-field gauge transformation,
\begin{align}
 B_2 \ \to \ B_2 - \frac{\rmd x^- \wedge \rmd z}{x^-\,z} \,. 
\end{align}
In the transformed background, the $B$-field is shifted
\begin{align}
\begin{split}
 \rmd s^2&= \frac{-2\,\rmd x^+\,\rmd x^- + \rmd z^2}{z^2} + \rmd s^2_{\rmS^3} + \rmd s_{\TT^4}^2 \,,
\\
 B_2&= \frac{\rmd x^- \wedge (x^-\,\rmd x^+ - z\,\rmd z)}{x^-\,z^2} + \frac{1}{4}\,\cos\theta\,\rmd \phi \wedge \rmd \psi \,, 
 \qquad \Exp{-2\Phi}=1\,,
\end{split}
\end{align}
we can check the isometries
\begin{align}
 \Lie_{\hat{K}_+}\CG_{mn}=\Lie_{\hat{K}_+}B_{mn}=\Lie_{\hat{K}_+}\Phi=0\,,\qquad 
 \Lie_{\hat{P}_+}\CG_{mn}=\Lie_{\hat{P}_+}B_{mn}=\Lie_{\hat{P}_+}\Phi=0\,.
\end{align}
Then, we can perform the usual $\beta$-deformation associated with $r=\frac{1}{8}\,K_+\wedge P_+$,
\begin{align}
\begin{split}
 \rmd s^2&= \frac{-2\,\rmd x^+\,\rmd x^- + \eta\,\rmd x^-\,(\rmd x^- - 2\,x^-\,z^{-1}\,\rmd z)}{z^2 + \eta\,(x^-)^2} + \frac{\rmd z^2}{z^2} + \rmd s^2_{\rmS^3} + \rmd s_{\TT^4}^2 \,,
\\
 B_2&= \frac{\rmd x^- \wedge (x^-\,\rmd x^+ - z\,\rmd z)}{x^-\,[z^2 + \eta\,(x^-)^2]} + \frac{1}{4}\,\cos\theta\,\rmd \phi \wedge \rmd \psi \,, 
 \qquad \Exp{-2\Phi}=\frac{z^2 + \eta\,(x^-)^2}{z^2}\,. 
\end{split}
\end{align}
Finally, we go back to the original coordinates, $\cH_{MN} = (\cF^{-1})_M{}^K\,(\cF^{-1})_N{}^L\,\cH_{KL}$, and obtain
\begin{align}
\begin{split}
 \rmd s^2&= \frac{-2\,\rmd x^+\,\rmd x^- + \eta\,\rmd x^-\,(\rmd x^- - 2\,x^-\,z^{-1}\,\rmd z)}{z^2 + \eta\,(x^-)^2} + \frac{\rmd z^2}{z^2} + \rmd s^2_{\rmS^3} + \rmd s_{\TT^4}^2 \,,
\\
 B_2&= \frac{\rmd x^- \wedge \rmd x^+ + \eta\,x^-\,z^{-1}\,\rmd x^- \wedge \rmd z}{z^2 + \eta\,(x^-)^2} + \frac{1}{4}\,\cos\theta\,\rmd \phi \wedge \rmd \psi \,, 
 \qquad \Exp{-2\Phi}=\frac{z^2 + \eta\,(x^-)^2}{z^2}\,. 
\end{split}
\label{eq:K_+wP_+}
\end{align}
This is a new solution of the usual supergravity. 

\subsubsection{General procedure}

In general, it is not easy to find a generalized coordinate transformation like \eqref{eq:gen-coord-trsf}, which removes the dual components of the generalized Killing vectors. 
However, in fact, it is not necessary to find such a coordinate transformation. 
As it is clear from the above procedure, for an $r$-matrix, $r=\frac{1}{2}\,r^{ij}\,\mathsf{T}_i\wedge \mathsf{T}_j$, associated with the generalized Killing vectors, the previous deformation is simply a transformation
\begin{align}
 \cH_{MN} \ \to \ \cH'_{MN} = h_M{}^K\,h_N{}^L\,\cH_{KL}\,,\qquad 
 h_{M}{}^N \equiv \delta_M^N - 2\,\eta\,r^{ij}\,\hat{\mathsf{T}}_{iM}\,\hat{\mathsf{T}}_j^N\,. 
\label{eq:general-Odd}
\end{align}
Requiring the generalized Killing vectors $\hat{\mathsf{T}}_i$ contained in the $r$-matrix to be mutually orthogonal (i.e.~$\eta_{MN}\,\hat{\mathsf{T}}_i^M\,\hat{\mathsf{T}}_j^N = 0$), we can easily see that the transformation matrix $h_{M}{}^N$ is an $\OO(D,D)$ matrix. 
In general, this $\OO(D,D)$ transformation is a combination of a $\beta$-transformation and diffeomorphisms, but in particular, when all of $\hat{\mathsf{T}}_{i}$ do not have the dual components, this $h_{M}{}^N$ reduces to the usual $\beta$-transformation matrix. 
We can easily check that the above solution \eqref{eq:K_+wP_+} can be obtained from the original background in the single step \eqref{eq:general-Odd}. 

When we consider a non-unimodular $r$-matrix, we suppose that the formula \eqref{eq:cI-formula} will be correct in a duality frame where $\hat{\mathsf{T}}_{i}$ take the form $(\hat{\mathsf{T}}_{i}^M)=(\hat{T}_{i}^m,\,0)$. 
Then, the deformed background will be a solution of modified DFT (mDFT) \cite{Sakatani:2016fvh} with
\begin{align}
 \bm{X}^M=\hat{\cI}^M \equiv \begin{pmatrix} \hat{\cI}^m\\ \hat{\cI}_m \end{pmatrix} 
 \equiv \eta\,r^{ij}\,[\hat{\mathsf{T}}_i,\,\hat{\mathsf{T}}_j]_{\rmC}^M \,. 
\end{align}
In terms of the GSE, it is a solution with $I^m=\hat{\cI}^m$ and $Z_m=\partial_m\Phi +I^n\,B_{nm} + \hat{\cI}_m$\,. 

\subsubsection{\texorpdfstring{$r=\frac{1}{2}\,\mathsf{K}_+\wedge \mathsf{K}_-$}{r=K\textplusinferior{\textwedge}K\textminusinferior}}

For an Abelian $r$-matrix $r=\frac{1}{2}\,\mathsf{K}_{+}\wedge \mathsf{K}_-$, we do not find a generalized coordinate system where dual components of both $\hat{\mathsf{K}}_+^M$ and $\hat{\mathsf{K}}_-^M$ vanish. 
However, from the general procedure \eqref{eq:general-Odd}, we can easily obtain the deformed background
\begin{align}
\begin{split}
 \rmd s^2 &= \frac{-2\,\rmd x^+\,\rmd x^- + \rmd z^2 + 2\,\eta\,[2\,(x^-\,\rmd x^+ + x^+ \,\rmd x^-) - (2\,x^+\,x^- + z^2)\,\frac{\rmd z}{z}]^2}{z^2+2\,\eta\,(z^2- 2\,x^- \, x^+)^2} 
\\
 &\quad+ \rmd s^2_{\rmS^3} + \rmd s_{\TT^4}^2\,, \qquad
 \Exp{-2\Phi} =\frac{z^2+2\,\eta\,(2\,x^+\,x^- -z^2)^2}{z^2}\,,
\\
 B_2&= \frac{\rmd x^-\wedge \rmd x^+ -4\,\eta\,(2\,x^+\,x^- - z^2) (\rmd x^- \wedge\rmd x^+ + (x^-\,\rmd x^+ - x^+\,\rmd x^-)\wedge\frac{\rmd z}{z}}{z^2+2\,\eta\,(z^2- 2\,x^- \, x^+)^2} 
\\
 &\quad + \frac{1}{4}\,\cos\theta\,\rmd \phi \wedge \rmd \psi \,. 
\end{split}
\end{align}
We can easily see that this is a solution of the usual supergravity. 

\subsubsection{\texorpdfstring{$r=\frac{1}{2}\,\mathsf{M}_{+-}\wedge \mathsf{K}_+$ or $r=\frac{1}{2}\,\mathsf{D}\wedge \mathsf{K}_+$}{r=M\textplusinferior\textminusinferior{\textwedge}{K\textplusinferior} or r=D{\textwedge}K\textplusinferior}}

Let us next consider a non-unimodular $r$-matrix $r=\frac{1}{2}\,\mathsf{M}_{+-}\wedge \mathsf{K}_+$, satisfying
\begin{align}
 \cI = \eta\, [\mathsf{M}_{+-},\,\mathsf{K}_+] = \eta\,\mathsf{K}_+ \,.
\end{align}
In this case, the deformed background
\begin{align}
 \rmd s^2 &= \frac{-2\,\rmd x^-\,\rmd x^+ + \rmd z^2 - 2\,\eta\, (\rmd x^- - x^-\,\frac{\rmd z}{z})\,[2\,x^+\,\rmd x^- + 2\,x^-\, \rmd x^+ - (2\,x^+\, x^- + z^2)\,\frac{\rmd z}{z}]}{z^2 - 2\,\eta\,x^-\, (2 x^- x^+ - z^2)}
\nn\\
 &\quad+ \rmd s^2_{\rmS^3} + \rmd s_{\TT^4}^2\,, \qquad
 \Exp{-2\Phi} = \frac{z^2-2\,\eta\,x^-\,(2\,x^+\,x^- -z^2)}{z^2} \,,
\nn\\
 B_2&= \frac{1}{4}\,\cos\theta\,\rmd \phi \wedge \rmd \psi
\nn\\
 &\quad + \frac{(1+2\,\eta\,x^-)\,\rmd x^-\wedge \rmd x^+ + \eta\, [\,2\,(x^-)^2\,\rmd x^+ - (4\,x^+\,x^- - z^2)\,\rmd x^-\,]\wedge \frac{\rmd z}{z}}{z^2 - 2\,\eta\,x^-\, (2\,x^+\,x^- -z^2)} \,,
\label{eq:MwedgeK+}
\end{align}
satisfies the equations of motion of mDFT with $\bm{X}^M =\eta\,\hat{\mathsf{K}}_+^M$\,. 

Similar to the example studied in section \ref{sec:AdS3-S3-T4MwedgeP}, we can freely rescale $\bm{X}^M$ as $\bm{X}^M\to \lambda\,\bm{X}^M$ $(\lambda\in\mathbb{R})$, and in a particular case $\lambda=0$, \eqref{eq:MwedgeK+} can be regarded as a solution of the usual supergravity. 

Interestingly, we can obtain the same background also by considering an $r$-matrix $r=\frac{1}{2}\,\mathsf{D}\wedge \mathsf{K}_+$, satisfying $\cI = \eta\, [\mathsf{D},\,\mathsf{K}_+] = -\eta\,\mathsf{K}_+$\,.
This also may be related to the degeneracy of $(\CG \pm B)_{mn}$ in the $\AdS3\times \rmS^3\times \TT^4$ background.

\subsubsection{Non-orthogonal case: \texorpdfstring{$r=\frac{1}{2}\,\mathsf{D} \wedge \mathsf{M}_{+-}$}{r=D{\textwedge}M\textplusinferior\textminusinferior}}

Let us finally comment on an Abelian $r$-matrix $r=\frac{1}{2}\,\mathsf{D} \wedge \mathsf{M}_{+-}$, which is a generalized version of the example considered in section \ref{sec:AdS3-DwM}. 
In this case, the associated generalized Killing vectors $\hat{\mathsf{D}}^M$ and $\hat{\mathsf{M}}_{+-}^N$ are not orthogonal to each other, $\eta_{MN}\,\hat{\mathsf{D}}^M\,\hat{\mathsf{M}}_{+-}^N\neq 0$, and $h_M{}^N$ in \eqref{eq:general-Odd} is not an $\OO(10,10)$ matrix. 
Accordingly, the deformed background is not a solution of the usual supergravity. 
In order to reproduce the example of section \ref{sec:AdS3-DwM} from the general procedure \eqref{eq:general-Odd}, we need to utilize trivial Killing vectors. 
Indeed, by introducing a trivial Killing vector, $\hat{\mathsf{N}}^M \equiv \partial^M \ln z$, $\hat{\mathsf{M}}'^M_{+-}\equiv \hat{\mathsf{M}}_{+-}^M -\hat{\mathsf{N}}^M$ takes the form $(\hat{\mathsf{M}}'^M_{+-})=(M_{+-}^m,\,0)$\,, and the transformation matrix $h_M{}^N$ associated with an $r$-matrix $r=\frac{1}{2}\,\mathsf{D} \wedge \hat{\mathsf{M}}'_{+-}$ is an $\OO(10,10)$ matrix. 
The deformed background reproduces the same background obtained in section \ref{sec:AdS3-DwM}. 

As this example indicates, the null condition $c_{ij}\equiv \eta_{MN}\,\hat{\mathsf{T}}_i^M\,\hat{\mathsf{T}}_j^N = 0$, is very important. 
Since a gauge transformation in DFT (i.e.~generalized diffeomorphism) is a local $\OO(D,D)$ rotation, in order to realize the deformation $h_M{}^N$ as a gauge symmetry in DFT, $h_M{}^N$ should be an $\OO(D,D)$ matrix. 
This requirement is equivalent to a condition
\begin{align}
 r^{ik}\,c_{kl}\,r^{lj}\,\hat{\mathsf{T}}_{i}^M\,\hat{\mathsf{T}}_{j}^N = 0\,. 
\end{align}
It is interesting to find a set of generalized Killing vectors $\hat{\mathsf{T}}_{i}^M$ and an $r$-matrix (satisfying CYBE) that satisfy the above condition with $c_{ij}\neq 0$, but here we simply require $c_{ij}=0$\,. 
Note that the same the null condition $c_{ij}=0$ is known in the context of the non-Abelian $T$-duality. 
As it has been studied in \cite{Hull:1989jk,Hull:1990ms}, the null condition played an important role in gauging non-Abelian isometries. 

\subsubsection{Short summary}

Let us summarize this subsection. 
Usually, we prepare a bi-vector $\hat{r}=\frac{1}{2}\,r^{ij}\,\hat{T}_i\wedge \hat{T}_j$ satisfying the homogeneous CYBE (or the Poisson condition)
\begin{align}
 [\hat{r},\,\hat{r}]_{\rmS} \equiv r^{ij}\,r^{kl}\, [\hat{T}_i,\,\hat{T}_k]\wedge T_j\wedge T_l
 = - r^{ij}\,r^{kl}\,f_{ik}{}^m\,\hat{T}_m\wedge \hat{T}_j\wedge \hat{T}_l = 0 \,,
\end{align}
where $[\cdot,\cdot]_{\rmS}$ is the Schouten bracket, and perform a local $\beta$-transformation
\begin{align}
 h_M{}^N = \begin{pmatrix} \bm{{1_{10}}} & \bm{{0_{10}}} \\ \bmr^{mn}(x) & \bm{{1_{10}}} \end{pmatrix},\qquad
 \bmr^{mn}(x) = -2\,\eta\,r^{ij}\,\hat{T}_i^m\, \hat{T}_j^n \,. 
\end{align}
In this subsection, in order to allow for the non-standard Killing vectors $\hat{K}_{\mu}$ satisfying \eqref{eq:K-B2-closed}, we have generalized the Killing vectors $\hat{T}_i$ into the generalized Killing vectors $\hat{\mathsf{T}}_i$\,. 
The generalized Killing vectors $\hat{\mathsf{T}}_i$ are defined such that their C-bracket satisfy the same commutation relations as those of $\hat{T}_i$\,. 
The homogeneous CYBE is generalized by replacing the usual Lie bracket with the C-bracket, and performing $\OO(D,D)$ transformations
\begin{align}
 h_{M}{}^N = \delta_M^N - 2\,\eta\,r^{ij}\,\hat{\mathsf{T}}_{iM}\,\hat{\mathsf{T}}_j^N\,,
\end{align}
we have obtained several new solutions of DFT. 
In particular, when all of the generalized Killing vectors $\hat{\mathsf{T}}_{i}^M$ do not have the dual components, this generalized transformation reduces to the usual local $\beta$-transformations. 

\section{Conclusions}
\label{sec:discussion}

In this paper, we have shown that, after suitable field redefinitions, a homogeneous YB-deformed $\AdS{5}\times \rmS^5$ superstring action associated with bosonic isometries can be always express as the usual GS superstring action up to quadratic order in fermions. 
The deformations were made only in the supergravity backgrounds and they were identified as local $\beta$-deformations. 
We have also found a DSM action that reproduces the GS type II superstring action up to quadratic order in fermions. 
After taking the diagonal gauge, the spacetime fermion is transformed as $\Theta_2\to \Omega\,\Theta_2$ under $\beta$-transformations. 
We found an explicit form of $\Omega$ in terms of the $\beta$-transformation parameter and the NS--NS fields. 
Moreover, $\beta$-deformations of $H$-fluxed AdS backgrounds were also studied and various solutions are obtained. 
In several examples of non-unimodular deformations, we unexpectedly obtained solutions of the usual supergravity. 

In this paper, we have mainly focused on the YB deformations of the $\AdS{5}\times \rmS^5$ superstring, but the equivalence between homogeneous YB deformations and local $\beta$-deformations will be shown also for other backgrounds.\footnote{See a recent paper \cite{Bakhmatov:2018apn} for a generalization of the ``YB deformations'' (or the open-closed string map) beyond the coset spaces.} 
Indeed, if we observe the NS--NS part of the deformed action discussed in section \ref{sec:YB-NS-NS}, it is clear that we have not used specific properties of the $\alg{psu}(2,2|4)$ algebra. 
At least when the $B$-field is absent, and the algebra $\alg{g}$ admits a projection $P$ to the bosonic coset generators $\sfP_{\sfa}$ and $\kappa_{\sfa\sfb}\equiv \str(\sfP_{\sfa}\,\sfP_{\sfb})$ is non-degenerate, the bosonic part of a coset YB sigma model associated with a skew-symmetric $R$-operator becomes
\begin{align}
 S_{(0)} &\sim \int \rmd^2\sigma\,\sqrt{-\gga}\,\Pg_{-}^{\WSa\WSb}\, \str\bigl[A_{\WSa}\, P\circ\cO_{-}^{-1}(A_{\WSb})\bigr]\bigr\rvert_{\theta=0}
 = \int \rmd^2\sigma\,\sqrt{-\gga}\,\Pg_{-}^{\WSa\WSb}\,e_{\WSa}{}^{\sfa}\,e_{\WSb}{}^{\sfb}\,k_{+\sfa\sfb} \,, 
\end{align}
where $P(A_{\WSa})\equiv e_{\WSa}{}^{\sfa}\,\sfP_{\sfa}$\,, $R_g(\sfP_{\sfa})\equiv \lambda_{\sfa}{}^{\sfb}\,\sfP_{\sfb}$\,, and $k_{+\sfa\sfb} \equiv [(\kappa^{-1}+\eta\,\kappa^{-1}\,\lambda)^{-1}]_{\sfa\sfb}$\,. 
According to the skew-symmetry $\lambda^{\sfa\sfb}=-\lambda^{\sfb\sfa}$ [$\lambda^{\sfa\sfb}\equiv (\kappa^{-1}\,\lambda)^{\sfa\sfb}$]\,, the YB deformation
\begin{align}
 E^{mn}=\OG^{mn} \ \to \ E'^{mn} = \OG^{mn} + \eta\, \lambda^{\sfa\sfb}\, e_{\sfa}{}^{m}\,e_{\sfb}{}^{n}\qquad 
 \bigl(\OG_{mn} \equiv e_{m}{}^{\sfa}\,e_{n}{}^{\sfb}\,\kappa_{\sfa\sfb}\bigr)\,,
\end{align}
can be regarded as a (local) $\beta$-deformation. 
The YB deformations of R--R fields are rather non-trivial, but we expect that the equivalence between YB deformations and local $\beta$-deformation will be shown in more general cases, as long as the $r$-matrix consists of bosonic generators. 
In a general analysis performed in \cite{Borsato:2016ose}, the YB sigma model action associated with a general $r$-matrix satisfying the homogeneous (or modified) CYBE has been expressed in the standard form of the GS superstring to all order in fermions. 
When the $r$-matrix contains fermionic generators, we expect the deformation of the target space can be no longer regarded as a local $\beta$-deformation. 
It will be an interesting future direction to clarify what kind of deformations are made in such cases.

As we showed in section \ref{sec:AdS3}, all of the $\beta$-deformed $\AdS3\times \rmS^3\times \TT^4$ backgrounds are solutions of (generalized) supergravity and string theory is well-defined in such backgrounds. 
In this sense, the local $\beta$-deformations are certain duality transformations in string theory. 
It will be interesting future work to check the integrability of string sigma model on the $\beta$-deformed $\AdS3\times \rmS^3\times \TT^4$ backgrounds obtained in this paper. 
It is also important to formulate the homogeneous YB deformations of type II superstring in $H$-fluxed background and show the equivalence to $\beta$-transformations. 

In this paper, we obtained the $\OO(10,10)$-invariant DSM action for type II superstring. 
Although we considered only up to the quadratic order in fermionic variables, it is important to obtain the complete action. 
In fact, a $T$-duality manifest GS superstring action has been proposed also in \cite{Hatsuda:2014aza,Hatsuda:2015cia}. 
In our approach, the R--R field strengths are contained in the $P$-$\brP$ or $\brP$-$P$ components of the generalized metric $\cM_{MN}$\,, but also in the approach of \cite{Hatsuda:2014aza,Hatsuda:2015cia}, they will appear in the ``left-right'' mixing terms. 
It is interesting future work to make the connection between the double-vielbein formalism and the approach of \cite{Hatsuda:2014aza,Hatsuda:2015cia} clearer, and obtain a $T$-duality manifest GS superstring action in arbitrary curved backgrounds. 
In the conventional GS superstring, from the requirement of the kappa invariance, the (generalized) supergravity equations of motion has been obtained \cite{TW}. 
By generalizing this analysis, it will be important to derive the type II DFT equations of motion from the $T$-duality manifest GS superstring action. 

It is also important to formulate the DSM that manifests the symmetry of non-Abelian $T$-dualities. 
When the target space is a group manifold, such DSM has been formulated in \cite{Klimcik:1995dy,Klimcik:1995ux} and its symmetry is clearly discussed in \cite{ReidEdwards:2010vp} (see also \cite{Hassler:2017yza,Lust:2018jsx} for relevant recent works). 
Its generalization to coset space will be important for a clearer understanding of YB deformations. 
In the usual DSM, the generalized vector $DX^M$ has the form
\begin{align}
 DX^M \equiv \begin{pmatrix} \rmd X^m \\ \rmd \tilde{X}_m -A_m \end{pmatrix} 
 = \begin{pmatrix} \rmd X^m \\ B_{mn}\, \rmd X^n + \CG_{mn}\,*_{\gga} \rmd X^n \end{pmatrix},
\end{align}
upon using the equations of motion. 
In order to keep the integrability of $\rmd X'^m$ after an $\OO(D,D)$ transformation
\begin{align}
 0= \rmd^2 X'^m = \rmd \bigl(\bms^{m}{}_{n}\,\rmd X^n + \bmr^{mn}\,B_{np}\, \rmd X^p + \bmr^{mn}\,\CG_{np}\,*_{\gga} \rmd X^p\bigr)\,, 
\label{eq:integrability}
\end{align}
we have to require the existence of a set of generalized Killing vectors \cite{Rennecke:2014sca}
\begin{align}
 V^{(m)N} \equiv \bigl(\bmr^{(m)n},\,\bms^{(m)}_n\bigr)\,,\qquad 
 \gLie_{V^{(m)}}\cH_{MN} = 0 \,. 
\label{eq:integrability-requirement}
\end{align}
As discussed in \cite{Rennecke:2014sca}, in the case of $\beta$-transformations ($\bms^{m}{}_{n}=\delta^m_n$ and $\bmr^{mn}=-\bmr^{nm}$), by requiring the set of generalized Killing vectors $V^{(m)}$ to form a closed algebra by means of the C-bracket, the homogeneous CYBE for the bi-vector $[\hat{r},\,\hat{r}]_{\rmS}=0$ ($\hat{r}\equiv \frac{1}{2}\,\bmr^{mn}\,\partial_m\wedge\partial_n$) is required. 
In the case of an Abelian $r$-matrix, we can find a coordinate system where all of the relevant Killing vectors $\hat{T}_i$ are constant vector, and the bi-vector $\bmr^{mn} = -2\,\eta\,r^{ij}\,\hat{T}_i^m\, \hat{T}_j^n$ automatically satisfies the requirement \eqref{eq:integrability-requirement} and we can perform the $\beta$-deformation without breaking the integrability. 
This is the usual constant $\beta$-shift in the presence of Abelian isometries, that can also realized as a TsT-transformation. 
In the case of non-Abelian $r$-matrices, since we cannot find a coordinate system where all of the Killing vectors $\hat{T}_i$ are constant vectors, the requirement \eqref{eq:integrability-requirement} is too restroctive. 
It will be interesting future work to relax the requirement \eqref{eq:integrability} by reformulating the DSM that manifests the symmetry of non-Abelian isometries. 

\section*{Acknowledgment}

We would like to thank Machiko Hatsuda, Jeong-Hyuck Park, Shozo Uehara, and Kentaroh Yoshida for valuable discussions. 
We also thank Eoin \'O.~Colg\'ain and Linus Wulff for useful comments on the first version of the manuscript. 
The work of J.S.\ was supported by the Japan Society for the Promotion of Science (JSPS). 

\appendix

\section{Conventions and Formulas}
\label{app:conventions}

\subsubsection*{Differential form and curvature}

The antisymmeterization is defined as
\begin{align}
 A_{[m_1\cdots m_n]} \equiv \frac{1}{n!}\,\bigl(A_{m_1\cdots m_n} \pm \text{permutations}\bigr) \,.
\end{align}
For conventions of differential forms, we use
\begin{align}
\begin{split}
 &\varepsilon^{01}= \frac{1}{\sqrt{-\gga}}\,,\qquad 
 \varepsilon_{01}= - \sqrt{-\gga} \,, \qquad 
 \rmd^{2}\sigma = \rmd \tau \wedge\rmd \sigma \,,
\\
 &(*_{\gga} \alpha_q)_{\WSa_1\cdots\WSa_{p+1-q}} =\frac{1}{q!}\,\varepsilon^{\WSb_1\cdots\WSb_q}{}_{\WSa_1\cdots\WSa_{p+1-q}}\,\alpha_{\WSb_1\cdots\WSb_q} \,,
\\
 & *_{\gga} (\rmd \sigma^{\WSa_1}\wedge \cdots \wedge \rmd \sigma^{\WSa_q}) 
 = \frac{1}{(p+1-q)!}\,\varepsilon^{\WSa_1\cdots\WSa_q}{}_{\WSb_1\cdots\WSb_{p+1-q}}\,\rmd \sigma^{\WSb_1}\wedge \cdots \wedge \rmd \sigma^{\WSb_{p+1-q}} \,,
\end{split}
\end{align}
on string worldsheet while in the spacetime, we define
\begin{align}
\begin{split}
 &\varepsilon^{1\cdots D}=-\frac{1}{\sqrt{-\CG}}\,,\qquad 
 \varepsilon_{1\cdots D}= \sqrt{-\CG} \,, \qquad 
 \epsilon^{1\cdots D} = - 1\,,\qquad 
 \epsilon^{1\cdots D} = 1 \,, 
\\
 &(* \alpha_q)_{m_1\cdots m_{p+1-q}} =\frac{1}{q!}\,\varepsilon^{n_1\cdots n_q}{}_{m_1\cdots m_{p+1-q}}\,\alpha_{n_1\cdots n_q} \,,\qquad 
 \rmd^{D}x = \rmd x^1\wedge\cdots\wedge\rmd x^D \,,
\\
 &* (\rmd \sigma^{m_1}\wedge \cdots \wedge \rmd \sigma^{m_q}) = \frac{1}{(p+1-q)!}\,\varepsilon^{m_1\cdots m_q}{}_{n_1\cdots n_{p+1-q}}\,\rmd x^{n_1}\wedge \cdots \wedge \rmd x^{n_{p+1-q}} \,,
\\
 &(\iota_v \alpha_n) = \frac{1}{(n-1)!}\,v^n\,\alpha_{nm_1\cdots m_{n-1}}\,\rmd x^{m_1}\wedge\cdots\wedge \rmd x^{m_{n-1}}\,. 
\end{split}
\end{align}

The spin connection is defined as
\begin{align}
 \omega_m{}^{\Loa\Lob} \equiv 2\,e^{n[\Loa}\,\partial_{[m} e_{n]}{}^{\Lob]} - e^{p[\Loa}\,e^{\Lob]q}\,\partial_{[p} e_{q]}{}^{\Loc}\,e_{m\Loc} \,,
\end{align}
which satisfies
\begin{align}
 \rmd e^{\Loa} + \omega^{\Loa}{}_{\Lob}\wedge e^{\Lob} = 0\,, 
\end{align}
where $e^{\Loa}\equiv e_m{}^{\Loa}\,\rmd x^m$ and $\omega^{\Loa}{}_{\Lob}\equiv \omega_m{}^{\Loa}{}_{\Lob}\,\rmd x^m$\,. 
The Riemann curvature tensor is defined as
\begin{align}
 R^{\Loa}{}_{\Lob} \equiv \frac{1}{2}\,R^{\Loa}{}_{\Lob\Loc\Lod}\,e^{\Loc}\wedge e^{\Lod} \equiv \rmd \omega^{\Loa}{}_{\Lob} + \omega^{\Loa}{}_{\Loc}\wedge \omega^{\Loc}{}_{\Lob} \,,\qquad 
 R^{\Loa}{}_{\Lob\Loc\Lod} = e_m{}^{\Loa}\,e_{\Lob}{}^n\,e_{\Loc}{}^p\,e_{\Lod}{}^q\,R^m{}_{npq} \,. 
\end{align}

\subsection*{(Generalized) supergravity}

Our conventions for the type II GSE \cite{Arutyunov:2015mqj,TW,Sakatani:2016fvh,Baguet:2016prz,Sakamoto:2017wor} are as follows:
\begin{align}
 &R_{mn}- \frac{1}{4}\,H_{mpq}\,H_n{}^{pq} + 2\,\sfD_m \partial_n \Phi + \sfD_m U_n +\sfD_n U_m = T_{mn} \,,
\nn\\
 &R + 4\,\sfD^m \partial_m \Phi - 4\,\abs{\partial \Phi}^2 - \frac{1}{2}\,\abs{H_3}^2 
  - 4\,\bigl(I^m I_m+U^m U_m + 2\,U^m\,\partial_m \Phi - \sfD_m U^m\bigr) =0 \,,
\nn\\
 &-\frac{1}{2}\,\sfD^k H_{kmn} + \partial_k\Phi\,H^k{}_{mn} + U^k\,H_{kmn} + \sfD_m I_n - \sfD_n I_m = \cK_{mn} \,,
\\
 &\rmd *\hat{F}_n -H_3\wedge * \hat{F}_{n+2} -\iota_I B_2 \wedge * \hat{F}_n -\iota_I * \hat{F}_{n-2} =0 \,,
\nn
\end{align}
where $\sfD_m$ is the usual covariant derivative associated with $\CG_{mn}$ and we have defined
\begin{align}
\begin{split}
 T_{mn} &\equiv \frac{1}{4}\Exp{2\Phi} \sum_p \biggl[ \frac{1}{(p-1)!}\, 
 \hat{F}_{(m}{}^{k_1\cdots k_{p-1}} \hat{F}_{n) k_1\cdots k_{p-1}} - \frac{1}{2}\, 
 \CG_{mn}\,\abs{\hat{F}_p}^2 \biggr] \,,
\\
 \cK_{mn}&\equiv \frac{1}{4}\Exp{2\Phi} \sum_p \frac{1}{(p-2)!}\, \hat{F}_{k_1\cdots k_{p-2}}\, 
 \hat{F}_{mn}{}^{k_1\cdots k_{p-2}} \,, 
\\
 \hat{F}_n&\equiv \rmd \hat{C}_{n-1} + H_3\wedge \hat{C}_{n-3} - \iota_I B_2 \wedge \hat{C}_{p-1} -\iota_I \hat{C}_{n+1} \,. 
\end{split}
\end{align}
The Killing vector $I=I^m\,\partial_m$ is defined to satisfy
\begin{align}
 \Lie_I \CG_{mn} = 0\,, \qquad 
 \Lie_I B_2 + \rmd \bigl(U -\iota_I B_2\bigr) = 0\,,\qquad 
 \Lie_I \Phi =0 \,, \qquad 
 I^m\,U_m = 0\,. 
\end{align}
When $I=0$, the GSE reduce to the usual supergravity equations of motion. 

When we describe the GSE as a special case of the modified DFT \cite{Sakatani:2016fvh}, the Killing vector $I^m$ and $U_m$ are packaged into a null generalized Killing vector
\begin{align}
 \bm{X}^M \equiv \begin{pmatrix} I^m \\ U_m -I^n B_{nm} \end{pmatrix} ,\quad 
 \gLie_{\bm{X}}\cH_{MN}=0\,,\quad 
 \gLie_{\bm{X}}d=0\,,\quad 
 \eta_{MN}\,\bm{X}^M\,\bm{X}^N=0\,. 
\end{align}
In a particular gauge $U_m = I^n B_{nm}$ (see \cite{Sakamoto:2017wor} for the details), the dual components of $\bm{X}^M$ vanish. 
We also define a generalized null vector
\begin{align}
 \bm{Y}^M \equiv \cH^{M}{}_N\,\bm{X}^N = \begin{pmatrix} U^m \\ I_m -U^n B_{nm} \end{pmatrix} ,\qquad 
 \eta_{MN}\,\bm{Y}^M\,\bm{Y}^N=0\,. 
\end{align}

As shown in \cite{Borsato:2016ose}, a YB deformed $\AdS5\times\rmS^5$ background associated with a non-unimodular $r$-matrix is a solution of the GSE. 
From various examples, we experimentally know that the Killing vector $I^m$ in the deformed background takes the form \eqref{eq:experimental}
\begin{align}
 I^m = \eta\,r^{ij}\,f_{ij}{}^k\,\hat{T}_k^m \,. 
\end{align}
Here, we derive the experimental formula for YB deformed $\AdS5\times\rmS^5$ backgrounds. 
In the case of the $\AdS5\times\rmS^5$ backgrounds, a general formula for $I$ was obtained in \cite{Borsato:2016ose}, and by neglecting contributions from fermionic generators, a simple expression
\begin{align}
\begin{split}
 I&= -\frac{\eta}{2}\, \kappa^{ij}\,\str\bigl\{[T_i,\,R(T_j)]\,\Ad_g\,(J_+^{(2)}+J_-^{(2)})\bigr\} \,,
\\
 (\kappa^{ij})&\equiv \kappa^{-1}\,,\qquad 
 \kappa\equiv (\kappa_{ij})\,,\qquad 
 \kappa_{ij} \equiv \str(T_i\,T_j)\,,
\end{split}
\end{align}
was given in a quite recent paper \cite{Wulff:2018aku}. 
This expression becomes
\begin{align}
 I &= -\frac{\eta}{2}\, \kappa^{ij}\, r^{kl}\,\kappa_{jl}\,(e_+^{\Loa}+e_-^{\Loa})\,\str\bigl([T_i,\,T_k]\,g\,\gP_{\Loa}\,g^{-1}\bigr)
\nn\\
 &= \frac{\eta}{2}\, r^{ik}\, (e_+^{\Loa}+e_-^{\Loa})\,f_{ik}{}^j\,[\Ad_g]_{\Loa}{}^l\, \kappa_{jl} 
 = \eta \, r^{ij}\, f_{ij}{}^k\,\bigl[\Ad_{g^{-1}}\bigr]_k{}^{\Lob}\, \,k_{-\Lob}{}^{\Loa}\,e'_{\Loa} \equiv I^{\Loa}\,e'_{\Loa}\,,
\end{align}
where we have used
\begin{align}
\begin{split}
 &[\Ad_g]_{\Loa}{}^k\, \kappa_{ki} = \str(g\,\gP_{\Loa}\,g^{-1}\,T_i) = \str(\gP_{\Loa}\,g^{-1}\,T_i\,g) = \bigl[\Ad_{g^{-1}}\bigr]_i{}^{\Loc}\, \eta_{\Loc\Loa}\,, 
\\
 &e_+^{\Loa}+e_-^{\Loa} = (k_{-}^{-1}\,k_+)_{\Lob}{}^{\Loa}\,e'^{\Lob} + e'^{\Lob} = [(2-k_{+}^{-1})\,k_+]_{\Lob}{}^{\Loa}\,e'^{\Lob} + e'^{\Lob}
 = 2 \,k_{+\Lob}{}^{\Loa}\,e'^{\Lob} = 2 \,k_{-}^{\Loa\Lob}\,e'_{\Lob}\,. 
\end{split}
\end{align}
Then, the curved components become
\begin{align}
 I^m = I^{\Loa}\,e'_{\Loa}{}^m = I^{\Loa}\,(k_{-}^{-1})_{\Loa}{}^{\Lob}\,e_{\Lob}{}^m = \eta \, r^{ij}\, f_{ij}{}^k\,\,\bigl[\Ad_{g^{-1}}\bigr]_k{}^{\Lob}\,e_{\Lob}{}^m = \eta \, r^{ij}\, f_{ij}{}^k\,\hat{T}_k^m \,,
\label{eq:I-derived}
\end{align}
and the formula \eqref{eq:experimental} is reproduced. 
Here, it is noted that, although the right-hand side of \eqref{eq:I-derived} is expressed by using the Killing vectors $\hat{T}_i^m$ on the undeformed background, the Killing vector $I^m$ on the left-hand side should be understood as a vector field defined on the YB-deformed $\AdS5\times\rmS^5$ background. 
Note also that if we use an identity $r^{ij}\,f_{ij}{}^k = \frac{1}{2}\,r^{kj}\,f_{ij}{}^i$\footnote{Here, indices $i,j,k$ are restricted to a subalgebra of $\alg{so}(2,4)\times\alg{so}(6)$ such that the $r$-matrix is invertible (see \cite{Borsato:2016ose} for more details). Accordingly, the indices in \eqref{eq:Im-trace} are also limited to the subalgebra.} (see (5.5) of \cite{Borsato:2016ose}), we can also express $I^m$ in terms of the trace of the structure constant
\footnote{As shown in \cite{Hoare:2016wsk,Borsato:2016pas,Hoare:2016wca,Borsato:2017qsx,Lust:2018jsx}, YB deformations can be also realized as non-Abelian $T$-dualities. In this context, a relation between the trace of the structure constant $f_{ki}{}^k$ and the Killing vector $I^m$ was noted in \cite{Hoare:2016wsk} (see also \cite{Hoare:2016wca}). This relation was further clarified in \cite{Borsato:2017qsx} and a simple expression $I^i=f_{ki}{}^k$ was obtained in \cite{Hong:2018tlp}.}
\begin{align}
 I^m = \frac{\eta}{2} \, r^{ij}\, f_{kj}{}^k\,\hat{T}_i^m \,. 
\label{eq:Im-trace}
\end{align}

\subsubsection*{Formulas for gamma matrices and spinors}

Products of antisymmetrized $32\times 32$ gamma matrices satisfy
\begin{align}
\begin{split}
 &\Gamma^{\Loa_1\cdots \Loa_p}\,\Gamma_{\Lob_1\cdots \Lob_q} 
 =\sum^{p+q}_{r=\abs{p-q}}\frac{(-1)^{\frac{u(u-1)}{2}}p!\,q!}{u!\,v!\,w!}\,\eta^{[\underline{\Loa_1}\Loc_1}\cdots\eta^{\underline{\Loa_v}\Loc_v}\,\delta_{[\Lob_1}^{\underline{\Loa_{v+1}}}\cdots \delta_{\Lob_u}^{\underline{\Loa_{p}}]}\,\Gamma_{|\Loc_1\cdots\Loc_v|\Lob_{u+1}\cdots \Lob_q]} 
\\
 &\Bigl[u\equiv \frac{1}{2}\,(p+q-r)\,,\quad
 v\equiv \frac{1}{2}\,(p-q+r)\,,\quad
 w\equiv \frac{1}{2}\,(-p+q+r)\Bigr]\,,
\label{eq:Gamma-p-q}
\end{split}
\end{align}
where the under-barred indices are totally antisymmetrized and the integer $r$ takes values
\begin{align}
 r=\abs{p-q}\,,\ \abs{p-q}+2\,,\ \dotsc \,,\ p+q-2\,,\ p+q\,,
\end{align}
and $u$, $v$ and $w$ are non-negative integers.
As particular cases, we obtain
\begin{align}
 &\Gamma^{\Loa_1\cdots \Loa_n}\,\Gamma^{\Lob}
 = \Gamma^{\Loa_1\cdots \Loa_n\Lob} + n\, \Gamma^{[\Loa_1\cdots \Loa_{n-1}}\,\eta^{\Loa_n]\Lob} \,. 
\label{eq:Gamma-n-1}
\\
 &\Gamma^{\Loa}\,\Gamma^{\Lob_1\cdots \Lob_n} 
 = \Gamma^{\Loa\Lob_1\cdots \Lob_n} + n\, \eta^{\Loa[\Lob_1} \,\Gamma^{\Lob_2\cdots \Lob_n]} \,. 
\end{align}

Arbitrary 32-component Majorana spinors $\Theta$ and $\Psi$ satisfy
\begin{align}
 &\brTheta\,\Gamma_{\Loa_1\Loa_2\cdots \Loa_n}\,\Psi = (-1)^{\frac{n(n+1)}{2}} \bar{\Psi}\,\Gamma_{\Loa_1\Loa_2\cdots \Loa_n}\,\Theta\,, 
\\
 &\brTheta\,\Gamma^{\Loa_1 \cdots \Loa_n}\,\Psi = 0 \qquad (n=1,\,2,\,5,\,6,\,9,\,10)\,. 
\end{align}
For spinors with a definite chirality, $\Gamma^{11}\,\Psi_{\pm}=\pm \Psi_{\pm}$ and $\Gamma^{11}\,\Theta_{\pm}=\pm \Theta_{\pm}$, we have
\begin{align}
\begin{split}
 \brTheta_{+}\,\Gamma_{\Loa_1\Loa_2\cdots \Loa_n}\,\Psi_{\pm} 
 &= \begin{cases}
 (-1)^{\frac{n(n+1)}{2}} \bar{\Psi}_{\pm}\,\Gamma_{\Loa_1\Loa_2\cdots \Loa_n}\,\Theta_{+} & (n\text{:odd/even})
\\
 0 & (n\text{:even/odd})
\end{cases}\,,
\\
 \brTheta_{-}\,\Gamma_{\Loa_1\Loa_2\cdots \Loa_n}\,\Psi_{\mp} 
 &= \begin{cases}
 (-1)^{\frac{n(n+1)}{2}} \bar{\Psi}_{\mp}\,\Gamma_{\Loa_1\Loa_2\cdots \Loa_n}\,\Theta_{-} & (n\text{:odd/even})
\\
 0 & (n\text{:even/odd})
\end{cases}\,,
\end{split}
\\
\begin{split}
 \brTheta_{+}\,\Gamma_{\Lob_1}\,\Gamma_{\Loa_1\Loa_2\cdots \Loa_n}\,\Gamma_{\Lob_2}\,\Psi_{\pm} 
 &= \begin{cases}
 (-1)^{\frac{n(n+1)}{2}} \bar{\Psi}_{\pm}\,\Gamma_{\Lob_2}\,\Gamma_{\Loa_1\Loa_2\cdots \Loa_n}\,\Gamma_{\Lob_1}\,\Theta_{+} & (n\text{:odd/even})
\\
 0 & (n\text{:even/odd})
\end{cases}\,,
\\
 \brTheta_{-}\,\Gamma_{\Lob_1}\,\Gamma_{\Loa_1\Loa_2\cdots \Loa_n}\,\Gamma_{\Lob_2}\,\Psi_{\mp} 
 &= \begin{cases}
 (-1)^{\frac{n(n+1)}{2}} \bar{\Psi}_{\mp}\,\Gamma_{\Lob_2}\,\Gamma_{\Loa_1\Loa_2\cdots \Loa_n}\,\Gamma_{\Lob_1}\,\Theta_{-} & (n\text{:odd/even})
\\
 0 & (n\text{:even/odd})
\end{cases}\,.
\end{split}
\label{eq:bTGGGT}
\end{align}

\section{\texorpdfstring{$\alg{psu}(2,2|4)$}{psu(2,2|4)} algebra}
\label{app:psu-algebra}

In this appendix, we collect our conventions and useful formulas on the $\alg{psu}(2,2|4)$ algebra (see for example \cite{Arutyunov:2009ga} for more details). 

\vspace{-\Pskip}
\subsection{Matrix realization}

\subsubsection*{$8 \times 8$ supermatrix representation}

The super Lie algebra $\alg{su}(2,2|4)$ can be realized by using $8 \times 8$ supermatrices $\cM$ satisfying $\str\cM =0$ and the reality condition
\begin{align}
 \cM^\dagger\, H+H\,\cM =0\,,\qquad 
 \cM = \begin{pmatrix} A & B \\ C & D \end{pmatrix} \,,
\label{eq:reality}
\end{align}
where $\str\cM\equiv \Tr A -\Tr D$ and the hermitian matrix $H$ is defined as
\begin{align}
 H\equiv \begin{pmatrix} \Sigma & \bm{0_4} \\ \bm{0_4} & \bm{1_4} \end{pmatrix} \,,\qquad
 \Sigma \equiv \begin{pmatrix} \bm{0_2} & -\ii\,\sigma_3 \\ \ii\,\sigma_3 & \bm{0_2} \end{pmatrix}=\sigma_2\otimes\sigma_3 \,. 
\end{align}
A trivial element satisfying the above requirement is the $\alg{u}(1)$ generator
\begin{align}
 \gZ = \ii \begin{pmatrix} \bm{1_4} & \bm{0_4} \\ \bm{0_4} & \bm{1_4} \end{pmatrix} \,,
\label{eq:gZ-def}
\end{align}
and the $\alg{psu}(2,2|4)$ is defined as the quotient $\alg{su}(2,2|4)/\alg{u}(1)$\,.

The $\alg{psu}(2,2|4)$ has an automorphism $\Omega$ defined as
\begin{align}
 \Omega(\cM)=-\cK \,\cM^{\ST}\,\cK^{-1}\,,\qquad
 \cK = \begin{pmatrix} K & \bm{0_4} \\ \bm{0_4} & K \end{pmatrix} \,,
\end{align}
where $K$ is a $4\times 4$ matrix
\begin{align}
 K\equiv {\footnotesize\begin{pmatrix} 0 & -1 & 0 & 0 \\ 1 & 0 & 0 & 0 \\ 0 & 0 & 0 & -1 \\ 0 & 0 & 1 & 0 \end{pmatrix}}\,,\qquad 
 K^{-1} = - K \,,
\end{align}
and $\cM^{\ST}$ represents the supertranspose of $\cM$ defined as
\begin{align}
 \cM^{\ST} = \begin{pmatrix} A^\rmT & -C^{\rmT} \\ B^{\rmT} & D^{\rmT} \end{pmatrix} \,. 
\end{align}
By using the automorphism $\Omega$ (of order four), we decompose $\alg{g}=\alg{psu}(2,2|4)$ as
\begin{align}
 \alg{g}=\alg{g}^{(0)}\oplus\alg{g}^{(1)}\oplus\alg{g}^{(2)}\oplus\alg{g}^{(3)}\,,
\end{align}
where $\Omega(\alg{g}^{(k)})=\ii^k\,\alg{g}^{(k)}$ ($k=0,1,2,3$) and the projector to each vector space $\alg{g}^{(k)}$ can be expressed as
\begin{align}
 P^{(k)}(\cM) \equiv \frac{1}{4}\,\bigl[\, \cM + \ii^{3k}\,\Omega(\cM)+\ii^{2k}\, \Omega^2(\cM) +\ii^k\,\Omega^3(\cM) \,\bigr]\,. 
\label{eq:P-i-projector}
\end{align}

\subsubsection*{Bosonic generators}

The bosonic generators of $\alg{psu}(2,2|4)$ algebra, $\gP_{\Loa}$ and $\gJ_{\Loa\Lob}$, can be represented by the following $8\times 8$ supermatrices:
\begin{align}
\begin{split}
 &\{\gP_{\Loa}\}\equiv \{\gP_{\check{\Loa}}\,, \gP_{\hat{\Loa}}\}\,,\qquad 
 \{\gJ_{\Loa\Lob}\}\equiv \{\gJ_{\check{\Loa}\check{\Lob}}\,, \gJ_{\hat{\Loa}\hat{\Lob}}\}\,,
\\
 &\gP_{\check{\Loa}} = 
 \begin{pmatrix}
 \frac{1}{2}\,\bm{\gamma}_{\check{\Loa}} & \bm{0_4} \\ 
 \bm{0_4} & \bm{0_4} 
 \end{pmatrix} \,, \qquad 
 \gJ_{\check{\Loa}\check{\Lob}} = 
 \begin{pmatrix}
 -\frac{1}{2}\,\bm{\gamma}_{\check{\Loa}\check{\Lob}} & \bm{0_4} \\ 
 \bm{0_4} & \bm{0_4} \end{pmatrix}
 \qquad (\check{\Loa},\,\check{\Lob}=0,\dotsc,4)\,, 
\\
 &\gP_{\hat{\Loa}} = 
 \begin{pmatrix}
  \bm{0_4} & \bm{0_4} \\ 
  \bm{0_4} & -\frac{\ii}{2}\,\bm{\gamma}_{\hat{\Loa}} 
 \end{pmatrix} \,, \qquad 
 \gJ_{\hat{\Loa}\hat{\Lob}} = 
 \begin{pmatrix}
  \bm{0_4} & \bm{0_4} \\ 
  \bm{0_4} & -\frac{1}{2}\, \bm{\gamma}_{\hat{\Loa}\hat{\Lob}} 
 \end{pmatrix}
 \qquad (\hat{\Loa},\,\hat{\Lob}=5,\dotsc,9)\,, 
\end{split}
\label{eq:P-J-super}
\end{align}
where we defined $4\times 4$ matrices $\bm{\gamma}_{\check{\Loa}} \equiv (\bm{\gamma}_{\check{\Loa}\check{i}}{}^{\check{j}})$ $(\check{i},\check{j}=1,\dotsc,4)$ and $\bm{\gamma}_{\check{\Loa}} \equiv (\bm{\gamma}_{\check{\Loa}\check{i}}{}^{\check{j}})$ $(\hat{i},\hat{j}=1,\dotsc,4)$
\begin{align}
 &\{\bm{\gamma}_{\check{\Loa}}\} \equiv \bigl\{\brgamma_0\,,\brgamma_1\,,\brgamma_2\,,\brgamma_3\,,\brgamma_5\,\bigr\} \,, \qquad 
 \{\bm{\gamma}_{\hat{\Loa}}\} \equiv \bigl\{-\brgamma_4\,,-\brgamma_1\,,-\brgamma_2\,,-\brgamma_3\,,-\brgamma_5\,\bigr\} \,,
\\[1mm]
 \begin{split}
 &\brgamma_1=
 {\footnotesize\begin{pmatrix}
  0 & 0 & 0 & -1\\
  0 & 0 & 1 & 0\\
  0 & 1 & 0 & 0\\
  -1& 0 & 0 & 0 
 \end{pmatrix}} , \quad 
 \brgamma_2=
 {\footnotesize\begin{pmatrix}
 0 & 0 & 0 & \ii \\
 0 & 0 & \ii & 0 \\
 0 & -\ii& 0 & 0 \\
 -\ii& 0 & 0 & 0
 \end{pmatrix}}, \quad 
 \brgamma_3=
 {\footnotesize\begin{pmatrix}
 0 & 0 & 1 & 0 \\
 0 & 0 & 0 & 1 \\
 1 & 0 & 0 & 0 \\
 0 & 1 & 0 & 0 
 \end{pmatrix}},
\\
 &\brgamma_0= -\ii\,\brgamma_4=
 {\footnotesize\begin{pmatrix}
  0 & 0 & 1 & 0 \\
  0 & 0 & 0 &-1 \\
  -1& 0 & 0 & 0 \\
  0 & 1 & 0 & 0 
 \end{pmatrix}}, \quad 
 \brgamma_5=\ii\,\brgamma_1\brgamma_2\brgamma_3\brgamma_0=
 {\footnotesize\begin{pmatrix}
  1 & 0 & 0 & 0 \\
  0 & 1 & 0 & 0 \\
  0 & 0 &-1 & 0 \\
  0 & 0 & 0 &-1
\end{pmatrix}},
\end{split}
\end{align}
and their antisymmeterizations $\bm{\gamma}_{\check{\Loa}\check{\Lob}} \equiv \bm{\gamma}_{[\check{\Loa}}\,\bm{\gamma}_{\check{\Lob}]}$ and $\bm{\gamma}_{\hat{\Loa}\hat{\Lob}} \equiv \bm{\gamma}_{[\hat{\Loa}}\,\bm{\gamma}_{\hat{\Lob}]}$. 
Here, $\brgamma_{\mu}$ ($\mu=0,\dotsc,3$) and $(\bm{\gamma}_{\Loa})\equiv (\bm{\gamma}_{\check{\Loa}},\,\bm{\gamma}_{\hat{\Loa}})$ satisfy
\begin{align}
 \{\brgamma_{\mu}\,, \brgamma_{\nu}\} = 2\,\eta_{\mu\nu}\,,\qquad (\eta_{\mu\nu}) \equiv \diag(-1,1,1,1)\,,\qquad
 (\bm{\gamma}_{\Loa})^{\rmT}= K\,\bm{\gamma}_{\Loa}\,K^{-1}\,. 
\end{align}

The conformal basis, $\{P_{\mu},\, M_{\mu\nu},\,D,\,K_{\mu}\}$\,, of a bosonic subalgebra $\alg{su}(2,2)\cong \alg{so}(2,4)$ that corresponds to the AdS isometries, can be constructed from $\gP_{\check{\Loa}}$ and $\gJ_{\check{\Loa}\check{\Lob}}$ as
\begin{align}
 P_\mu \equiv \gP_\mu + \gJ_{\mu 4}\,,\qquad K_\mu \equiv \gP_\mu - \gJ_{\mu 4}\,,\qquad M_{\mu\nu}\equiv \gJ_{\mu\nu}\,,\qquad D\equiv \gP_4\,,
\end{align}
where $P_{\mu}$, $M_{\mu\nu}$, $D$, and $K_{\mu}$ represent the translation generators, the Lorentz generators, the dilatation generator, and the special conformal generators, respectively. 
On the other hand, a bosonic subalgebra $\alg{su}(4)\cong\alg{so}(6)$ that corresponds to the isometries of $\rmS^5$ are generated by $\gP_{\hat{\Loa}}$ and $\gJ_{\hat{\Loa}\hat{\Lob}}$\,. 
We choose the Cartan generators of $\alg{su}(4)$ as follows
\begin{align}
 h_1 \equiv \gJ_{57}\,,\qquad h_2 \equiv \gJ_{68}\,,\qquad h_3 \equiv \gP_9\,.
\end{align}

For later convenience, let us also define $16\times 16$ matrices $\gamma_{\Loa}$, $\hat{\gamma}_{\Loa}$, and $\gamma_{\Loa\Lob}$ as
\begin{align}
\begin{split}
 (\gamma_{\Loa})&\equiv(\gamma_{\check{\Loa}},\, \gamma_{\hat{\Loa}}) 
 =(\bm{\gamma}_{\check{\Loa}}\otimes \bm{1_4},\, \bm{1_4}\otimes \bm{\gamma}_{\hat{\Loa}})\,,
\\
 (\hat{\gamma}_{\Loa})&\equiv(\hat{\gamma}_{\check{\Loa}},\, \hat{\gamma}_{\hat{\Loa}}) 
 =(\bm{\gamma}_{\check{\Loa}}\otimes \bm{1_4},\, \bm{1_4}\otimes \ii\,\bm{\gamma}_{\hat{\Loa}})\,,
\\
 (\gamma_{\Loa\Lob})&\equiv (\gamma_{\check{\Loa}\check{\Lob}},\,\gamma_{\hat{\Loa}\hat{\Lob}})
 =(\bm{\gamma}_{\check{\Loa}\check{\Lob}}\otimes \bm{1_4}\,, \bm{1_4}\otimes\bm{\gamma}_{\hat{\Loa}\hat{\Lob}}) \,,
\end{split}
\label{eq:gamma-16}
\end{align}
which satisfy
\begin{align}
\begin{split}
 &(\gamma_{\check{\Loa}})^\dagger = \gamma_{\check{0}}\,\gamma_{\check{\Loa}}\,\gamma_{\check{0}} \,,\qquad 
 (\gamma_{\hat{\Loa}})^\dagger = -\gamma_{\check{0}}\,\gamma_{\hat{\Loa}}\,\gamma_{\check{0}} \,,\qquad 
 (\gamma_{\Loa})^\rmT = (K\otimes K)^{-1}\, \gamma_{\Loa}\,(K\otimes K) \,, 
\\
 &(\hat{\gamma}_{\check{\Loa}})^\dagger = \gamma_{\check{0}}\,\hat{\gamma}_{\check{\Loa}}\,\gamma_{\check{0}} \,,\qquad 
 (\hat{\gamma}_{\Loa})^\rmT = (K\otimes K)^{-1}\, \hat{\gamma}_{\Loa}\,(K\otimes K)\,,
\\
 &\{\gamma_{\Loa},\,\gamma_{\Lob}\}=2\,\eta_{\Loa\Lob}\,,\qquad
 \{\hat{\gamma}_{\check{\Loa}},\,\hat{\gamma}_{\check{\Lob}}\}=2\,\eta_{\check{\Loa}\check{\Lob}}\,,\qquad
 \{\hat{\gamma}_{\hat{\Loa}},\,\hat{\gamma}_{\hat{\Lob}}\}=-2\,\delta_{\hat{\Loa}\hat{\Lob}}\,.
\end{split}
\end{align}
We can easily see $\gamma_{\check{\Loa}\check{\Lob}}= \gamma_{[\check{\Loa}}\,\gamma_{\check{\Lob}]}$ and $\gamma_{\hat{\Loa}\hat{\Lob}}=\gamma_{[\hat{\Loa}}\, \gamma_{\hat{\Lob}]}$\,. 
If we also define $\hat{\gamma}_{\check{\Loa}\check{\Lob}}\equiv \hat{\gamma}_{[\check{\Loa}}\,\hat{\gamma}_{\check{\Lob}]}$ and $\hat{\gamma}_{\hat{\Loa}\hat{\Lob}}\equiv \hat{\gamma}_{[\hat{\Loa}}\,\hat{\gamma}_{\hat{\Lob}]}$\,, they satisfy
\begin{align}
 \hat{\gamma}_{\Loa\Lob} = -\frac{1}{2}\,R_{\Loa\Lob}{}^{\Loc\Lod}\, \gamma_{\Loc\Lod}\,,
\end{align}
where $R_{\Loa\Lob}{}^{\Loc\Lod}$ are the tangent components of Riemann tensor in $\AdS{5}\times\rmS^5$, whose non-vanishing components are
\begin{align}
 R_{\check{\Loa}\check{\Lob}}{}^{\check{\Loc}\check{\Lod}} = -2\, \delta_{[\check{\Loa}}^{[\check{\Loc}}\,\delta_{\check{\Lob}]}^{\check{\Lod}]} \,,\qquad 
 R_{\hat{\Loa}\hat{\Lob}}{}^{\hat{\Loc}\hat{\Lod}} = 2\, \delta_{[\hat{\Loa}}^{[\hat{\Loc}}\,\delta_{\hat{\Lob}]}^{\hat{\Lod}]} \,.
\end{align}

\subsubsection*{Fermionic generators}

The fermionic generators $(\gQ^I)^{\check{\SPa}\hat{\SPa}}$ $(\check{\SPa}, \hat{\SPa}=1,\dotsc, 4)$ are given by
\begin{align}
 (\gQ^1)^{\check{\SPa}\hat{\SPa}} &=
 \begin{pmatrix}
  \bm{0_4} & \ii\,\delta^{\check{\SPa}}_{\check{i}}\,K^{\hat{j}\hat{\SPa}} \\
  -\delta_{\hat{i}}^{\hat{\SPa}}\, K^{\check{\SPa}\check{j}} & \bm{0_4}
 \end{pmatrix} ,
\qquad 
 (\gQ^2)^{\check{\SPa}\hat{\SPa}} =
 \begin{pmatrix}
  \bm{0_4} & - \delta^{\check{\SPa}}_{\check{i}} \,K^{\hat{j}\hat{\SPa}} \\
  \ii\,\delta_{\hat{i}}^{\hat{\SPa}}\, K^{\check{\SPa}\check{j}}& \bm{0_4}
 \end{pmatrix}\,. 
\label{eq:Q-matrix}
\end{align}
As discussed in \cite{Arutyunov:2015qva}, these matrices do not satisfy the reality condition \eqref{eq:reality} but rather their redefinitions $\mathcal{Q}^I$ do. 
The choice, $\gQ^I$ or $\mathcal{Q}^I$, is a matter of convention, and we here employ $\gQ^I$ by following \cite{Arutyunov:2015qva}. 
We also introduce Grassmann-odd coordinates $\theta_I\equiv (\theta_{\check{\SPa}\hat{\SPa}})_I$ which are 16-component Majorana--Weyl spinors satisfying
\begin{align}
 (\gQ^I\,\theta_I)^{\dagger}\, H+H\,(\gQ^I\,\theta_I) = 0 \,. 
\label{eq:Q-theta-reality}
\end{align}
Since the matrices $\gQ^I$ satisfy
\begin{align}
\begin{split}
 &(\gQ^I)^{\dagger}_{\check{\SPa}\hat{\SPa}}
 =-\ii \,K^{-1}_{\check{\SPa}\check{\SPb}}\,(\gQ^I)^{\check{\SPb}\hat{\SPb}}\,K^{-1}_{\hat{\SPb}\hat{\SPa}} \,,
\\
 &H\,(\gQ^I)^{\check{\SPa}\hat{\SPa}}\,H^{-1}=\ii\,(\bm{\gamma}^0)_{\check{\SPb}}{}^{\check{\SPa}}\,(\gQ^I)^{\check{\SPb}\hat{\SPa}}\,,
\end{split}
\end{align}
the condition \eqref{eq:Q-theta-reality} is equivalent to the Majorana condition
\begin{align}
 \brtheta_I \equiv\theta^\dagger_I\, \gamma^0 = \theta_I^{\rmT}\,(K\otimes K)\,,
\end{align}
or more explicitly,
\begin{align}
 \brtheta_I^{\check{\SPa}\hat{\SPa}} = \theta_{I\check{\SPb}\hat{\SPb}}\, K^{\check{\SPb}\check{\SPa}}\,K^{\hat{\SPb}\hat{\SPa}}\,. 
\end{align}

\subsubsection*{Commutation relations}

The generators of $\alg{su}(2,2|4)$ algebra, $\gP_{\Loa}$, $\gJ_{\Loa\Lob}$, $\gQ^I$, and $\gZ$ satisfy the following commutation relations:
\begin{align}
\begin{split}
 [\gP_{\Loa},\,\gP_{\Lob}] &= \frac{1}{2}\,R_{\Loa\Lob}{}^{\Loc\Lod}\,\gJ_{\Loc\Lod}\,, 
\qquad 
 [\gJ_{\Loa\Lob},\,\gP_{\Loc}] = \eta_{\Loc\Loa}\,\gP_{\Lob} - \eta_{\Loc\Lob}\,\gP_{\Loa} \,, 
\\
 [\gJ_{\Loa\Lob},\,\gJ_{\Loc\Lod}] 
 &= \eta_{\Loa\Loc}\,\gJ_{\Lob\Lod}-\eta_{\Loa\Lod}\,\gJ_{\Lob\Loc}-\eta_{\Lob\Loc}\,\gJ_{\Loa\Lod}+\eta_{\Lob\Lod}\,\gJ_{\Loa\Loc} \,, 
\\
 [\gQ^I\,\theta_I,\,\gP_{\Loa}] &= \frac{\ii}{2}\,\epsilon^{IJ}\,\gQ^J\,\hat{\gamma}_{\Loa}\,\theta_I\,, \qquad 
 [\gQ^I\,\theta_I,\,\gJ_{\Loa\Lob}] = \frac{1}{2}\,\delta^{IJ}\,\gQ^I\,\gamma_{\Loa\Lob}\,\theta_J\,,
\\
 [\gQ^I\,\theta_I,\,\gQ^J\,\psi_J] 
 &= -\ii\, \delta^{IJ}\, \brtheta_I\,\hat{\gamma}^{\Loa}\,\psi_J\, \gP_{\Loa} 
   - \frac{1}{4}\,\epsilon^{IJ}\, \brtheta_I\,\gamma^{\Loa\Lob}\,\psi_J\,R_{\Loa\Lob}{}^{\Loc\Lod}\, \gJ_{\Loc\Lod} 
   - \frac{1}{2}\, \delta^{IJ}\, \brtheta_I\,\psi_J\,\gZ \,,
\end{split}
\label{eq:su(2,2|4)}
\end{align}
and the $\alg{psu}(2,2|4)$ algebra is obtained by dropping the last term proportional to $\gZ$\,. 

On the other hand, the bosonic generators $\{P_{\mu},\, M_{\mu\nu},\,D,\,K_{\mu}\}$ satisfy the $\alg{so}(2,4)$ algebra,
\begin{align}
\begin{split}
 [P_\mu,\, K_\nu]&= 2\,\bigl(\eta_{\mu\nu}\, D - M_{\mu\nu}\bigr)\,,\quad 
 [D,\, P_{\mu}]= P_\mu\,,\quad [D,\,K_\mu]= -K_\mu\,,
\\
 [M_{\mu\nu},\, P_\rho] &= \eta_{\mu\rho}\, P_\nu-\eta_{\nu\rho}\, P_\mu \,,\quad 
 [M_{\mu\nu},\, K_\rho] = \eta_{\mu\rho}\,K_\nu-\eta_{\nu\rho}\,K_\mu\,, 
\\
 [M_{\mu\nu},\,M_{\rho\sigma}]&= \eta_{\mu\rho}\,M_{\nu\sigma}-\eta_{\mu\sigma}\,M_{\nu\rho} - \eta_{\nu\rho}\,M_{\mu\sigma}+\eta_{\nu\sigma}\,M_{\mu\rho}\,.
\end{split}
\label{eq:so(2-4)-algebra}
\end{align}

\subsubsection*{Supertrace and Projections}

For generators of the $\alg{psu}(2,2|4)$ algebra, the supertrace become
\begin{align}
\begin{split}
 &\str(\gP_{\Loa}\,\gP_{\Lob})=\eta_{\Loa\Lob}\,,\qquad
 \str(\gJ_{\Loa\Lob}\,\gJ_{\Loc\Lod}) =R_{\Loa\Lob\Loc\Lod}\,,
\\
 &\str(\gQ^I\theta_I\,\gQ^J\lambda_J) =-2\,\epsilon^{IJ}\,\brtheta_I\,\lambda_J\,, 
\end{split}
\end{align}
where $R_{\Loa\Lob\Loc\Lod}\equiv R_{\Loa\Lob}{}^{\Loe\Lof}\,\eta_{\Loe\Loc}\,\eta_{\Lod\Lof}$ and
\begin{align}
 \eta_{\Loa\Lob} \equiv \begin{pmatrix} \eta_{\check{\Loa}\check{\Lob}} & 0 \\ 0 & \eta_{\hat{\Loa}\hat{\Lob}} \end{pmatrix}\,, \quad 
 \eta_{\check{\Loa}\check{\Lob}} \equiv \diag (-1,1,1,1,1)\,,\quad 
 \eta_{\hat{\Loa}\hat{\Lob}} \equiv \diag (1,1,1,1,1)\,. 
\end{align}

Each $\mathbb{Z}_4$-component $\alg{g}^{(i)}$ is spanned by the following generators:
\begin{align}
 \alg{g}^{(0)}\! = \Span_{\mathbb{R}}\{\gJ_{\Loa\Lob}\}\,,\quad 
 \alg{g}^{(1)}\! = \Span_{\mathbb{R}}\{\gQ^1\}\,,\quad 
 \alg{g}^{(2)}\! = \Span_{\mathbb{R}}\{\gP_{\Loa}\}\,,\quad 
 \alg{g}^{(3)}\! = \Span_{\mathbb{R}}\{\gQ^2\}\,. 
\end{align}
Then, from the definition of $d_{\pm}$ \eqref{eq:dpm},
\begin{align}
 d_{\pm} \equiv \mp P^{(1)}+2\,P^{(2)}\pm P^{(3)}\,.
\end{align}
we obtain
\begin{align}
 d_\pm(\gP_{\Loa}) = 2\, \gP_{\Loa}\,,\qquad d_\pm(\gJ_{\Loa\Lob}) =0\,,\qquad 
 d_\pm(\gQ^I) = \mp \sigma_3^{IJ}\,\gQ^J \,. 
\end{align}

\vspace{-\Pskip}
\subsection{Connection to ten-dimensional quantities}

By using the $16\times 16$ matrices $\gamma_{\Loa}$ defined in \eqref{eq:gamma-16}, the $32\times 32$ gamma matrices $(\Gamma_{\Loa})^{\SPa}{}_{\SPb}$ are realized as
\begin{align}
 (\Gamma_{\Loa}) \equiv \bigl(\Gamma_{\check{\Loa}},\, \Gamma_{\hat{\Loa}}\bigr) 
 \equiv \bigl(\sigma_1\otimes \gamma_{\check{\Loa}},\, \sigma_2\otimes \gamma_{\hat{\Loa}}\bigr)\,. 
\end{align}
We can also realize the charge conjugation matrix as
\begin{align}
 C= \ii\,\sigma_2\otimes K \otimes K \,. 
\end{align}

The $32$-component Majorana--Weyl fermions $\Theta_I$ expressed as
\begin{align}
 \Theta_I = \begin{pmatrix} 1 \\ 0\end{pmatrix}\otimes \theta_I \,,
\label{eq:Theta-theta}
\end{align}
which satisfies the chiral conditions
\begin{align}
 \Gamma^{11}\,\Theta_I = \Theta_I\,. 
\end{align}
The Majorana condition is given by
\begin{align}
 \brTheta_I = \Theta_{I}^\rmT\,C = \begin{pmatrix} 0 & 1 \end{pmatrix}\otimes \brtheta_I \,.
\label{eq:brTheta-brtheta}
\end{align}
This decomposition leads to the following relations between $32$- and $8$-component fermions:
\begin{align}
 &\brtheta_I \hat{\gamma}_{\Loa} \theta_J = \brTheta_I \Gamma_{\Loa} \Theta_J \,,
\label{eq:lift-32-AdS5-1}
\\
 &\brtheta_I\,\hat{\gamma}_{\Loa}\,\hat{\gamma}_{\Lob}\,\theta_J 
 = -\ii\,\brTheta_I\,\Gamma_{\Loa}\, \Gamma_{01234}\,\Gamma_{\Lob}\,\Theta_J 
 = \ii\,\brTheta_I\,\Gamma_{\Loa}\, \Gamma_{56789}\,\Gamma_{\Lob}\,\Theta_J \,,
\\
&\ii\,\sigma_1\otimes \bm{1_4}\otimes \bm{1_4} = \Gamma_{01234}\,,\qquad 
 \sigma_2\otimes \bm{1_4}\otimes \bm{1_4} = \Gamma_{56789} \,,
\end{align}
The second relation plays an important role for a supercoset construction of the $\AdS{5} \times \rmS^5$ background since the R--R bispinor in the $\AdS{5}\times \rmS^5$ background takes the form
\begin{align}
 \bisF_5 
 =\frac{1}{5!}\,\bisF_{\Loa_1\cdots \Loa_5}\,\Gamma^{\Loa_1\cdots \Loa_5}
 =4\,(\Gamma^{01234}+\Gamma^{56789})\,.
\end{align}
Indeed, we obtain
\begin{align}
 \brtheta_I\,\hat{\gamma}_{\Loa}\,\hat{\gamma}_{\Lob}\theta_J
 =\frac{\ii}{8}\,\brTheta_I\,\Gamma_{\Loa}\,\bisF_5\,\Gamma_{\Lob}\,\Theta_J\,. 
\label{eq:lift-32-AdS5-2}
\end{align}
We can also show the following relations:\footnote{Recall that $\gamma_{\Loa\Lob}$ has only the components $(\gamma_{\Loa\Lob})=(\gamma_{\check{\Loa}\check{\Lob}},\,\gamma_{\hat{\Loa}\hat{\Lob}})$.}
\begin{align}
 &\brtheta_I\,\hat{\gamma}_{\Loa}\, \gamma_{\Lob\Loc}\,\theta_J = \brTheta_I\,\Gamma_{\Loa}\, \Gamma_{\Lob\Loc} \,\Theta_J \,, 
\label{eq:lift-32-AdS5-3}
\\
 &\brtheta_I\, \gamma_{\Loa\Lob} \,\theta_J
 = -\ii\,\brTheta_I\,\Gamma_{01234}\, \Gamma_{\Loa\Lob} \,\Theta
 = -\ii\,\brTheta_I\,\Gamma_{56789}\, \Gamma_{\Loa\Lob} \,\Theta_J \,, 
\\
 & \brtheta_I\, \gamma_{\Loa\Lob}\,\gamma_{\Loc\Lod}\,\theta_J
 = -\ii\,\brTheta_I\,\Gamma_{01234}\, \Gamma_{\Loa\Lob}\,\Gamma_{\Loc\Lod} \,\Theta_J 
 = -\ii\,\brTheta_I\,\Gamma_{56789}\, \Gamma_{\Loa\Lob}\,\Gamma_{\Loc\Lod} \,\Theta_J \,. 
\end{align}

\section{Geometry of reductive homogeneous space}
\label{app:homogeneous-space}

In this appendix, we review geometry of reductive homogeneous spaces (see for example \cite{Castellani:1991et,Ortin:2004ms} for more details). 

\vspace{-\Pskip}
\subsection{Generalities}

Let us consider a homogeneous space $G/H$ and decompose the Lie algebra as a direct sum of vector spaces, $\alg{g}=\alg{h}\oplus\alg{k}$. 
If $[\alg{k},\,\alg{h}]\subset \alg{k}$ is satisfied, $G/H$ is called reductive, and if $[\alg{k},\,\alg{k}]\subset \alg{h}$ is further satisfied, $G/H$ is called symmetric. 
We denote the basis of $\alg{h}$ as $\{\sfJ_{\sfi}\}$ $(\sfi=1,\dotsc,\dim G-\dim H)$ and those of $\alg{k}$ as $\{\sfP_{\sfa}\}$ $(\sfa=1,\dotsc,\dim G-\dim H)$\,. 

We choose a gauge where the coset representative $g(x)$ is expanded only in terms of $\sfK_{m}$, like $g(x)=\exp(x^{m}\,\sfK_{m})$ ($m=1,\dotsc,\dim G-\dim H$)\,. 
Here, $\{\sfK_{m}\}$ is arbitrary as long as $\{\sfK_{m}\}$ and $\{\sfJ_{\sfi}\}$ span the vector spaces $\alg{g}$\,. 
An obvious choice is $\{\sfK_{m}\}=\{\sfP_{\sfa}\}$, but it is not necessary to choose in that way. 
Once we fix the set $\{\sfK_{m}\}$, in order to maintain the gauge choice under a left multiplication $g(x)\to g_L\,g(x)$, we need to simultaneously perform a local right multiplication,
\begin{align}
 g(x) \ \to \ g(x') = g_L\,g(x)\,h^{-1}(x)\qquad (h\in H) \,.
\label{eq:left-G}
\end{align}
Then, if we expand the left-invariant Maurer--Cartan 1-form as
\begin{align}
 A\equiv g^{-1}\,\rmd g = e^{\sfa}\,\sfP_{\sfa} - \Omega^{\sfi}\,\sfJ_{\sfi} \,, 
\label{eq:MC-1-form-decomp}
\end{align}
we obtain the following transformation laws under the left multiplication \eqref{eq:left-G}:
\begin{align}
 e^{\sfa}(x)\ \to \ e'^{\sfa}(x') = \Lambda^{\sfa}{}_{\sfb}\,e^{\sfb}(x) \,,\qquad 
 \Omega^{\sfi}(x)\ \to \ \Omega'^{\sfi}(x') = \bigl[\Lambda^{\sfi}{}_{\sfj} \, \Omega^{\sfj} - (h^{-1}\,\rmd h)^{\sfi}\bigr](x) \,,
\end{align}
where we have defined $h\,\sfP_{\sfa}\,h^{-1}=\Lambda^{\sfb}{}_{\sfa}\,\sfP_{\sfb}$ and $h\,\sfJ_{\sfi}\,h^{-1}\equiv \Lambda^{\sfj}{}_{\sfi}\,\sfP_{\sfj}$\,. 
This shows that $\Omega^{\sfi}$ behaves as a connection of $H$\,. 
From the decomposition \eqref{eq:MC-1-form-decomp}, the Maurer--Cartan equations become
\begin{align}
\begin{split}
 0 =\rmd A + A\wedge A 
 &= \bigl(\rmd e^{\sfa} - \Omega^{\sfi} \wedge e^{\sfb}\, f_{\sfi\sfb}{}^{\sfa} 
 + \frac{1}{2}\, e^{\sfb}\wedge e^{\sfc}\,f_{\sfb\sfc}{}^{\sfa}\bigr)\,\sfP_{\sfa}
\\
 &\quad - \Bigl(\rmd \Omega^{\sfi} - \frac{1}{2}\,\Omega^{\sfj}\wedge \Omega^{\sfk}\,f_{\sfj\sfk}{}^{\sfi} - \frac{1}{2}\, e^{\sfb}\wedge e^{\sfc}\,f_{\sfb\sfc}{}^{\sfi} \Bigr) \,\sfJ_{\sfi}\,. 
\end{split}
\end{align}
If we regard $e^{\sfa}$ as the vielbein on $G/H$ and suppose the absence of torsion
\begin{align}
 T^{\sfa} \equiv \rmd e^{\sfa} + \omega^{\sfa}{}_{\sfb}\wedge e^{\sfb} = 0\,,
\end{align}
the Maurer--Cartan equations show that the spin connection can be expressed as
\begin{align}
 \omega^{\sfa}{}_{\sfb} = - \Omega^{\sfi} \, f_{\sfi\sfb}{}^{\sfa} + \frac{1}{2}\, e^{\sfc} \,f_{\sfc\sfb}{}^{\sfa} \,.
\label{eq:spin-connection-Omega}
\end{align}
Moreover, the associated Riemann curvature tensor is expressed as
\begin{align}
\begin{split}
 R^{\sfa}{}_{\sfb} &\equiv \frac{1}{2}\,e^{\sfc}\wedge e^{\sfd}\,R_{\sfc\sfd}{}^{\sfa}{}_{\sfb} 
 \equiv \rmd \omega^{\sfa}{}_{\sfb} + \omega^{\sfa}{}_{\sfc}\wedge \omega^{\sfc}{}_{\sfb} 
\\
 &= - \frac{1}{2}\, e^{\sfe}\wedge e^{\sff}\,\Bigl(f_{\sfe\sff}{}^{\sfi}\, f_{\sfi\sfb}{}^{\sfa} +\frac{1}{2}\,f_{\sfe\sff}{}^{\sfc}\,f_{\sfc\sfb}{}^{\sfa}-\frac{1}{2}\,f_{\sfe\sfd}{}^{\sfa}\,f_{\sff\sfb}{}^{\sfd} \Bigr) \,,
\\
 R^{\sfa}{}_{\sfb\sfc\sfd} &= - \Bigl(f_{\sfc\sfd}{}^{\sfi}\, f_{\sfi\sfb}{}^{\sfa} +\frac{1}{2}\,f_{\sfc\sfd}{}^{\sfe}\,f_{\sfe\sfb}{}^{\sfa}-\frac{1}{2}\,f_{\sfc\sfe}{}^{\sfa}\,f_{\sfd\sfb}{}^{\sfe} \Bigr)\,.
\end{split}
\label{eq:Riemann-reductive}
\end{align}

In order to obtain the Killing vectors on $G/H$, let us consider an infinitesimal left multiplication
\begin{align}
 g_L=1+\epsilon^i\,T_i\,,\qquad h=1-\epsilon^i\,W_i{}^{\sfi}\,\sfJ_{\sfi}\,, 
\end{align}
under which the coordinates are supposed to transform as
\begin{align}
 x'^{m} = x^{m} + \epsilon^i\,\hat{T}_i^m \,. 
\end{align}
We obtain
\begin{align}
\begin{split}
 \epsilon^i\,\bigl( T_i\,g + g\,W_i{}^{\sfi}\,\sfJ_{\sfi}\bigr) 
 &= \delta_{\epsilon} g = g(x+ \epsilon^i\,\hat{T}_i)-g(x)
 = \epsilon^i\, \hat{T}_i^m\, \partial_m g 
\\
 &= \epsilon^i\, \hat{T}_i^m\, g\,\bigl(e_m{}^{\sfa}\,\sfP_{\sfa} - \Omega_m{}^{\sfi}\,\sfJ_{\sfi}\bigr)\,,
\end{split}
\end{align}
and this leads to
\begin{align}
 \bigl[\Ad_{g^{-1}}\bigr]_i{}^j\,T_j \equiv g^{-1}\,T_i\,g = \hat{T}_i^{\sfa}\,\sfP_{\sfa} - \bigl(\hat{T}_i^m\,\Omega_{m}{}^{\sfi} + W_i{}^{\sfi}\bigr)\,\sfJ_{\sfi} \,,
\end{align}
where $\hat{T}_i^{\sfa} \equiv \hat{T}_i^m\, e_m{}^{\sfa}$\,. 
We thus obtain the following expression:
\begin{align}
 \hat{T}_i^{\sfa} = \bigl[\Ad_{g^{-1}}\bigr]_i{}^{\sfa} \,,\qquad 
 W_i{}^{\sfi} = - \hat{T}_i^m\,\Omega_{m}{}^{\sfi} - \bigl[\Ad_{g^{-1}}\bigr]_i{}^{\sfi} \,.
\label{eq:Killing-formula}
\end{align}
Under the same variation, we obtain
\begin{align}
 \delta_\epsilon A&= \epsilon^i\,\bigl[e^{\sfa}\,W_i{}^{\sfi}\,f_{\sfa\sfi}{}^{\sfb}\,\sfP_{\sfb} + (\rmd W_i{}^{\sfi}-\Omega^{\sfj}\,W_i{}^{\sfk}\,f_{\sfj\sfk}{}^{\sfi})\,\sfJ_{\sfi}\bigr]\,,
\\
 \delta_\epsilon e^{\sfa}&=\epsilon^i\, e^{\sfb}\,W_i{}^{\sfi}\,f_{\sfb\sfi}{}^{\sfa}\,,\qquad 
 \delta_\epsilon \Omega^{\sfi} = \epsilon^i\,\bigl(\Omega^{\sfj}\,W_i{}^{\sfk}\,f_{\sfj\sfk}{}^{\sfi} - \rmd W_i{}^{\sfi}\bigr) \,.
\end{align}
If we define the metric on $G/H$ as
\begin{align}
 g_{mn} \equiv e_m{}^{\sfa}\,e_n{}^{\sfb}\,\kappa_{\sfa\sfb} \,,
\end{align}
by using a constant matrix $\kappa_{\sfa\sfb}$ satisfying
\begin{align}
 f_{\sfi(\sfa}{}^{\sfc}\,\kappa_{\sfb)\sfc} = 0 \,,
\end{align}
the metric is invariant under the variation,
\begin{align}
 \delta_\epsilon g_{mn} = -2\,\epsilon^i\, e_{(m}{}^{\sfa}\,e_{n)}{}^{\sfb}\,W_i{}^{\sfi}\,f_{\sfi(\sfa}{}^{\sfc}\,\kappa_{\sfb)\sfc} = 0 \,. 
\end{align}
We can check that the variation is the same as the Lie derivative,
\begin{align}
 \delta_\epsilon e_m{}^{\sfa} = \epsilon^i\,\Lie_{\hat{T}_i} e_m{}^{\sfa} = \epsilon^i\,\bigl(\hat{T}_i^n\,\partial_n e_m{}^{\sfa} + \partial_m \hat{T}_i^n\,e_n{}^{\sfa}\bigr) \,,
\end{align}
and the invariance of the metric indicates that $\hat{T}_i^m$ are Killing vectors associated with the generator $T_i$\,. 

From $T_i\,g = \hat{T}_i^m\, \partial_m g -g\,W_i{}^{\sfi}\,\sfJ_{\sfi}$\,, we can calculate commutators of two variations as
\begin{align}
 [T_i,\,T_j]\,g = -(\Lie_{\hat{T}_i} \hat{T}_j)^m\, \partial_m g 
 + \bigl(\hat{T}_i^n\,\partial_n W_j{}^{\sfi} - \hat{T}_j^n\,\partial_n W_i{}^{\sfi}+ W_i{}^{\sfj}\, \,W_j{}^{\sfk}\,f_{\sfj\sfk}{}^{\sfi} \bigr)\, g\,\sfJ_{\sfi} \,. 
\end{align}
On the other hand, from $[T_i,\,T_j]=f_{ij}{}^k\,T_k$\,, we can also express the left-hand side as
\begin{align}
 [T_i,\,T_j]\,g = f_{ij}{}^k\,T_k\,g = f_{ij}{}^k\,\hat{T}_k^m\, \partial_m g -f_{ij}{}^k\,g\,W_k{}^{\sfi}\,\sfJ_{\sfi}\,,
\end{align}
and by comparing these, we obtain
\begin{align}
\begin{split}
 &[\hat{T}_i,\, \hat{T}_j]^m = (\Lie_{\hat{T}_i} \hat{T}_j)^m = - f_{ij}{}^k\,\hat{T}_k^m\,,
\\
 &\hat{T}_i^n\,\partial_n W_j{}^{\sfi} - \hat{T}_j^n\,\partial_n W_i{}^{\sfi}+ W_i{}^{\sfj}\, \,W_j{}^{\sfk}\,f_{\sfj\sfk}{}^{\sfi} = -f_{ij}{}^k\,g\,W_k{}^{\sfi}\,. 
\end{split}
\end{align}

\vspace{-\Pskip}
\subsection{\texorpdfstring{$\AdS{5}\times \rmS^5$}{AdS\textfiveinferior{x}S\textfivesuperior}}

In the case of the (bosonic) coset
\begin{align}
 \AdS{5}\times \rmS^5 = \frac{\SO(2,4)}{\SO(1,4)}\times \frac{\SO(6)}{\SO(5)}\,,
\end{align}
the two sets of generators are given by
\begin{align}
 \{\sfP_{\sfa}\} = \bigl\{\gP_{\Loa}\bigr\}\,,\qquad 
 \{\sfJ_{\sfi}\} = \bigl\{\gJ_{\Loa\Lob}/\sqrt{2!} \bigr\}\,,
\end{align}
and it is a symmetric coset space ($f_{\Loa\Lob}{}^{\Loc}=0$). 
The normalization $\tfrac{1}{\sqrt{2!}}$ is introduced to prevent overcounting coming from the summation of antisymmetrized indices. 
Quantities with the index $\sfi$ always contains the factor $\tfrac{1}{\sqrt{2!}}$ and, for example, the Maurer--Cartan 1-form \eqref{eq:MC-1-form-decomp} is expressed as
\begin{align}
 A = e^{\sfa}\,\sfP_{\sfa} - \Omega^{\sfi}\,\sfJ_{\sfi}\qquad\leftrightarrow\qquad
 A = e^{\Loa}\,\gP_{\Loa} - \frac{1}{2}\,\Omega^{\Loa\Lob}\,\gJ_{\Loa\Lob} \,. 
\end{align}
From \eqref{eq:spin-connection-Omega} and $f_{\Loa\Lob}{}^{\Loc}=0$, the spin connection becomes
\begin{align}
 \omega^{\Loa}{}_{\Lob} = - \frac{1}{2}\, \Omega^{[\Loc\Lod]} \, f_{[\Loc\Lod]\Lob}{}^{\Loa} 
 = \Omega^{\Loa\Loc} \, \eta_{\Loc\Lob} \,,
\label{eq:index-convention}
\end{align}
where we used $f_{[\Loc\Lod]\Lob}{}^{\Loa}=2\,\eta_{\Lob[\Loc}\,\delta_{\Lod]}^{\Loa}$ (see $[\gJ,\,\gP]$-commutator of \eqref{eq:su(2,2|4)}) and we obtain
\begin{align}
 A = e^{\Loa}\,\gP_{\Loa} - \frac{1}{2}\,\omega^{\Loa\Lob}\,\gJ_{\Loa\Lob} \,,
\end{align}
independent of the explicit parameterization of $g$ like \eqref{eq:group-parameterization}. 

From \eqref{eq:Riemann-reductive} and $f_{\Loa\Lob}{}^{\Loc}=0$, the Riemann curvature tensor becomes
\begin{align}
 R^{\Loa}{}_{\Lob\Loc\Lod} = - \frac{1}{2}\, f_{\Loc\Lod}{}^{[\Loe\Lof]}\, f_{[\Loe\Lof]\Lob}{}^{\Loa} 
 = f_{\Loc\Lod}{}^{[\Loa\Loe]}\, \eta_{\Loe\Lob} \,.
\end{align}
This explains why the $[\gP,\,\gP]$-commutator in \eqref{eq:su(2,2|4)} is expressed in terms of the Riemann tensor; $f_{\Loc\Lod}{}^{[\Loa\Lob]}=R^{\Loa\Lob}{}_{\Loc\Lod}=R_{\Loc\Lod}{}^{\Loa\Lob}$\,. 

\section{Equivalence of (\ref{eq:R-R-relation1}) and (\ref{eq:R-R-relation2})}
\label{app:formula-proof}

In this appendix, we prove that the relation \eqref{eq:R-R-relation1},
\begin{align}
 \hat{\cF} = \Exp{\Phi-\tilde{\phi}}\Exp{-B_2\wedge} \Exp{-\beta\vee} \check{\cF} \,,
\end{align}
is equivalent to the relation \eqref{eq:R-R-relation2},
\begin{align}
 \bisF = \check{\bm{\cF}}\,\Omega_0^{-1}\,,\qquad 
 \Omega^{-1}_0 =(\det \Einv_{\Loa}{}^{\Lob})^{-\frac{1}{2}} \text{\AE}\bigl(-\tfrac{1}{2}\,\beta^{\Loa\Lob}\,\Gamma_{\Loa\Lob}\bigr) \,.
\end{align}
Since $\Exp{\Phi-\tilde{\phi}} = (\det \Einv_{\Loa}{}^{\Lob})^{-\frac{1}{2}}$ from \eqref{eq:two-dilatons}, we here show the equivalence of two relations,
\begin{align}
 \hat{F} = \Exp{-B_2\wedge} \Exp{-\beta\vee} \check{F} \qquad \Leftrightarrow\qquad 
 \hat{\bmF} = \check{\bmF}\,\brOmega_0^{-1} \,,\quad \brOmega_0^{-1} \equiv \text{\AE}\bigl(-\tfrac{1}{2}\,\beta^{\Loa\Lob}\,\Gamma_{\Loa\Lob}\bigr) \,,
\end{align}
where $\hat{\bmF}\equiv \Exp{-\Phi}\bisF$ and $\check{\bmF}\equiv \Exp{-\tilde{\phi}}\check{\bm{\cF}}$\,. 

\vspace{-\Pskip}
\subsection{Evaluation of \texorpdfstring{$\hat{F} =\Exp{-B_2\wedge}\Exp{-\beta\vee}\check{F}$}{\^{F}=exp(-B\textwedge)exp(-\textbeta\textvee)\v{F}}}

Let us first evaluate $\Exp{-B_2\wedge}\Exp{-\beta\vee}\check{F}$\,. 
This can be expanded as
\begin{align}
 \Exp{-B_2\wedge}\Exp{-\beta\vee}\check{F}
 &=\sum_{k:\Atop{\text{even}}{\text{odd}}} \sum_{t=0}^{[\frac{k}{2}]} \sum_{s=0}^{t+[\frac{D-k}{2}]} \frac{(-1)^{s}}{2^{s+t}\,s!\,t!\,(k-2\,t)!}\,B_{m_1m_2}\cdots B_{m_{2s-1}m_{2s}}\,\beta^{n_1n_2}\cdots\beta^{n_{2t-1}n_{2t}}
\nn\\
 &\quad\times
 \check{F}_{n_1\cdots n_{2t}m_{2s+1}\cdots m_{2s+k-2t}}\, \rmd x^{m_1}\wedge\cdots\wedge\rmd x^{m_{2s+k-2t}}
\nn\\
 &=\sum_{r:\Atop{\text{even}}{\text{odd}}} \sum_{s=0}^{[\frac{r}{2}]} \sum_{t=0}^{s+[\frac{D-r}{2}]} \frac{(-1)^s}{2^{s+t}\,s!\,t!\,(r-2s)!}\,\beta^{\Loc_1\Loc_2}\cdots\beta^{\Loc_{2t-1}\Loc_{2t}}\,
 \beta_{\Lob_1\Lob_2}\cdots \beta_{\Lob_{2s-1}\Lob_{2s}}
\nn\\
 &\quad\times e_{m_1}{}^{\Lob_1}\cdots e_{m_{2s}}{}^{\Lob_{2s}}\,
 \check{F}_{\Loc_1\cdots \Loc_{2t}m_{2s+1}\cdots m_r}\,\rmd x^{m_1}\wedge\cdots\wedge\rmd x^{m_r} \,,
\end{align}
where the square bracket $[n]$ denotes the integral part of $n$\,, and in the second equality, we have used relations \eqref{eq:dual-RR-flat-components} and \eqref{eq:g-G-B-beta}. 
Then, $\hat{F}$ with flat indices becomes
\begin{align}
 \hat{F}_{\Loa_1\cdots \Loa_k} &\equiv e_{\Loa_1}{}^{m_1}\cdots e_{\Loa_k}{}^{m_k}\,\hat{F}_{m_1\cdots m_k} 
\nn\\
 &=\sum_{s=0}^{[\frac{k}{2}]} \sum_{t=0}^{s+[\frac{D-k}{2}]} \frac{(-1)^{s}\,k!}{2^{s+t}\,s!\,t!\,(k-2s)!}\,
 \beta^{\Loc_{1}\Loc_{2}}\cdots\beta^{\Loc_{2t-1}\Loc_{2t}} 
\nn\\
&\quad\times
 \beta_{[\Loa_1\Loa_2}\cdots \beta_{\Loa_{2s-1}\Loa_{2s}}\, (\Einv^\rmT)_{\Loa_{2s+1}}{}^{\Lob_{2s+1}}\cdots (\Einv^\rmT)_{\Loa_{k}]}{}^{\Lob_{k}}\,
 \check{F}_{\Lob_{2s+1}\cdots \Lob_{k} \Loc_1\cdots \Loc_{2t}}
\nn\\
 &=\sum_{s=0}^{[\frac{k}{2}]} \sum_{t=0}^{s+[\frac{D-k}{2}]} \sum_{u=0}^{k-2s} \frac{(-1)^{s}\,k!}{2^{s+t}\,s!\,t!\,u!\,(k-2s-u)!}\,
 \beta^{\Loc_{1}\Loc_{2}}\cdots\beta^{\Loc_{2t-1}\Loc_{2t}} 
\nn\\
 &\quad\times \beta_{[\Loa_1\Loa_2}\cdots \beta_{\Loa_{2s-1}\Loa_{2s}}\,
 \beta_{\Loa_{2s+1}}{}^{\Lob_{1}}\cdots \beta_{\Loa_{2s+u}}{}^{\Lob_{u}}\,
 \check{F}_{|\Lob_{1}\cdots \Lob_{u}|\Loa_{2s+u+1}\cdots \Loa_k] \Loc_1\cdots \Loc_{2t}}\,,
\label{eq:Fhat-Fcheck-flat}
\end{align}
where we used $e_{\Loa}{}^m = (\Einv^{\rmT})_{\Loa}{}^{\Lob}\, \tilde{e}_{\Lob}{}^m$ and $(\Einv^\rmT)_{\Loa}{}^{\Lob}=\delta_{\Loa}^{\Lob}+\beta_{\Loa}{}^{\Lob}$\,. 

\vspace{-\Pskip}
\subsection{Evaluation of \texorpdfstring{$\hat{\bmF} =\check{\bmF} \,\brOmega_0^{-1}$}{\^{F}=\v{F} \={\textOmega}\textzeroinferior\textminussuperior\textonesuperior}}

Next, by using the definitions,
\begin{align}
 \check{\bmF} = \sum_{k:\Atop{\text{even}}{\text{odd}}}\frac{1}{k!}\, \check{F}_{\Loa_1\cdots \Loa_k}\,\Gamma^{\Loa_1\cdots \Loa_k} \,,\qquad
 \brOmega_0^{-1} = \sum_{\ell=0}^{[\frac{D}{2}]} \frac{(-1)^\ell}{2^\ell\,\ell!}\,
 \beta^{\Lob_1 \Lob_2}\cdots\beta^{\Lob_{2\ell-1} \Lob_{2\ell}}\,\Gamma_{\Lob_1\cdots \Lob_{2\ell}} \,,
\end{align}
let us expand the right-hand side of $\hat{\bmF}=\check{\bmF} \,\brOmega_0^{-1}$ as
\begin{align}
 \hat{\bmF} 
 &= \sum^{[\frac{D}{2}]}_{\ell=0}\sum_{k:\Atop{\text{even}}{\text{odd}}}\frac{(-1)^\ell}{2^\ell\,\ell!\,k!}\,
 \check{F}_{\Loa_1\cdots \Loa_k}\,
 \beta^{\Lob_1\Lob_2}\cdots\beta^{\Lob_{2\ell-1}\Lob_{2\ell}}\,
 \Gamma^{\Loa_1\cdots \Loa_k}\,\Gamma_{\Lob_1\cdots \Lob_{2\ell}}
\nn\\
 &= \sum^{[\frac{D}{2}]}_{\ell=0}\sum_{k:\Atop{\text{even}}{\text{odd}}}\sum^{2\ell+k}_{s=\abs{2\ell-k}} \frac{(-1)^{\frac{r(r-1)}{2}}\,{}_{s}\mathrm{C}_{r}\,(2\ell)!}{2^\ell\,\ell!\,s!\,(2\ell-r)!}\,
 \beta^{[\Lob_1\Lob_2}\cdots \beta^{\Lob_{2\ell-1} \Lob_{2\ell}]}\,
 \eta_{\Lob_{1}\Loc_{1}}\cdots \eta_{\Lob_{r}\Loc_{r}}\,
 \check{F}_{\Lob_{r+1}\cdots \Lob_{2\ell} \Loc_{r+1}\cdots \Loc_{s}}\,
 \Gamma^{\Loc_1\cdots \Loc_s}\,,
\end{align}
where we used the formula \eqref{eq:Gamma-p-q} and defined $r\equiv \frac{2\ell-k+s}{2}$\,. 
Then, the R--R field strength $\hat{F}_{\Loa_1\cdots \Loa_k}$ with flat indices becomes
\begin{align}
 \hat{F}_{\Loa_1\cdots \Loa_k}
 =\sum_{\ell,\,r} \frac{(-1)^{\frac{r(r-1)}{2}}\,{}_k\mathrm{C}_r\,(2\ell)!}{2^\ell\,\ell!\,(2\ell-r)!}\, \beta^{[\Lob_1 \Lob_2}\cdots\beta^{\Lob_{2\ell-1} \Lob_{2\ell}]}
 \,\eta_{\Lob_1[\underline{\Loa_{1}}}\cdots \eta_{\Lob_{r}\underline{\Loa_{r}}}\, \check{F}_{\Lob_{r+1}\cdots\Lob_{2\ell}\underline{\Loa_{r+1}\cdots\Loa_{k}}]} \,,
\label{eq:hatF-flat}
\end{align}
where the under-barred indices are totally antisymmetrized and non-negative integers $\ell$ and $r$ run over the region where the following relations are satisfied:
\begin{align}
 0\leq 2\,\ell-r \,,\qquad 0\leq k-r \,, \qquad 0\leq k+2\,\ell-2\,r\leq D\,.
\end{align}
We can further expand the right-hand side of \eqref{eq:hatF-flat} as
\footnote{We used the identity for arbitrary totally antisymmetric tensors $C^{\Loa\Lob}$\,, $A_{\Loa_1\cdots \Loa_r}$\,, and $B_{\Loa_{r+1}\cdots \Loa_{2\ell}}$,
\begin{align*}
 &C^{[\Loa_1\Loa_2}\cdots C^{\Loa_{2\ell-1}\Loa_{2\ell}]}\, A_{\Loa_1\cdots \Loa_r}\,B_{\Loa_{r+1}\cdots \Loa_{2\ell}}
\\
 &=\sum_{u}\frac{(-1)^{\frac{u(u-1)}{2}}\,2^u\,\ell!}{s!\,t!\,u!}\,\frac{r!\,(2\ell-r)!}{(2\ell)!}\, C^{\Lob_1\Lob_2}\cdots C^{\Lob_{2t-1}\Lob_{2t}} 
\\
 &\quad\times \bigl(C^{\Loa_1\Loa_2}\cdots C^{\Loa_{2s-1}\Loa_{2s}}\bigr)
 \bigl(C^{\Loc_1\Lod_1}\cdots C^{\Loc_u\Lod_u}\bigr)
 \,A_{\Loa_1\cdots \Loa_{2s}\Loc_1\cdots \Loc_u}\,B_{\Lod_1\cdots \Lod_u\Lob_1\cdots \Lob_{2t}}\,,
\end{align*}
where $0\leq r \leq 2 \ell$,\ $2 s\equiv r-u$, and $2 t\equiv 2 \ell-r-u$.
The range of the summation over $u$ is as follows:
\begin{align*}
 u= \begin{cases}
    0\,, 2\,, \dotsc\,, \min (r,2\ell-r) & \text{for $r$ even}\,, \\
    1\,, 3\,, \dotsc\,, \min (r,2\ell-r) & \text{for $r$ odd}\,.
  \end{cases}
\end{align*}}
\begin{align}
 \hat{F}_{\Loa_1\cdots \Loa_k} 
 &=\sum_{\ell,\,r,\,u} \frac{(-1)^{s}\,2^u\,k!}{2^\ell\,s!\,t!\,u!\,(k-r)!} \,
 \beta^{\Lob_{1} \Lob_{2}}\cdots\beta^{\Lob_{2t-1}\Lob_{2t}}
\nn\\
 &\quad\times 
 \beta_{[\underline{\Loa_{1}\Loa_{2}}}\cdots\beta_{\underline{\Loa_{2s-1}\Loa_{2s}}} \,
 \beta_{\underline{\Loa_{2s+1}}}{}^{\Loc_{1}}\cdots \beta_{\underline{\Loa_{r}}}{}^{\Loc_{u}}\,
 \check{F}_{|\Loc_{1}\cdots\Loc_{u}| \underline{\Loa_{r+1}\cdots\Loa_k}]\Lob_{1}\cdots\Lob_{2t}} \,,
\end{align}
where $s$ and $t$ are defined as
\begin{align}
 2\,s\equiv r-u \,,\qquad 2\,t\equiv 2\,\ell-r-u \,,
\end{align}
and non-negative integers $\ell$, $r$, and $u$ run over the region where
\begin{align}
 0\leq s\,,\quad 0\leq t\,,\quad r\leq k \,,\quad 0\leq k+2\,t-2\,s \leq D \,,
\end{align}
are satisfied. 
If we change the variables, we obtain a more explicit expression,
\begin{align}
 \hat{F}_{\Loa_1\cdots \Loa_k} 
 &=\sum_{s=0}^{[\frac{k}{2}]} \sum_{t=0}^{s+[\frac{D-k}{2}]} \sum_{u=0}^{k-2s} \frac{(-1)^{s}\,k!}{2^{s+t}\,s!\,t!\,u!\,(k-2s-u)!} \,
 \beta^{\Loc_{1}\Loc_{2}}\cdots\beta^{\Loc_{2t-1}\Loc_{2t}}
\nn\\
 &\times 
 \beta_{[\underline{\Loa_{1}\Loa_{2}}}\cdots\beta_{\underline{\Loa_{2s-1}\Loa_{2s}}} \,
 \beta_{\underline{\Loa_{2s+1}}}{}^{\Lod_{1}}\cdots \beta_{\underline{\Loa_{r}}}{}^{\Lod_{u}}\,
 \check{F}_{|\Lod_{1}\cdots\Lod_{u}| \underline{\Loa_{r+1}\cdots\Loa_k}]\Loc_{1}\cdots\Loc_{2t}} \,. 
\end{align}
This precisely matches with \eqref{eq:Fhat-Fcheck-flat} and the equivalence has been proven. 

\section{The spinor rotation \texorpdfstring{$\Omega$}{\textOmega}}
\label{app:omega}

In this Appendix, we prove the formula \eqref{eq:Hassan-formula} in an arbitrary even dimension $D$. 
Namely, we prove that the spinor representation of a local Lorentz transformation $\Lambda^{\Loa}{}_{\Lob} \equiv (O_+^{-1}\,O_-)^{\Loa}{}_{\Lob}$ ($O_\pm \equiv \delta^{\Loa}_{\Lob} \pm a^{\Loa}{}_{\Lob}$, $a^{\Loa\Lob}=-a^{\Lob\Loa}$) is given by
\begin{align}
 \Omega_{(a)} = (\det O_{\pm})^{-\frac{1}{2}} \text{\AE}\bigl(-\tfrac{1}{2}\,a_{\Loa\Lob}\,\Gamma^{\Loa\Lob}\bigr)\,,
 \qquad 
 \Omega_{(a)}^{-1} =(\det O_{\pm})^{-\frac{1}{2}} \text{\AE}\bigl(\tfrac{1}{2}\,a_{\Loa\Lob}\,\Gamma^{\Loa\Lob}\bigr) \,.
\end{align}

If we define matrices
\begin{align}
 \Omega_\pm \equiv \text{\AE}\bigl(\pm\tfrac{1}{2}\,a_{\Loa\Lob}\,\Gamma^{\Loa\Lob}\bigr)
 = \sum^{5}_{p=0}\frac{(\pm 1)^p}{2^p\,p!}\, a_{\Lob_1 \Lob_2}\cdots a_{\Lob_{2p-1}\Lob_{2p}}\,\Gamma^{\Lob_1\Lob_2\cdots \Lob_{2p-1}\Lob_{2p}}\,,
\end{align}
we can easily show the identity,
\begin{align}
 (O_\mp)^{\Loa}{}_{\Lob}\,\Gamma^{\Lob}\,\Omega_\pm - (O_{\pm})^{\Loa}{}_{\Lob}\,\Omega_\pm\,\Gamma^{\Lob}
 = [\Gamma^{\Loa},\,\Omega_\pm] \mp a^{\Loa}{}_{\Lob}\,\{\Gamma^{\Lob},\,\Omega_\pm\} =0\,,
\end{align}
and this leads to
\begin{align}
 \Omega^{-1}_\pm\,\Gamma^{\Loa}\,\Omega_\pm = (O_{\mp}^{-1}\,O_{\pm})^{\Loa}{}_{\Lob}\,\Gamma^{\Lob} \,. 
\label{eq:Omega-pm}
\end{align}
Choosing the lower sign, we obtain the desired relation,
\begin{align}
 \Omega^{-1}_-\,\Gamma^{\Lob}\,\Omega_- = \Lambda^{\Loa}{}_{\Lob}\,\Gamma^{\Lob} \,.
\end{align}
In the following, we rescale $\Omega_-$ and define $\Omega_{(a)}$ such that $\Omega_{(a)}^{-1}=\Omega_{(-a)}$\,. 
The relation \eqref{eq:Omega-pm} implies that $\Omega^{-1}_\pm$ is proportional to $\Omega_{\mp}$\,, and we denote their relation as
\begin{align}
 \Omega^{-1}_- = \frac{1}{\abs{\Omega}^2}\,\Omega_+ \,. 
\end{align}
We shall show $\abs{\Omega}^2 =\det (\delta_{\Loa}^{\Lob} - a_{\Loa}{}^{\Lob})= \det O_- \, (= \det O_+)$\,, and then we find that $\Omega_{(a)} \equiv \abs{\Omega}^{-1}\,\Omega_-$ satisfies the relation $\Omega_{(a)}^{-1}=\Omega_{(-a)}$\,. 

We can compute $\abs{\Omega}^2 = \Omega_-\,\Omega_+$ as
\begin{align}
 \abs{\Omega}^2
 &= \sum_{p=0}^{D/2}\, \frac{(2p)!}{2^{2p}\,(p!)^2}\, a^{[\Lob_1 \Lob_2}\cdots a^{\Lob_{2p-1} \Lob_{2p}]}\, a_{[\Lob_{1} \Lob_{2}}\cdots a_{\Lob_{2p-1} \Lob_{2p}]}
\nn\\
 &=\sum_{p=0}^{D/2}\ \sum_{0\leq \Lob_1<\cdots<\Lob_{2p}\leq D-1}\ \sum_{\sigma\in S_{2p}}
 \Bigl(\frac{\sgn(\sigma)}{2^p\,p!}\,
 a_{\Lob_{\sigma(1)}\Lob_{\sigma(2)}}\cdots a_{\Lob_{\sigma(2p-1)} \Lob_{\sigma(2p)}}\Bigr)
\nn\\
 &\qquad\qquad\qquad\times \sum_{\sigma'\in S_{2p}}\Bigl(\frac{\sgn(\sigma')}{2^p\,p!}\,
 a^{\Lob_{\sigma'(1)}\Lob_{\sigma'(2)}}\cdots a^{\Lob_{\sigma'(2p-1)} \Lob_{\sigma'(2p)}}\Bigr)
\nn\\
 &=\sum_{p=0}^{D/2}\ \sum_{0\leq \Lob_1<\cdots<\Lob_{2p}\leq D-1} \varepsilon_{\Lob_1}\,\Pf\bigl[a(\Lob_1,\dotsc,\Lob_{2p})\bigr]^2\,,
\end{align}
where $S_{2p}$ is the symmetric group on a set of $2p$ indices, $\sgn(\sigma)$ is the sign of a permutation $\sigma\in S_{2p}$ and $\varepsilon_{\Lob_1} \equiv \eta_{\Lob_1\Lob_1}$ is $-1$ for $\Lob_1=0$ and $+1$ for $\Lob_1\geq 1$. 
The Pfaffian
\begin{align}
 \Pf[A(\Lob_1,\dotsc,\Lob_{2p})]
 \equiv \sum_{\sigma\in S_{2p}} \biggl(\frac{\sgn(\sigma)}{2^p\,p!}
 A_{\Lob_{\sigma(1)}\Lob_{\sigma(2)}}\cdots A_{\Lob_{\sigma(2p-1)} \Lob_{\sigma(2p)}}\biggr)\,,
\end{align}
is the polynomial in matrix elements of the antisymmetric matrix $A(\Lob_1,\dotsc,\Lob_{2p})$ which is defined by
\begin{align}
 A(\Lob_1,\dotsc,\Lob_{2p})=
 \begin{pmatrix} 0~& A_{{\Lob}_1{\Lob}_2} & \ldots & A_{{\Lob}_1{\Lob}_{2p-1}}& A_{{\Lob}_1{\Lob}_{2p}} \\
      -A_{{\Lob}_1{\Lob}_2} & 0 & \ldots & A_{{\Lob}_2{\Lob}_{2p-1}}& A_{{\Lob}_2{\Lob}_{2p}} \\
      \vdots & \vdots & \ddots & \vdots& \vdots \\
      -A_{{\Lob}_1{\Lob}_{2p-1}} & -A_{{\Lob}_2{\Lob}_{2p-1}} & \ldots & 0& A_{{\Lob}_{2p-1}{\Lob}_{2p}} \\
      -A_{{\Lob}_{1}{\Lob}_{2p}} &-A_{{\Lob}_{2}{\Lob}_{2p}} & \ldots & -A_{{\Lob}_{2p-1}{\Lob}_{2p}} & 0
 \end{pmatrix} \,.
\end{align}
As it is well known, the square of the Pfaffian $\Pf[A(\Lob_1,\dotsc,\Lob_{2p})]^2$ coincides with $\det [A(\Lob_1,\dotsc,\Lob_{2p})]$\,.

If we define a matrix function $p_A(x)$ $(x\in \mathbb{R})$ of a $D\times D$ antisymmetric matrix $A$ as
\begin{align}
 p_A(x) =-\det (x\,\eta_{\Loa\Lob}-A_{\Loa\Lob})\,,
\end{align}
its Taylor series around $x=0$ is
\begin{align}
 p_A(x) = x^{D} + c_2\, x^{D-2} + \cdots + c_{D-2}\,x^{2}+c_{D}\,,
\end{align}
where the coefficients $c_{2p}$ $(p=0\,,1\,,\dotsc,\,D/2)$ are given by
\begin{align}
 c_{2p}=\frac{1}{(D-2p)!}\,p^{(D-2p)}_A(0) =\sum_{0\leq \Lob_1<\cdots<\Lob_{2p}\leq D-1} \varepsilon_{\Lob_1}\,\det \bigl[A(\Lob_1,\dotsc,\Lob_{2p})\bigr]\,.
\end{align}
From this, we finally obtain
\begin{align}
 \abs{\Omega}^2 
 &=\sum_{p=0}^{D/2}\ \sum_{0\leq \Lob_1<\cdots<\Lob_{2p}\leq D-1} \varepsilon_{\Lob_1}\,\det \bigl[a(\Lob_1,\dotsc,\Lob_{2p})\bigr]
\nn\\
 &=p_a(1)= -\det \bigl(\eta_{\Loa\Lob}-a_{\Loa\Lob}\bigr) = \det \bigl(\delta_{\Loa}^{\Lob}-a_{\Loa}{}^{\Lob}\bigr) = \det O_- \,.
\end{align}

\section{Expansion of \texorpdfstring{$\cO^{-1}_{\pm}$}{O\textpm\textminussuperior\textonesuperior}}
\label{app:expansion-O}

In this appendix, we expand the operators $\cO^{-1}_{\pm}\equiv (1\pm\eta\,R_g\circ d_\pm)^{-1}$ in terms of $\theta$\,. 
To this end, we first use the parameterization $g=g_{\bos}\cdot g_{\fer}$, and expand $R_g(X)$ as
\begin{align}
 R_g(X) &= g_{\fer}^{-1}\,g_{\bos}^{-1}\,R(g_{\bos}\,g_{\fer}\,X\,g_{\fer}^{-1}\,g_{\bos}^{-1})\,g_{\bos}\,g_{\fer}
\nn\\
 &= R_{g_{\bos}}(X) - [\chi,\,R_{g_{\bos}}(X)] + R_{g_{\bos}}([\chi,\,X])
\nn\\
 &\quad + \frac{1}{2}\,R_{g_{\bos}}([\chi,\,[\chi,\,X]]) + \frac{1}{2}\,[\chi,\,[\chi,\,R_{g_{\bos}}(X)]] - [\chi,\,R_{g_{\bos}}([\chi,\,X])] + \cO(\theta^3)\,, 
\end{align}
where $R_{g_{\bos}}(X) \equiv g_{\bos}^{-1}\,R(g_{\bos}\,X\,g_{\bos}^{-1})\,g_{\bos}$ and $\chi\equiv \gQ^I\, \theta_I$\,. 
We can then expand $\cO_{\pm}$ as
\begin{align}
\begin{split}
 \cO_{\pm}&=1\pm \eta\, R_g\circ d_\pm =\cO_{\pm(0)} + \cO_{\pm(1)} + \cO_{\pm(2)} + \cO(\theta^3)\,,
\\
 \cO_{(0)}(X)&= 1 \pm \eta\, R_{g_{\bos}}(d_\pm(X)) \,,
\\
 \cO_{\pm(1)}(X)&=\pm \eta\, R_{g_{\bos}}([\chi,\, d_{\pm}(X)])\mp \eta\,[\chi,\,R_{g_{\bos}}\circ d_{\pm}(X)]\,,
\\
 \cO_{\pm(2)}(X)&=\mp\frac{\eta}{2}\,\bigl([\chi,\,[\chi,\,R_{g_{\bos}}\circ d_{\pm}(X)]]-R_{g_{\bos}}([\chi,\,[\chi,\,d_{\pm}(X)]])\bigr) -[\chi,\,\cO_{\pm(1)}(X)]\,.
\end{split}
\end{align}
The inverses can be also expanded as
\begin{align}
\begin{split}
 \cO_{\pm}^{-1}&=\frac{1}{1\pm\eta\,R_g\circ d_\pm}
 = \cO_{\pm(0)}^{-1}+\cO_{\pm(1)}^{-1}+\cO_{\pm(2)}^{-1}+\cO(\theta^3)\,,
\\
 \cO_{\pm(0)}^{-1}&= \frac{1}{1\pm \eta\, R_{g_{\bos}}\circ d_{\pm}}\,,
\\
 \cO_{\pm(1)}^{-1}&= -\cO_{\pm(0)}^{-1}\circ\cO_{\pm(1)}\circ\cO_{\pm(0)}^{-1}\,,
\\
 \cO_{\pm(2)}^{-1}&= -\cO_{\pm(0)}^{-1}\circ\cO_{\pm(2)}\circ\cO_{\pm(0)}^{-1} -\cO_{\pm(1)}^{-1}\circ\cO_{\pm(1)}\circ\cO_{\pm(0)}^{-1}\,.
\end{split}
\end{align}

\paragraph*{\underline{Order $\cO(\theta^0)$:}}

The leading order part $\cO^{-1}_{\pm(0)}$ of the inverse operators act as
\begin{align}
\begin{split}
 \cO^{-1}_{\pm(0)}(\gP_{\Loa})
 &=k_{\pm\Loa}{}^{\Lob}\,\gP_{\Lob} \mp \eta\,k_{\pm\Loa}{}^{\Lob}\,\lambda_{\Lob}{}^{\Loc\Lod}\,\gJ_{\Loc\Lod}\,,
\\
 \cO^{-1}_{\pm(0)}(\gJ_{\Loa\Lob})&=\gJ_{\Loa\Lob}\,,\qquad
 \cO^{-1}_{\pm(0)}(\gQ^I)=\gQ^I\,,
\end{split}
\label{eq:Oinv-0}
\end{align}
where we have used \eqref{eq:Rg-operation} and defined $k_{\pm\Loa}{}^{\Lob}$ as
\begin{align}
 k_{\pm\Loa}{}^{\Lob} \equiv \bigl[(1\pm 2\,\eta\,\lambda)^{-1}\bigr]{}_{\Loa}{}^{\Lob} \,. 
\end{align}
Note that $k_{\pm\Loa}{}^{\Lob}$ satisfies $k_{\pm\Loa\Lob}\equiv k_{\pm\Loa}{}^{\Loc}\,\eta_{\Loc\Lob}=k_{\mp\Lob\Loa}$ due to the antisymmetry of $\lambda_{\Loa\Lob}$ given in \eqref{eq:lambda-properties}. 

\paragraph*{\underline{Order $\cO(\theta^1)$:}}

At the next order, we obtain
\begin{align}
\begin{split}
 \cO_{\pm(1)}(\gP_{\Loa})
 &=\pm \eta\,\gQ^I\,\Bigl(\ii\,\epsilon^{IJ}\,\lambda_{\Loa}{}^{\Lob}\,\hat{\gamma}_{\Lob}-\frac{1}{2}\,\delta^{IJ}\,\lambda_{\Loa}{}^{\Lob\Loc}\,\gamma_{\Lob\Loc}\Bigr)\,\theta_J\,, \qquad
 \cO_{\pm(1)}(\gJ_{\Loa\Lob}) =0\,,
\\
 \cO_{\pm(1)}(\gQ^I\psi_I) 
 &=-\frac{\ii}{2}\,\eta\,\brtheta_I\,
 \bigl(2\,\sigma_3^{IJ}\,\lambda_{\Lob}{}^{\Loc}\,\hat{\gamma}_{\Loc}+\ii\,\sigma_1^{IJ}\,\lambda_{\Lob}{}^{\Loc\Lod}\,\gamma_{\Loc\Lod}\bigr)\,\psi_J\,\eta^{\Lob\Loa}\,\gP_{\Loa} + (\gJ\text{-term})\,,
\end{split}
\end{align}
where ``$(\gJ\text{-term})$'' represents terms proportional to $\gJ_{\Loa\Lob}$ that are not relevant to our computation. 
Then, the operations of the inverse operators are
\begin{align}
\begin{split}
 \cO^{-1}_{\pm(1)}(\gP_{\Loa})
 &=\mp\eta\,\gQ^I\, k_{\pm\Loa}{}^{\Lob}
 \Bigl(\ii\,\epsilon^{IJ}\,\lambda_{\Lob}{}^{\Loc}\,\hat{\gamma}_{\Loc}
 -\frac{1}{2}\,\delta^{IJ}\,\lambda_{\Lob}{}^{\Loc\Lod}\,\gamma_{\Loc\Lod}\Bigr)\,\theta_J\,, \qquad
 \cO^{-1}_{\pm(1)}(\gJ_{\Loa\Lob}) =0\,,
\\
 \cO^{-1}_{\pm(1)}(\gQ^I\psi_I)
 &=\frac{\ii}{2}\eta\,\brtheta_I\, k_\pm^{\Lob\Loa}
 \bigl(2\,\sigma_3^{IJ}\,\lambda_{\Lob}{}^{\Loc}\,\hat{\gamma}_{\Loc}+\ii\,\sigma_1^{IJ}\,\lambda_{\Lob}{}^{\Loc\Lod}\,\gamma_{\Loc\Lod}\bigr)\,\psi_J\gP_{\Loa}
 +(\gJ\text{-term})\,.
\end{split}
\end{align}

\paragraph*{\underline{Order $\cO(\theta^2)$:}}

Finally, the operators at the quadratic order are given by
\begin{align}
\begin{split}
 \cO_{\pm(2)}(\gP_{\Loa})&=\pm\frac{\ii}{2}\,\eta\,\brtheta_I\,\hat{\gamma}^{\Lob}
 \Bigl(\ii\,\epsilon^{IJ}\,\lambda_{\Loa}{}^{\Loc}\,\hat{\gamma}_{\Loc}-\frac{1}{2}\,\delta^{IJ}\,\lambda_{\Loa}{}^{\Loc\Lod}\,\gamma_{\Loc\Lod}\Bigr)\,\theta_J\,\gP_{\Lob}
\nn\\
 &\quad\mp\frac{\ii}{2}\,\eta\,\brtheta_I\,
 \Bigl(\ii\,\epsilon^{IJ}\,\lambda^{\Lob}{}_{\Loc}\,\hat{\gamma}^{\Loc}-\frac{1}{2}\,\delta^{IJ}\,\eta^{\Lob\Loe}\,\lambda_{\Loe}{}^{\Loc\Lod}\,\gamma_{\Loc\Lod}\Bigr)\,\hat{\gamma}_{\Loa}\,\theta_J\,\gP_{\Lob} +(\gJ\text{-term}) \,,
\\
 \cO_{\pm(2)}(\gJ_{\Loa\Lob})&=0\,. 
\end{split}
\end{align}
The inverses are
\begin{align}
 \cO^{-1}_{\pm(2)}(\gP_{\Loa})
 &= \mp\frac{\ii}{2}\,\brtheta_I\,k_{\pm}^{\Lob\Loh}\,k_{\pm\Loa}{}^{\Lod}\,
 \biggl[ (\delta^{IJ}\,\delta_{\Lob}^{\Loc}+2\,\eta\,\sigma_3^{IJ}\,\lambda_{\Lob}{}^{\Loc})\,\hat{\gamma}_{\Loc}\,\Bigl(-\frac{\eta}{2}\,\lambda_{\Lod}{}^{\Loe\Lof}\, \gamma_{\Loe\Lof}\Bigr)
\nn\\
 &\qquad+ \Bigl(\frac{\eta}{2}\,\lambda_{\Lob}{}^{\Loe\Lof}\, \gamma_{\Loe\Lof}\Bigr)\,(\delta^{IJ}\,\delta_{\Lod}^{\Loc}+2\,\eta\,\sigma_3^{IJ}\,\lambda_{\Lod}{}^{\Loc})\,\hat{\gamma}_{\Loc}
\nn\\
 &\qquad+\frac{\ii}{2}\,\epsilon^{IJ}\,\bigl[\hat{\gamma}_{\Lob}\,(2\,\eta\, \lambda_{\Lod}{}^{\Loc}\,\hat{\gamma}_{\Loc})
 -(2\,\eta\,\lambda_{\Lob}{}^{\Loc}\,\hat{\gamma}_{\Loc})\,\hat{\gamma}_{\Lod}\bigr]
\nn\\
 &\qquad
 +\ii\,\sigma_1^{IJ}\,(\eta\,\lambda_{\Lob}{}^{\Loe\Lof}\, \gamma_{\Loe\Lof})\,\Bigl(-\frac{\eta}{2}\,\lambda_{\Lod}{}^{\Lof\Log}\, \gamma_{\Lof\Log}\Bigr)
\nn\\
 &\qquad+\frac{\ii}{2}\,\sigma_1^{IJ}\,(2\,\eta\,\lambda_{\Lob}{}^{\Lod}\,\hat{\gamma}_{\Lod})\,(2\,\eta\,\lambda_{\Lod}{}^{\Loc}\,\hat{\gamma}_{\Loc})
 \biggr]\,\theta_J\,\gP_{\Loh} +(\gJ\text{-term})\,.
\end{align}
Operators of $\cO^{-1}_{\pm(2)}$ on other generators are not necessary for the computation of the action.

\section{Deformed torsionful spin connections}
\label{app:torsionful-spin-connections}

In this appendix, we show that two torsionful spin connections $W_{\pm}^{\Loa\Lob}$ introduced in \eqref{eq:e-torsionful-spin-pm} satisfy the following relations (we basically follow the discussion of \cite{Borsato:2016ose}):
\begin{align}
 \omega'^{\Loa\Lob}_+ &\equiv \omega'^{\Loa\Lob} + \frac{1}{2}\,e'_{\Loc}\,H'^{\Loc\Loa\Lob} =W_{+}^{\Loa\Lob}\,,
\label{eq:Lorntz-connection1}
\\
 \omega'^{\Loa\Lob}_- &\equiv \omega'^{\Loa\Lob}-\frac{1}{2}\,e'_{\Loc}\,H'^{\Loc\Loa\Lob}
  =\Lambda^{\Loa}{}_{\Loc}\,\Lambda^{\Lob}{}_{\Lod}\,W_{-}^{\Loc\Lod} + (\Lambda\,\rmd\Lambda^{-1})^{\Loa\Lob} \,,
\label{eq:Lorntz-connection2}
\end{align}
which are assumed in \eqref{eq:torsionful-spin}. 

\vspace{-\Pskip}
\subsection{Two expressions of the deformed \texorpdfstring{$H$}{H}-flux}
\label{app:deformed-H-flux}

In order to show \eqref{eq:Lorntz-connection1} and \eqref{eq:Lorntz-connection2}, we here obtain two expressions for the deformed $H$-flux.
Let us begin by considering two expressions of the deformed $B$-field [recall \eqref{eq:G-B-prime} and \eqref{eq:e-torsionful-spin-pm}]
\begin{align}
\begin{split}
 B_2'&= -\eta\,\lambda_{\Loa\Lob}\,e_{+}^{\Loa}\wedge e_{+}^{\Lob} =\eta\,\str \bigl[J_+^{(2)}\wedge R_g(J_+^{(2)})\bigr]\bigr\rvert_{\theta=0} 
\\
 &= - \eta\,\lambda_{\Loa\Lob}\,e_{-}^{\Loa}\wedge e_{-}^{\Lob} =\eta\,\str \bigl[J_-^{(2)}\wedge R_g(J_-^{(2)})\bigr]\bigr\rvert_{\theta=0} \,,
\end{split}
\label{eq:B-YB-2}
\end{align}
where $J_{\pm}$ are defined in \eqref{eq:J-O-inv-A} and $J_{\pm}^{(n)}\equiv P^{(n)}\,J_{\pm}$\,. 
Since we are only interested in the $B$-field at order $\cO(\theta^0)$\,, in the following computation, we ignore terms involving the grade-$1$ and $3$ components of $A$ and $J_{\pm}$ (where we have $d_{\pm}\sim 2\,P^{(2)}$). 

The exterior derivatives of the two expressions in Eq.~\eqref{eq:B-YB-2} become
\begin{align}
 H'_3&\equiv \frac{1}{3!}\,H'_{\Loa\Lob\Loc}\,e'^{\Loa}\wedge e'^{\Lob}\wedge e'^{\Loc}\equiv \rmd B'_2
\nn\\
 &=\eta\,\rmd\,\str \bigl[J_{\pm}^{(2)}\wedge R_g(J_{\pm}^{(2)})\bigr]\bigr\rvert_{\theta=0}
\nn\\
 &=2\,\eta\,\str \bigl[\rmd J_{\pm}^{(2)}\wedge R_g(J_{\pm}^{(2)})+J_{\pm}^{(2)}\wedge \{A,\,R_g(J_{\pm}^{(2)})\}\bigr]\bigr\rvert_{\theta=0}
\nn\\
 &=2\,\eta\,\str \bigl[\rmd J_{\pm}^{(2)}\wedge R_g(J_{\pm}^{(2)})+J_{\pm}^{(2)}\wedge \{J_{\pm}^{(0)}+J_{\pm}^{(2)},\,R_g(J_{\pm}^{(2)})\}\bigr]\bigr\rvert_{\theta=0}
\nn\\
 &\quad \pm 4\,\eta^2\,\str \bigl[J_{\pm}^{(2)}\wedge \{R_g(J_{\pm}^{(2)}),\,R_g(J_{\pm}^{(2)})\}\bigr]\bigr\rvert_{\theta=0} \,.
\label{eq:H3-eq1}
\end{align}
Here, in the third line, we have used a relation
\begin{align}
 \rmd \bigl[ R_g(B) \bigr] = R_g(\rmd B) -\{A,\,R_g(B)\}+R_g(\{A,\,B\}) \qquad 
 \bigl[\,\{B,\,C\} \equiv B\wedge C + C\wedge B \,\bigr]\,. 
\end{align}
for $\alg{g}$-valued 1-forms $B$ and $C$, and in the last equality, we have used the relation
\begin{align}
 A\rvert_{\theta=0} = \cO_{\pm} (J_{\pm})\bigr\rvert_{\theta=0} = J_{\pm}^{(0)}+J^{(2)}_{\pm} \pm 2\,\eta\,R_g(J_{\pm}^{(2)})\bigr\rvert_{\theta=0}\,. 
\end{align}
It is easy to see that the last term in \eqref{eq:H3-eq1} vanishes by using the cyclic property of the supertrace and the homogeneous CYBE
\begin{align}
 \{R_g(J_{\pm}^{(2)}),\, R_g(J_{\pm}^{(2)})\}-2\,R_g\{R_g(J_{\pm}^{(2)}),\, J_{\pm}^{(2)}\}=0\,.
\end{align}
Now, we utilize the deformed structure equation \cite{Borsato:2016ose}
\begin{align}
 \rmd J_\pm &= \rmd\bigl(\cO_{\pm}^{-1}\,A\bigr) = -\cO_{\pm}^{-1}\,\rmd\cO_{\pm} \,\cO_{\pm}^{-1}\,A + \cO_{\pm}^{-1}\,(\rmd A)
\nn\\
 &= \mp \eta\,\cO_{\pm}^{-1}\,(\rmd R_g)\,d_{\pm}\,J_{\pm} - \cO_{\pm}^{-1}\,(A\wedge A)
\nn\\
 &= \mp \eta\,\cO_{\pm}^{-1}\,\bigl[-\{A,\,R_g(d_{\pm}\,J_{\pm})\} + R_g\,\{A,\,d_{\pm}\,J_{\pm}\} \bigr]\,d_{\pm}\,J_{\pm} - \frac{1}{2}\,\cO_{\pm}^{-1}\,\{A,\, A\}
\nn\\
 &= - \frac{1}{2}\,\cO_{\pm}^{-1}\,\{J_{\pm},\, J_{\pm}\} \mp \eta\,\cO_{\pm}^{-1}\,R_g\,\{J_{\pm},\,d_{\pm}\,J_{\pm}\}
\nn\\
 &\quad + \frac{\eta^2}{2}\,\cO_{\pm}^{-1}\,\bigl[\{R_g(d_{\pm}\,J_{\pm}),\,R_g(d_{\pm}\,J_{\pm})\}-2\,R_g\bigl(\{R_g(d_{\pm}\,J_{\pm}),\,d_{\pm}\,J_{\pm}\}\bigr)\bigr]
\nn\\
 &= - \frac{1}{2}\,\cO_{\pm}^{-1}\,\{J_{\pm},\, J_{\pm}\} \mp \eta\,\cO_{\pm}^{-1}\,R_g\,\{J_{\pm},\,d_{\pm}\,J_{\pm}\} \,,
\end{align}
where we have repeatedly used $A=\cO_{\pm}(J_{\pm})$\,, and in the last equality, we have used the homogeneous CYBE. 
In the following computation, since terms involving $J_{\pm}^{(1)}$ or $J_{\pm}^{(3)}$ are irrelevant, we have
\begin{align}
 \rmd J_{\pm} =-\frac{1}{2}\,\{J_{\pm},\,J_{\pm}\} \mp 2\,\eta\,\cO_{\pm}^{-1}\,R_g\{J_{\pm}^{(2)},\,J_{\pm}^{(2)}\} \,,
\label{eq:dJ+-}
\end{align}
and then \eqref{eq:H3-eq1} is simplified as
\begin{align}
 H'_3&=2\,\eta\,\str\bigl[\bigl(\{J_{\pm},\,J_{\pm}^{(2)}\}-\frac{1}{2}\,P^{(2)}\,\{J_{\pm},\,J_{\pm}\}
  \mp 2\,\eta\, P^{(2)}\,\cO_{\pm}^{-1}\, R_g\{J_{\pm}^{(2)},\,J_{\pm}^{(2)}\} \bigr)\wedge R_g(J_{\pm}^{(2)})\bigr]\bigr\rvert_{\theta=0}
\nn\\
 &=2\,\eta\,\str\bigl[ \{J_{\pm}^{(2)},\,J_{\pm}^{(2)}\}\wedge \cO_{\mp}^{-1}\,R_g(J_{\pm}^{(2)})\bigr]\bigr\rvert_{\theta=0} \,,
\label{eq:H3-eq2}
\end{align}
where, in the last equality, we have used relations\footnote{The transpose of an operator $\mathsf{O}$ is defined as $\str[\,A\,\mathsf{O}(B)\,]=\str[\,\mathsf{O}^{\rmT}(A)\,B\,]$\,. Since $R_g$ is defined to be antisymmetric, $R_g^\rmT=-R_g$, and $d_{\pm}$ satisfies $d_{\pm}^\rmT=d_{\mp}$ [see \eqref{eq:dpm-transpose}], we can for example show $\cO_{\pm}^\rmT=1\mp\eta\,d_{\mp}\,R_g$, $d_{\pm}\,\cO_{\pm}=\cO_{\mp}^\rmT\,d_{\pm}$, and $d_{\pm}\,\cO_{\pm}^{-1}=\cO_{\mp}^{-\rmT}\,d_{\pm}$\,.}
\begin{align}
\begin{split}
 &\pm\eta\,d_{\pm}\,\cO_{\pm}^{-1}\,R_g = \pm\eta\,\cO_{\mp}^{-\rmT}\,d_{\pm}\,R_g = 1-\cO_{\mp}^{-\rmT} \,,
\\
 &\{J_+,\,J_+^{(2)}\} - \frac{1}{2}\,P^{(2)}\{J_+,\,J_+\} = \{J^{(2)}_+,\,J^{(2)}_+\} \,.
\end{split}
\end{align}
We can further rewrite the expression \eqref{eq:H3-eq2} by using the operator $M=\cO_-^{-1}\,\cO_+$ and its inverse $M^{-1}=\cO_+^{-1}\,\cO_-$\,. 
From
\begin{align}
 P^{(0)}\,M^{\pm1}\,P^{(2)} = P^{(0)}\,\cO_{\mp}^{-1}\,(\cO_{\mp}\pm 4\,\eta \, R_{g})\,P^{(2)}=\pm 4\,\eta\, P^{(0)}\,\cO_{\mp}^{-1}\,R_g\, P^{(2)}\,,
\end{align}
we can rewrite \eqref{eq:H3-eq2} as
\begin{align}
 H'_3 = \pm \frac{1}{2}\,\str\bigl[\{J_{\pm}^{(2)},\,J_{\pm}^{(2)}\}\wedge M^{\pm1}(J_{\pm}^{(2)})\bigr]\bigr\rvert_{\theta=0} \,.
\end{align}
Finally, by introducing a notation
\begin{align}
\begin{split}
 &M(\gP_{\Loc})\rvert_{\theta=0} = (\Lambda^{-1})_{\Loc}{}^{\Loa} \, \gP_{\Loa} + \frac{1}{2}\,M_{\Loc}{}^{\Loa\Lob}\,\gJ_{\Loa\Lob} \,,\quad
 M^{-1}(\gP_{\Loc})\rvert_{\theta=0} = \Lambda_{\Loc}{}^{\Loa} \, \gP_{\Loa} + \frac{1}{2}\,(M^{-1})_{\Loc}{}^{\Loa\Lob}\,\gJ_{\Loa\Lob} \,,
\\
 &\bigl[\,\Lambda_{\Loa}{}^{\Lob}= (k_-^{-1})_{\Loa}{}^{\Loc}\, k_{+\Loc}{}^{\Lob}\,,\quad 
 M_{\Loc}{}^{\Loa\Lob} \equiv 4\,\eta\,k_{-\Loc}{}^{\Lod}\,\lambda_{\Lod}{}^{\Loa\Lob} \,,\quad
 (M^{-1})_{\Loc}{}^{\Loa\Lob} \equiv -4\,\eta\,k_{+\Loc}{}^{\Lod}\,\lambda_{\Lod}{}^{\Loa\Lob}\,\bigr] \,,
\end{split}
\end{align}
and using $e_+^{\Loa}=\Lambda_{\Lob}{}^{\Loa}\,e'^{\Lob}$\,, we obtain two expressions for the deformed $H$-flux
\begin{align}
 H'_3 &= \frac{1}{2}\,\Lambda_{[\Loc}{}^{\Log}\,\Lambda_{\Loa}{}^{\Loe}\,\Lambda_{\Lob]}{}^{\Lof}\,M_{\Log,\Loe\Lof}\,e'^{\Loa}\wedge e'^{\Lob}\wedge e'^{\Loc}\qquad 
 \bigl[\,M_{\Loc,\Loa\Lob} \equiv M_{\Loc}{}^{\Loe\Lof}\,\eta_{\Loe\Loa}\,\eta_{\Lof\Lob}\,\bigr]
\label{eq:Hprime1}
\\
  &=-\frac{1}{2}\,M_{[\Loc,\Loa\Lob]}^{-1} \,e'^{\Loa}\wedge e'^{\Lob}\wedge e'^{\Loc}\qquad 
 \bigl[\,M^{-1}_{\Loc,\Loa\Lob} \equiv (M^{-1})_{\Loc}{}^{\Loe\Lof}\,\eta_{\Loe\Loa}\,\eta_{\Lof\Lob}\,\bigr]\,.
\label{eq:Hprime2}
\end{align}

\vspace{-\Pskip}
\subsection{Deformed torsionful spin connections}

By considering the leading order part $\cO(\theta^{0})$ of \eqref{eq:dJ+-}, we obtain
\begin{align}
 \rmd e_{\pm}^{\Loa}+(\omega_{[\pm]})^{\Loa}{}_{\Lob}\wedge e_{\pm}^{\Lob}=0\,,
\label{eq:de+-}
\end{align}
where the spin connections $(\omega_{[\pm]})_{\Loa\Lob}$ associated with the deformed vielbeins $e_{\pm}^{\Loa}$ have the form
\begin{align}
 \omega_{[\pm]\Loa\Lob} = W_{\pm\Loa\Lob}+\frac{1}{2}\,e_{\pm}^{\Loc}\,\bigl(M^{\pm1}_{\Loa,\Lob\Loc}+M^{\pm1}_{\Lob,\Loc\Loa} - M^{\pm1}_{\Loc,\Loa\Lob}\bigr)\,.
\label{eq:deformed-spin-2}
\end{align}
In particular, for the spin connection $\omega'^{\Loa\Lob}\equiv\omega_{[-]}^{\Loa\Lob}$ associated with the deformed vielbeins $e'^{\Loa}=e_{-}^{\Loa}$, using the formula $H'_{\Loa\Lob\Loc} =-3\,M_{[\Loc,\Loa\Lob]}^{-1}$ in \eqref{eq:Hprime2}, we obtain the first relation \eqref{eq:Lorntz-connection1} as
\begin{align}
 &\omega'^{\Loa\Lob}+\frac{1}{2}\,e'_{\Loc}\,H'^{\Loc\Loa\Lob}
 =\omega'^{\Loa\Lob}-\frac{1}{2}\,e'_{\Loc}\,\bigl[(M^{-1})_{\Loa}{}^{\Lob\Loc}+(M^{-1})_{\Lob}{}^{\Loc\Loa}+(M^{-1})_{\Loc}{}^{\Loa\Lob}\bigr]
\nn\\
 &=W_{-}^{\Loa\Lob}-e'^{\Loc}\,(M^{-1})_{\Loc}{}^{\Loa\Lob} = W_{-}^{\Loa\Lob} + 4\,\eta\,e'^{\Loc}\,k_{+\Loc}{}^{\Lod}\,\lambda_{\Lod}{}^{\Loa\Lob}
  =W_{+}^{\Loa\Lob}\,,
\end{align}
where in the last equality we have used\footnote{In order to show the relation $W_{+}^{\Loa\Lob}=W_{-}^{\Loa\Lob}-e'^{\Loc}\,(M^{-1})_{\Loc}{}^{\Loa\Lob}$\,, it will be easier to observe the bosonic part of the relation $J_{+}^{(0)}=P^{(0)}\,M^{-1}(J_{-})$\,.}
\begin{align}
 W_{+}^{\Loa\Lob} &= W_{-}^{\Loa\Lob} + 2\,\eta\,(e_+^{\Loc}+e_-^{\Loc})\,\lambda_{\Loc}{}^{\Loa\Lob}
 = W_{-}^{\Loa\Lob} + 2\,\eta\,e'^{\Loc}\,(\Lambda_{\Loc}{}^{\Lod}+\delta_{\Loc}^{\Lod})\,\lambda_{\Lod}{}^{\Loa\Lob}
\nn\\
 &= W_{-}^{\Loa\Lob} + 2\,\eta\,e'^{\Loc}\,[(k_-^{-1})_{\Loc}{}^{\Loe}+(k_+^{-1})_{\Loc}{}^{\Loe}]\,k_{+\Loe}{}^{\Lod}\,\lambda_{\Lod}{}^{\Loa\Lob}
  = W_{-}^{\Loa\Lob} + 4\,\eta\,e'^{\Loc}\,k_{+\Loc}{}^{\Lod}\, \lambda_{\Lod}{}^{\Loa\Lob} \,. 
\end{align}

On the other hand, if we take the upper sign in \eqref{eq:de+-}, from $e_+^{\Loa}=\Lambda_{\Lob}{}^{\Loa}\,e'^{\Lob}$, we obtain
\begin{align}
 \rmd e'^{\Loa} + \bigl[(\Lambda^{-1})_{\Loc}{}^{\Loa} \,\rmd \Lambda_{\Lob}{}^{\Loc} + \Lambda^{\Loa\Lod}\, \omega_{+\Lod\Loe}\, \Lambda_{\Loc}{}^{\Loe}\bigr]\wedge e'^{\Loc}=0\,. 
\label{eq:deprime-2}
\end{align}
From the upper sign of \eqref{eq:deformed-spin-2}, $H'_{\Loa\Lob\Loc} = 3\,\Lambda_{\Loa}{}^{\Lod}\,\Lambda_{\Lob}{}^{\Loe}\,\Lambda_{\Loc}{}^{\Lof}\,M_{[\Lod,\Loe\Lof]}$ in \eqref{eq:Hprime1}, and the identity $\Lambda_{\Loa}{}^{\Lod}\,M_{\Lod,\Lob\Loc}=-(M^{-1})_{\Loa,\Lob\Loc}$\,, we can show
\begin{align}
 \Lambda_{\Loa}{}^{\Lod} \,\omega_{+\Lod\Loe}\,\Lambda_{\Loc}{}^{\Loe} 
 &= \Lambda_{\Loa}{}^{\Lod} \,W_{+\Lod\Loe}\,\Lambda_{\Loc}{}^{\Loe} + \frac{1}{2}\,\Lambda_{\Loa}{}^{\Lod} \, \Lambda_{\Loc}{}^{\Loe}\,\Lambda_{\Lob}{}^{\Lof}\,e'^{\Lob}\,\bigl(M_{\Lod,\Loe\Lof}+M_{\Loe,\Lof\Lod} - M_{\Lof,\Lod\Loe}\bigr)
\nn\\
 &= \Lambda_{\Loa}{}^{\Lod} \,\Lambda_{\Loc}{}^{\Loe} \, \bigl(W_{+\Lod\Loe} - \Lambda_{\Lob}{}^{\Lof}\,e'^{\Lob}\, M_{\Lof,\Lod\Loe} \bigr)
 + \frac{1}{2}\,e'^{\Lob}\,H'_{\Lob\Loa\Loc} 
\nn\\
 &= \Lambda_{\Loa}{}^{\Lod} \,\Lambda_{\Loc}{}^{\Loe} \, W_{-\Lod\Loe} + \frac{1}{2}\,e'^{\Lob}\,H'_{\Lob\Loa\Loc} \,. 
\end{align}
This together with \eqref{eq:deprime-2} shows the second relation \eqref{eq:Lorntz-connection2},
\begin{align}
 \omega'^{\Loa\Lob} = \Lambda^{\Loa}{}_{\Lod} \,\Lambda^{\Lob}{}_{\Loe} \, W_{-}^{\Lod\Loe} + (\Lambda\,\rmd \Lambda^{-1})^{\Loa\Lob} + \frac{1}{2}\,e'_{\Loc}\,H'^{\Loc\Loa\Lob} \,. 
\end{align}

\section{Deformed Lagrangian at order \texorpdfstring{$\cO(\theta^2)$}{O(\texttheta\texttwosuperior)}}
\label{app:deformed-Lagrangian}

In this appendix, we show that the YB-deformed sigma model action can be rewritten as the conventional GS superstring action (up to quadratic order in fermions) by performing suitable field redefinitions. 

\vspace{-\Pskip}
\subsection{A derivation of the deformed Lagrangian at \texorpdfstring{$\cO(\theta^2)$}{O(\texttheta\texttwosuperior)}}

Let us start with a straightforward computation of the deformed Lagrangian $\cL_{(2)}$ by using the results obtained in Appendix \ref{app:expansion-O}. 
For convenience, we decompose $\cL_{(2)}$ as
\begin{align}
 \cL_{(2)}=\cL_{(2,0,0)}+\cL_{(0,0,2)}+\cL_{(1,1,0)}+\cL_{(0,1,1)}+\cL_{(0,2,0)}+\cL_{(1,0,1)}+\cO(\theta^4)\,,
\end{align}
where we have defined
\begin{align}
 \cL_{(a,b,c)} \equiv -\frac{\dlT}{2}\,\Pg_-^{\WSa\WSb}\,\str[A_{\WSa(a)}\,d_-\circ \cO^{-1}_{-(b)}(A_{\WSb(c)})] \,, 
\end{align}
and $A_{\WSa(a)}$ ($a=0,1,2$) have the following form as we can see from \eqref{eq:A-AdS5xS5}:
\begin{align}
\begin{split}
 A_{\WSa(0)} &\equiv e_{\WSa}^{\Loa}\,\gP_{\Loa} -\frac{1}{2}\,\omega_{\WSa}{}^{\Loa\Lob} \,\gJ_{\Loa\Lob}\,,\qquad 
 A_{\WSa(1)} \equiv \gQ^I\,D_{\WSa}^{IJ}\theta_J \,,
\\
 A_{\WSa(2)} &\equiv \frac{\ii}{2}\,\brtheta_I\,\hat{\gamma}^{\Loa}\,D_{\WSa}^{IJ}\theta_J \,\gP_{\Loa} + \frac{1}{8}\,\epsilon^{IK}\,\brtheta_I\,\gamma^{\Loc\Lod}\,R_{\Loc\Lod}{}^{\Loa\Lob}\,D_{\WSa}^{KJ}\theta_J \,\gJ_{\Loa\Lob} \,.
\end{split}
\end{align}
Each part is given by
\begin{align}
 &\cL_{(2,0,0)}+\cL_{(0,0,2)}
 = -\ii\,\frac{\dlT}{2}\,\Pg_-^{\WSa\WSb}\,\brtheta_I\,\bigl(e_{+\WSa}{}^{\Loa}\,\hat{\gamma}_{\Loa}\,D^{IJ}_{\WSb}+e_{-\WSb}{}^{\Loa}\,\hat{\gamma}_{\Loa}\,D^{IJ}_{\WSa}\bigr)\theta_J\,.
\label{eq:part-1}
\\
 &\cL_{(1,1,0)}= -\ii\,\frac{\dlT}{2}\,\Pg_-^{\WSa\WSb}\, \brtheta_I\,
 \biggl[\Bigl(2\,\eta\,\sigma_3^{IJ}\,e_{-\WSb}{}^{\Loa}\,\lambda_{\Loa}{}^{\Lob}\,\hat{\gamma}_{\Lob}
 -\frac{\ii}{2}\,\sigma_1^{IJ}\,\delta W_{-\WSb}{}^{\Lob\Loc}\,\gamma_{\Lob\Loc}\Bigr)\,D_{\WSa}\theta_J
\nn\\
 &\qquad\qquad\qquad +\frac{\ii}{2}\, \sigma_1^{IJ}\,\bigl(2\,\eta\,e_{-\WSb}{}^{\Loa}\,\lambda_{\Loa}{}^{\Loc}\,\hat{\gamma}_{\Loc}\bigr)\bigl(e_{\WSa}{}^{\Lod}\,\hat{\gamma}_{\Lod}\bigr)\, \theta_J
 -\Bigl(\frac{1}{4}\,\delta W_{-\WSb}{}^{\Lob\Loc}\,\gamma_{\Lob\Loc}\Bigr) \bigl(\sigma_3^{IJ}\,e_{\WSa}{}^{\Lod}\,\hat{\gamma}_{\Lod}\bigr)\, \theta_J \biggr]\,.
\label{eq:part-2}
\\
 &\cL_{(0,1,1)} =-\ii\,\frac{\dlT}{2}\,\Pg_-^{\WSa\WSb}\, \brtheta_I\,
 \biggl[\Bigl(2\,\eta\,\sigma_3^{IJ}\,e_{+\WSa}{}^{\Loa}\,\lambda_{\Loa}{}^{\Lob}\,\hat{\gamma}_{\Lob}
 +\frac{\ii}{2}\,\sigma_1^{IJ}\,\delta W_{+\WSa}{}^{\Lob\Loc}\,\gamma_{\Lob\Loc}\Bigr)\,D_{\WSb}\theta_J
\nn\\
 &\qquad\qquad\qquad +\frac{\ii}{2}\,\sigma_1^{IJ}\,\bigl(2\,\eta\,e_{+\WSa}{}^{\Loa}\,\lambda_{\Loa}{}^{\Loc}\,\hat{\gamma}_{\Loc}\bigr)\,\bigl(e_{\WSb}{}^{\Lod}\,\hat{\gamma}_{\Lod}\bigr)\,\theta_J
 +\Bigl(\frac{1}{4}\,\delta W_{+\WSa}{}^{\Lob\Loc}\,\gamma_{\Lob\Loc}\Bigr)\bigl(\sigma_3^{IJ}\,e_{\WSb}{}^{\Lod}\,\hat{\gamma}_{\Lod}\bigr)\,\theta_J \biggr]\,,
\label{eq:part-3}
\\
 &\cL_{(0,2,0)}=-\ii\,\frac{\dlT}{2}\,\Pg_-^{\WSa\WSb}\,\brtheta_I\,\biggl[
 e_{+\WSa}{}^{\Loa}\,\bigl(\delta^{IJ}\,\delta_{\Loa}^{\Lob}+2\,\eta\,\sigma_3^{IJ}\,\lambda_{\Loa}{}^{\Lob}\bigr)\,\hat{\gamma}_{\Lob}\,
 \Bigl(\frac{1}{4}\,\delta W_{-\WSb}{}^{\Loc\Lod}\,\gamma_{\Loc\Lod}\Bigr)
\nn\\
 &\qquad\qquad\qquad\qquad\qquad +\Bigl(\frac{1}{4}\,\delta W_{+\WSa}{}^{\Loc\Lod}\,\gamma_{\Loc\Lod}\Bigr)
 e_{-\WSb}{}^{\Loa}\,\bigl(\delta^{IJ}\,\delta_{\Loa}^{\Lob}+2\,\eta\,\sigma_3^{IJ}\,\lambda_{\Loa}{}^{\Lob}\bigr)\,\hat{\gamma}_{\Lob}
\nn\\
 &\qquad\qquad\qquad\qquad\qquad +\frac{\ii}{2}\,\epsilon^{IJ}\,\bigl[\bigl(e_{+\WSa}{}^{\Loa}\,\hat{\gamma}_{\Loa}\bigr)\bigl(2\,\eta\,e_{-\WSb}{}^{\Loc}\,\lambda_{\Loc}{}^{\Lod}\,\hat{\gamma}_{\Lod}\bigr)
 -\bigl(2\,\eta\,e_{+\WSa}{}^{\Loa}\,\lambda_{\Loa}{}^{\Lob}\,\hat{\gamma}_{\Lob}\bigr)\bigl(e_{-\WSb}{}^{\Loc}\,\hat{\gamma}_{\Loc}\bigr)\bigr]
\nn\\
 &\qquad\qquad\qquad\qquad\qquad +\frac{\ii}{8}\,\sigma_1^{IJ}\,\bigl(\delta W_{+\WSa}{}^{\Loa\Lob}\,\gamma_{\Loa\Lob}\bigr)\bigl(\delta W_{-\WSb}{}^{\Loc\Lod}\,\gamma_{\Loc\Lod}\bigr)
\nn\\
 &\qquad\qquad\qquad\qquad\qquad +\frac{\ii}{2}\,\sigma_1^{IJ}\,\bigl(2\,\eta\,e_{+\WSa}{}^{\Loa}\,\lambda_{\Loa}{}^{\Lob}\,\hat{\gamma}_{\Lob}\bigr)\bigl(2\,\eta\,e_{-\WSb}{}^{\Loc}\,\lambda_{\Loc}{}^{\Lod}\,\hat{\gamma}_{\Lod}\bigr) \biggr]\,\theta_J\,.
\label{eq:part-4}
\\
 &\cL_{(1,0,1)}=-\ii\,\frac{\dlT}{2}\,\Pg_-^{\WSa\WSb}\,\bigl(-2\,\sigma_3^{IK}\,\brtheta_I\,e_{[\WSa}{}^{\Loa}\,\hat{\gamma}_{\Loa}\,D_{\WSb]}^{KJ}\theta_J\bigr)\,.
\label{eq:part-5}
\end{align}
Here, we have defined $\delta W_{\pm}^{\Loa\Lob}$ as
\begin{align}
 \delta W_{\pm}^{\Loa\Lob}=\pm 2\,\eta\,e_{\pm}^{\Loc}\, \lambda_{\Loc}{}^{\Loa\Lob}\,,
\end{align}
which are parts of the torsionful spin connections
\begin{align}
 W_{\pm}^{\Loa\Lob}=\omega^{\Loa\Lob}+\delta W_{\pm}^{\Loa\Lob}\,.
\end{align}

Gathering the results \eqref{eq:part-1}--\eqref{eq:part-5}, we can calculate $\cL_{(2)}$\,. 
In the following computation, it may be useful to use the following identities:
\begin{align}
\begin{split}
 \brtheta_I\,\hat{\gamma}_{\Loa}\,\gamma_{\Loc\Lod}\,\theta_J
 &=\brtheta_J\,\gamma_{\Loc\Lod}\,\hat{\gamma}_{\Loa}\,\theta_I\,,\qquad
 \brtheta_I\hat{\gamma}_{\Loa}\,\hat{\gamma}_{\Lob}\theta_J
 =-\brtheta_J\,\hat{\gamma}_{\Lob}\,\hat{\gamma}_{\Loa}\,\theta_I\,,
\\
 \brtheta_I\,\gamma_{\Loa\Lob}\,\theta_J
 &=\brtheta_J\,\gamma_{\Loa\Lob}\,\theta_I\,,\qquad
 \brtheta_I\,\gamma_{\Loa\Lob}\,\gamma_{\Loc\Lod}\,\theta_J
 = - \brtheta_J \gamma_{\Loc\Lod}\,\gamma_{\Loa\Lob}\,\theta_I\,.
\label{eq:thetaflip}
\end{split}
\end{align}
The result is
\begin{align}
 \cL_{(2)}&=-\ii\,\frac{\dlT}{2}\,\Pg_-^{\WSa\WSb}\,\brtheta_I\,
 \biggl\{ \bigl[\sigma^{IJ}_3\,e_{\WSb}{}^{\Loa}+e_{-\WSb}{}^{\Lob}\,(\delta^{IJ}\,\delta_{\Lob}^{\Loa}+2\,\eta\,\sigma_3^{IJ}\,\lambda_{\Lob}{}^{\Loa})\bigr]\,\hat{\gamma}_{\Loa}\,D_{+\WSa}
\nn\\
 &\qquad +\bigl[-\sigma^{IJ}_3\,e_{\WSa}{}^{\Loa} + e_{+\WSa}{}^{\Lob}\,(\delta^{IJ}\,\delta_{\Lob}^{\Loa}+2\,\eta\,\sigma_3^{IJ}\,\lambda_{\Lob}{}^{\Loa})\bigr]\,\hat{\gamma}_{\Loa}\,D_{-\WSb}
\nn\\
 &\qquad +\frac{\ii}{2}\,\epsilon^{IJ}\,\Bigl[\bigl(e_{+\WSa}{}^{\Loa}\,\hat{\gamma}_{\Loa}\bigr)\bigl(e_{\WSb}{}^{\Lob}\,\hat{\gamma}_{\Lob}+2\,\eta\,e_{-\WSb}{}^{\Loc}\,\lambda_{\Loc}{}^{\Lod}\,\hat{\gamma}_{\Lod}\bigr)
 +\bigl(e_{-\WSb}{}^{\Loa}\,\hat{\gamma}_{\Loa}\bigr)\bigl(e_{\WSa}{}^{\Lob}\,\hat{\gamma}_{\Lob}-2\,\eta\,e_{+\WSa}{}^{\Loa}\,\lambda_{\Loa}{}^{\Lob}\,\hat{\gamma}_{\Lob}\bigr) \Bigr]
\nn\\
 &\qquad +\frac{\ii}{2}\,\sigma_1^{IJ}\,\Bigl[\bigl(2\,\eta\,e_{-\WSb}{}^{\Loa}\,\lambda_{\Loa}{}^{\Loc}\,\hat{\gamma}_{\Loc}\bigr)\bigl(e_{\WSa}{}^{\Lod}\,\hat{\gamma}_{\Lod}\bigr)
 +\bigl(2\,\eta\,e_{+\WSa}{}^{\Loa}\,\lambda_{\Loa}{}^{\Loc}\,\hat{\gamma}_{\Loc}\bigr)\bigl(e_{\WSb}{}^{\Lod}\,\hat{\gamma}_{\Lod}\bigr)
\nn\\
 &\qquad\qquad\qquad +\bigl(2\,\eta\,e_{+\WSa}{}^{\Loa}\,\lambda_{\Loa}{}^{\Lob}\,\hat{\gamma}_{\Lob}\bigr)\bigl(2\,\eta\,e_{-\WSb}{}^{\Loc}\,\lambda_{\Loc}{}^{\Lod}\,\hat{\gamma}_{\Lod}\bigr)\Bigr]
\nn\\
 &\qquad +\frac{\ii}{2}\,\sigma_1^{IJ}\,
  \Bigl[\delta W_{+\WSa}{}^{\Loa\Lob}\,\gamma_{\Loa\Lob}\,D_{\WSb}
       -\delta W_{-\WSb}{}^{\Loa\Lob}\,\gamma_{\Loa\Lob}\,D_{\WSa}
       +\frac{1}{4}\,\bigl(\delta W_{+\WSa}{}^{\Loa\Lob}\,\gamma_{\Loa\Lob}\bigr)\bigl(\delta W_{-\WSb}{}^{\Loc\Lod}\,\gamma_{\Loc\Lod}\bigr)\Bigr]\biggr\}\,\theta_J
\nn\\
 &=-\ii\,\frac{\dlT}{2}\,\Pg_-^{\WSa\WSb}\,\brtheta_I\,
 \biggl\{2\,\pi^{IJ}_+\,e_{-\WSb}{}^{\Loa}\,\hat{\gamma}_{\Loa}\,D_{+\WSa}
      +2\,\pi^{IJ}_-\,e_{+\WSa}{}^{\Loa}\,\hat{\gamma}_{\Loa}\,D_{-\WSb}
      +\ii\,\epsilon^{IJ}\,e_{-\WSb}{}^{\Loa}\,\hat{\gamma}_{\Loa}\,e_{+\WSa}{}^{\Lob}\,\hat{\gamma}_{\Lob}
\nn\\
 &\qquad\qquad +\frac{\ii}{2} \,\sigma_1^{IJ}\,\Bigl[\bigl(e_{+\WSa}{}^{\Loa}\,\hat{\gamma}_{\Loa}\bigr)\bigl(e_{-\WSb}{}^{\Lob}\,\hat{\gamma}_{\Lob}\bigr) -\bigl(e_{\WSa}{}^{\Loa}\,\hat{\gamma}_{\Loa}\bigr)\bigl(e_{\WSb}{}^{\Lob}\,\hat{\gamma}_{\Lob}\bigr)
\nn\\
 &\qquad\qquad +\delta W_{+\WSa}{}^{\Loa\Lob}\, \gamma_{\Loa\Lob}\,D_{\WSb}
  -\delta W_{-\WSb}{}^{\Loa\Lob}\,\gamma_{\Loa\Lob}\,D_{\WSa}
  +\frac{1}{4}\,\bigl(\delta W_{+\WSa}{}^{\Loa\Lob}\,\gamma_{\Loa\Lob}\bigr)\bigl(\delta W_{-\WSb}{}^{\Loc\Lod}\,\gamma_{\Loc\Lod}\bigr)\Bigr]\biggr\}\,\theta_J\,.
\end{align}
In the second equality, we have used
\begin{align}
\begin{split}
 2\,\pi^{IJ}_+\, e_{-\WSa}{}^{\Loa}=& \sigma^{IJ}_3\,e_{\WSb}{}^{\Loa}+e_{-\WSb}{}^{\Lob}\,\bigl(\delta^{IJ}\,\delta_{\Lob}^{\Loa}+2\,\eta\,\sigma_3^{IJ}\,\lambda_{\Lob}{}^{\Loa}\bigr)\,,
\\
 2\,\pi^{IJ}_-\, e_{+\WSa}{}^{\Loa}=&-\sigma^{IJ}_3\,e_{\WSa}{}^{\Loa}+e_{+\WSa}{}^{\Lob}\,\bigl(\delta^{IJ}\,\delta_{\Lob}^{\Loa}+2\,\eta\,\sigma_3^{IJ}\,\lambda_{\Lob}{}^{\Loa}\bigr)\,,
\end{split}
\end{align}
where the projection operators $\pi_{\pm}^{IJ}$ are defined by
\begin{align}
 \pi_{\pm}^{IJ}=\frac{\delta^{IJ} \pm \sigma_3^{IJ}}{2} \,.
\end{align}
Now, we decompose the deformed Lagrangian into two parts
\begin{align}
 \cL_{(2)} =\cL_{(2)}^{\rmc}+\delta\cL_{(2)} \,,
\end{align}
where $\cL_{(2)}^{\rmc}$ takes the form of the canonical GS Lagrangian after taking the diagonal gauge (see section \ref{sec:YB-RR}) while $\delta\cL_{(2)}$ is the remaining part. 
The explicit form of $\cL_{(2)}^{\rmc}$ is given by
\begin{align}
 \cL_{(2)}^{\rmc} &=-\ii\,\dlT\,\brtheta_I\,\biggl[
   \Pg_-^{\WSa\WSb}\,\pi_+^{IJ}\,e_{-\WSb}{}^{\Loa}\,\hat{\gamma}_{\Loa}\,D_{+\WSa}
   +\Pg_-^{\WSa\WSb}\,\pi_-^{IJ}\,e_{+\WSa}{}^{\Loa}\,\hat{\gamma}_{\Loa}\,D_{-\WSb}
   +\frac{\ii}{2}\,\epsilon^{IJ}\,e_{-\WSb}{}^{\Loa}\,\hat{\gamma}_{\Loa}\,e_{+\WSa}{}^{\Lob}\,\hat{\gamma}_{\Lob}\biggr]\,\theta_J
\nn\\
  &=-\ii\,\dlT\,\biggl[
   \Pg_+^{\WSa\WSb}\,e_{-\WSa}{}^{\Loa}\,\brtheta_1\,\hat{\gamma}_{\Loa}\,D_{+\WSb}\theta_1
   +\Pg_-^{\WSa\WSb}\,e_{+\WSa}{}^{\Loa}\,\brtheta_2\,\hat{\gamma}_{\Loa}\,D_{-\WSb}\theta_2 
   +\ii\, \Pg_+^{\WSa\WSb}\,\epsilon^{IJ}\,\brtheta_1\,e_{-\WSa}{}^{\Loa}\,\hat{\gamma}_{\Loa}\,e_{+\WSb}{}^{\Lob}\,\hat{\gamma}_{\Lob}\,\theta_2\biggr] \,.
\end{align}
On the other hand, the remaining term $\delta\cL_{(2)}$ has the form
\begin{align}
 \delta\cL_{(2)}
 &= \frac{\dlT}{4}\,\Pg_-^{\WSa\WSb}\,\sigma_1^{IJ}\,\brtheta_I\,
  \biggl[\delta W_{+\WSa}{}^{\Loa\Lob}\,\hat{\gamma}_{\Loa\Lob}\,D_{\WSb}
  -\delta W_{-\WSb}{}^{\Loa\Lob}\,\hat{\gamma}_{\Loa\Lob}\,D_{\WSa}
\nn\\
 &\qquad +\frac{1}{4}\,\bigl(\delta W_{+\WSa}{}^{\Loa\Lob}\,\gamma_{\Loa\Lob}\bigr)\bigl(\delta W_{-\WSb}{}^{\Loc\Lod}\,\gamma_{\Loc\Lod}\bigr)
 +\bigl(e_{+\WSa}{}^{\Loa}\,\hat{\gamma}_{\Loa}\bigr)\bigl(e_{-\WSb}{}^{\Lob}\,\hat{\gamma}_{\Lob}\bigr)
 -\bigl(e_{\WSa}{}^{\Loa}\,\hat{\gamma}_{\Loa}\bigr)\bigl(e_{\WSb}{}^{\Lob}\,\hat{\gamma}_{\Lob}\bigr)\biggr]\,\theta_J\,.
\end{align}
By using \eqref{eq:thetaflip}, this can be rewritten as
\begin{align}
 \delta\cL_{(2)} &= \frac{\dlT}{4}\,\Pg_-^{\WSa\WSb}\,\sigma_1^{IJ}\,\brtheta_I\,
 \biggl[\bigl(\delta W_{+\WSa}{}^{\Loa\Lob}\,\partial_{\WSb}-\delta W_{-\WSb}{}^{\Loa\Lob}\,\partial_{\WSa}\bigr)\,\gamma_{\Loa\Lob}
\nn\\
 &\quad\qquad\qquad\qquad +\frac{1}{8}\,\bigl(\delta W_{+\WSa}{}^{\Loa\Lob}\,\omega_{\WSb}{}^{\Loc\Lod}
 -\delta W_{-\WSb}{}^{\Loa\Lob}\,\omega_{\WSa}{}^{\Loc\Lod}+\delta W_{+\WSa}{}^{\Loa\Lob}\,\delta W_{-\WSb}{}^{\Loc\Lod}\bigr)\, \bigl[\gamma_{\Loa\Lob},\,\gamma_{\Loc\Lod}\bigr]
\nn\\
 &\quad\qquad\qquad\qquad +\frac{1}{2}\,(e_{+\WSa}{}^{\Loa}\,e_{-\WSb}{}^{\Lob}
 -e_{\WSa}{}^{\Loa}\,e_{\WSb}{}^{\Lob})\,[\hat{\gamma}_{\Loa},\hat{\gamma}_{\Lob}] \biggr]\,\theta_J
\nn\\
 &=\frac{\dlT}{4}\,\Pg_-^{\WSa\WSb}\,\sigma_1^{IJ}\,
 \brtheta_I\,\biggl[
 \delta W_{+\WSa}{}^{\Loa\Lob}\,\partial_{\WSb}-\delta W_{-\WSb}{}^{\Loa\Lob}\,\partial_{\WSa}
 +(\delta W_{+\WSa})^{\Loa}{}_{\Loc}\,\omega_{\WSb}{}^{\Loc\Lob}
 -(\delta W_{-\WSb})^{\Loa}{}_{\Loc}\,\omega_{\WSa}{}^{\Loc\Lob}
\nn\\
 &\quad\qquad\qquad\qquad +(\delta W_{+\WSa})^{\Loa}{}_{\Loc}\,\delta W_{-\WSb}{}^{\Loc\Lob}
     -\frac{1}{2}\,(e_{+\WSa}{}^{\Loc}\,e_{-\WSb}{}^{\Lod}
     -e_{\WSa}{}^{\Loc}\, e_{\WSb}{}^{\Lod})\,R_{\Loc\Lod}{}^{\Loa\Lob}\biggr]\,\gamma_{\Loa\Lob}\theta_J\,,
\label{eq:delta-cL2}
\end{align}
where we used
\begin{align}
 \hat{\gamma}_{\Loa\Lob} =-\frac{1}{2}\,R_{\Loa\Lob}{}^{\Loc\Lod}\, \gamma_{\Loc\Lod}\,,\qquad 
 [\gamma_{\Loa\Lob},\,\gamma_{\Loc\Lod}]
 =-2\,\bigl(\eta_{\Loa\Loc}\,\gamma_{\Lob\Lod}-\eta_{\Lob\Loc}\,\gamma_{\Loa\Lod}-\eta_{\Loa\Lod}\,\gamma_{\Lob\Loc}+\eta_{\Lob\Lod}\,\gamma_{\Loa\Loc}\bigr)\,.
\end{align}
In the following, we show that $\delta\cL_{(2)}$ are completely canceled by performing an appropriate redefinition of the bosonic fields $X^m$\,. 

\vspace{-\Pskip}
\subsection{Bosonic shift}
\label{app:bosonic-shift}

We consider the redefinition of $X^m$\,,
\begin{align}
 X^m\ \to\ X^m + \xi^m\,,\qquad \xi^m\equiv \frac{\eta}{4}\,\sigma_1^{IJ}\, e^{\Loc m}\,\lambda_{\Loc}{}^{\Loa\Lob}\,\brtheta_I\, \gamma_{\Loa\Lob}\,\theta_J +\cO(\theta^4)\,. 
\label{eq:shift}
\end{align}
This was originally considered in \cite{Arutyunov:2015qva,Kyono:2016jqy} such that the unwanted terms involving $\sigma_1^{IJ}\, \brtheta_I\, \gamma_{\Loa\Lob}\,\partial_{\WSa}\theta_J$ in \eqref{eq:delta-cL2} are canceled out by the deviation of the Lagrangian under the shift \eqref{eq:shift}, $\delta\cL_{\YB}=\delta\cL_{(0)}+\cO(\theta^2)$ where
\begin{align}
\begin{split}
 &\cL_{(0)}\equiv -\dlT\,\Pg_{-}^{\WSa\WSb}\,E'_{mn}\,\partial_{\WSa} X^m\,\partial_{\WSb} X^n 
\\
 &\bigl(E'_{mn}=\CG'_{mn}+B'_{mn} = e_m{}^{\Loa}\,e_n{}^{\Lob}\,k_{+\Loa\Lob} = \str [A_{m}\,\cO_{-}^{-1}(A_{n})]\rvert_{\theta=0}\bigr)\,. 
\end{split}
\end{align}
As we show below, in fact, $\delta\cL_{(2)}$ is completely canceled out under the redefinition \eqref{eq:shift} when the $r$-matrix satisfies the homogeneous CYBE, which has been checked for specific examples in the previous works. 

For simplicity, we introduce a shorthand notation
\begin{align}
 \frac{1}{\sqrt{2!}}\,\gJ_{\Loa\Lob} \ \rightarrow \ \gJ_{\sfi} \,,
\end{align}
with combinatoric factors discussed around \eqref{eq:index-convention}. 
In this notation, the commutation relations of bosonic generators $\{\gP_{\Loa},\, \gJ_{\sfi}\}$ and matrices $\{\hat{\gamma}_{\Loa},\,\gamma_{\sfi}\}$ become
\begin{align}
\begin{split}
 [\gP_{\Loa},\,\gP_{\Lob}] &=f_{\Loa\Lob}{}^{\sfi}\,\gJ_{\sfi}\,,\qquad
 [\gJ_{\sfi},\,\gJ_{\sfj}]= f_{\sfi\sfj}{}^{\sfk}\, \gJ_{\sfk}\,,\qquad
 [\gP_{\Loa},\,\gP_{\sfi}]=f_{\Loa\sfi}{}^{\Lob}\, \gP_{\Lob}\,,
\\
 [\hat{\gamma}_{\Loa},\,\hat{\gamma}_{\Lob}] &=-2\,f_{\Loa\Lob}{}^{\sfi}\, \gamma_{\sfi}\,,\qquad
 [\gamma_{i},\, \gamma_{j}]=-2\,f_{\sfi\sfj}{}^{\sfk}\, \gamma_{\sfk} \,.
\end{split}
\end{align}
Then, $\delta\cL_{(2)}$ in \eqref{eq:delta-cL2} can be expressed as
\begin{align}
 \delta\cL_{(2)}
 &=\frac{\dlT}{2}\,\Pg_-^{\WSa\WSb}\sigma_1^{IJ}
 \brtheta_I\, \gamma_{\sfk}\,\bigl(\delta W_{+\WSa}{}^{\sfk}\,\partial_{\WSb}\theta_J-\delta W_{-\WSb}{}^{\sfk}\,\partial_{\WSa}\theta_J\bigr) 
\nn\\
 &\quad -\frac{\dlT}{4}\,\Pg_-^{\WSa\WSb}\,\biggl[\bigl(\delta W_{+\WSa}{}^{\sfi}\,\omega_{\WSb}{}^{\sfj}
 +\omega_{\WSa}{}^{\sfi}\,\delta W_{-\WSb}{}^{\sfj}+\delta W_{+\WSa}{}^{\sfi}\,\delta W_{-\WSb}{}^{\sfj}\bigr)\, f_{\sfi\sfj}{}^{\sfk}
\nn\\
 &\quad\qquad\qquad +\bigl(e_{+\WSa}{}^{\Loa}\,e_{-\WSb}{}^{\Lob}-e_{\WSa}{}^{\Loa}\,e_{\WSb}{}^{\Lob}\bigr)\,f_{\Loa\Lob}{}^{\sfk} \biggr]\,\sigma_1^{IJ}\,\brtheta_I\, \gamma_{\sfk}\,\theta_J\,.
\label{eq:L2-shift}
\end{align}

\subsubsection*{A computation of $\delta \cL_{(0)}$}

A straightforward computation shows
\begin{align}
 \delta \cL_{(0)}&= -\dlT\,\Pg_{-}^{\WSa\WSb}\,\Lie_{\xi}E'_{mn}\,\partial_{\WSa} X^m\,\partial_{\WSb} X^n
\nn\\
 &=-\dlT\,\Pg_-^{\WSa\WSb}\,\bigl( \xi^p\,\partial_p E'_{mn} + \partial_m \xi^p\, E'_{pn}
 +\partial_{n}\xi^p \,E'_{mp}\bigr)\,\partial_{\WSa}X^{m}\,\partial_{\WSb}X^n
\nn\\
 &=-\frac{\dlT}{2}\,\Pg_-^{\WSa\WSb}\,\sigma_1^{IJ}\, \brtheta_I\,\gamma_{\sfk}\,\bigl(\delta W_{+m}{}^{\sfk}\,\partial_n\theta_J -\delta W_{-n}{}^{\sfk}\,\partial_m\theta_J\bigr)\,\partial_{\WSa} X^m\,\partial_{\WSb} X^n 
\nn\\
 &\quad -\frac{\eta\,\dlT}{2}\,\Pg_-^{\WSa\WSb}\,\sigma_1^{IJ}\,\brtheta_I\,\gamma_{\sfk}\,\theta_J\,\Bigl[\lambda_{\Loc}{}^{\sfk}\,e_m{}^{\Loa}\,e_n{}^{\Lob}\,e^{\Loc p}\,\partial_p k_{+\Loa\Lob} + \partial_m\lambda_{\Loa}{}^{\sfk}\,e_{-n}{}^{\Loa} + \partial_n\lambda_{\Loa}{}^{\sfk}\,e_{+m}{}^{\Loa} 
\nn\\
 &\quad\qquad\qquad\qquad\qquad\qquad 
  + 2\,e^{\Loc p}\,\lambda_{\Loc}{}^{\sfk}\,\bigl(\partial_{[p}e_{m]\Loa}\,e_{-n}{}^{\Loa}+ \partial_{[p}e_{n]\Loa}\,e_{+m}{}^{\Loa}\bigr)\Bigr]\,\partial_{\WSa} X^m\,\partial_{\WSb} X^n \,,
\end{align}
where we have used $e_{\pm m}{}^{\Loa}=k_{\pm \Lob}{}^{\Loa}\,e_m{}^{\Lob}$ and $\delta W_{\pm m}{}^{\sfk}=\pm 2\,\eta\,e_{\pm m}{}^{\Loc}\,\lambda_{\Loc}{}^{\sfk}$\,. 
From this, we can easily see that the terms involving $\sigma_1^{IJ}\, \brtheta_I\, \gamma_{\sfk}\,\partial_{\WSa}\theta_J$ in $\delta \cL_{(0)}$ and $\delta\cL_{(2)}$ cancel each other out. 

We can further compute $\partial_p k_{+\Loa\Lob}$ and $\partial_m\lambda_{\Loa}{}^{\sfk}$ as follows by recalling their original definitions. 
The $\partial_m\lambda_{\Loa}{}^{\sfk}$ can be obtained from
\begin{align}
 \partial_{m}\lambda_{\Loc}{}^{\sfi}\,\gJ_{\sfi} &=\partial_{m}\bigl[P^{(0)}R_{g_{\bos}}(\gP_{\Loc})\bigr]\bigr\rvert_{\theta=0}
 =-P^{(0)} [A_{m},\,R_{g_{\bos}}(\gP_{\Loc})]\bigr\rvert_{\theta=0}+P^{(0)}R_{g_{\bos}}\bigl([A_{m},\,\gP_{\Loc}]\bigr)\bigr\rvert_{\theta=0}
\nn\\
 &=\bigl[e_m{}^{\Loa}\,(\lambda_{\sfk}{}^{\sfi}\,f_{\Loa\Loc}{}^{\sfk} - \lambda_{\Loc}{}^{\Lob}\,f_{\Loa\Lob}{}^{\sfi})
 +\omega_m{}^{\sfj}\,(\lambda_{\Loc}{}^{\sfk}\,f_{\sfj\sfk}{}^{\sfi} - \lambda_{\Lod}{}^{\sfi}\,f_{\sfj\Loc}{}^{\Lod})\bigr]\,\gJ_{\sfi}\,.
\end{align}
where we have used
\begin{align}
\begin{split}
 &\partial_{m}R_g(\,\cdot\,)=-[A_{m},\,R_g(\cdot)\,]+R_g([A_{m},\,\cdot\,]) \,, \qquad 
 A_m\rvert_{\theta=0}=e_m{}^{\Loa}\,\gP_{\Loa}-\omega_m{}^{\sfk}\,\gJ_{\sfk} \,,
\\
 &R_g(\gP_{\Loa})\rvert_{\theta=0}=\lambda_{\Loa}{}^{\Lob}\,\gP_{\Lob}-\lambda_{\Loa}{}^{\sfk}\,\gJ_{\sfk} \,,\qquad 
 R_g(\gJ_{\sfi})\rvert_{\theta=0}=\lambda_{\sfi}{}^{\Lob}\,\gP_{\Lob}-\lambda_{\sfi}{}^{\sfk}\,\gJ_{\sfk} \,. 
\end{split}
\end{align}
Similarly, we obtain
\begin{align}
 \partial_{m}k_{+\Loa\Lob}&=\partial_{m}\str\bigl[\gP_{\Lob}\,\cO_+^{-1}(\gP_{\Loa})\bigr]\bigr\rvert_{\theta=0}
 = -\frac{1}{2}\,\str\bigl[d_-(\gP_{\Lob})\,\cO_+^{-1}\circ\partial_{m}\cO_+\circ\cO_+^{-1}(\gP_{\Loa})\bigr]\bigr\rvert_{\theta=0}
\nn\\
 &= -\frac{1}{2}\,\str\bigl[\cO_+^{-\rmT}\circ d_-(\gP_{\Lob})\, \partial_{m}\cO_+\circ\cO_+^{-1}(\gP_{\Loa})\bigr]\bigr\rvert_{\theta=0}
\nn\\
 &= -\frac{\eta}{2}\,\str\bigl[d_-\,\cO_-^{-1}(\gP_{\Lob})\, \partial_{m} R_g\circ d_+\circ\cO_+^{-1}(\gP_{\Loa})\bigr]\bigr\rvert_{\theta=0}
\nn\\
 &= -2\,\eta\,k_{+\Loa}{}^{\Loc}\,k_{-\Lob}{}^{\Lod}\,\str\bigl[\gP_{\Lod}\,\bigl\{-[A_m,\,R_g(\gP_{\Loc})] + R_g([A_m,\,\gP_{\Loc}]) \bigr\}\bigr]\bigr\rvert_{\theta=0}
\nn\\
 &=-2\,\eta\,k_{+\Loa}{}^{\Loc}\,k_{-\Lob\Lod}\,\bigl[e_m{}^{\Loe}\,(\lambda_{\sfk}{}^{\Lod}\,f_{\Loe\Loc}{}^{\sfk}-\lambda_{\Loc}{}^{\sfk}\,f_{\Loe\sfk}{}^{\Lod}) +\omega_m{}^{\sfk}\,(\lambda_{\Loc}{}^{\Loe}\,f_{\sfk\Loe}{}^{\Lod}-\lambda_{\Loe}{}^{\Lod}\,f_{\sfk\Loc}{}^{\Loe})\bigr] \,.
\end{align}
By using several identities, such as
\begin{align}
 e_m{}^{\Loa}=e_{\pm m}{}^{\Lob}\,\bigl(\delta_{\Lob}^{\Loa}\pm 2\,\eta\,\lambda_{\Lob}{}^{\Loa}\bigr)\,,\qquad
 \partial_{[m}e_{n]}{}^{\Loa}=-\omega_{[m}{}^{\Loa\Lob}\,e_{n]\Lob}\,,
\end{align}
and the explicit forms of the structure constants, we can straightforwardly obtain
\begin{align}
 \delta \cL_{(0)}&=-\frac{\dlT}{2}\,\Pg_-^{\WSa\WSb}\,\sigma_1^{IJ}\,\brtheta_I\,\gamma_{\sfk}\,\bigl(\delta W_{+m}{}^{\sfk}\,\partial_{n}\theta_J -\delta W_{-n}{}^{\sfk}\,\partial_{m}\theta_J\bigr) \,\partial_{\WSa}X^{m}\,\partial_{\WSb}X^n
\nn\\
 &\quad +\frac{\dlT}{4}\,\Pg_-^{\WSa\WSb}\,\Bigl[\bigl(\delta W_{+m}{}^{\sfi}\,\omega_{n}{}^{\sfj} +\omega_m{}^{\sfi}\,\delta W_{-n}{}^{\sfj}\bigr)\,f_{\sfi\sfj}{}^{\sfk}
  +2\,\eta\,\bigl(e_{m}{}^{\Loa}\,e_{-n}{}^{\Loc}\,\lambda_{\Loc}{}^{\Lob}- e_{+m}{}^{\Loc}\,\lambda_{\Loc}{}^{\Loa}\,e_{n}{}^{\Lob}\bigr)\,f_{\Loa\Lob}{}^{\sfk}
\nn\\
 &\qquad +4\,\eta^2\,\bigl(e_{+m}{}^{\Loc}\,\lambda_{\Loc}{}^{\sfj}\,e_{-n}{}^{\Lob}- e_{+m}{}^{\Lob}\,e_{-n}{}^{\Loc}\,\lambda_{\Loc}{}^{\sfj}\bigr)\,
  f_{\Lob\sfj}{}^{\Loa}\,\lambda_{\Loa}{}^{\sfk}
\nn\\
 &\qquad -4\,\eta^2\,\bigl(e_{+m}{}^{\Loc}\,\lambda_{\Loc}{}^{\Loa}\,e_{-n}{}^{\Lob}+e_{+m}{}^{\Loa}\,e_{-n}{}^{\Loc}\,\lambda_{\Loc}{}^{\Lob}\bigr)\,f_{\Loa\Lob}{}^{\sfj}\,\lambda_{\sfj}{}^{\sfk} \Bigr]\,\sigma_1^{IJ}\,\brtheta_I\,\gamma_{\sfk}\,\theta_J\,\partial_{\WSa}X^{m}\,\partial_{\WSb}X^n\,.
\label{eq:L0-shift}
\end{align}

\subsubsection*{A computation of $\delta \cL_{(0)}+\delta\cL_{(2)}$}

Now, the sum of \eqref{eq:L0-shift} and \eqref{eq:L2-shift} becomes
\begin{align}
 &\delta \cL_{(0)}+\delta\cL_{(2)}
\nn\\
 &= \frac{\dlT}{4}\,\Pg_-^{\WSa\WSb}\,\biggl[ 4\,\eta^2\,(e_{+m}{}^{\Loa}\,\lambda_{\Loa}{}^{\sfi})(e_{-n}{}^{\Lob}\,\lambda_{\Lob}{}^{\sfj})\,f_{\sfi\sfj}{}^{\sfk}
\nn\\
 &\quad +2\,\eta\,\bigl(e_{m}{}^{\Loa}\,e_{-n}{}^{\Loc}\,\lambda_{\Loc}{}^{\Lob}- e_{+m}{}^{\Loc}\,\lambda_{\Loc}{}^{\Loa}\,e_{n}{}^{\Lob}\bigr)\,f_{\Loa\Lob}{}^{\sfk}
 -\bigl(e_{+m}{}^{\Loa}\,e_{-n}{}^{\Lob}-e_{m}{}^{\Loa}\,e_{n}{}^{\Lob}\bigr)\,f_{\Loa\Lob}{}^{\sfk}
\nn\\
 &\quad +4\,\eta^2\,\bigl(e_{+m}{}^{\Loc}\,\lambda_{\Loc}{}^{\sfj}\,e_{-n}{}^{\Lob}- e_{+m}{}^{\Lob}\,e_{-n}{}^{\Loc}\,\lambda_{\Loc}{}^{\sfj}\bigr)\,f_{\Lob\sfj}{}^{\Loa}\,\lambda_{\Loa}{}^{\sfk}
\nn\\
 &\quad -4\,\eta^2\,\bigl(e_{+m}{}^{\Loc}\,\lambda_{\Loc}{}^{\Loa}\,e_{-n}{}^{\Lob}+e_{+m}{}^{\Loa}\,e_{-n}{}^{\Loc}\,\lambda_{\Loc}{}^{\Lob}\bigr)\,f_{\Loa\Lob}{}^{\sfj}\,\lambda_{\sfj}{}^{\sfk}\biggr]\,\sigma_1^{IJ}\,\brtheta_I\,\gamma_{\sfk}\,\theta_J\,\partial_{\WSa}X^{m}\,\partial_{\WSb}X^n\,,
\end{align}
where we used $\delta W_{\pm}^{\sfi}=\pm 2\,\eta\,e_{\pm}^{\Loc}\, \lambda_{\Loc}{}^{\sfi}$ in the first line.
Moreover, the second line is simplified as
\begin{align}
 2\,\eta\,\bigl(e_{m}{}^{\Loa}\, e_{-n}{}^{\Loc}\,\lambda_{\Loc}{}^{\Lob}- e_{+m}{}^{\Loc}\,\lambda_{\Loc}{}^{\Loa}\,e_{n}{}^{\Lob}\bigr)
 -\bigl(e_{+m}{}^{\Loa}\,e_{-n}{}^{\Lob}-e_{m}{}^{\Loa}\,e_{n}{}^{\Lob}\bigr)
 =4\,\eta^2\,e_{+m}{}^{\Loc}\,\lambda_{\Loc}{}^{\Loa}\,e_{-n}{}^{\Lod}\,\lambda_{\Lod}{}^{\Lob}\,,
\end{align}
and we obtain
\begin{align}
\begin{split}
 \delta \cL_{(0)}+\delta\cL_{(2)}&=\eta^2\,\dlT\,\Pg_-^{\WSa\WSb}\,
 \Bigl[(e_{+m}{}^{\Loc}\,\lambda_{\Loc}{}^{\Loa})(e_{-n}{}^{\Lod}\,\lambda_{\Lod}{}^{\Lob})\,f_{\Loa\Lob}{}^{\sfk}
 +(e_{+m}{}^{\Loa}\,\lambda_{\Loa}{}^{\sfi})(e_{-n}{}^{\Lob}\,\lambda_{\Lob}{}^{\sfj})\,f_{\sfi\sfj}{}^{\sfk}
\\
 &\quad +\bigl(e_{+m}{}^{\Loc}\,\lambda_{\Loc}{}^{\sfj}\,e_{-n}{}^{\Lob} - e_{+m}{}^{\Lob}\, e_{-n}{}^{\Loc}\,\lambda_{\Loc}{}^{\sfj}\bigr)\,f_{\Lob\sfj}{}^{\Loa}\,\lambda_{\Loa}{}^{\sfk}
\\
 &\quad -\bigl(e_{+m}{}^{\Loc}\,\lambda_{\Loc}{}^{\Loa}\,e_{-n}{}^{\Lob}+e_{+m}{}^{\Loa}\,e_{-n}{}^{\Loc}\,\lambda_{\Loc}{}^{\Lob}\bigr)\,f_{\Loa\Lob}{}^{\sfj}\,\lambda_{\sfj}{}^{\sfk}\Bigr]\,\sigma_1^{IJ}\,\brtheta_I\,\gamma_{\sfk}\,\theta_J\,\partial_{\WSa}X^{m}\,\partial_{\WSb}X^n\,.
\end{split}
\label{eq:sum-shift2}
\end{align}
Remarkably, the quantities in the square bracket of \eqref{eq:sum-shift2} are precisely the grade-$0$ component of $\CYBE_g(J_{+m}^{(2)},\,J_{-n}^{(2)})$\,,
\begin{align}
\begin{split}
 \bigl[\CYBE^{(0)}_g(J_{+m}^{(2)},\,J_{-n}^{(2)})\bigr]^{\sfk}
 &= (e_{+m}{}^{\Loc}\,\lambda_{\Loc}{}^{\Loa})(e_{-n}{}^{\Lod}\,\lambda_{\Lod}{}^{\Lob})\,f_{\Loa\Lob}{}^{\sfk}
  +(e_{+m}{}^{\Loa}\,\lambda_{\Loa}{}^{\sfi})(e_{-n}{}^{\Lob}\,\lambda_{\Lob}{}^{\sfj})\,f_{\sfi\sfj}{}^{\sfk}
\\
 &\quad +\bigl(e_{+m}{}^{\Loc}\,\lambda_{\Loc}{}^{\sfj}\,e_{-n}{}^{\Lob}- e_{+m}{}^{\Lob}\,e_{-n}{}^{\Loc}\,\lambda_{\Loc}{}^{\sfj}\bigr)\, f_{\Lob\sfj}{}^{\Loa}\,\lambda_{\Loa}{}^{\sfk}
\\
 &\quad -(e_{+m}{}^{\Loc}\,\lambda_{\Loc}{}^{\Loa}\,e_{-n}{}^{\Lob}+e_{+m}{}^{\Loa}\,e_{-n}{}^{\Loc}\,\lambda_{\Loc}{}^{\Lob})\,f_{\Loa\Lob}{}^{\sfj}\,\lambda_{\sfj}{}^{\sfk}\,,
\end{split}
\end{align}
where we have used $J_{\pm m}^{(2)}=e_{\pm m}{}^{\Loa}\,\gP_{\Loa}$ [see \eqref{eq:Jpm-expansion}]. 
Therefore, we obtain 
\begin{align}
 \delta \cL_{(0)}+\delta\cL_{(2)}
 =\eta^2\,\dlT\,\Pg_-^{\WSa\WSb}\,\sigma_1^{IJ}\,
 \bigl[\CYBE^{(0)}_g(J_{+m}^{(2)},\,J_{-n}^{(2)})\bigr]^{\sfi}\, \brtheta_I\,\hat{\gamma}_{\sfi}\,\theta_J\,\partial_{\WSa}X^{m}\,\partial_{\WSb}X^n\,,
\end{align}
which shows that $\delta \cL_{(0)}+\delta\cL_{(2)}$ vanishes when the $r$-matrix satisfies the homogeneous CYBE.

\section{The \texorpdfstring{$\kappa$}{\textkappa}-symmetry transformation}
\label{app:kappa}

In fact, the YB sigma model action defined in \eqref{eq:YBsM} is invariant under the following $\kappa$-symmetry variations (see \cite{Kawaguchi:2014qwa} for the details):
\begin{align}
 \cO_-^{-1}\,g^{-1}\,\delta_{\kappa}g
 &= \Pg_-^{\WSa\WSb}\,\bigl\{\gQ^1\,\kappa_{1\WSa},\,J_{-\WSb}^{(2)}\bigr\}
  + \Pg_+^{\WSa\WSb}\,\bigl\{\gQ^2\,\kappa_{2\WSa},\,J_{+\WSb}^{(2)}\bigr\} \,, 
\label{eq:kappa1}
\\
 \delta_\kappa\bigl(\sqrt{-\gga}\,\gga^{\WSa\WSb}\bigr)
 &= \frac{1}{4}\,\sqrt{-\gga}\,\str\Bigl[\Upsilon\,\Bigl(\bigl[\gQ^1\kappa_{1(+)}^{\WSa},\,J_{+(+)}^{(1)\WSb}\bigr]
 + \bigl[\gQ^2\,\kappa^{\WSa}_{2(-)},\,J_{-(-)}^{(3)\WSb}\bigr]\Bigr) + (\WSa \leftrightarrow \WSb)\Bigr]\,,
\label{eq:kappa2}
\end{align}
where $\kappa_{I\WSa}$ $(I=1,2)$ are local fermionic parameters and we have defined $\Upsilon \equiv \diag (\bm{1_4},\,-\bm{1_4})$\,. 
Here and hereafter, worldsheet vectors with $(\pm)$ are projected with the projection operator $\Pg_{\pm\WSa}{}^{\WSb}$, like $J_{\pm(\pm)}^{\WSa}\equiv \Pg_{\pm}^{\WSa\WSb}\,J_{\pm\WSb}$\,.

In this appendix, following the procedure of \cite{Arutyunov:2015qva}, we rewrite the $\kappa$-variations \eqref{eq:kappa1} and \eqref{eq:kappa2} as the standard $\kappa$-variations in the GS type IIB superstring \cite{Grisaru:1985fv},
\begin{align}
 \delta_\kappa X^m &=-\frac{\ii}{2}\,e'^{\Loa m}\,\brTheta_I\,\Gamma_{\Loa}\,\delta_\kappa\Theta'_I + \cO(\Theta'^3)\,, 
\label{eq:kappaX1}
\\
 \delta_\kappa\Theta'_I
 &= \frac{1}{4}\,(\delta^{IJ}\,\gga^{\WSa\WSb}-\sigma_3^{IJ}\,\epsilon^{\WSa\WSb})\,e'_{\WSb}{}^{\Loa}\,\Gamma_{\Loa}\,K'_{J\WSa} + \cO(\Theta'^2)\,,
\label{eq:kappath1}
\\
 \frac{1}{\sqrt{-\gga}}\,\delta_{\kappa}\bigl(\sqrt{-\gga}\,\gga^{\WSa\WSb}\bigr)
 &= -2\,\ii\,\bar{K}'^{(\WSa}_{1(+)}\,D'^{\WSb)}_{+(+)}\Theta'_1 - 2\,\ii\,\bar{K}'^{(\WSa}_{2(-)}\,D'^{\WSb)}_{-(-)}\Theta'_2
\nn\\
 &\quad + \frac{\ii}{8}\,\biggl[\bar{K}'^{(\WSa}_{1(+)}\,\bisF'\,e'^{\WSb)\Loa}_{(+)}\,\Gamma_{\Loa}\,\Theta'_2 -\brTheta'_1\,e'^{(\WSb|\Loa}_{(-)}\,\Gamma_{\Loa}\,\bisF'\,K'^{|\WSa)}_{2(-)}\biggr] + \cO(\Theta'^3)\,,
\label{eq:kappaga1}
\end{align}
where the detailed notations are explained below. 
In the course of the rewriting, we need to identify the supergravity background $(e'_m{}^{\Loa},\,B'_{mn},\,\bisF')$ as the $\beta$-deformed $\AdS{5}\times\rmS^5$ background. 
In this sense, the following computation serves as a non-trivial check of the equivalence between YB deformations and local $\beta$-deformations. 

\subsubsection*{Bosonic fields}

We first consider the $\kappa$-symmetry transformation of the bosonic fields $X^m$\,, which can be extracted from the grade-$2$ component of \eqref{eq:kappa1}.
From \eqref{eq:kappa1}, we can easily see
\begin{align}
 P^{(2)}\circ\cO_-^{-1}\,g^{-1}\,\delta_{\kappa}g=0\,.
\end{align}
The left-hand side can be expanded as 
\begin{align}
 P^{(2)}\circ\cO_-^{-1}\,g^{-1}\,\delta_{\kappa}g
 &=P^{(2)}\,\biggl[\Bigl(e_{m}{}^{\Loa}\,\delta_{\kappa}X^m+\frac{\ii}{2}\,\brtheta_I\,\hat{\gamma}^{\Loa}\,\delta_{\kappa}\theta_I \Bigr)\,\cO_{-(0)}^{-1}(\gP_{\Loa})
 +\cO_{-(1)}^{-1}\bigl(\gQ^{I}\,\delta_{\kappa}\theta_{I}\bigr) +\cO(\theta^3) \biggr]
\nn\\
 &=\biggl[e_{m\Lob}\,\delta_{\kappa}X^m +\frac{\ii}{2}\,\bigl(\delta^{IJ}\,\delta_{\Lob}^{\Loc} + 2\,\eta\,\sigma_3^{IJ}\,\lambda_{\Lob}{}^{\Loc}\bigr)\,\brtheta_I\,\hat{\gamma}_{\Loc}\,\delta_\kappa\,\theta_J\biggr]\,k_-^{\Lob\Loa}\,\gP_{\Loa}
\nn\\
 &\quad -\frac{\eta}{2}\,\sigma_1^{IJ}\,\brtheta_I\,\lambda_{\Lob}{}^{\Loc\Lod}\,\gamma_{\Loc\Lod}\,\delta_\kappa\theta_J\,k_{-}^{\Lob\Loa}\,\gP_{\Loa} + \cO(\theta^3) \,.
\end{align}
Now, by performing the redefinition \eqref{eq:bosonic-shift} of $X^m$\,, the term proportional to $\sigma^{IJ}_1$ disappears and we obtain
\begin{align}
 0=\Bigl[ e_{m \Lob}\,\delta_{\kappa}X^m + \frac{\ii}{2}\,\bigl(\delta^{IJ}\,\delta_{\Lob}^{\Loc}+2\,\eta\,\sigma_3^{IJ}\,\lambda_{\Lob}{}^{\Loc}\bigr)\,\brtheta_I\,\hat{\gamma}_{\Loc}\, \delta_\kappa\theta_J\Bigr]\,k_-^{\Lob\Loa}\,\gP_{\Loa}+\cO(\theta^3)\,.
\label{eq:X-kappa-expansion}
\end{align}
Then, solving the equation \eqref{eq:X-kappa-expansion} for $\delta_{\kappa}X^{m}$\,, we obtain
\begin{align}
 \delta_{\kappa}X^{m}
 &= -\frac{\ii}{2}\,(k_-^{-1})^{\Loa}{}_{\Lob}\,e^{\Lob m}\,\brtheta_1\,\hat{\gamma}_{\Loa}\,\delta_{\kappa}\theta_1
 -\frac{\ii}{2}\,(k_{+}^{-1})^{\Loa}{}_{\Lob}\,e^{\Lob m}\,\brtheta_2\,\hat{\gamma}_{\Loa}\,\delta_{\kappa}\theta_2 + \cO(\theta^3)
\nn\\
 &=-\frac{\ii}{2}\,e_-^{\Loa m}\,\brTheta_1\,\Gamma_{\Loa}\,\delta_{\kappa}\Theta_1 - \frac{\ii}{2}\,e_+^{\Loa m}\,\brTheta_2\,\Gamma_{\Loa}\,\delta_{\kappa}\Theta_2+\cO(\Theta^3)\,,
\label{eq:kappa-X}
\end{align}
where it is note that the inverse of $e_{\pm m}{}^{\Loa}=e_{m}{}^{\Lob}\,k_{\pm \Lob}{}^{\Loa}$ is $e_{\pm \Loa}{}^{m}=(k_\pm^{-1})_{\Loa}{}^{\Lob}\,e_{\Lob}{}^{m}$\,. 
Finally, by using $\Lambda_{\Loa}{}^{\Lob} \, \Gamma_{\Lob}=\Omega^{-1} \,\Gamma_{\Loa}\,\Omega$ and the redefined fermions $\Theta'_I$ given in \eqref{eq:fermi-redef}, \eqref{eq:kappa-X} becomes
\begin{align}
 \delta_{\kappa}X^{m} &=-\frac{\ii}{2}\,e'^{\Loa m}\,\brTheta_1\,\Gamma_{\Loa}\,\delta_{\kappa}\Theta_1 -\frac{\ii}{2}\,e'^{\Loa m}\,\brTheta_2\,\Omega^{-1}\,\Gamma_{\Loa}\,\Omega\,\delta_{\kappa}\Theta_2+\cO(\Theta^3)
\nn\\
 &=-\frac{\ii}{2}\,e'^{\Loa m}\,\brTheta'_I\,\Gamma_{\Loa}\,\delta_{\kappa}\Theta'_I + \cO(\Theta^3) \,,
\end{align}
which is the usual $\kappa$-variation \eqref{eq:kappaX1} of $X^m$\,.

\subsubsection*{Fermionic fields}

Next, let us consider the $\kappa$-variations of fermionic variables.
These can be found from
\begin{align}
\begin{split}
 P^{(1)}\,\cO^{-1}_-\,g^{-1}\,\delta_{\kappa}g&=\Pg^{\WSa\WSb}_-\,\{\gQ^1\,\kappa_{1\WSa},\,J_{-\WSb}^{(2)}\}\,,
\\
 P^{(3)}\,\cO^{-1}_-\,g^{-1}\,\delta_{\kappa}g&=\Pg^{\WSa\WSb}_+\,\{\gQ^2\,\kappa_{2\WSa},\,J_{+\WSb}^{(2)}\}\,.
\end{split}
\label{eq:P1P3Ogdg}
\end{align}
Indeed, the left-hand side gives
\begin{align}
 P^{(1)}\,\cO^{-1}_-\,g^{-1}\,\delta_{\kappa}g = \gQ^1\,\delta_{\kappa}\theta_1 + \cO(\theta^2)\,,\qquad
 P^{(3)}\,\cO^{-1}_-\,g^{-1}\,\delta_{\kappa}g = \gQ^2\,\delta_{\kappa}\theta_2 + \cO(\theta^2)\,. 
\end{align}
In order to evaluate the right-hand side, we use the following relations \cite{Arutyunov:2015qva}:
\begin{align}
 \gQ^I\,\gP_{\check{\Loa}}+\gP_{\check{\Loa}}\,\gQ^I=\frac{1}{2}\,\gQ^I\,\hat{\gamma}_{\check{\Loa}}\,,\qquad
 \gQ^I\,\gP_{\hat{\Loa}}+\gP_{\hat{\Loa}}\,\gQ^I=-\frac{1}{2}\,\gQ^I\,\hat{\gamma}_{\hat{\Loa}} \,,
\end{align}
which can be verified by using the matrix representations of $\gP_{\Loa}$ and $\gQ_I$ given in \eqref{eq:P-J-super} and \eqref{eq:Q-matrix}. 
Then, the transformations \eqref{eq:P1P3Ogdg} become
\begin{align}
\begin{split}
 \gQ^1\delta_{\kappa}\theta_1&=\frac{1}{2}\,\Pg_-^{\WSa\WSb}\,\gQ^1\,\bigl(e_{-\WSb}{}^{\check{\Loa}}\,\hat{\gamma}_{\check{\Loa}}- e_{-\WSb}{}^{\hat{\Loa}}\,\hat{\gamma}_{\hat{\Loa}}\bigr)\,\kappa_{1\WSa} + \cO(\theta^2)\,,
\\
 \gQ^2\delta_{\kappa}\theta_2&=\frac{1}{2}\,\Pg_+^{\WSa\WSb}\,\gQ^2\,\bigl(e_{+\WSb}{}^{\check{\Loa}}\,\hat{\gamma}_{\check{\Loa}}- e_{+\WSb}{}^{\hat{\Loa}}\,\hat{\gamma}_{\hat{\Loa}}\bigr)\,\kappa_{2\WSa} + \cO(\theta^2)\,.
\end{split}
\end{align}
By using relations \eqref{eq:Theta-theta} and \eqref{eq:brTheta-brtheta}, these can be rewritten as
\begin{align}
\begin{split}
 \delta_{\kappa}\Theta_1&=\frac{1}{2}\,\Pg^{\WSa\WSb}_-\,e_{-\WSb}{}^{\Loa}\,\Gamma_{\Loa}\,K_{1\WSa} + \cO(\Theta^2)\,,
\\
 \delta_{\kappa}\Theta_2&=\frac{1}{2}\,\Pg^{\WSa\WSb}_+\,e_{+\WSb}{}^{\Loa}\,\Gamma_{\Loa}\,K_{2\WSa} + \cO(\Theta^2)\,,
\end{split}
\end{align}
where we have introduced 
\begin{align}
 K_I\equiv \begin{pmatrix} 0\\ 1 \end{pmatrix} \otimes \kappa_I\,,\qquad
 \bar{K}_I\equiv \begin{pmatrix} 1 & 0 \end{pmatrix} \otimes \bar{\kappa}_I\,,
\end{align}
and used
\begin{align}
 \Gamma_{\check{\Loa}}\,K_{I} = \begin{pmatrix} 1\\ 0 \end{pmatrix} \otimes \hat{\gamma}_{\check{\Loa}}\,\kappa_I \,,\qquad 
 \Gamma_{\hat{\Loa}}\,K_{I} = - \begin{pmatrix} 1\\ 0 \end{pmatrix} \otimes \hat{\gamma}_{\hat{\Loa}}\,\kappa_I \,.
\end{align}
Finally, using the redefined fermions $\Theta'_I$ and considering redefinitions of $K_I$\,,
\begin{align}
 K'_1=K_1\,,\qquad K'_2=\Omega\,K_2 \,,
\label{eq:fermi-redef-K}
\end{align}
we obtain the standard $\kappa$-variations of the fermions \eqref{eq:kappath1}.

\subsubsection*{Worldsheet metric}

Finally, we rewrite the $\kappa$-variation \eqref{eq:kappa2} of $\gga_{\WSa\WSb}$ into the standard form \eqref{eq:kappaga1}.

By using the expansion \eqref{eq:Jpm-expansion} of the deformed currents $J_{\pm}$\,, the variation \eqref{eq:kappa2} can be expanded as
\begin{align}
 \frac{1}{\sqrt{-\gga}}\,\delta_{\kappa}\bigl(\sqrt{-\gga}\,\gga^{\WSa\WSb}\bigr) 
 = -\ii\,\bar{\kappa}_{1(+)}^{\WSa}\,D^{\WSb\,1J}_{+(+)}\theta_J
  -\ii\,\bar{\kappa}_{2(-)}^{\WSa}\,D^{\WSb\,2J}_{-(-)}\theta_J
  +(\WSa\leftrightarrow \WSb)+\cO(\theta^3)\,,
\end{align}
where we have used commutation relations of $\alg{su}(2,2|4)$ algebra
\begin{align}
\begin{split}
 \bigl[\gQ^1\,\kappa^{\WSa}_{1(+)},\,J_{+(+)}^{(1)\WSb}\bigr]
 &=\Bigl(-\frac{1}{2}\,\bar{\kappa}^{\WSa}_{1(+)}\,D^{\WSb\,1J}_{+(+)}\theta_J + \cO(\theta^3)\Bigr)\,\gZ + (\gP\text{-term}) \,,
\\
 \bigl[\gQ^2\kappa^{\WSa}_{2(-)},\,J_{-(-)}^{(3)\WSb}\bigr]
 &=\Bigl(-\frac{1}{2}\,\bar{\kappa}^{\WSa}_{2(-)}\,D^{\WSb\,2J}_{-(-)}\theta_J + \cO(\theta^3)\Bigr)\,\gZ + (\gP\text{-term})\,,
\end{split}
\end{align}
and supertrace formulas
\begin{align}
 \str\bigl[\Upsilon\,\gZ\bigr]=8\,\ii\,,\qquad \str\bigl[\Upsilon\,(\text{other generators})\bigr]=0\,.
\end{align}
It is noted that the redefinition \eqref{eq:bosonic-shift} of $X^m$ does not affect the variation of the worldsheet metric at the leading order in $\theta$.

Then, by using the $32\times 32$ gamma matrices, the variation can be expressed as
\begin{align}
 \frac{1}{\sqrt{-\gga}}\,\delta_{\kappa}\bigl(\sqrt{-\gga}\,\gga^{\WSa\WSb}\bigr)
 &= -\ii\,\bar{K}_{1(+)}^{\WSa}\,D^{\WSb}_{+(+)}\Theta_1 -\ii\,\bar{K}_{2(-)}^{\WSa}\,D^{\WSb}_{-(-)}\Theta_2
\nn\\
 &\quad +\frac{\ii}{16}\,\biggl[\bar{K}_{1(+)}^{\WSa}\,\bisF_5\,e^{\WSb\Loa}_{+(+)}\,\Gamma_{\Loa}\,\Theta_2 - \bar{K}_{2(-)}^{\WSa}\,\bisF_5\,e^{\WSb\Loa}_{-(-)}\,\Gamma_{\Loa}\,\Theta_1 \biggr]
\nn\\
 &\quad +(\WSa \leftrightarrow \WSb) + \cO(\Theta^3)\,,
\end{align}
where we have used relations
\begin{align}
 \bar{\kappa}_{I(\pm)}^{\WSa}\,\gamma_{\Loa\Lob}\,\theta_J =\bar{K}_{I(\pm)}^{\WSa}\,\Gamma_{\Loa\Lob}\Theta_J \,,\qquad
 \bar{\kappa}_{I(\pm)}^{\WSa}\,\hat{\gamma}_{\Loa}\,\theta_J =\frac{\ii}{8}\,\bar{K}_{I(\pm)}^{\WSa}\,\bisF_5\,\Gamma_{\Loa}\,\Theta_J\,.
\end{align}
Finally, we perform the redefinitions \eqref{eq:fermi-redef} and \eqref{eq:fermi-redef-K}. 
By using relations \eqref{eq:torsionful-spin} and \eqref{eq:YBRR0}, the variation of the worldsheet metric becomes
\begin{align}
 \frac{1}{\sqrt{-\gga}}\,\delta_\kappa\bigl(\sqrt{-\gga}\,\gga^{\WSa\WSb}\bigr)
 &= -\ii\,\bar{K}'^{\WSa}_{1(+)}\,D'^{\WSb}_{+(+)}\Theta'_1 - \ii\,\bar{K}'^{\WSa}_{2(-)}\,D'^{\WSb}_{-(-)}\Theta'_2
\nn\\
 &\quad +\frac{\ii}{16}\,\biggl[\bar{K}'^{\WSa}_{1(+)}\,\bigl(\bisF_5\,\Omega^{-1}\bigr)\,e'^{\WSb \Loa}_{(+)}\,\Gamma_{\Loa}\,\Theta'_2
 -\brTheta'_1\,e'^{\WSb\Loa}_{(-)}\,\Gamma_{\Loa}\,\bigl(\bisF_5\,\Omega^{-1}\bigr)\,K'^{\WSa}_{2(-)}\biggr]
\nn\\
 &\quad +(\WSa \leftrightarrow \WSb)+\cO(\Theta'^3)
\nn\\
 &=-\ii\,\bar{K}'^{\WSa}_{1(+)}\,D'^{\WSb}_{+(+)}\Theta'_1 - \ii\,\bar{K}'^{\WSa}_{2(-)}\,D'^{\WSb}_{-(-)}\Theta'_2
\nn\\
 &\quad +\frac{\ii}{16}\,\biggl[\bar{K}'^{\WSa}_{1(+)}\,\bisF'\,e'^{\WSb\Loa}_{(+)}\,\Gamma_{\Loa}\,\Theta'_2 - \brTheta'_1\,e'^{\WSb\Loa}_{(-)}\,\Gamma_{\Loa}\,\bisF'\,K'^{\WSa}_{2(-)}\biggr]
\nn\\
 &\quad +(\WSa \leftrightarrow \WSb)+\cO(\Theta'^3) \,.
\end{align}
In this way, we have obtained the standard $\kappa$-variation of the worldsheet metric \eqref{eq:kappaga1}.

\end{document}